\documentclass[twocolumn,twocolappendix]{aastex631}
\usepackage{amsmath}

\usepackage{ulem}

\received{April 13, 2021}
\revised{June 11, 2021}
\accepted{Jul, 2021}

\shorttitle{}
\shortauthors{Yamada et al.}
\graphicspath{{./}{figures/}}

\begin{document}

\title{Comprehensive Broadband X-ray and Multiwavelength Study of 
Active Galactic Nuclei in Local 57 Ultra/luminous Infrared Galaxies Observed with NuSTAR and/or Swift/BAT}

\correspondingauthor{Satoshi Yamada}
\email{styamada@kusastro.kyoto-u.ac.jp}

\author[0000-0002-9754-3081]{Satoshi Yamada}
\affiliation{Department of Astronomy, Kyoto University, Kitashirakawa-Oiwake-cho, Sakyo-ku, Kyoto 606-8502, Japan}

\author[0000-0001-7821-6715]{Yoshihiro Ueda}
\affiliation{Department of Astronomy, Kyoto University, Kitashirakawa-Oiwake-cho, Sakyo-ku, Kyoto 606-8502, Japan}

\author[0000-0002-0114-5581]{Atsushi Tanimoto}
\affiliation{Department of Physics, The University of Tokyo, Tokyo 113-0033, Japan}

\author[0000-0001-6186-8792]{Masatoshi Imanishi}
\affiliation{National Astronomical Observatory of Japan, Osawa, Mitaka, Tokyo 181-8588, Japan}
\affiliation{Department of Astronomical Science, Graduate University for Advanced Studies (SOKENDAI), 2-21-1 Osawa, Mitaka, Tokyo 181-8588, Japan}

\author[0000-0002-3531-7863]{Yoshiki Toba}
\affiliation{Department of Astronomy, Kyoto University, Kitashirakawa-Oiwake-cho, Sakyo-ku, Kyoto 606-8502, Japan}
\affiliation{Research Center for Space and Cosmic Evolution, Ehime University, 2-5 Bunkyo-cho, Matsuyama, Ehime 790-8577, Japan}

\author[0000-0001-5231-2645]{Claudio Ricci}
\affiliation{N\'ucleo de Astronom\'{\i}a de la Facultad de Ingenier\'{\i}a, Universidad Diego Portales, Av. Ej\'ercito Libertador 441, Santiago, Chile}
\affiliation{Kavli Institute for Astronomy and Astrophysics, Peking University, Beijing 100871, People's Republic of China}
\affiliation{George Mason University, Department of Physics \& Astronomy, MS 3F3, 4400 University Drive, Fairfax, VA 22030, USA}

\author[0000-0003-3474-1125]{George~C. Privon}
\affiliation{National Radio Astronomy Observatory, 520 Edgemont Rd, Charlottesville,
VA 22903, USA}



\begin{abstract}

We perform a systematic X-ray spectroscopic analysis of 
57 local ultra/luminous infrared galaxy systems 
(containing 84 individual galaxies) 
observed with Nuclear Spectroscopic Telescope Array and/or Swift/BAT.
Combining soft X-ray data obtained with Chandra, XMM-Newton, Suzaku 
and/or Swift/XRT, 
we identify 40 hard ($>$10~keV)
X-ray detected active galactic nuclei (AGNs) 
and constrain their torus parameters with the X-ray clumpy torus model
XCLUMPY \citep{Tanimoto2019}.
Among the AGNs at $z < 0.03$, for which sample biases are minimized, the fraction of Compton-thick
($N_{\rm H} \geq 10^{24}$~cm$^{-2}$) AGNs reaches 64$^{+14}_{-15}$\% 
(6/9 sources) in late mergers, while 24$^{+12}_{-10}$\%
(3/14 sources) in early mergers,
consistent with the tendency reported by \citet{Ricci2017bMNRAS}. We
find that the bolometric AGN luminosities derived from the infrared
data increase, but the X-ray 
to bolometric luminosity ratios decrease, with merger stage. The X-ray weak AGNs in
late mergers ubiquitously show massive outflows at sub-pc to
kpc scales.  Among them, the most luminous AGNs ($L_{\rm bol,AGN} \sim
10^{46}$~erg~s$^{-1}$) have relatively small column densities of
$\lesssim$10$^{23}$~cm$^{-2}$ and almost super-Eddington ratios
($\lambda_{\rm Edd} \sim$~1.0). Their torus
covering factors ($C_{\rm T}^{\rm (22)} \sim 0.6$) are larger than
those of Swift/BAT selected AGNs with similarly high Eddington ratios. 
These results suggest 
a scenario that, in the final stage of mergers, 
multiphase strong outflows are produced due to chaotic quasi-spherical
inflows and the AGN becomes extremely X-ray weak and deeply buried due to
obscuration by inflowing and/or outflowing material.

\end{abstract}



\section{Introduction} 
\label{S1_intro}

For understanding the coevolution of galaxy and supermassive black hole
(SMBH), gas-rich galaxy mergers (i.e., major mergers) are considered as
a key phenomenon. Many theoretical studies \citep[e.g.,][]{Hopkins2006}
suggest that major mergers trigger both intense star formation in the 
galaxies and mass
accretion onto the SMBHs, being enshrouded by a huge amount of gas and dust. 
This makes them appear as luminous and ultraluminous infrared
galaxies (U/LIRGs; \citealt{Sanders1996} for a review)\footnote{LIRGs
and ULIRGs are defined by using the total 8--1000~$\mu$m luminosities of
$L_{\rm IR}$ = 10$^{11}$--10$^{12}L_{\odot}$ and $>$10$^{12}L_{\odot}$,
respectively.} containing obscured active galactic nuclei (AGNs).
ULIRGs may be in a prephase of quasi stellar objects (quasars) 
where star forming activities become relatively weak. 
Thus, studies of ULIRGs are indispensable to 
reveal the whole history of star formation and SMBH growth
triggered by major mergers.

Recent numerical simulations of galaxy mergers
\citep[e.g.,][]{Kawaguchi2020} predict that the mass accretion rates
can exceed the Eddington limits as the separation of the
two SMBHs decreases.  
However, the nuclear environments of
accreting SMBHs in U/LIRGs have been still veiled 
in mystery. This is because the
central regions in the final phase of mergers become deeply ``buried'' with
large covering factors, where even the direction of the lowest
dust column-density can be opaque to the ionizing UV photons
\citep[e.g.,][]{Imanishi2006,Imanishi2008,Yamada2019}.  To shed light on the
properties of these hidden AGNs, hard X-rays ($>$10~keV) are useful
thanks to their high penetrating power against
obscuration.\footnote{Considering the relation between $V$-band extinction and 
hydrogen column density of the Galactic interstellar medium, 
$N_{\rm H}$/$A_V$ = $2 \times 10^{21}$~cm$^{-2}$~mag$^{-1}$ \citep{Draine2003}, 
the penetration depth of 10~keV X-rays corresponds to at least $A_V \sim 500$~mag.} 
In fact,
\citet{Ricci2017bMNRAS} identity 25 AGNs in 30 interacting U/LIRGs with
the Nuclear Spectroscopic Telescope Array (NuSTAR;
\citealt{Harrison2013}) and/or Swift/BAT, which cover 
the 3--79~keV and 14--195~keV bands, respectively. 
They also report that the fraction of Compton-thick (CT; with hydrogen
column densities $N_{\rm H} \geq 10^{24}$~cm$^{-2}$) AGNs is 
significantly larger ($65^{+12}_{-13}$\%) in late mergers than in early ones 
($35^{+13}_{-12}$\%).
This indicates that the AGNs in late stage mergers are indeed deeply
buried by CT material.

To best constrain the properties of obscuring material (``torus'') in AGNs
through X-ray observations, construction of spectral models
based on the realistic geometries of AGN tori is crucial.
Since many studies indicate that the torus structure is clumpy
\citep[e.g.,][]{Krolik1988,Wada2002,Honig2007,Laha2020},
\citet{Tanimoto2019} develop the X-ray clumpy torus
model called XCLUMPY, which adopts
the same geometry as that of the infrared (IR) CLUMPY model
\citep{Nenkova2008a,Nenkova2008b}. 
The model is applied to the broadband X-ray spectra
of nearby AGNs to constrain their torus covering factors
\citep[e.g.,][]{Miyaji2019,Tanimoto2020,Tanimoto2021,Toba2020,Ogawa2021,Uematsu2021}.


To understand the connection between mass accretion 
and obscuration in mergers, investigation of differences in the AGN
structure between mergers and nonmergers may provide us with a clue.
\citet{Yamada2020} apply the XCLUMPY model to the spectra of two
``nonmerging LIRGs'' and find that their AGNs are not buried, unlike
those in late mergers. This implies that accreting environments are 
different between major mergers and secular processes.  From a
statistical study of Swift/BAT selected AGNs, whose majority are
nonmergers, \citet{Ricci2017cNature} suggest that the torus
structure is regulated by radiation pressure from the AGN.
Thus, studies of AGNs in U/LIRGs 
will enable us to answer if the same physics works or not 
in deeply buried AGNs in major mergers.  

Comparison of X-ray luminosities\footnote{In 
this paper, we define ``absorption-corrected X-ray luminosities''
of an AGN in the 2--10~keV and 10--50~keV bands, 
$L_{2-10}$ and $L_{10-50}$, 
as those corrected for line-of-sight absorption 
that is recognized in the X-ray spectra.
It remains possible, however, that 
a part of the X-ray emission from the innermost region 
is completely blocked by optically thick material and hence the true 
X-ray luminosity could be underestimated, 
as discussed in Section~\ref{subsub6-1-6_Unified-view}.}
with bolometric ones inferred from
infrared observations also provide us with information on the AGN
physics in U/LIRGs. 
\citet{Teng2015} report that the bolometric-to-X-ray
luminosity ratio ($\kappa_{\rm bol,X}$) of the AGN in an ULIRG is much larger
(i.e., X-ray weak), $\kappa_{\rm bol,X} \sim$ $10^2$--$10^4$, 
than that of a normal AGN, for
which typically $\kappa_{\rm bol,X} \sim 20$ \citep[e.g.,][]{Vasudevan2007}. 
The reason for the difference
is not clear. To answer this question, comprehensive multiwavelength
investigation of a large number of AGNs in U/LIRGs is required.

In this work, we carry out a systematic X-ray spectral study of a
sample of 57 local U/LIRG systems (containing 84 individual galaxies)
observed with NuSTAR and/or Swift/BAT by combining the available soft X-ray
data of Chandra, XMM-Newton, Suzaku, and/or Swift/XRT.  The paper is
structured as follows.  Section~\ref{S2_sample} describes our sample and merger
classification.  Details of the X-ray observations and data reduction
are presented in Section~\ref{S3_reduction}. 
Section~\ref{S4_analysis} describes our X-ray spectral analysis. 
In Section~\ref{S5_results} we compare our X-ray results with 
those of previous multiwavelength observations.
In Section~\ref{S6_discussion} we discuss the origin of the X-ray
weakness, torus structure,
and the connection between AGN and starburst activities.
Section~\ref{S7_conclusion} summarizes our main conclusions. 
In two companion papers (\citealt{Ricci2021b}; C. Ricci et al., in preparation) 
we will discuss: i) the evolution of obscuration along the merger sequence, 
and ii) the relation between X-ray emission and multiwavelength proxies of 
AGN activity.
The cosmological parameters adopted are $H_{\rm 0}$ = 70 km s$^{-1}$
Mpc$^{-1}$, $\Omega_{\rm M}$ = 0.3, and $\Omega_{\rm \Lambda}$ = 0.7.
All uncertainties are quoted at the 90\% confidence level
unless otherwise stated.

\section{Sample Description}
\label{S2_sample}
\subsection{Sample Selection}
\label{sub2-1_selection}

The IRAS Revised Bright Galaxy Sample (RBGS; \citealt{Sanders2003}) is
an all-sky survey of IR galaxies having the IRAS 60 $\mu$m flux
densities above 5.24 Jy and Galactic latitude $|b| \geq 5\degr$.  Our
targets are taken from the GOALS sample \citep{Armus2009}, which is a
complete subset of the RBGS with an 8--1000~$\mu$m luminosity threshold
of $L_{\rm IR} \geq 10^{11} L_{\odot}$ consisting of 
180 LIRGs and 22 ULIRGs at redshifts $z < 0.088$.

Most of the U/LIRGs in GOALS are interacting systems
\citep{Haan2011a,Stierwalt2013}. In total 244 and 246 nuclei are observed
with the Spitzer SH and LH IRS modules, respectively, within the 202 U/LIRG
systems \citep{Inami2013}. In X-rays, the presence of AGNs in the
individual nuclei is investigated with Chandra for 106 systems (C-GOALS;
\citealt{Iwasawa2011}; \citealt{Torres-Alba2018}).
\citet{Ricci2017bMNRAS} report the nuclear AGN/starburst 
activities of 30 interacting 
systems in GOALS, basically using NuSTAR observations.
In this work, we select all 57 U/LIRG systems (containing 84 galaxies) in
the GOALS sample observed with NuSTAR by 2020 April and/or 
detected with Swift/BAT in the 105-month data.
NGC~232/NGC~235 is the only system that is detected with
Swift/BAT but not has been observed with NuSTAR.
Our sample consists of 46 merging and 11 nonmerging U/LIRG systems.

\subsection{Merger Stages}
\label{sub2-2_stage}

We adopt the visually derived merger stage classification of
\citet{Stierwalt2013}, which is mainly based on the Hubble Space
Telescope (HST) and Spitzer/IRAC 3.6~$\mu$m images.  According to their
results and recent works \citep[e.g.,][]{Ricci2017bMNRAS}, we divide the
morphologies into five designations: ``stage-A'' (galaxy pairs prior to
a first encounter); ``stage-B'' (post-first-encounter with galaxy disks
still symmetric and in tact but with signs of tidal tails); ``stage-C''
(showing amorphous disks, tidal tails, and other signs of merger
activity); ``stage-D'' (two nuclei in a common envelop); and ``stage-N''
(no signs of merger activity or massive neighbors).  Hereafter, stage-A
and B galaxies are assigned as ``early mergers'', stage-C and D sources
are ``late mergers'', and stage-N objects are ``nonmergers''.

There remain possibilities that some stage-N sources, such as NGC 7130
\citep{Davies2014}, are past mergers. In general, however, it is very
difficult to distinguish whether a stage-N source is a past merger or
not. Also, it is reported that they show different properties of star
formation and AGN obscuration from those of stage-D mergers
\citep[e.g.,][]{Shangguan2019,Yamada2019}. Thus, we treat these objects
as stage-N in this work. Moreover, we re-categorize the merger stages
for the seven systems by the following reasons:

\begin{enumerate}
\item NGC 1068: the galaxy is a LIRG in the GOALS
sample but is not classified by \citet{Stierwalt2013}.
Although NGC 1068 may show tidally induced structures of a past minor
merger \citep{Tanaka2017}, the host galaxy is morphologically similar to
a normal spiral galaxy. As the case of NGC 7130, we treat it as a stage-N source.

\item ESO 203-1: the system is assigned as a stage-D merger 
by \citet{Stierwalt2013}. The HST images show, however, that it is an
early merger with two nuclei separated by $\sim$8~kpc
\citep{Haan2011a,Ricci2021b}. We thus regard it as a stage-B merger.

\item NGC 4418 and MCG+00-32-013: the brighter galaxy, NGC~4418, is classified 
as a stage-N LIRG, but recent works \citep{Varenius2017,Boettcher2020}
suggest that the central starburst activity would be enhanced 
by the nearby dwarf galaxy MCG+00-32-013 with the separation of 29.4~kpc.
Since no clear feature of galactic interaction is evident, we classify it as a stage-A merger.

\item MCG--03-34-064: the source is classified as a
stage-A merger consisting of MCG--03-34-064 and its nearby galaxy MCG--03-34-063, whose angular separation is 106\farcs2. 
However, they have different redshifts, 0.0165 and 0.0213, respectively.
\citet{Alonso-Herrero2012} also report this point. Thus, we re-classify the brighter IR source MCG--03-34-064 as a stage-N object.

\item IC 4518A and IC 4518B: the system is a LIRG in
the GOALS sample, but is not classified by \citet{Stierwalt2013}.
We treat it as a stage-B merger according to \citet{Ricci2017bMNRAS},
who report signs of tidal interactions and the separation of $\sim$12~kpc
in the optical and IR images.

\item IRAS F18293--3413: the galaxy is assigned as a stage-C merger by \citet{Stierwalt2013}. As noted by \citet{Ricci2017bMNRAS}, however, the HST imaging shows that the two sources do not share a common envelop, and the object seems to be in a past merger \citep{Haan2011a}. Considering the much smaller size of the companion source, we re-categorized it as a stage-N object.
\end{enumerate}

\clearpage
\startlongtable
\begin{deluxetable*}{llllcccccc}
\label{T1_properties}
\tablecaption{Basic Information of our Sample}
\tabletypesize{\footnotesize}
\tablehead{
\colhead{IRAS Name} &
\colhead{Object Name} &
\colhead{R.A.} &
\colhead{Decl.} &
\colhead{$z$} &
\colhead{M} &
\colhead{$D_{12}$} &
\colhead{$D_{12}$} &
\colhead{log($L_{\rm IR}$)} &
\colhead{IRS}\\ 
& & \ \ (J2000) & \ \ \ \ (J2000) & & & (arcsec) & (kpc) & ($L_{\odot}$) &
}
\decimalcolnumbers
\startdata 
F00085--1223   & NGC 34 & 00:11:06.61 & $-$12:06:28.33 & 0.0196 & D & S & S & 11.49 & Y\\
F00163--1039 N & MCG--02-01-052 & 00:18:50.15 & $-$10:21:41.49 & 0.0273 & B & 56.1 & 30.7 & [11.48] & n\\
F00163--1039 S & MCG--02-01-051 & 00:18:50.90 & $-$10:22:36.49 & 0.0271 & B & 56.1 & 30.7 & 11.48 & Y\\
F00344--3349   & ESO 350-38 & 00:36:52.49 & $-$33:33:17.23 & 0.0206 & C & 5.2$^{g}$ & 2.2$^{g}$ & 11.28 & Y\\
F00402--2349 W & NGC 232 & 00:42:45.81 & $-$23:33:40.69 & 0.0226 & B & 120.8 & 54.7 & [11.44]: & Y\\
F00402--2349 E & NGC 235 & 00:42:52.81 & $-$23:32:27.71 & 0.0222 & B & 120.8 & 54.7 & [11.44]: & Y\\
F00506+7248 E & MCG+12-02-001 & 00:54:03.94 & +73:05:05.23 & 0.0157 & C & 0.9$^{h}$ & 0.3$^{h}$ & 11.50 & Y\\
F01053--1746 W & IC 1623A & 01:07:46.49 & $-$17:30:22.50$^{a}$ & 0.0201 & C & 15.7 & 6.4 & [11.71] & n\\
F01053--1746 E & IC 1623B & 01:07:47.57 & $-$17:30:25.04 & 0.0203 & C & 15.7 & 6.4 & 11.71 & Y\\
F02071--1023 W2& NGC 833 & 02:09:20.85 & $-$10:07:59.11 & 0.0129 & A & 56.4 & 15.3 & [11.05]: & Y\\
F02071--1023 W & NGC 835 & 02:09:24.61 & $-$10:08:09.31 & 0.0136 & A & 56.4 & 15.3 & [11.05]: & Y\\
F02071--1023 E & NGC 838 & 02:09:38.56 & $-$10:08:46.12 & 0.0128 & A & 148.8 & 39.1 & [11.05]: & Y\\
F02071--1023 S & NGC 839 & 02:09:42.73 & $-$10:11:01.64 & 0.0129 & A & 148.8 & 39.1 & [11.05]: & Y\\
F02401--0013   & NGC 1068 & 02:42:40.77 & $-$00:00:47.84 & \,\,\,0.0038$^\dagger$ & N & n & n & 11.40 & Y\\
F03117+4151 N & UGC 2608 & 03:15:01.41 & +42:02:08.39 & 0.0233 & N & n & n & 11.41 & Y\\
F03117+4151 S & UGC 2612 & 03:15:14.60 & +41:58:50.51 & 0.0318 & N & n & n & [11.41] & Y\\
F03164+4119   & NGC 1275 & 03:19:48.16 & +41:30:42.11 & 0.0176 & N & n & n & 11.26 & Y\\
F03316--3618   & NGC 1365 & 03:33:36.46 & $-$36:08:26.37 & \,\,\,0.0055$^\dagger$ & N & n & n & 11.00 & Y\\
F04454--4838   & ESO 203-1 & 04:46:49.54 & $-$48:33:29.90$^{a}$ & 0.0529 & B & 7.5$^{a}$ & 7.7$^{a}$ & 11.86 & Y\\
F05054+1718 W & CGCG 468-002W & 05:08:19.71 & +17:21:48.09 & 0.0175 & B & 29.5 & 10.3 & [11.22]: & Y\\
F05054+1718 E & CGCG 468-002E & 05:08:21.21 & +17:22:08.34 & 0.0168 & B & 29.5 & 10.3 & [11.22]: & Y\\
F05189--2524   & IRAS F05189--2524 & 05:21:01.40 & $-$25:21:45.27 & 0.0426 & D & S & S & 12.16 & Y\\
F06076--2139 N & IRAS F06076--2139 & 06:09:45.79 & $-$21:40:23.52 & 0.0374 & C & 8.3 & 6.2 & 11.65 & Y\\
F06076--2139 S & 2MASS 06094601--2140312 & 06:09:46.01 & $-$21:40:31.26 & 0.0374 & C & 8.3 & 6.2 & [11.65] & n\\
F08354+2555   & NGC 2623 & 08:38:24.02 & +25:45:16.29 & 0.0185 & D & S & S & 11.60 & Y\\
F08520--6850 W & ESO 060-IG016 West & 08:52:29.36 & $-$69:02:01.07 & 0.0451 & B & 15.4 & 13.6 & [11.82] & n\\
F08520--6850 E & ESO 060-IG016 East & 08:52:32.05 & $-$69:01:55.74 & 0.0451 & B & 15.4 & 13.6 & 11.82 & Y\\
F08572+3915   & IRAS F08572+3915 & 09:00:25.36 & +39:03:54.23 & 0.0580 & D & \ \,4.4$^{h}$ & \ \,5.6$^{h}$ & 12.16 & Y\\
F09320+6134   & UGC 5101 & 09:35:51.69 & +61:21:10.52 & 0.0394 & D & S & S & 12.01 & Y\\
F09333+4841 W & MCG+08-18-012 & 09:36:30.86 & +48:28:10.46 & 0.0252 & A & 65.3 & 33.6 & [11.34] & n\\
F09333+4841 E & MCG+08-18-013 & 09:36:37.19 & +48:28:27.86 & 0.0259 & A & 65.3 & 33.6 & 11.34 & Y\\
F10015--0614 S & MCG--01-26-013 & 10:03:57.03 & $-$06:29:47.12 & 0.0161 & A & 108.8 & 36.5 & [11.37] & n\\
F10015--0614 N & NGC 3110 & 10:04:02.12 & $-$06:28:29.12 & 0.0169 & A & 108.8 & 36.5 & 11.37 & Y\\
F10038--3338   & ESO 374-IG032 & 10:06:04.58 & $-$33:53:05.55 & 0.0340 & D & S & S & 11.78 & Y\\
F10257--4339   & NGC 3256 & 10:27:51.28 & $-$43:54:13.55 & 0.0094 & D & \ \,5.1$^{h}$ & \ \,1.0$^{h}$ & 11.64 & Y\\
F10565+2448   & IRAS F10565+2448 & 10:59:18.13 & +24:32:34.74 & 0.0431 & D & \ \,7.4$^{h}$ & \ \,6.7$^{h}$ & 12.08 & Y\\
F11257+5850 W & NGC 3690 West & 11:28:31.00 & +58:33:41.20$^{b}$ & 0.0102 & C & 22.0 & 4.6 & [11.93]: & Y\\
F11257+5850 E & NGC 3690 East & 11:28:33.70 & +58:33:47.20$^{b}$ & 0.0104 & C & 22.0 & 4.6 & [11.93]: & Y\\
F12043--3140 N & ESO 440-58 & 12:06:51.80 & $-$31:56:47.00 & 0.0232 & B & 11.8 & 5.5 & [11.43]: & Y\\
F12043--3140 S & MCG--05-29-017 & 12:06:51.87 & $-$31:56:58.75 & 0.0228 & B & 11.8 & 5.5 & [11.43]: & Y\\
F12112+0305   & IRAS F12112+0305 & 12:13:46.11 & +02:48:41.50 & 0.0733 & D & \ \,3.0$^{i}$ & \ \,4.1$^{i}$ & 12.36 & Y\\
F12243--0036 NW& NGC 4418 & 12:26:54.63 & $-$00:52:39.51 & \,\,\,0.0073$^\dagger$ & A & 179.9 & 29.4 & 11.19 & Y\\
F12243--0036 SE& MCG+00-32-013 & 12:27:04.53 & $-$00:54:21.14 & \,\,\,0.0074$^\dagger$ & A & 179.9 & 29.4 & [11.19] & n\\
F12540+5708   & Mrk 231 & 12:56:14.23 & +56:52:25.24 & 0.0422 & D & S & S & 12.57 & Y\\
F12590+2934 S & NGC 4922S & 13:01:24.51 & +29:18:30.15 & 0.0239 & C & 22.1 & 10.6 & [11.38] & n\\
F12590+2934 N & NGC 4922N & 13:01:25.27 & +29:18:49.88 & 0.0236 & C & 22.1 & 10.6 & 11.38 & Y\\
F13126+2453   & IC 860 & 13:15:03.51 & +24:37:07.80 & 0.0112 & N & n & n & 11.14 & Y\\
13120--5453    & IRAS 13120--5453 & 13:15:06.38 & $-$55:09:22.60 & 0.0308 & D & S & S & 12.32 & Y\\
F13188+0036   & NGC 5104 & 13:21:23.12 & +00:20:33.38 & 0.0186 & N & n & n & 11.27 & Y\\
F13197--1627   & MCG--03-34-064 & 13:22:24.48 & $-$16:43:42.09 & 0.0165 & N & n & n & 11.28 & Y\\
F13229--2934   & NGC 5135 & 13:25:44.06 & $-$29:50:01.24 & 0.0137 & N & n & n & 11.30 & Y\\
F13362+4831 S & Mrk 266B & 13:38:17.35 & +48:16:31.90 & 0.0276 & B & 10.1 & 5.6 & [11.56]: & Y\\
F13362+4831 N & Mrk 266A & 13:38:17.78 & +48:16:41.02 & 0.0279 & B & 10.1 & 5.6 & [11.56]: & Y\\
F13428+5608   & Mrk 273 & 13:44:42.07 & +55:53:13.17 & 0.0378 & D & \ \,0.9$^{h}$ & \ \,0.7$^{h}$ & 12.21 & Y\\
F14348--1447   & IRAS F14348--1447 & 14:37:38.32 & $-$15:00:23.97 & 0.0827 & D & \ \,3.4$^{j}$ & \ \,5.2$^{j}$ & 12.39 & Y\\
F14378--3651   & IRAS F14378--3651 & 14:40:59.01 & $-$37:04:31.94 & 0.0681 & D & S & S & 12.23 & Y\\
F14544--4255 W & IC 4518A & 14:57:41.18 & $-$43:07:55.49$^{c}$ & 0.0163 & B & 35.5 & 11.5 & 11.23 & Y\\
F14544--4255 E & IC 4518B & 14:57:44.41 & $-$43:07:52.69 & 0.0155 & B & 35.5 & 11.5 & [11.23] & n\\
F15250+3608   & IRAS F15250+3608 & 15:26:59.43 & +35:58:37.23 & 0.0552 & D & \ \,0.7$^{k}$ & 0.8 & 12.08 & Y\\
F15327+2340 W & Arp 220W & 15:34:57.22 & +23:30:11.49$^{d}$ & 0.0181 & D & 1.0 & 0.4 & [12.28]: & Y(u)\\
F15327+2340 E & Arp 220E & 15:34:57.29 & +23:30:11.34$^{d}$ & 0.0181 & D & 1.0 & 0.4 & [12.28]: & Y(u)\\
F16504+0228 S & NGC 6240S & 16:52:58.90 & +02:24:03.36$^{e}$ & 0.0245 & D & 1.7 & 0.8 & [11.93]: & Y(u)\\
F16504+0228 N & NGC 6240N & 16:52:58.92 & +02:24:05.03$^{e}$ & 0.0245 & D & 1.7 & 0.8 & [11.93]: & Y(u)\\
F16577+5900 N & NGC 6285 & 16:58:24.02 & +58:57:21.23 & 0.0190 & B & 91.0 & 34.5 & [11.37] & Y\\
F16577+5900 S & NGC 6286 & 16:58:31.38 & +58:56:10.21 & 0.0183 & B & 91.0 & 34.5 & 11.37 & Y\\
F17138--1017   & IRAS F17138--1017 & 17:16:35.70 & $-$10:20:38.00 & 0.0173 & D & S & S & 11.49 & Y\\
F18293--3413   & IRAS F18293--3413 & 18:32:41.18 & $-$34:11:27.46 & 0.0182 & N & S & S & 11.88 & Y\\
F19297--0406   & IRAS F19297--0406 & 19:32:22.30 & $-$04:00:01.79 & 0.0857 & D & S & S & 12.45 & Y\\
F20221--2458 SW& NGC 6907 & 20:25:06.94 & $-$24:48:38.16 & 0.0106 & B & 43.9 & 9.4 & 11.11 & Y\\
F20221--2458 NE& NGC 6908 & 20:25:08.97 & $-$24:48:04.11$^{f}$ & 0.0102 & B & 43.9 & 9.4 & [11.11] & n\\
20264+2533 W  & NGC 6921 & 20:28:28.84 & +25:43:24.19 & 0.0145 & A & 91.4 & 26.5 & [11.11] & n\\
20264+2533 E  & MCG+04-48-002 & 20:28:35.06 & +25:44:00.18 & 0.0139 & A & 91.4 & 26.5 & 11.11 & Y\\
F20550+1655 NW& II Zw 096 & 20:57:23.63 & +17:07:44.60 & 0.0355 & C & 11.6 & 8.1 & [11.94]: & Y\\
F20550+1655 SE& IRAS F20550+1655 SE & 20:57:24.08 & +17:07:34.96 & 0.0350 & C & 11.6 & 8.1 & [11.94]: & Y\\
F20551--4250   & ESO 286-19 & 20:58:26.82 & $-$42:38:59.42 & 0.0430 & D & S & S & 12.06 & Y\\
F21453--3511   & NGC 7130 & 21:48:19.49 & $-$34:57:04.73 & 0.0162 & N & n & n & 11.42 & Y\\
F23007+0836 S & NGC 7469 & 23:03:15.67 & +08:52:25.28 & 0.0163 & A & 79.7 & 26.2 & 11.65 & Y\\
F23007+0836 N & IC 5283 & 23:03:18.00 & +08:53:37.13 & 0.0160 & A & 79.7 & 26.2 & [11.65] & n\\
F23128--5919   & ESO 148-2 & 23:15:46.77 & $-$59:03:15.94 & 0.0446 & C & \ \,4.5$^{l}$ & 3.9 & 12.06 & Y\\
F23157+0618   & NGC 7591 & 23:18:16.27 & +06:35:09.11 & 0.0165 & N & n & n & 11.12 & Y\\
F23254+0830 W & NGC 7674 & 23:27:56.70 & +08:46:44.24 & 0.0289 & A & 33.3 & 19.5 & 11.56 & Y\\
F23254+0830 E & MCG+01-59-081 & 23:27:58.75 & +08:46:57.88 & 0.0295 & A & 33.3 & 19.5 & [11.56] & n\\
\tablebreak
23262+0314 W  & NGC 7679 & 23:28:46.67 & +03:30:40.99 & 0.0171 & A & 269.8 & 93.8 & 11.11 & Y\\
23262+0314 E  & NGC 7682 & 23:29:03.90 & +03:32:00.00 & 0.0171 & A & 269.8 & 93.8 & [11.11] & n\\
\enddata
\tablecomments{Columns: 
(1) IRAS name; 
(2) object name; 
(3--4) the best available source right ascension and declination (J2000) in CDS Portal (based on SIMBAD; \url{http://cdsportal.u-strasbg.fr/}); 
(5) redshift in NASA/IPAC Extragalactic Database (NED);
(6) merger stage as classified from the Hubble Space Telescope (HST) and IRAC 3.6 $\mu$m imaging \citep{Stierwalt2013}; 
(7--8) the separation between the two nuclei in arcsec and kpc. S and n mean that a single nucleus is observed in merging and nonmerging U/LIRGs, respectively;
(9) logarithmic total IR luminosity in units of $L_{\odot}$ \citep{Armus2009}. 
Values in brackets should be upper limits due to contamination from nearby much brighter (or equally bright with the suffix ``:'') IR sources.
(10) Y and n mark detection and nondetection with Spitzer/IRS in the 10--20 $\mu$m (SH) or 19--38 $\mu$m (LH) band, respectively \citep{Alonso-Herrero2012,Mazzarella2012,Inami2013}. The (u) means that the two nuclei are not clearly divided.\\
\textbf{References:} The source positions in Column (3--4) and the separations in Column (7--8) are adopted from the references: 
(a) \citet{Haan2011a}; (b) \citet{Zezas2003}; (c) \citet{Alonso-Herrero2020}; 
(d) \citet{Scoville2017}; (e) \citet{Puccetti2016}; (f) NED; (g) \citet{Kunth2003};
(h) \citet{Ricci2017bMNRAS}; (i) \citet{Imanishi2020}; (j) \citet{Imanishi2014}; 
(k) \citet{Scoville2000}; (l) \citet{Zenner1993}.
}
\tablenotetext{\dagger}{We utilize redshift-independent measurements of the distance 
for the closest objects (at $z < 0.01$), NGC 1068 (14.4 Mpc; \citealt{Tully1988,Bauer2015}), NGC 1365 (17.3 Mpc; \citealt{Venturi2018}), and NGC 4418/MCG+00-32-013 (34 Mpc; e.g., \citealt{Ohyama2019}).}
\end{deluxetable*}

\begin{enumerate}

\item[7] NGC 6907 and NGC 6908: NGC 6907 is classified as
stage-N but \citet{Madore2007} suggest that NGC 6907 and the dwarf
galaxy NGC 6908 are interacting and show tidal signatures.
The separation is 9.4 kpc. Thus, we treated them as a stage-B merger.

\end{enumerate}

\subsection{Basic Information on Individual Objects}
\label{sub2-3_bias}
To obtain the accurate galaxy separations in mergers, we refer to the
central positions not only from the SIMBAD but also from 
the literature.
We finally confirm them via visual inspection in the optical images
(observed with HST, PanSTARRS, SDSS, and/or DSS2), the mid-IR images
(with Spitzer/IRAC 1--4), and/or the soft X-ray images (with Chandra
and/or XMM-Newton). 
The basic properties of our targets and their
references are summarized in Table~\ref{T1_properties}.

In this work, we utilize the hard X-ray observations
with NuSTAR to identify the AGNs in local U/LIRGs.
To determine which galaxy in the system contains an AGN, 
we firstly check the peak position 
in the NuSTAR 8--24~keV image against
the galaxy positions (see Section~\ref{sub4-1_counts}).
In many cases, however, NuSTAR cannot resolve the merging systems because
their angular separations are smaller than the angular resolution
(a full-width at half-maximum (FWHM) of 18\arcsec\ and a half power
diameter of 58\arcsec). In such cases, it is also possible
that a system actually contain more than one AGNs. 
Thus, we also 
refer to the soft X-ray (i.e., Chandra and/or XMM-Newton) 
and other multiwavelength data to confirm 
the presence of AGN(s) in each system as reported 
in Appendix~\ref{Appendix-B}.

\begin{figure}
    \epsscale{1.30}
    \plotone{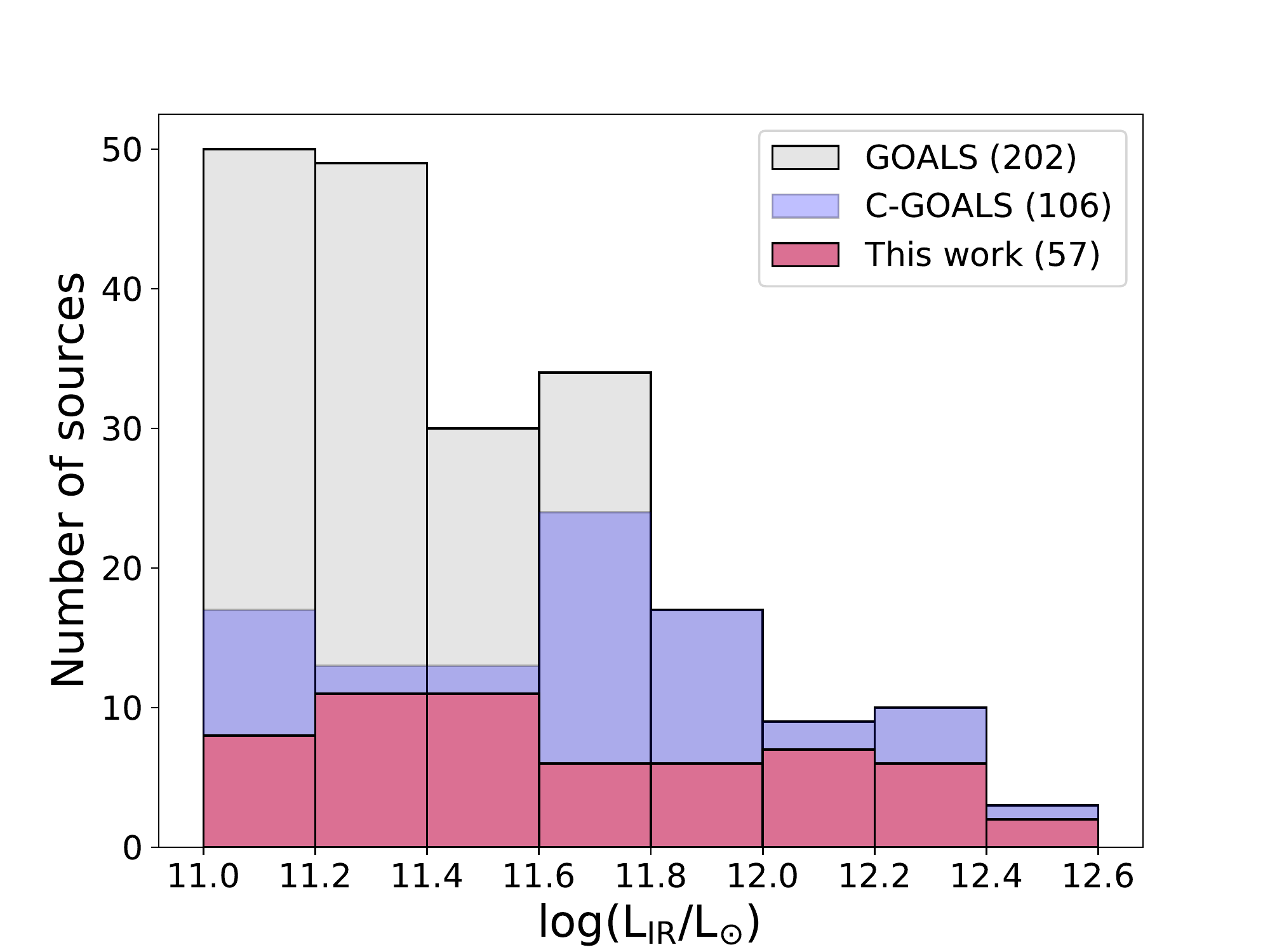}
    \plotone{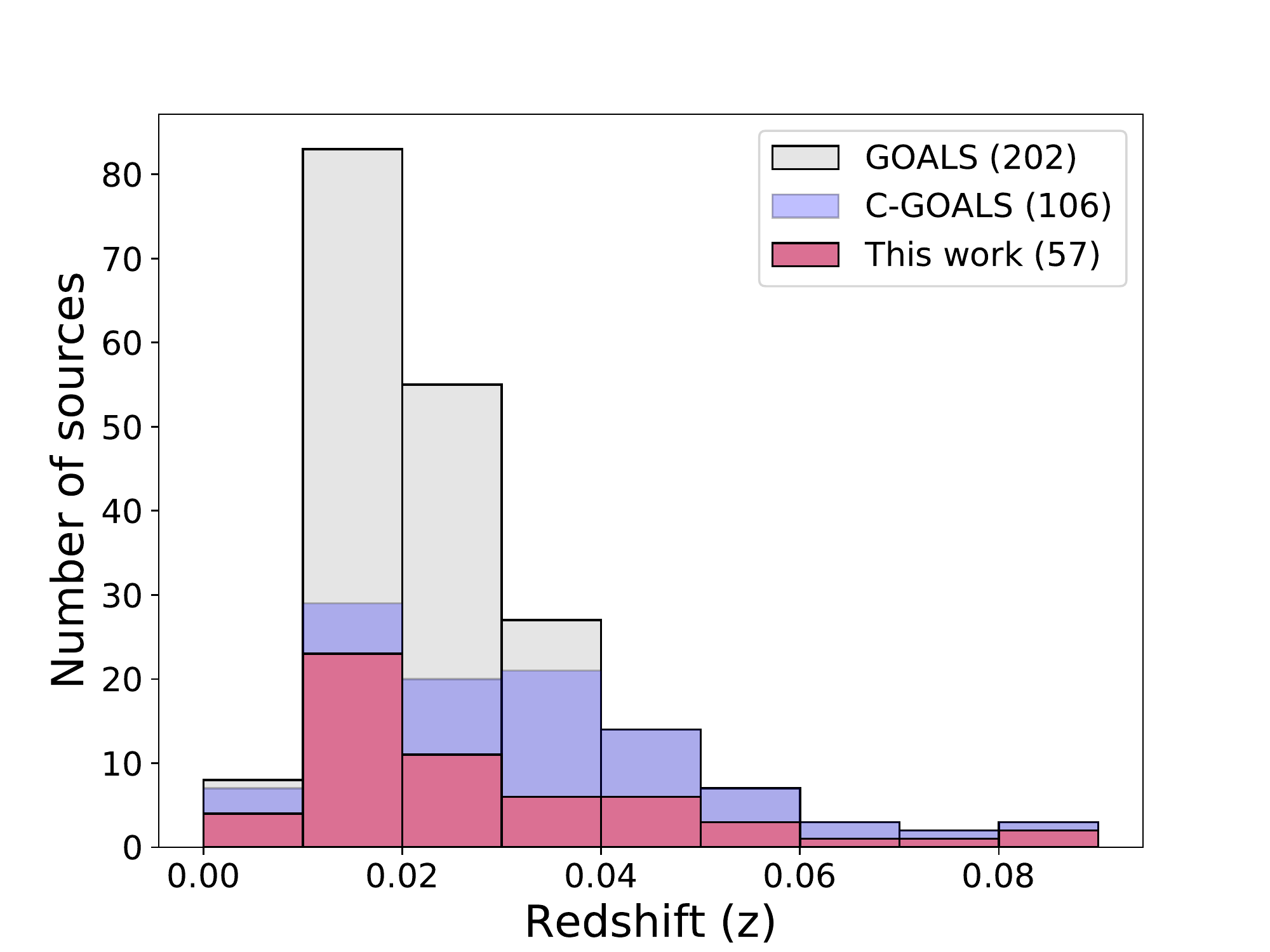}
    \plotone{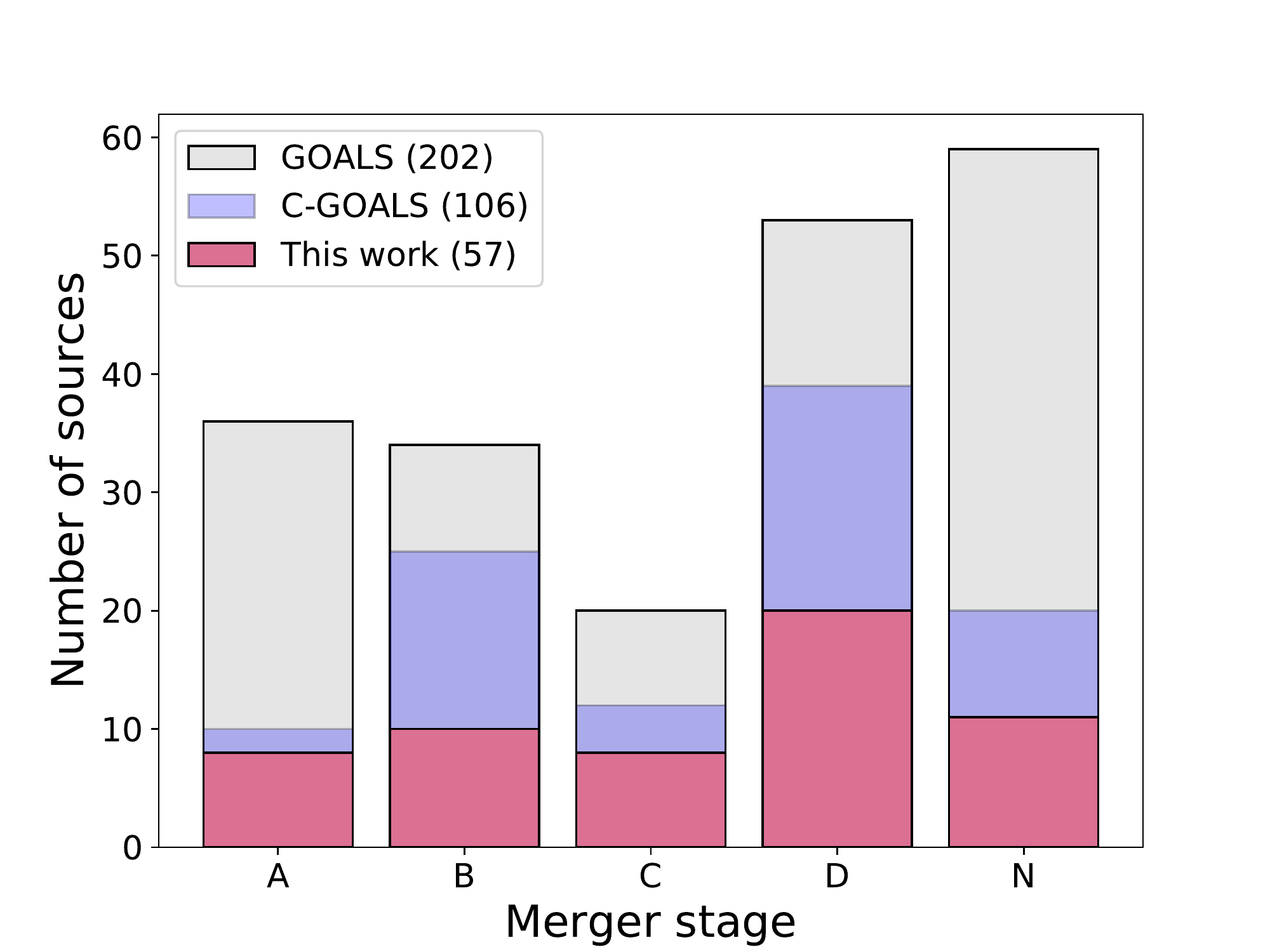}
        \caption{Upper panel: 
histogram of IR luminosity ($L_{\rm IR}$) 
        for the whole GOALS sample (gray) compared with the C-GOALS
 sample (blue) and our sample (red).
        Middle panel: histogram of redshift (taken from NED).
        Bottom panel: histogram of merger stage.\\
        \label{F1_hist}}
\end{figure}

\subsection{Characteristics of Our Sample}
\label{sub2-4_bias}

Figure~\ref{F1_hist} shows the histograms of the 8--1000 $\mu$m IR luminosity, 
redshift, and merger stage for the U/LIRGs in the GOALS, C-GOALS, and our samples.  
As seen from the upper and middle panels,
our sample covers the full ranges of the parent GOALS
sample over $L_{\rm IR} = 10^{11.0}$--$10^{12.6} L_{\odot}$ and $z$ = 0.001--0.088. 
A Kolmogorov-Smirnov (KS) test indicates that 
the difference in the IR
luminosity distribution between our targets and the rest 
of the GOALS sample is significant at a $>$99\% confidence level, while
the difference in the redshift is not ($p$-value = 0.214). Thus, we
may regard our sample as a good representative subsample of the local
U/LIRGs, although it is weakly biased toward higher IR luminosities
compared with the parent GOALS sample.

As shown in the bottom panel, the numbers of the 
five merger stages in our sample are 8, 10, 8,
20, and 11 for stage-A, B, C, D, and N, whereas those in GOALS are 36,
34, 20, 53, and 59, respectively.\footnote{We treat ESO 550-25 as a
stage-B merger based on the optical image, which is not classified by
\citet{Stierwalt2013}.} 
A chi-square test of independence suggests that the difference 
of the distribution between our targets and the rest of the GOALS sample 
is not significant ($p$-value = 0.132).
We note a possible bias that objects that show
AGN features in the soft X-ray and/or IR band may have
been preferentially selected to be observed with NuSTAR. The
smaller fraction of stage-N (nonmergers) in our sample than in the GOALS
sample may be affected by this bias. We have to keep it mind when we
discuss any statistical properties (e.g., AGN fraction) of U/LIRGs using
our sample (see e.g., Section~\ref{sub5-2_CTfraction}).

\ifnum0=1
Figure~\ref{F1_hist} shows the histograms of the 8--1000 $\mu$m IR luminosity, 
redshift and merger stage for the U/LIRGs in GOALS, C-GOALS, and our sample.  
As seen from the upper and middle panels,
our sample covers the full ranges of the parent GOALS
sample over $L_{\rm IR} = 10^{11.0}$--$10^{12.6} L_{\odot}$ and $z$ = 0.001--0.088. 
Thus, although our sample is far from being complete, we
may regard our sample as a good representative subsample of the GOALS
galaxies. 
As shown in the bottom panel, the numbers of the 
five merger stages in our sample are 8, 10, 8,
20, and 11 for stage-A, B, C, D, and N, whereas those in GOALS are 36,
34, 20, 53, and 59, respectively.\footnote{We treat ESO 550-25 as a
stage-B merger based on the optical image, which is not classified by
\citet{Stierwalt2013}.} We note a possible bias that objects that show
AGN features in the soft X-ray and/or IR band may have
been preferentially selected to be observed with NuSTAR. The
smaller fraction of stage-N (nonmergers) in our sample than in the GOALS
sample may be affected by this bias. We have to keep it mind when we
discuss any statistical properties (e.g., AGN fraction) of U/LIRGs using
our sample (see e.g., Section~\ref{sub5-2_CTfraction}).
\fi

\subsection{Targets with Complex Spectral Features}
\label{sub2-5_Xray-ref}

For the eight well-studied U/LIRG systems (hosting 10 AGNs), NGC 1068,
NGC 1275, NGC 1365, Mrk 266, NGC 6240, UGC 2608, NGC 5135, and NGC 7469,
we refer to
the X-ray spectral results of previous works based on the NuSTAR data.
The broadband spectra of NGC 1068, NGC 1275, and NGC 1365 are reproduced
with a multi-component reflector model \citep{Bauer2015}, a jet-dominant
AGN model \citep{Rani2018}\footnote{NGC 1275 is a
radio galaxy in the Perseus cluster hosting a low-luminosity AGN
\citep{Hitomi-Collaboration2018}. The contamination of the jet
emission makes it difficult to constrain the nuclear structure.
Thus, we only discuss its X-ray flux and band ratio in this paper.}, and
a multi-layer variable absorber model \citep{Rivers2015},
respectively. 
The dual AGN systems of Mrk 266B/Mrk 266A
\citep{Iwasawa2020} and NGC 6240S/NGC 6240N \citep{Puccetti2016} are
observed with Chandra by separating the individual AGN components, 
and their hard X-ray emissions are 
detected with NuSTAR. The absorption hydrogen 
column densities of these five U/LIRG systems taken from the literature are
summarized in Table~\ref{T2_previousXref}.\footnote{The 
hydrogen column densities of NGC~1068, NGC~1365,
Mrk~266B/Mrk~266A and NGC~6240S/NGC~6240N are derived with 
other torus models than XCLUMPY. We note that the column 
densities obtained with XCLUMPY are consistent with 
the values obtained with smooth-torus models for the other AGNs
in our sample (see also Appendix~\ref{Appendix-B}).}
We also refer to the NuSTAR results of UGC 2608,
NGC~5135 (heavily CT AGNs in stage-N LIRGs; \citealt{Yamada2020}), and
NGC 7469 (unobscured AGN whose spectrum is subject to ionized warm
absorbers; \citealt{Ogawa2021}), for which the XCLUMPY torus model is
applied; the best-fitting results are shown in 
Table~\ref{T5_bestpara-AGN}.
In Appendix~\ref{Appendix-B}, we present detailed information obtained
from previous X-ray and multiwavelength studies for all 84 galaxies
including these 8 U/LIRG systems.

\begin{deluxetable*}{lccc}
\label{T2_previousXref}
\tablecaption{Previous NuSTAR Works for AGNs with Complex Spectral Features}
\tablewidth{0pt}
\tabletypesize{\footnotesize}
\tablehead{
\colhead{Object} &
\colhead{$N^{\rm LoS}_{\rm H}$} &
\colhead{Model} &
\colhead{Ref.}
}
\decimalcolnumbers
\startdata
NGC 1068 & $\sim$1000 & Multi-component reflector model & \citet{Bauer2015} \\
NGC 1275 & \nodata & Jet-dominant AGN model in the Perseus Cluster  & \citet{Rani2018} \\
NGC 1365 & $\sim$10 & Multi-layer variable absorber model & \citet{Rivers2015} \\
Mrk 266B/Mrk 266A & 700/6.8 & Dual-AGN model & \citet{Iwasawa2020} \\
NGC 6240S/NGC 6240N & 147$^{+21}_{-17}$/155$^{+72}_{-23}$  & Dual-AGN model & \citet{Puccetti2016}
\enddata
\tablecomments{Columns:
(1) object name;
(2) line-of-sight hydrogen column density in units of 10$^{22}$ cm$^{-2}$;
(3--4) X-ray spectral model and their references.}
\end{deluxetable*}

\section{Observations and Data Reduction}
\label{S3_reduction}

For the 49 U/LIRG systems other than the 8 systems mentioned in Section~\ref{sub2-5_Xray-ref}, 
we analyzed all the available data of NuSTAR (Section~\ref{sub3-1_nustar}), 
Chandra (Section~\ref{sub3-2_chandra}), XMM-Newton (Section~\ref{sub3-3_xmm}), 
and Suzaku (Section~\ref{sub3-4_suzaku}),
observed by 2020 April. We also analyzed their Swift/XRT data when no
other soft X-ray data were available, and utilized the Swift/BAT spectra
in the 105 month catalog if detected (Section~\ref{sub3-5_swift}). For the other 8
systems, we reduced the same data as used in their references (see
Section~\ref{sub2-5_Xray-ref}) and obtained the count rates, in order to treat the results
in a uniform manner for all the 57 systems. A log of the X-ray
observations of our sample is provided in Table~\ref{T3_log}.  In the following, we
describe the data reduction for each satellite.

\subsection{NuSTAR}
\label{sub3-1_nustar}

NuSTAR \citep{Harrison2013}, the first satellite capable of focusing
hard X-rays of 3--79~keV, observed 56 U/LIRG systems in the GOALS sample
by Cycle-4. 
Ten systems were observed in Cycle-1 (PI: F.E. Bauer) and
other 14 systems are in Cycle-3 (PIs: C. Ricci; G. C. Privon) as campaigns of observing
local U/LIRGs in different merger stages. 
We analyzed all these data including those of the other 32 systems
in the archive.

We reduced the data using the NuSTAR Data Analysis Software
(\textsc{nustardas}) v1.8.0 within HEASOFT v6.25, adopting the
calibration files released on 2019 May 13.  Calibrated and cleaned event
files were produced using the \textsc{nupipeline} script with the
\textsc{saamode=optimized} and \textsc{tentacle=yes} options.  The
\textsc{nuproducts} task was used to extract the source spectra, background
spectra, and response files.  We adopted circular source regions of
45\arcsec\ radii or those of 30\arcsec\ or smaller radii\footnote{We
adopted circular regions of 25\arcsec\ radii for MCG--02-01-052, NGC~833, 
and IRAS F14348--1447.} when there are contaminating
nearby sources close to the target. 
We also chose 30\arcsec\ radii when the signal-to-noise (S/N) ratios of
the 8--24~keV count rates obtained from the 45\arcsec\ radius regions were
lower than $<$3$\sigma$. 
The background spectra were taken from the annuluses centered on the
X-ray sources with inner and outer radii of 90\arcsec\ and 150\arcsec,
respectively, by avoiding regions of contamination from nearby sources
and CCD gaps. Net exposure times for these targets range from $\sim$10~ks
to $\sim$200~ks. After confirming that the spectra of the two focal
plane modules (FPMA and FPMB; FPM) agreed with each other within the
statistical uncertainties, we coadded them with the \textsc{addascaspec}
task.

\subsection{Chandra}
\label{sub3-2_chandra}

We analyzed the Chandra/ACIS \citep{Garmire2003} imaging data of 46
U/LIRG systems, most of which were obtained by the project to follow up
GOALS objects (PI: D. Sanders, see \citealt{Iwasawa2011} and
\citealt{Torres-Alba2018}). The data reduction was performed with the
Chandra data analysis package \textsc{ciao} v4.11 and the Calibration database
(CALDB) v4.8.4.1.  The event files were reprocessed with the
\textsc{chandra\_repro} tool. The source spectrum was extracted from 
the circular region of a typically 10\arcsec\ radius 
centered on the optical position or the X-ray peak position for a bright source.
We adopt a radius large enough to include the extended
emission of the whole galaxy, between 3\arcsec\ (for a galaxy in a close merging system, ESO 440-58) and 50\arcsec\ (for the nearest source, NGC 1068).
The background spectrum was taken from a nearby source-free 
circular region with the same, or 10\arcsec\ radius when the source radius is $<$10\arcsec.

\subsection{XMM-Newton}
\label{sub3-3_xmm}

The XMM-Newton \citep{Jansen2001} data of 35 U/LIRG systems were
utilized in this work.  We reduced the EPIC/MOS (MOS1, MOS2) and EPIC/pn
data with the standard Science Analysis System (\textsc{sas}) v17.0.0
\citep{Gabriel2004}, and then processed their raw data files by the
\textsc{emproc} and \textsc{epproc} scripts, respectively. We filtered
out periods of high background activity. The threshold is set to
be 1.5 times the averaged count rates obtained from periods with
$\leq$0.5 counts s$^{-1}$ in the energy range above 10~keV.
The source spectrum was extracted from a circular region with a
radius of typically 25\arcsec, whereas the background was from a
source-free circular region with a radius of 40\arcsec\ in the same CCD
chips. We generated the ARF with \textsc{arfgen} and RMF with
\textsc{rmfgen}. The spectra and response files of EPIC/MOS1 and MOS2
were combined.

\subsection{Suzaku}
\label{sub3-4_suzaku}

For 18 U/LIRG systems, we utilized the data of the Suzaku
\citep{Mitsuda2007} observations. Suzaku carried the X-ray CCD cameras
called the X-ray Imaging Spectrometers (XIS), covering the 0.2--12~keV
band. XIS-0, XIS-2, and XIS-3 are frontside-illuminated cameras
(XIS-FI) and XIS-1 is a backside-illuminated one (XIS-BI). XIS-2 data
were not available after 2006 November due to the malfunction.
We generated the cleaned event files with the Suzaku calibration
database released on 2018 October 23. The source spectrum was selected from
a circular region of 1\farcm7 radius, while the background was from a
partial annulus with inner and outer radii of 3\farcm0 and 5\farcm7
within the field of view by excluding regions of nearby sources.
The ARFs and RMFs were generated using \textsc{xisarfgen} and
\textsc{xisrmfgen} tasks \citep{Ishisaki2007}, respectively.  The
spectra and responses of the XIS-FI detectors were merged. 

Suzaku also carried the Hard X-ray Detector (HXD), a non-imaging
instrument consisting of the silicon PIN diodes (covering 10--70~keV)
and GSO scintillators (40--600~keV). We did not use the GSO data,
because our targets were too faint to be detected with GSO. The HXD-PIN
data were reduced with \textsc{aepipeline} task. We utilized the
``tuned'' background event files \citep{Fukazawa2009} to produce the non
X-ray background spectra of HXD-PIN, to which the simulated cosmic
X-ray background spectrum was added.

\subsection{Swift}
\label{sub3-5_swift}

We analyzed Swift/XRT data covering the $\sim$0.3--10~keV band
for three U/LIRG systems that were neither observed with Chandra, 
XMM-Newton, nor Suzaku: CGCG 468-002W/CGCG
468-002E, NGC~5104, and NGC 6907/NGC 6908.\footnote{The Swift/XRT data of
ObsID 33513001 and 33513002 performed on 2014 November 3 as ToO observations
were not used since they may be largely affected by 
the emission from the supernova in NGC~6908 \citep{Margutti2014}.}
We also analyzed those of ESO~148-2 and NGC~7679, which showed large
time variability between the epoch of NuSTAR and Chandra/XMM-Newton
observations (see Appendix~\ref{Appendix-B}).

The data reduction was carried out with the \textsc{xrtpipeline} v0.13.4
script. The source spectrum was taken from a circular region with a
radius of 45\arcsec, and the background was from a partial annulus with
inner and outer radii of 90\arcsec\ and 150\arcsec, respectively.  All
spectra of each object were combined to improve the photon statistics.

To better constrain the broadband spectra, we also utilized the
time-averaged Swift/BAT spectra in the 105-month catalog
\citep{Oh2018}\footnote{\url{https://swift.gsfc.nasa.gov/results/bs105mon/}}, 
which cover the 14--195~keV band. 
To identify the counterparts,
we matched our targets with the Swift/BAT catalog 
within error radii defined as:
\begin{align}
& \rm{Error\ radius\ (arcmin)} = \sqrt{\left(\frac{30.5}{\rm S/N}\right)^2 + 0.1^2},
\end{align}
where S/N represents the significance level of of the BAT detection
\citep[see][]{Oh2018}.
Thus, we adopted the BAT spectra for 11 galaxies (NGC 235, CGCG
468--002W, IRAS F05189--2524, UGC 5101, MCG--03-34-064, Mrk~273, IC
4518A, MCG+04-48-002, NGC 7130, NGC 7674, and NGC 7679) except for
the dual AGN system NGC~833/NGC~835 and 5 systems for which 
we refer to the previous NuSTAR results (NGC 1068, NGC 1275, NGC 1365,
NGC~6240S/NGC~6240N, and NGC 7469; see Section~\ref{sub2-5_Xray-ref}).
\\

\section{X-Ray Spectral Analysis}
\label{S4_analysis}
\subsection{X-ray Counts and NuSTAR Counterparts}
\label{sub4-1_counts}

Since many of the U/LIRGs in our sample are interacting systems, 
we have to identify which of the two galaxies or both contributes to the
hard X-ray emission observed with NuSTAR. First, we check the source
peak position in the NuSTAR 8--24~keV band image, where the AGN
component is expected to dominate. If the NuSTAR position is consistent
with one galaxy but not with the other, then we regard the former as the
counterpart of the hard X-ray source. Here we take into account both the
statistical ($<$17\arcsec; i.e., the value of the radius with 1$\sigma$ enclosed fraction of $\approx$51\arcsec\ divided by the S/N ratio) and systematic 
(8\arcsec; 90\% confidence) errors in
the astrometry of NuSTAR \citep{Harrison2013}. 
If the NuSTAR error circle contains both
galaxies, then we refer to the previous work based on Chandra and/or
other wavelength data to identify AGNs in either or both galaxies
(Appendix~\ref{Appendix-B}). We regard the galaxy that contains an AGN as the
counterpart of the NuSTAR source. Among our sample, only two systems,
Mrk~266B/Mrk~266A and NGC~6240S/NGC~6240N, are identified as
``dual-AGNs'' where both nuclei significantly contribute to the hard
X-ray emission observed with NuSTAR.

\clearpage
\startlongtable
\begin{deluxetable*}{llccc}
\label{T3_log}
\tablecaption{X-ray Observation Log}
\tabletypesize{\footnotesize}
\tablehead{
\colhead{Object} &
\colhead{Satellite} &
\colhead{ObsID} &
\colhead{Observation Date} &
\colhead{Net Exp.}
}
\decimalcolnumbers
\startdata
NGC 34 
& NuSTAR     & 60101068002 & 2015-07-31 & 21.0/21.0 \\
& Chandra    &       15061 & 2015-07-31 & 14.8 \\
& XMM-Newton &  0150480501 & 2002-12-22 & 21.8/21.8/14.5 \\
\hline
MCG--02-01-052/MCG--02-01-051 
& NuSTAR     & 60101069002 & 2015-10-29 & 20.3/20.5 \\
& Chandra    &       13823 & 2011-10-30 & 29.6 \\
\hline
ESO 350-38 
& NuSTAR     & 60374008002 & 2018-01-15 & 22.4/22.4 \\
& Chandra    &        8175 & 2006-10-28 & (u; time-v) \\
& Chandra    &       16695 & 2015-11-29 & 24.7 \\
& Chandra    &       16696 & 2016-09-12 & 24.7 \\
& Chandra    &       16697 & 2017-11-24 & 23.8 \\
\hline
NGC 232/NGC 235 
& Chandra    &       12872 & 2011-01-17 & 6.9 \\
& Chandra    &       15066 & 2013-01-04 & 14.8 \\
& Suzaku     &   708026010 & 2013-12-10 & 19.8/17.2 \\
\hline
MCG+12-02-001
& NuSTAR     & 60101070002 & 2015-08-03 & 25.8/25.5 \\
& Chandra    &       15062 & 2012-11-22 & 14.3 \\
\hline
IC 1623A/IC 1623B 
& NuSTAR     & 50401001002 & 2019-01-19 & 197.9/198.8 \\
& NuSTAR     & 60374003002 & 2018-01-29 & 20.9/21.0 \\
& Chandra    &        7063 & 2005-10-20 & 59.4 \\
& XMM-Newton &  0025540101 & 2001-06-26 & 11.4/11.4/2.7 \\
& XMM-Newton &  0830440101 & 2019-01-10 & 31.2/31.2/25.8 \\
\hline
NGC 833/NGC 835/NGC 838/NGC 839 
& NuSTAR     & 60061346002 & 2015-09-13 & 18.1/18.1 \\
& Chandra    &       15181 & 2013-07-16 & 49.5 \\
& Chandra    &       15666 & 2013-07-18 & 29.7 \\
& Chandra    &       15667 & 2013-07-21 & 58.3 \\
& Chandra    &         923 & 2000-11-16 & (u; time-v) \\
& Chandra    &       10394 & 2008-11-23 & (u; time-v) \\
& XMM-Newton &  0115810301 & 2000-01-23 & 52.3/47.8/44.7 \\
& Suzaku     &   709009010 & 2014-08-04 & (u; cont) \\
\hline
NGC 1068
& NuSTAR     & 60002030002 & 2012-12-18 & 57.9/57.8 \\
& NuSTAR     & 60002030004 & 2012-12-20 & 48.6/48.5 \\
& NuSTAR     & 60002030006 & 2012-12-21 & 19.5/19.4 \\
& XMM-Newton &  0111200101 & 2000-07-29 & 39.1/--/35.1 \\
& XMM-Newton &  0111200201 & 2000-07-30 & 39.0/--/35.3 \\
\hline
UGC 2608
& NuSTAR     & 60001161002 & 2014-10-08 & 22.1/22.2 \\
& XMM-Newton &  0002942401 & 2002-01-28 & 4.8/4.8/2.1 \\
\hline
UGC 2612
& NuSTAR     & 60001161002 & 2014-10-08 & 22.1/22.2 \\
& Chandra    &       17280 & 2015-12-11 & 4.7 \\
& XMM-Newton &  0002942401 & 2002-01-28 & (u; faint) \\
& XMM-Newton &  0002942501 & 2002-01-28 & (u; faint) \\
& Suzaku     &   701007010 & 2006-08-02 & (u; faint) \\
& Suzaku     &   701007020 & 2007-02-04 & (u; cont) \\
\hline
NGC 1275
& NuSTAR     & 90202046002 & 2017-02-01 & 20.3/19.5 \\
& NuSTAR     & 90202046004 & 2017-02-04 & 28.2/28.1 \\
& XMM-Newton &  0085110101 & 2001-01-30 & 53.1/53.1/24.6 \\
& XMM-Newton &  0305780101 & 2006-01-29 & 123.3/123.3/76.1 \\
\hline
NGC 1365
& NuSTAR     & 60002046002 & 2012-07-25 & 36.2/36.0 \\
& NuSTAR     & 60002046003 & 2012-07-26 & 40.6/40.5 \\
& NuSTAR     & 60002046005 & 2012-12-24 & 66.2/66.3 \\
& NuSTAR     & 60002046007 & 2013-01-23 & 73.0/73.6 \\
& NuSTAR     & 60002046009 & 2013-02-12 & 69.7/69.5 \\
& XMM-Newton &  0692840201 & 2012-07-25 & 133.9/133.9/118.2 \\
& XMM-Newton &  0692840301 & 2012-12-24 & 121.6/121.6/107.8 \\
& XMM-Newton &  0692840401 & 2013-01-23 & 125.0/125.0/92.8 \\
& XMM-Newton &  0692840501 & 2013-02-12 & 122.4/122.5/115.8 \\
\hline
ESO 203-1 
& NuSTAR     & 60374001002 & 2018-05-25 & 21.2/21.1 \\
& Chandra    &        7802 & 2008-01-17 & 14.8 \\
\hline
CGCG 468-002W/CGCG 468-002E 
& NuSTAR     & 60006011002 & 2012-07-23 & 15.5/15.5 \\
& Swift/XRT  &49706(001--) & 2013-04-17 & 81.5 \\
&  & \ \ \ \ \ \ \ (--011) & 2014-11-11 &\\
\hline
IRAS F05189--2524 
& NuSTAR     & 60002027002 & 2013-02-20 & 22.9/23.2 \\
& NuSTAR     & 60002027004 & 2013-10-02 & 21.2/21.2 \\
& NuSTAR     & 60002027005 & 2013-10-02 & 7.9/8.0 \\
& NuSTAR     & 60201022002 & 2016-09-05 & 144.0/143.8 \\
& Chandra    &        2034 & 2001-10-30 & 19.7 \\
& Chandra    &        3432 & 2002-01-30 & 14.9 \\
& XMM-Newton &  0085640101 & 2001-03-17 & 11.6/11.6/8.1 \\
& XMM-Newton &  0722610101 & 2013-10-02 & 36.1/36.1/30.8 \\
& XMM-Newton &  0790580101 & 2016-09-06 & 96.4/96.4/83.9 \\
& Suzaku     &   701097010 & 2006-04-10 & 78.2/48.7 \\
\hline
IRAS F06076--2139/2MASS 06094601--2140312 
& NuSTAR     & 60370004002 & 2018-01-29 & 22.7/23.1 \\
& Chandra    &       15052 & 2012-12-12 & 14.8 \\
\hline
NGC 2623 
& NuSTAR     & 60374010002 & 2018-05-24 & 38.7/38.6 \\
& Chandra    &        4059 & 2003-01-03 & 19.8 \\
& XMM-Newton &  0025540301 & 2001-04-27 & 11.9/11.9/5.7 \\
\hline
ESO 060-IG016 West/ESO 060-IG016 East 
& NuSTAR     & 60101053002 & 2015-12-01 & 41.1/41.4 \\
& Chandra    &        7888 & 2007-05-31 & 14.7 \\
\hline
IRAS F08572+3915 
& NuSTAR     & 50401004002 & 2019-04-04 & 211.2/209.3 \\
& NuSTAR     & 60001088002 & 2013-05-23 & 23.7/23.8 \\
& Chandra    &        6862 & 2006-01-26 & 14.9 (u; faint) \\
& XMM-Newton &  0200630101 & 2004-04-13 & 28.2/28.3/23.5 \\
& XMM-Newton &  0830420101 & 2019-04-05 & 57.6/57.5/51.0 \\
& XMM-Newton &  0830420201 & 2019-04-07 & 61.2/61.2/51.9 \\
& Suzaku     &   701053010 & 2006-04-14 & 77.2/58.6 \\
\hline
UGC 5101 
& NuSTAR     & 60001068002 & 2014-03-21 & 17.6/17.4 \\
& Chandra    &        2033 & 2001-05-28 & 49.3 \\
& XMM-Newton &  0085640201 & 2001-11-12 & 33.9/34.0/26.5 \\
& Suzaku     &   701002010 & 2006-03-31 & 49.0/41.4 \\
\tablebreak
& Suzaku     &   701002020 & 2006-10-31 & 42.2/31.6 \\
\hline
MCG+08-18-012/MCG+08-18-013 
& NuSTAR     & 60101071002 & 2015-09-25 & 20.2/20.0 \\
& Chandra    &       15067 & 2013-06-03 & 13.8 \\
\hline
MCG--01-26-013/NGC 3110 
& NuSTAR     & 60101072002 & 2015-11-12 & 18.9/18.9 \\
& Chandra    &       15069 & 2013-02-02 & 14.9 \\
& XMM-Newton &  0550460201 & 2008-06-04 & 35.2/33.0/24.7 \\
& XMM-Newton &  0550461001 & 2008-06-04 & 4.2/4.2/2.3 \\
\hline
ESO 374-IG032
& NuSTAR     & 60101055002 & 2016-01-14 & 51.1/51.7 \\
& Chandra    &        7807 & 2007-03-07 & 14.4 \\
\hline
NGC 3256 
& NuSTAR     & 50002042002 & 2014-08-24 & 170.7/171.2 \\
& Chandra    &         835 & 2000-01-05 & 27.8 \\
& Chandra    &        3569 & 2003-05-23 & 27.2 \\
& Chandra    &       16026 & 2014-08-27 & 15.4 \\
& XMM-Newton &  0112810201 & 2001-12-15 & 16.1/16.1/11.6 \\
& XMM-Newton &  0300430101 & 2005-12-06 & 131.0/131.1/115.6 \\
\hline
IRAS F10565+2448 
& NuSTAR     & 60001090002 & 2013-05-22 & 25.2/24.6 \\
& Chandra    &        3952 & 2003-10-23 & 28.9 \\
& XMM-Newton &  0150320201 & 2003-06-17 & 30.7/30.7/26.4 \\
& Suzaku     &   702115010 & 2007-11-06 & 39.4/33.1 \\
\hline
NGC 3690 West/NGC 3690 East
& NuSTAR     & 50002041002 & 2013-03-12 & 8.9/9.4 \\
& NuSTAR     & 50002041003 & 2013-03-13 & 59.8/59.7 \\
& Chandra    &        1641 & 2001-07-13 & 24.3 \\
& Chandra    &        6227 & 2005-02-14 & 10.2 \\
& Chandra    &       15077 & 2013-03-13 & 51.9 \\
& Chandra    &       15619 & 2013-03-12 & 38.5 \\
& XMM-Newton &  0112810101 & 2001-05-06 & 21.3/21.3/15.4 \\
& XMM-Newton &  0679381101 & 2011-12-15 & 11.2/11.2/8.5 \\
\hline
ESO 440-58/MCG--05-29-017 
& NuSTAR     & 60101073002 & 2015-08-23 & 27.6/27.6 \\
& Chandra    &       15064 & 2013-03-20 & 14.8 \\
\hline
IRAS F12112+0305 
& NuSTAR     & 60374005002 & 2018-01-17 & 18.9/18.8 \\
& Chandra    &        4110 & 2003-04-15 & (u; faint) \\
& Chandra    &        4934 & 2004-07-17 & (u; faint) \\
& XMM-Newton &  0081340801 & 2001-12-30 & 22.3/22.3/17.9 \\
\hline
NGC 4418/MCG+00-32-013
& NuSTAR     & 60101052002 & 2015-07-03 & 47.2/47.0 \\
& Chandra    &        4060 & 2003-03-10 & 19.8 \\
& Chandra    &       10391 & 2009-02-20 & 5.7 \\
& Suzaku     &   701001010 & 2006-07-13 & 68.4/53.4 \\
\hline
Mrk 231
& NuSTAR     & 60002025002 & 2012-08-26 & 40.2/39.4 \\
& NuSTAR     & 60002025004 & 2013-05-09 & 27.5/28.2 \\
& NuSTAR     & 80302608002 & 2017-10-19 & 78.7/78.9 \\
& NuSTAR     & 90102001002 & 2015-04-02 & 31.1/31.2 \\
& NuSTAR     & 90102001004 & 2015-04-19 & 27.7/27.8 \\
& NuSTAR     & 90102001006 & 2015-05-28 & 30.4/30.3 \\
& Chandra    &        1031 & 2000-10-19 & 39.2 \\
& Chandra    &        4028 & 2003-02-03 & 39.7 \\
& Chandra    &        4029 & 2003-02-11 & 38.6 \\
& Chandra    &        4030 & 2003-02-20 & 36.0 \\
& Chandra    &       11851 & 2010-07-11 & 4.8 \\
& Chandra    &       13947 & 2012-08-24 & 177.7 \\
& Chandra    &       13948 & 2012-08-27 & 173.8 \\
& Chandra    &       13949 & 2012-08-23 & 40.0 \\
& XMM-Newton &  0081340201 & 2001-06-07 & 21.4/21.4/17.2 \\
& XMM-Newton &  0770580401 & 2015-04-25 & 22.6/22.6/19.0 \\
& XMM-Newton &  0770580501 & 2015-05-28 & 24.9/24.9/20.9 \\
& Suzaku     &   706037010 & 2011-04-27 & 197.5/75.2 \\
\hline
NGC 4922S/NGC 4922N
& NuSTAR     & 60101074002 & 2015-11-17 & 19.9/19.7 \\
& Chandra    &        4775 & 2004-11-02 & 3.8 \\
& Chandra    &       15065 & 2013-11-02 & 14.9 \\
& Chandra    &       18201 & 2016-03-06 & 5.8 \\
\hline
IC 860 
& NuSTAR     & 60301024002 & 2018-02-01 & 71.0/72.1 \\
& Chandra    &       10400 & 2009-03-24 & 19.2 \\
\hline
IRAS 13120--5453 
& NuSTAR     & 60001091002 & 2013-02-25 & 26.2/26.2 \\
& Chandra    &        7809 & 2006-12-01 & 14.7 \\
& XMM-Newton &  0693520201 & 2013-02-20 & 125.5/125.9/111.3 \\
\hline
NGC 5104 
& NuSTAR     & 60370003002 & 2018-01-12 & 20.4/20.4 \\
& Swift/XRT  &36966(001--) & 2007-07-24 & 8.2 \\
&  & \ \ \ \ \ \ \ (--004) & 2008-01-05 &\\
\hline
MCG--03-34-064 
& NuSTAR     & 60101020002 & 2016-01-17 & 76.6/78.1 \\
& Chandra    &        7373 & 2006-07-31 & 7.1 \\
& XMM-Newton &  0206580101 & 2005-01-24 & 44.1/44.1/38.3 \\
& XMM-Newton &  0506340101 & 2008-01-24 & 88.4/88.4/77.0 \\
& XMM-Newton &  0763220201 & 2016-01-17 & 137.0/139.6/123.3 \\
& Suzaku     &   704022010 & 2009-07-01 & (u; cont) \\
& Suzaku     &   707027010 & 2013-01-04 & (u; cont) \\
\hline
NGC 5135
& NuSTAR     & 60001153002 & 2015-01-14 & 33.4/32.5 \\
& Chandra    &        2187 & 2001-09-04 & 29.3 \\
\hline
Mrk 266B/Mrk 266A
& NuSTAR     & 60465005002 & 2019-02-08 & 29.7/30.1 \\
& Chandra    &        2044 & 2001-11-02 & 19.7 \\
\hline
Mrk 273 
& NuSTAR     & 60002028002 & 2013-11-04 & 69.2/69.6 \\
& Chandra    &       18177 & 2016-09-06 & 59.9 \\
& Chandra    &       18178 & 2017-02-16 & 30.4 \\
& Chandra    &       19783 & 2017-02-18 & 31.8 \\
& Chandra    &       19784 & 2017-02-26 & 33.6 \\
& Chandra    &       20009 & 2017-02-14 & 34.3 \\
& Chandra    &         809 & 2000-04-19 & (u; time-v) \\
& XMM-Newton &  0722610201 & 2013-11-04 & 21.2/21.2/17.5 \\
& XMM-Newton &  0101640401 & 2002-05-07 & (u; time-v) \\
& XMM-Newton &  0651360301 & 2010-05-13 & (u; time-v) \\
& XMM-Newton &  0651360501 & 2010-05-15 & (u; time-v) \\
& XMM-Newton &  0651360601 & 2010-05-17 & (u; time-v) \\
& XMM-Newton &  0651360701 & 2010-06-26 & (u; time-v) \\
& Suzaku     &   701050010 & 2006-07-07 & 79.9/77.4 \\
\hline
IRAS F14348--1447 
& NuSTAR     & 60374004002 & 2018-01-27 & 21.1/21.0 \\
& Chandra    &        6861 & 2006-03-12 & 14.7 \\
& XMM-Newton &  0081341401 & 2002-07-29 & 21.4/21.4/17.9 \\
\hline
IRAS F14378--3651 
& NuSTAR     & 60001092002 & 2013-02-28 & 24.5/24.4 \\
& Chandra    &        7889 & 2007-06-25 & 13.9 \\
\hline
IC 4518A/IC 4518B 
& NuSTAR     & 60061260002 & 2013-08-02 & 7.8/7.8 \\
& XMM-Newton &  0401790901 & 2006-08-07 & 11.5/11.5/8.8 \\
& XMM-Newton &  0406410101 & 2006-08-15 & 24.4/--/21.2 \\
& Suzaku     &   706012010 & 2012-02-02 & 64.4/53.7 \\
\hline
IRAS F15250+3608 
& NuSTAR     & 60374009002 & 2018-01-17 & 18.0/18.3 \\
& Chandra    &        4112 & 2003-08-27 & 9.8 \\
& XMM-Newton &  0081341101 & 2002-02-22 & 19.2/19.2/15.3 \\
\hline
Arp 220W/Arp 220E
& NuSTAR     & 60002026002 & 2013-08-13 & 66.1/66.0 \\
& Chandra    &       16092 & 2014-04-30 & 169.2 \\
& Chandra    &       16093 & 2014-05-07 & 66.7 \\
& Chandra    &         869 & 2000-06-24 & (u; time-v) \\
& XMM-Newton &  0101640801 & 2002-08-11 & 13.5/13.5/10.8 \\
& XMM-Newton &  0101640901 & 2003-01-15 & 14.4/14.4/8.3 \\
& XMM-Newton &  0722610301 & 2013-08-13 & 34.3/34.3/29.5 \\
& XMM-Newton &  0205510201 & 2005-01-14 & (u; highbg) \\
& XMM-Newton &  0205510401 & 2005-02-19 & (u; highbg) \\
& XMM-Newton &  0205510501 & 2005-02-27 & (u; highbg) \\
& Suzaku     &   700006010 & 2006-01-07 & 90.8/87.4 \\
\hline
NGC 6240S/NGC 6240N
& NuSTAR     & 60002040002 & 2015-03-18 & 30.9/30.8 \\
& Chandra    &       12713 & 2011-05-31 & 145.4 \\
& Chandra    &        1590 & 2001-07-29 & 36.7 \\
\hline
NGC 6285/NGC 6286 
& NuSTAR     & 60101075002 & 2015-05-29 & 17.2/17.2 \\
& Chandra    &       10566 & 2009-09-18 & 14.0 \\
& XMM-Newton &  0203390701 & 2004-02-12 & 7.7/7.7/5.7 \\
& XMM-Newton &  0203391201 & 2004-03-25 & (u; highbg) \\
\hline
IRAS F17138--1017 
& NuSTAR     & 60101076002 & 2015-06-26 & 24.8/25.2 \\
& Chandra    &       15063 & 2013-07-12 & 14.8 \\
\hline
IRAS F18293--3413 
& NuSTAR     & 60101077002 & 2016-02-20 & 21.2/21.0 \\
& Chandra    &        7815 & 2007-02-25 & 14.0 \\
& Chandra    &       21379 & 2019-08-08 & 79.0 \\
& XMM-Newton &  0670300701 & 2012-03-16 & 21.3/21.3/17.7 \\
\hline
IRAS F19297--0406 
& NuSTAR     & 60374007002 & 2018-03-03 & 20.0/20.0 \\
& Chandra    &        7890 & 2007-06-18 & 16.4 \\
\hline
NGC 6907/NGC 6908 
& NuSTAR     & 60370001002 & 2018-05-13 & 22.0/21.6 \\
& Swift/XRT  &33513(003--) & 2014-11-10 & 10.4 \\
&  & \ \ \ \ \ \ \ (--006) & 2014-11-19 &\\
\hline
NGC 6921/MCG+04-48-002 
& NuSTAR     & 60061300002 & 2013-05-18 & 19.2/19.2 \\
\tablebreak
& XMM-Newton &  0312192301 & 2006-04-23 & 13.7/13.7/10.8 \\
& Suzaku     &   702081010 & 2007-04-18 & (u; cont) \\
\hline
II Zw 096/IRAS F20550+1655 SE 
& NuSTAR     & 60374002002 & 2018-03-28 & 29.0/29.1 \\
& Chandra    &        7818 & 2007-09-10 & 14.6 \\
& XMM-Newton &  0670140101 & 2011-10-28 & 75.7/75.7/66.1 \\
\hline
ESO 286-19 
& NuSTAR     & 60101054002 & 2015-07-30 & 44.5/44.0 \\
& Chandra    &        2036 & 2001-10-31 & 44.9 \\
& XMM-Newton &  0081340401 & 2001-04-21 & 20.9/20.9/16.4 \\
\hline
NGC 7130 
& NuSTAR     & 60061347002 & 2014-08-17 & 21.2/21.1 \\
& NuSTAR     & 60261006002 & 2016-12-15 & 40.2/39.9 \\
& Chandra    &        2188 & 2001-10-23 & 38.6 \\
& Suzaku     &   703012010 & 2008-05-11 & 44.5/34.2 \\
\hline
NGC 7469
& NuSTAR     & 60101001002 & 2015-06-12 & 21.6/21.5 \\
& XMM-Newton &  0760350201 & 2015-06-12 &  --/--/62.6 \\
\hline
IC 5283 
& NuSTAR     & 60101001002 & 2015-06-12 & (u; cont) \\
& NuSTAR     & 60101001004 & 2015-11-24 & (u; cont) \\
& NuSTAR     & 60101001006 & 2015-12-15 & (u; cont) \\
& NuSTAR     & 60101001008 & 2015-12-22 & (u; cont) \\
& NuSTAR     & 60101001010 & 2015-12-25 & (u; cont) \\
& NuSTAR     & 60101001012 & 2015-12-27 & (u; cont) \\
& NuSTAR     & 60101001014 & 2015-12-28 & (u; cont) \\
& XMM-Newton &  0112170101 & 2000-12-26 & 17.8/17.8/12.3 \\
& XMM-Newton &  0112170301 & 2000-12-26 & 23.2/23.2/16.2 \\
& XMM-Newton &  0760350201 & 2015-06-12 &  --/--/62.6 \\
& Suzaku     &   703028010 & 2008-06-24 & (u; cont) \\
\hline
ESO 148-2 
& NuSTAR     & 60374006002 & 2018-03-06 & 27.1/26.8 \\
& NuSTAR     & 60374006004 & 2018-03-29 & 19.1/19.0 \\
& Chandra    &        2037 & 2001-09-30 & 49.3 \\
& XMM-Newton &  0081340301 & 2002-11-19 & 23.0/23.0/17.8 \\
& XMM-Newton &  0109463601 & 2001-10-17 & 5.1/5.3/-- \\
& XMM-Newton &  0079940101 & 2005-10-22 & 18.4/18.4/-- \\
& Swift/XRT  &35164(001--) & 2005-12-08 & 23.6 \\
&  & \ \ \ \ \ \ \ (--008) & 2006-06-27 &\\
\hline
NGC 7591 
& NuSTAR     & 60370002002 & 2018-01-09 & 19.3/19.2 \\
& Chandra    &       10264 & 2009-07-05 & 4.9 \\
\hline
NGC 7674/MCG+01-59-081 
& NuSTAR     & 60001151002 & 2014-09-30 & 52.0/51.9 \\
& XMM-Newton &  0200660101 & 2004-06-02 & 10.1/10.1/8.1 \\
& Suzaku     &   708023010 & 2013-12-08 & 52.2/45.7 \\
\hline
NGC 7679/NGC 7682 
& NuSTAR     & 60368002002 & 2017-10-06 & 22.4/21.7 \\
& XMM-Newton &  0301150501 & 2005-05-27 & 19.7/19.8/16.2 \\
& Swift/XRT  &    88108002 & 2017-10-06 & 1.9 \\
\enddata
\tablecomments{Columns: 
(1) object name; 
(2) satellite name; 
(3) observation ID; 
(4) observation date in YYYY/MM/DD; 
(5) net exposure (FPMA and FPMB for NuSTAR; ACIS-S or ACIS-I for Chandra; MOS1, MOS2, and pn for XMM-Newton; each of available XIS 0, 1, 2, 3 and HXD-PIN for Suzaku; and XRT for Swift).
The (u) means that the data are unused due to large time
 variability among the soft X-ray observations (time-v), too faint
 source or short exposure time (faint), contamination from brighter
 nearby sources (cont), or data with high background flaring throughout
 the observation (highbg).
}
\end{deluxetable*}
\clearpage

We find that NuSTAR detects hard X-ray emission from 37 galaxies and the
two inseparable dual-AGN systems in the 8--24~keV band (or in the
10--40~keV band for MCG+12-02-001 and IRAS F14348--1447) at $>$3$\sigma$
significance. These are considered to be good candidates of AGNs.
Table~\ref{T4_count-rate} summarizes the NuSTAR net count rates (NCR$^{\rm Nu}$) in the full (3--24~keV), soft
(3--8~keV), and hard (8--24~keV) bands, and the band ratio (the 
count-rate ratio between the 8--24~keV and 3--8~keV bands) for all
objects. The band ratio, BR$^{\rm Nu}$ (= NCR$^{\rm Nu}_{\rm 8-24}$/NCR$^{\rm Nu}_{\rm 3-8}$), is useful to identify obscured
AGNs with a threshold of BR$^{\rm Nu} > 1.7$, corresponding to an
effective photon index of $\Gamma_{\rm eff} \lesssim 0.6$
\citep{Lansbury2017}. Table~\ref{T4_count-rate} also lists the observed 
2--7~keV fluxes. 
These fluxes are calculated from the broadband spectral 
analysis described in Section~\ref{sub4-2_model}. For the sources mentioned
in Section~\ref{sub2-5_Xray-ref}, we calculate the fluxes
converted from the 
count rates of Chandra, XMM-Newton, and/or Swift/XRT by assuming a power-law 
photon index of 1.8.
For the undetected sources, 
we constrain the 3$\sigma$ upper limits by adopting the same conversion.

\subsection{X-ray Spectral Model}
\label{sub4-2_model}

We utilize \textsc{xspec} v12.10.1 \citep{Arnaud1996} for spectral
fitting. Galactic absorption is considered with the \textsf{phabs} model
whose hydrogen column density is fixed at the value of \citet{Willingale2013}.  
The solar
abundances of \citet{Anders1989} are assumed. The flux cross-calibration
uncertainties among different instruments are taken into account by
multiplying constant factors.
We set the constant values for NuSTAR/FPM and XMM-Newton/MOS to unity
as references, whose relative flux calibration is known to be accurate
within a few percent \citep{Madsen2015,Madsen2017}, whereas those of the
other instruments are allowed to vary between 0.8 and 1.2. 
If a source is not detected with NuSTAR in any bands, 
we adopt Chandra, 
XMM-Newton (MOS, pn), or Swift/XRT as the calibration reference
in this priority order.
We link that of Suzaku/HXD-PIN to 1.16 (1.18) times that of XIS-FI for
the data observed at the XIS (HXD) nominal pointing position, on the
basis of the calibration with the Crab Nebula.
The XIS energy range of 1.6--1.9~keV is ignored for NGC 7674 due to
the large calibration error. 

To obtain the highest quality spectrum from each instrument, we combine
the spectra of all observations unless they show large time variability
above $\sim$1~keV (listed in Table~\ref{T3_log} with the mark
``time-v'').\footnote{When time variability is found only at lower
energies, we use the combined spectra above $\sim$1--2~keV.}
The time
variability among the spectra of different instruments are
taken into account by multiplying a constant factor to the AGN
transmitted component (Section~\ref{subsub4-2-1_AGNmodel}).
For the sources whose NuSTAR total (source and background) counts are 
$<$200 counts or the S/N ratios in the 8--24~keV band
are $<$3$\sigma$, we employ the Cash statistics
\citep{Cash1979}\footnote{The Cash statistic is also adopted for IRAS
F08572+3915 where the detection significance in the NuSTAR 3--8~keV band 
is $<$3$\sigma$ level.}, whereas we use $\chi^2$ statistics for the 
other objects.

Excluding the 8 U/LIRG systems mentioned in
Section~\ref{sub2-5_Xray-ref}, we analyze the broadband spectra of the 31 
systems that are significantly ($>$3$\sigma$) detected with NuSTAR in
the 8--24~keV or 10--40~keV band, and those of NGC 235, which is
detected with Swift/BAT.
Among the 32 hard X-ray detected sources, 
we apply an AGN model (Section~\ref{subsub4-2-1_AGNmodel}) to 
30 sources that show hard X-ray power law components. 
Whereas, we apply a starburst-dominant model (Section~\ref{subsub4-2-2_SBmodel}) 
to the other
two sources, IC~1623A and NGC~3256, since their spectra show high energy
cutoff features around $\sim$10~keV and hence are likely 
dominated by X-ray binary (XRB) populations.
The latter model is also adopted for the soft X-ray spectra
of the 27 sources that are undetected with NuSTAR in the 8--24~keV
band. For 9 sources out of these 59 sources, we introduce additional
emission lines (Section~\ref{subsub4-2-3_lines}) on the AGN or starburst-dominant model to
improve the spectral fit.

\subsubsection{AGN Model}
\label{subsub4-2-1_AGNmodel}

For the 30 AGNs detected in the hard X-ray band (see above), we apply
the XCLUMPY model \citep{Tanimoto2019}, which is a Monte Carlo-based
spectral model from clumpy tori in AGNs. The torus geometry and
geometrical parameters are the same as those in the CLUMPY
\citep{Nenkova2008a,Nenkova2008b} model used for IR studies.  As
mentioned in Section~\ref{S1_intro}, many IR observations suggest that AGN tori
are not smooth but have dusty clumpy structure
\citep[e.g.,][]{Krolik1988,Wada2002,Honig2007}. Recent X-ray spectral
works also support the clumpy nature of AGN tori, as such models can
explain a large amount of unabsorbed reflection components 
often observed in obscured AGNs
\citep[e.g.,][]{LiuYang2014,Furui2016,Tanimoto2018,Tanimoto2019}.
Studies of the variability of the line-of-sight column density
using X-ray monitoring data also support the clumpy structure
\citep[e.g.,][]{Lamer2003,Puccetti2007,Risaliti2009,Risaliti2011,Maiolino2010,Markowitz2014,Laha2020}.
Thus, we regard this model as one of the most realistic models.

\clearpage
\startlongtable
\begin{deluxetable*}{lrcrrrrlc}
\label{T4_count-rate}
\tablecaption{X-ray Count Rates and NuSTAR Band Ratios in Our Sample}
\tabletypesize{\footnotesize}
\tablehead{
\colhead{Object} &
\colhead{$F^{\rm obs}_{2-7}$} &
\colhead{Facility} &
\colhead{NCR$^{\rm Nu}_{3-24}$} &
\colhead{NCR$^{\rm Nu}_{3-8}$} &
\colhead{NCR$^{\rm Nu}_{8-24}$} &
\colhead{BR$^{\rm Nu}$} &
\colhead{$>$3$\sigma$ Det.} &
\colhead{$N^{\rm Gal}_{\rm H}$}
}
\decimalcolnumbers
\startdata
NGC 34 & 1.7e-13 & C & 6.96 $\pm$ 0.49 & 3.04 $\pm$ 0.32 & 3.91 $\pm$ 0.37 & 1.29 $\pm$ 0.18 &  F S H & 2.32\\
MCG--02-01-052 & 8.3e-15 & C & 0.14 $\pm$ 0.20 & 0.20 $\pm$ 0.15 & $<$0.34\ \ \ \ \ & $<$1.70\ \ \ \ \ & \nodata & 3.51\\  
MCG--02-01-051 & 5.7e-14 & C & 0.20 $\pm$ 0.19 & 0.17 $\pm$ 0.14 & 0.03 $\pm$ 0.12 & 0.18 $\pm$ 0.72 & \nodata & 3.51\\  
ESO 350-38 & 9.7e-14 & C & 0.62 $\pm$ 0.20 & 0.58 $\pm$ 0.16 & 0.04 $\pm$ 0.13 & 0.07 $\pm$ 0.22 &  F S  & 2.52\\  
NGC 232 & 3.8e-14 & C & \nodata\ \ \ \ \ & \nodata\ \ \ \ \ & \nodata\ \ \ \ \ & \nodata\ \ \ \ \ & \nodata & 1.49\\
NGC 235$^{a}$ & 2.1e-12 & C & \nodata\ \ \ \ \ & \nodata\ \ \ \ \ & \nodata\ \ \ \ \ & \nodata\ \ \ \ \ & \nodata & 1.48\\
MCG+12-02-001 & 1.0e-13 & C & 1.37 $\pm$ 0.23 & 0.92 $\pm$ 0.17 & 0.46 $\pm$ 0.15 & 0.50 $\pm$ 0.19 &  F S (H)$^{b}$ & 36.1\\  
IC 1623A & 1.9e-13 & C & 2.88 $\pm$ 0.12 & 2.22 $\pm$ 0.09 & 0.66 $\pm$ 0.07 & 0.30 $\pm$ 0.03 &  F S H & 1.45\\
IC 1623B & (cont) & C & (cont)\ \ \ \ \ & (cont)\ \ \ \ \ & (cont)\ \ \ \ \ & \nodata\ \ \ \ \ & \nodata & 1.45\\
NGC 833 & 1.5e-13 & C & 11.87 $\pm$ 0.60 & 5.37 $\pm$ 0.41 & 6.42 $\pm$ 0.44 & 1.20 $\pm$ 0.12 &  F S H & 2.73\\
NGC 835 & 6.0e-13 & C & 19.24 $\pm$ 0.76 & 6.63 $\pm$ 0.45 & 12.50 $\pm$ 0.61 & 1.89 $\pm$ 0.16 &  F S H & 2.74\\
NGC 838 & 5.9e-14 & C & 1.16 $\pm$ 0.26 & 0.66 $\pm$ 0.18 & 0.45 $\pm$ 0.18 & 0.68 $\pm$ 0.33 &  F S  & 2.78\\  
NGC 839 & 7.3e-14 & C & 0.37 $\pm$ 0.21 & 0.27 $\pm$ 0.14 & 0.08 $\pm$ 0.15 & 0.30 $\pm$ 0.58 & \nodata & 2.80\\  
NGC 1068 & 2.5e-12 & XPN & 132.40 $\pm$ 0.74 & 79.21 $\pm$ 0.57 & 52.86 $\pm$ 0.47 & 0.67 $\pm$ 0.01 &  F S H & 3.22\\
UGC 2608 & 1.5e-13 & X & 9.55 $\pm$ 0.58 & 3.15 $\pm$ 0.38 & 6.41 $\pm$ 0.44 & 2.03 $\pm$ 0.28 &  F S H & 20.3\\
UGC 2612 & $<$1.5e-14 & C & $<$0.31\ \ \ \ \ & $<$0.42\ \ \ \ \ & $<$0.14\ \ \ \ \ & \nodata\ \ \ \ \ & \nodata & 20.0\\  
NGC 1275 & 3.4e-11 & X & 905.70 $\pm$ 3.58 & 637.10 $\pm$ 3.06 & 265.80 $\pm$ 1.84 & 0.42 $\pm$ 0.01 &  F S H & 20.7\\
NGC 1365 & 7.0e-12 & X & 445.93 $\pm$ 1.80 & 246.43 $\pm$ 1.34 & 198.18 $\pm$ 1.20 & 0.80 $\pm$ 0.01 &  F S H & 1.39\\
ESO 203-1 & $<$1.1e-14 & C & $<$0.05\ \ \ \ \ & $<$0.11\ \ \ \ \ & $<$0.16\ \ \ \ \ & \nodata\ \ \ \ \ & \nodata & 1.42\\  
CGCG 468-002W & 5.5e-12 & XRT & 214.10 $\pm$ 2.68 & 123.60 $\pm$ 2.03 & 89.91 $\pm$ 1.74 & 0.73 $\pm$ 0.02 &  F S H & 31.1\\
CGCG 468-002E & (cont) & XRT & (cont)\ \ \ \ \ & (cont)\ \ \ \ \ & (cont)\ \ \ \ \ & \nodata\ \ \ \ \ & \nodata & 31.1\\
IRAS F05189--2524 & 1.7e-12 & C & 72.86 $\pm$ 0.45 & 46.06 $\pm$ 0.35 & 26.54 $\pm$ 0.27 & 0.58 $\pm$ 0.01 &  F S H & 1.80\\
IRAS F06076--2139 & 6.3e-14 & C & 3.46 $\pm$ 0.39 & 1.46 $\pm$ 0.25 & 1.99 $\pm$ 0.29 & 1.36 $\pm$ 0.31 &  F S H & 9.56\\
2MASS 06094601--2140312 & $<$1.1e-14 & C & (cont)\ \ \ \ \ & (cont)\ \ \ \ \ & (cont)\ \ \ \ \ & \nodata\ \ \ \ \ & \nodata & 9.56\\
NGC 2623 & 5.9e-14 & C & 1.98 $\pm$ 0.26 & 1.13 $\pm$ 0.18 & 0.84 $\pm$ 0.19 & 0.74 $\pm$ 0.21 &  F S H & 3.54\\
ESO 060-IG016 West & $<$8.5e-15 & C & (cont)\ \ \ \ \ & (cont)\ \ \ \ \ & (cont)\ \ \ \ \ & \nodata\ \ \ \ \ & \nodata & 7.26\\
ESO 060-IG016 East & 1.1e-13 & C & 3.04 $\pm$ 0.26 & 1.68 $\pm$ 0.19 & 1.34 $\pm$ 0.18 & 0.80 $\pm$ 0.14 &  F S H & 7.24\\
IRAS F08572+3915 & 1.2e-14 & X & 0.28 $\pm$ 0.06 & 0.11 $\pm$ 0.04 & 0.17 $\pm$ 0.04 & 1.55 $\pm$ 0.67 &  F H & 2.17\\  
UGC 5101 & 8.4e-14 & C & 15.02 $\pm$ 0.73 & 3.08 $\pm$ 0.37 & 11.85 $\pm$ 0.63 & 3.85 $\pm$ 0.51 &  F S H & 3.33\\
MCG+08-18-012 & $<$1.2e-14 & C & 0.06 $\pm$ 0.18 & 0.09 $\pm$ 0.13 & $<$0.36\ \ \ \ \ & $<$4.00\ \ \ \ \ & \nodata & 1.74\\  
MCG+08-18-013 & 3.7e-14 & C & 0.27 $\pm$ 0.18 & 0.33 $\pm$ 0.14 & $<$0.27\ \ \ \ \ & $<$0.82\ \ \ \ \ & \nodata & 1.75\\  
MCG--01-26-013 & 5.9e-15 & C & $<$0.28\ \ \ \ \ & $<$0.20\ \ \ \ \ & $<$0.24\ \ \ \ \ & \nodata \ \ \ \ \ & \nodata & 3.90\\  
NGC 3110 & 9.4e-14 & C & 0.43 $\pm$ 0.21 & 0.22 $\pm$ 0.14 & 0.22 $\pm$ 0.15 & 1.00 $\pm$ 0.93 & \nodata & 3.89\\  
ESO 374-IG032 & 1.2e-14 & C & 0.24 $\pm$ 0.12 & 0.21 $\pm$ 0.09 & 0.04 $\pm$ 0.08 & 0.19 $\pm$ 0.39 & \nodata & 11.9\\  
NGC 3256 & 4.5e-13 & C & 6.36 $\pm$ 0.17 & 4.92 $\pm$ 0.14 & 1.42 $\pm$ 0.09 & 0.29 $\pm$ 0.02 &  F S H & 13.2\\
IRAS F10565+2448 & 3.8e-14 & C & 0.23 $\pm$ 0.19 & 0.41 $\pm$ 0.15 & $<$0.15\ \ \ \ \ & $<$0.37\ \ \ \ \ & \nodata & 1.12\\  
NGC 3690 West & 8.7e-13 & C & 26.03 $\pm$ 0.46 & 11.90 $\pm$ 0.31 & 14.08 $\pm$ 0.34 & 1.18 $\pm$ 0.04 &  F S H & 0.936\\
NGC 3690 East & (cont) & C & (cont)\ \ \ \ \ & (cont)\ \ \ \ \ & (cont)\ \ \ \ \ & \nodata\ \ \ \ \ & \nodata & 0.935\\
ESO 440-58 & 1.7e-14 & C & 0.07 $\pm$ 0.15 & 0.07 $\pm$ 0.11 & 0.01 $\pm$ 0.10 & 0.14 $\pm$ 1.45 & \nodata & 7.10\\  
MCG--05-29-017 & 1.3e-14 & C & (cont)\ \ \ \ \ & (cont)\ \ \ \ \ & (cont)\ \ \ \ \ & \nodata\ \ \ \ \ & \nodata & 7.11\\
\tablebreak
IRAS F12112+0305 & 1.3e-14 & X & 0.08 $\pm$ 0.19 & 0.22 $\pm$ 0.15 & $<$0.22\ \ \ \ \ & $<$1.00\ \ \ \ \ & \nodata & 1.93\\  
NGC 4418 & 1.4e-14 & C & 0.31 $\pm$ 0.12 & 0.19 $\pm$ 0.09 & 0.12 $\pm$ 0.09 & 0.63 $\pm$ 0.56 & \nodata & 2.09\\  
MCG+00-32-013 & (cont) & C & (cont)\ \ \ \ \ & (cont)\ \ \ \ \ & (cont)\ \ \ \ \ & \nodata\ \ \ \ \ & \nodata & 2.09\\
Mrk 231 & 5.3e-13 & C & 22.50 $\pm$ 0.23 & 11.43 $\pm$ 0.17 & 11.01 $\pm$ 0.16 & 0.96 $\pm$ 0.02 &  F S H & 0.992\\
NGC 4922S & 4.1e-15 & C & (cont)\ \ \ \ \ & (cont)\ \ \ \ \ & (cont)\ \ \ \ \ & \nodata\ \ \ \ \ & \nodata & 0.966\\
NGC 4922N & 8.0e-14 & C & 3.49 $\pm$ 0.42 & 1.59 $\pm$ 0.28 & 1.84 $\pm$ 0.30 & 1.16 $\pm$ 0.28 &  F S H & 0.966\\
IC 860 & $<$2.2e-14 & C & $<$0.27\ \ \ \ \ & $<$0.12\ \ \ \ \ & 0.03 $\pm$ 0.07 & $>$0.25\ \ \ \ \ & \nodata & 1.05\\  
IRAS 13120--5453 & 1.7e-13 & C & 5.55 $\pm$ 0.41 & 2.92 $\pm$ 0.29 & 2.64 $\pm$ 0.29 & 0.90 $\pm$ 0.13 &  F S H & 35.1\\
NGC 5104 & 2.6e-14 & XRT & 0.50 $\pm$ 0.21 & 0.15 $\pm$ 0.13 & 0.33 $\pm$ 0.16 & 2.20 $\pm$ 2.18 & \nodata & 1.95\\  
MCG--03-34-064 & 1.6e-12 & C & 86.39 $\pm$ 0.77 & 29.47 $\pm$ 0.45 & 56.78 $\pm$ 0.62 & 1.93 $\pm$ 0.04 &  F S H & 6.44\\
NGC 5135 & 2.6e-13 & C & 13.48 $\pm$ 0.51 & 5.27 $\pm$ 0.33 & 8.19 $\pm$ 0.39 & 1.55 $\pm$ 0.12 &  F S H & 6.02\\
Mrk 266B/Mrk 266A$^{c}$ & 1.7e-13 & C & 8.25 $\pm$ 0.44 & 4.29 $\pm$ 0.31 & 3.93 $\pm$ 0.31 & 0.92 $\pm$ 0.10 &  F S H & 1.75\\
Mrk 273 & 2.0e-13 & C & 26.62 $\pm$ 0.47 & 9.48 $\pm$ 0.29 & 17.01 $\pm$ 0.37 & 1.79 $\pm$ 0.07 &  F S H & 0.909\\
IRAS F14348--1447 & 2.5e-14 & C & 1.25 $\pm$ 0.32 & 0.79 $\pm$ 0.23 & 0.46 $\pm$ 0.22 & 0.58 $\pm$ 0.33 &  F S (H)$^{b}$ & 10.8\\  
IRAS F14378--3651 & 1.5e-14 & C & 0.28 $\pm$ 0.18 & 0.26 $\pm$ 0.13 & 0.02 $\pm$ 0.12 & 0.08 $\pm$ 0.46 & \nodata & 8.22\\  
IC 4518A & 1.0e-12 & X & 143.00 $\pm$ 3.10 & 66.47 $\pm$ 2.11 & 76.07 $\pm$ 2.26 & 1.14 $\pm$ 0.05 &  F S H & 13.1\\
IC 4518B & 8.4e-15 & XPN & (cont)\ \ \ \ \ & (cont)\ \ \ \ \ & (cont)\ \ \ \ \ & \nodata\ \ \ \ \ & \nodata & 13.0\\
IRAS F15250+3608 & 8.7e-15 & X & 0.03 $\pm$ 0.19 & 0.09 $\pm$ 0.13 & $<$0.33\ \ \ \ \ & $<$3.67\ \ \ \ \ & \nodata & 1.57\\  
Arp 220W & 7.1e-14 & C & 2.03 $\pm$ 0.19 & 1.54 $\pm$ 0.15 & 0.49 $\pm$ 0.12 & 0.32 $\pm$ 0.08 &  F S H & 4.58\\
Arp 220E & (cont) & C & (cont)\ \ \ \ \ & (cont)\ \ \ \ \ & (cont)\ \ \ \ \ & \nodata\ \ \ \ \ & \nodata & 4.58\\
NGC 6240S/NGC 6240N$^{c}$ & 1.0e-12 & C & 108.00 $\pm$ 1.36 & 26.96 $\pm$ 0.69 & 80.72 $\pm$ 1.17 & 2.99 $\pm$ 0.09 &  F S H & 6.26\\
NGC 6285 & 1.4e-14 & C & 0.28 $\pm$ 0.20 & 0.42 $\pm$ 0.16 & $<$0.22\ \ \ \ \ & $<$0.52\ \ \ \ \ & \nodata & 1.87\\  
NGC 6286 & 5.0e-14 & C & 3.31 $\pm$ 0.43 & 0.93 $\pm$ 0.27 & 2.38 $\pm$ 0.33 & 2.56 $\pm$ 0.82 &  F S H & 1.87\\
IRAS F17138--1017 & 8.8e-14 & C & 2.72 $\pm$ 0.34 & 1.65 $\pm$ 0.25 & 1.02 $\pm$ 0.23 & 0.62 $\pm$ 0.17 &  F S H & 31.0\\
IRAS F18293--3413 & 1.2e-13 & C & 1.76 $\pm$ 0.27 & 1.33 $\pm$ 0.22 & 0.44 $\pm$ 0.16 & 0.33 $\pm$ 0.13 &  F S  & 14.3\\  
IRAS F19297--0406 & 1.2e-14 & C & 0.38 $\pm$ 0.21 & 0.19 $\pm$ 0.14 & 0.20 $\pm$ 0.15 & 1.05 $\pm$ 1.11 & \nodata & 28.2\\  
NGC 6907 & 5.5e-14 & XRT & 1.06 $\pm$ 0.24 & 0.62 $\pm$ 0.17 & 0.42 $\pm$ 0.16 & 0.68 $\pm$ 0.32 &  F S  & 6.59\\  
NGC 6908 & (cont) & XRT & (cont)\ \ \ \ \ & (cont)\ \ \ \ \ & (cont)\ \ \ \ \ & \nodata\ \ \ \ \ & \nodata & 6.59\\
NGC 6921 & 1.8e-13 & X & 22.27 $\pm$ 0.91 & 4.47 $\pm$ 0.44 & 17.75 $\pm$ 0.80 & 3.97 $\pm$ 0.43 &  F S H & 34.4\\
MCG+04-48-002 & 1.1e-12 & X & 34.81 $\pm$ 1.00 & 10.00 $\pm$ 0.55 & 24.72 $\pm$ 0.84 & 2.47 $\pm$ 0.16 &  F S H & 34.5\\
II Zw 096 & 8.7e-15 & C & (cont)\ \ \ \ \ & (cont)\ \ \ \ \ & (cont)\ \ \ \ \ & \nodata\ \ \ \ \ & \nodata & 9.16\\
IRAS F20550+1655 SE & 5.8e-14 & C & 1.22 $\pm$ 0.21 & 0.83 $\pm$ 0.16 & 0.39 $\pm$ 0.14 & 0.47 $\pm$ 0.19 &  F S  & 9.16\\  
ESO 286-19 & 4.3e-14 & C & 0.64 $\pm$ 0.14 & 0.42 $\pm$ 0.11 & 0.20 $\pm$ 0.10 & 0.48 $\pm$ 0.27 &  F S  & 3.76\\  
NGC 7130 & 1.5e-13 & C & 10.75 $\pm$ 0.34 & 2.89 $\pm$ 0.19 & 7.86 $\pm$ 0.28 & 2.72 $\pm$ 0.20 &  F S H & 2.11\\
NGC 7469 & 1.9e-11 & XPN & 588.90 $\pm$ 3.75 & 365.60 $\pm$ 2.95 & 221.60 $\pm$ 2.30 & 0.61 $\pm$ 0.01 &  F S H & 5.39\\
IC 5283 & $<$1.1e-14 & X & (cont)\ \ \ \ \ & (cont)\ \ \ \ \ & (cont)\ \ \ \ \ & \nodata\ \ \ \ \ & \nodata & 5.39\\
ESO 148-2 & 1.3e-13 & C & 0.86 $\pm$ 0.15 & 0.55 $\pm$ 0.11 & 0.31 $\pm$ 0.10 & 0.56 $\pm$ 0.21 &  F S H & 1.68\\  
NGC 7591 & $<$4.4e-14 & C & $<$0.53\ \ \ \ \ & $<$0.33\ \ \ \ \ & 0.03 $\pm$ 0.14 & $>$0.09\ \ \ \ \ & \nodata & 7.52\\  
NGC 7674 & 3.5e-13 & X & 24.22 $\pm$ 0.52 & 10.96 $\pm$ 0.35 & 13.15 $\pm$ 0.38 & 1.20 $\pm$ 0.05 &  F S H & 5.20\\
MCG+01-59-081 & $<$1.6e-14 & X & (cont)\ \ \ \ \ & (cont)\ \ \ \ \ & (cont)\ \ \ \ \ & \nodata\ \ \ \ \ & \nodata & 5.19\\
\tablebreak
NGC 7679 & 4.2e-13 & X & 48.01 $\pm$ 1.08 & 28.53 $\pm$ 0.83 & 19.34 $\pm$ 0.70 & 0.68 $\pm$ 0.03 &  F S H & 6.25\\
NGC 7682 & 1.5e-13 & X & 12.07 $\pm$ 0.60 & 5.09 $\pm$ 0.39 & 6.88 $\pm$ 0.45 & 1.35 $\pm$ 0.14 &  F S H & 6.40\\
\enddata
\tablecomments{Columns: 
(1) object name; 
(2--3) observed 2--7~keV flux corrected for Galactic absorption in units
of erg s$^{-1}$ cm$^{-2}$, and the facility of the soft X-ray data. 
Fluxes are calculated from the best-fitting spectral 
model described in Section~\ref{sub4-2_model}. For the sources 
mentioned in Section~\ref{sub2-5_Xray-ref} and the 
undetected sources (given as upper limits), they are
converted from the X-ray count rates with the HEASARC tool WebPIMMS (v4.11a;
\url{https://heasarc.gsfc.nasa.gov/cgi-bin/Tools/w3pimms/w3pimms.pl})
assuming a power-law of $\Gamma$ = 1.8 with Galactic absorption. 
The adopted soft X-ray observations of Chandra (C), XMM-Newton/MOS (X), pn (XPN),
and Swift/XRT (XRT) are listed in Column (3).
(4--6) background-subtracted net count rates of combined NuSTAR observations in the full (3--24~keV), soft (3--8~keV) and hard (8--24~keV) bands in units of 10$^{-3}$ cts s$^{-1}$;
(7) NuSTAR Band ratio (i.e., the ratio of 8--24~keV and 3--8~keV count rates);
(8) NuSTAR $>$3$\sigma$ detection in the full (F), soft (S), and hard (H) bands;
(9) hydrogen column density of Galactic absorption in units of 10$^{20}$ cm$^{-2}$ \citep{Willingale2013}.
The (cont) means that the value is smaller than that of the interacting galaxy, and is difficult to estimate due to its contamination.
The error is 1$\sigma$, and the upper and lower limits are 3$\sigma$ levels.
}
\tablenotetext{a}{NuSTAR has not observed NGC 235, but Swift/BAT detects it in the 14--195~keV band at a 25.51$\sigma$ level \citep{Oh2018}.}
\tablenotetext{b}{For MCG+12-02-001 and IRAS F14348--1447, the NuSTAR
 net count rates in the 8--24~keV band are 2.98$\sigma$ and 2.08$\sigma$ levels, while
 those in the 10--40~keV band are 3.37$\sigma$ and 3.06$\sigma$ levels, respectively.}
\tablenotetext{c}{The system hosting two bright AGNs whose emissions
 could not be spatially resolved with NuSTAR.}
\end{deluxetable*}

The spectral model in the \textsc{xspec} terminology is represented as: 
\begin{align}
&\mathsf{const1 * phabs * zphabs *} \notag\\
&\mathsf{(const2 *zphabs * cabs * zcutoffpl + const3 * zcutoffpl} \notag\\
&\mathsf{+\ atable\{xclumpy\_R.fits\} + atable\{xclumpy\_L.fits\}} \notag\\
&\mathsf{+\ apec[\times 1\rm{-}3])}.
\end{align}
In the fist term, 
\textsf{const1} and \textsf{phabs} represent the instrumental 
cross-normalization factor and Galactic absorption, respectively. The
\textsf{zphabs} accounts for an absorption of the host galaxy when 
required (only for NGC 3690W, UGC 5101, and IRAS 13120--5453). The first
term represents the direct component modeled by an absorbed power-law
with an exponential cutoff, which is fixed at 300~keV
\citep[e.g.,][]{Dadina2008}. The photon index ($\Gamma_{\rm AGN}$) is
set to be a free parameter between 1.5 and 2.5. When it cannot be well
constrained, we fix it at 1.8 as a typical value in AGNs
\citep[e.g.,][]{Ueda2014,Ricci2017dApJS}. We consider Compton scattering
effect out of the line of sight (\textsf{cabs}), whose hydrogen column
density ($N_{\rm H}$) is tied to that of photometric absorption
(\textsf{zphabs}).  The \textsf{const2} factor ($C^{\rm Time}$) is a
constant to take account of time variability among different
instruments. This is allowed to vary within 0.1--10 (i.e., a factor of
$<$10 in variability) in order to avoid unrealistic solutions.

The second term, an unabsorbed cutoff power-law, represents 
the scattering component from the AGN 
and/or the emission from XRBs in the host galaxy.
The normalization is linked to that of the direct
component, while the photon index ($\Gamma_{\rm bin}$) is allowed to
vary between 1.5 and 3.0.\footnote{This is because in reality the
``scattered'' component may contain many soft X-ray emission lines from
a photoionized plasma that are hard to be resolved with CCD spectra,
making the apparent slope (i.e., that obtained with a simple power law
model) steeper than that of the direct component.} The \textsf{const3}
is multiplied, which gives a fraction of the unabsorbed emission ($f_{\rm
bin}$) relative to the transmitted AGN component at 1~keV, 
containing both the XRB and scattered AGN components. 
Its range is restricted within 0\%--20\%.\footnote{The 
typical scattering fraction is $\lesssim$5\% for the obscured AGNs with 
$N_{\rm H}^{\rm LoS} \geq 10^{22}$ cm$^{-2}$ from the 70-month Swift/BAT
catalog \citep{Gupta2021}, while
the $f_{\rm bin}$ for AGNs in stage-D mergers are $\sim$10\%, indicating
that the component of XRBs may significantly affect the values
(see Section~\ref{sub6-3_X-SFR}).}
For the three stage-D mergers whose \textsf{const3} values cannot be
constrained (IRAS F08572+3915, Arp~220W, and IRAS F17138--1017), we
adopt 10\% as a typical value obtained from the other stage-D mergers.

The third and fourth terms are the table models of XCLUMPY, providing
the reflection continuum (xclumpy\_R.fits) and fluorescence emission
lines (xclumpy\_L.fits), respectively.
The free parameters are the hydrogen column density along the equatorial
plane ($N_{\rm H}^{\rm Equ}$: 10$^{23}$--10$^{26}$ cm$^{-2}$), torus
angular width ($\sigma$: 10\degr--70\degr), and inclination of the
observer ($i$: 20\degr--87\degr).\footnote{The inner
($r_{\rm in}$ = 0.05~pc) and outer radii ($r_{\rm out}$ = 1.00~pc),
the radius of each clump ($R_{\rm clump}$ = 0.002~pc), 
the number of the clumps along the equatorial plane 
($N^{\rm Equ}_{\rm clump}$ = 10), and
the index of radial clumpy distribution ($q$ = 0.5) are fixed.}  When
these values are not well constrained, we first fix the inclination to
either 30\degr, 60\degr, or 80\degr\ that gives the 
smallest $\chi^2$ or C value.
Furthermore, if $\sigma$ is not well determined, we fix it to be
20\degr, a typical value obtained from a local AGN sample
\citep{Ogawa2021}.\footnote{The averaged value of $\sigma$
that we can constrain, $\sim$22\degr,
is consistent with the adopted typical value (20\degr). 
We note that, even if different values of $\sigma$, 10\degr\ and 30\degr, are adopted,
the main results are almost unchanged except for the torus covering
factors. Thus, the torus covering factors are not calculated when
$\sigma$ is fixed (Section~\ref{subsub5-3-4_Edd}).}
We link the line-of-sight column density ($N_{\rm
H}^{\rm LoS}$) in the first term to the value converted from these torus
parameters by following Equation (3) in \citet{Tanimoto2019}.  
When $N_{\rm H}^{\rm LoS}$ is found to be variable
among the observation epochs of different satellites, it is set 
to be a free parameter.  
The normalizations and photon indices are fixed to those of the direct component.
The fifth term is optically-thin thermal emission in the host galaxy
modeled by the \textsf{apec} code \citep{Smith2001}. Whenever
necessary, we consider two or three components with different temperatures.

The best-fit spectra of the 30 AGNs are shown in Figures~\ref{C1-F}--\ref{C5-F}. The
best-fit parameters, observed X-ray fluxes, observed
X-ray luminosities, and the equivalent width (EW) of the Fe K$\alpha$ line are
summarized in Table~\ref{T5_bestpara-AGN}, including the the results of UGC~2608, NGC~5135
\citep{Yamada2020}, and NGC~7469 \citep{Ogawa2021} analyzed with the
XCLUMPY model. Thus, we finally obtain X-ray spectral results of 40
AGNs, consisting of 33 AGNs analyzed with XCLUMPY and 7 AGNs for which
we refer to the literature (NGC~1068, NGC~1275, NGC~1365,
Mrk~266B/Mrk~266A, and NGC~6240S/NGC~6240N). Comparison of our results
with previous works for individual sources is summarized in Appendix~\ref{Appendix-B}.

\subsubsection{Starburst-dominant Model}
\label{subsub4-2-2_SBmodel}

We apply a starburst-dominant model to the
two sources detected with NuSTAR in the 8--24~keV band (IC~1623A and
NGC~3256), and to the 27 sources not detected with NuSTAR in this band.
Hereafter, we call these two and 27 sources as ``starburst-dominant sources''
and ``hard X-ray undetected sources'', respectively.
The two starburst-dominant sources show the smallest values of the band ratio,
BR$^{\rm Nu} \approx$ 0.3, among the hard X-ray detected sources.  This
means that their spectra are steeper than those of AGNs in the 3--24~keV
range (see also Section~\ref{sub5-1_Xcolor}). Actually, their broadband spectra cannot
be well reproduced with the AGN model because of the high energy cutoff
features around 10~keV.\footnote{Although Arp~220W also shows BR$^{\rm
Nu} \approx$ 0.3, it does not show such a cutoff feature at $\sim$10
keV. If we fit the spectrum with \textsf{zcutoffpl}, we obtain $E_{\rm
cut} > 92$~keV (see Appendix~\ref{Appendix-B}).}
Moreover, there are little signatures of AGNs in these starburst-dominant and
hard X-ray undetected sources based on multiwavelength observations as
discussed in Appendix~\ref{Appendix-B}.\footnote{Although IC~1623B might have an AGN,
its X-ray emission is much weaker than that of IC~1623A
\citep[e.g.,][]{Grimes2006}.}

The starburst-dominant model is described as
\begin{align}
&\mathsf{const1 * phabs * zphabs * (zcutoffpl + apec[\times 1\rm{-}2])}.
\end{align}
The cross-normalization (\textsf{const1}) and Galactic absorption (\textsf{phabs}) 
are the same as those in the AGN model. 
The absorption by the host galaxy (\textsf{zphabs}) is adopted only for
NGC~3256.  The first term represents emission from 
XRBs in the host galaxy. We note that this component would
include a scattered component from the AGN if present. The photon index
($\Gamma_{\rm bin}$) is allowed to vary between 1.5 and 3.0.
The second term is optically-thin thermal emission modeled by \textsf{apec}.

The best-fit spectra are shown in Figures~\ref{C6-F}--\ref{C10-F}. The best-fit
parameters, observed fluxes, and observed luminosities are listed in
Table~\ref{T6_bestpara-SB}. Appendix~\ref{Appendix-B} compares 
the results of our spectral analysis 
with those of previous studies for the individual sources.

\subsubsection{Additional lines}
\label{subsub4-2-3_lines}

The XCLUMPY model contains fluorescence lines from elements with atomic
numbers up to $Z=30$, among which the Fe~K$\alpha$ line at 6.4~keV is
the strongest. Besides them, we also include Gaussian components
(\textsf{zgauss} in \textsc{xspec}) to model other prominent emission
lines when required. In particular, emission lines from highly ionized
iron ions (\ion{Fe}{25} and \ion{Fe}{26}) at $\sim$6.7--7.0~keV
are often detected from C-GOALS objects, which are most probably
produced by highly ionized interstellar medium energized by starburst
activities \citep[see e.g.,][]{Iwasawa2009,Iwasawa2011,Lindner2012,Gilli2014}. Also, 
\ion{Si}{13} line at 1.86~keV is often observed, which originates from
AGN photoionized gas and/or thermal emission from the starburst
region. Accordingly, we add these highly-ionized Fe and Si K$\alpha$ lines if
they are found to significantly improve the fit. When necessary, we also
consider other emission lines (\ion{O}{8} K$\alpha$ at 0.654~keV,
\ion{Mg}{11} K$\alpha$ at 1.342~keV, and Ni K$\alpha$ at
7.48~keV). The best-fit values of these additional lines, adopted for
nine sources, are listed in Table~\ref{T7_bestpara-lines}.

\clearpage
\begin{longrotatetable}
\begin{deluxetable*}{lcccccccccc}
\label{T5_bestpara-AGN}
\tablecaption{Summary of the Best-fit Spectral Parameters for AGNs}
\tablewidth{700pt}
\tabletypesize{\small}
\tablehead{
\colhead{Object} &
\colhead{$N^{\rm LOS}_{\rm H}$} &
\colhead{$N^{\rm Host}_{\rm H}$} &
\colhead{$N^{\rm Equ}_{\rm H}$} &
\colhead{$\sigma$} &
\colhead{$i$} &
\colhead{$\Gamma_{\rm AGN}$} &
\colhead{$A_{\rm AGN}$} &
\colhead{$\Gamma_{\rm bin}$} &
\colhead{$f_{\rm bin}$} &
\colhead{$N^{\rm Time}_{\rm H}$} \\
\ \ \ \ \ \ \ \ \ \ (1) & (2) & (3) & (4) & (5) & (6) & (7) & (8) & (9) & (10) & (11)\\
 & $kT_1$ & $kT_2$ & $kT_3$ & $C_{\rm ACIS}$ & $C_{\rm pn}$ & $C_{\rm FI}$& $C_{\rm BI}$ & $C_{\rm XRT}$ & $C_{\rm BAT}$ & $C^{\rm Time}$\\
 & (12) & (13) & (14) & (15) & (16) & (17) & (18) & (19) & (20) & (21)\\
 & log$F^{\rm obs}_{0.5-2}$ & log$F^{\rm obs}_{2-10}$ & log$F^{\rm obs}_{10-50}$ & log$L^{\rm apec}_{\rm 0.5-2}$ & log$L^{\rm bin}_{\rm 0.5-2}$ & log$L^{\rm obs}_{\rm 0.5-2}$ & log$L^{\rm obs}_{2-10}$ & log$L^{\rm obs}_{10-50}$ & EW & $\chi^2$/$\chi^2_{\nu}$ (C/C$_{\nu}$)\\
 & (22) & (23) & (24) & (25) & (26) & (27) & (28) & (29) & (30) & (31)
}
\startdata
\multicolumn{11}{c}{Stage-A}\\
\hline 
NGC 833 & 26.1$^{+1.2}_{-1.1}$ & \nodata & 32.0$^{+68.0}_{-22.9}$ & 13.7$^{+2.2}_{-1.5}$ & 60$^{\rm (f)}$ & 1.98$^{+0.24}_{-0.22}$ & 10.2$^{+8.0}_{-4.3}$ & 2.49$^{+0.25}_{-0.25}$ & 0.6$^{+0.5}_{-0.3}$ & \nodata\\
 & 0.64$^{+0.10}_{-0.09}$ & \nodata & \nodata & 0.80$^{+0.11}$ & 1.20$_{-0.07}$ & \nodata & \nodata & \nodata & \nodata & 0.44$^{\rm +0.08[C]}_{-0.08}$/0.58$^{\rm +0.08[X]}_{-0.07}$\\
 & $-$13.65 & $-$12.16 & $-$11.62 & 39.53 & 39.67 & 39.92 & 41.41 & 41.95 & 49 & 185.9/1.18 [$\chi^2$]\\
\hline 
NGC 835 & 29.8$^{+0.8}_{-1.5}$ & \nodata & 100.0$_{-73.4}$ & 12.4$^{+1.7}_{-0.1}$ & 60$^{\rm (f)}$ & 1.50$^{+0.11}$ & 5.1$^{+1.8}_{-0.4}$ & 2.27$^{+0.14}_{-0.15}$ & 2.3$^{+0.3}_{-0.6}$ & \nodata\\
 & 0.89$^{+0.06}_{-0.09}$ & 0.41$^{+0.07}_{-0.04}$ & \nodata & 0.81$^{+0.05}_{-0.01}$ & 0.95$^{+0.06}_{-0.05}$ & \nodata & \nodata & \nodata & \nodata & 1.95$^{\rm +0.19[C]}_{-0.21}$/0.11$^{\rm +0.02[X]}_{-0.01}$\\
 & $-$13.10 & $-$12.10 & $-$11.28 & 40.34 & 40.02 & 40.52 & 41.51 & 42.33 & 60 & 338.3/1.05 [$\chi^2$]\\
\hline 
NGC 6921 & 173.1$^{+27.8}_{-31.9}$ & \nodata & 2.2$^{+2.5}_{-0.8}$ & 20.0$^{+43.4}_{-9.7}$ & 80$^{\rm (f)}$ & 1.67$^{+0.21}_{-0.17}$ & 33$^{+45}_{-21}$ & 3.00$_{-1.09}$ & 0.6$^{+1.2}_{-0.5}$ & \nodata\\
 & 0.36$^{+0.49}_{-0.16}$ & \nodata & \nodata & \nodata & 1.20$_{-0.20}$ & \nodata & \nodata & \nodata & \nodata & 0.45$^{\rm +0.77[X]}_{-0.45}$\\
 & $-$13.07 & $-$12.46 & $-$11.09 & 40.22 & 40.37 & 40.60 & 41.19 & 42.58 & 1039 & 42.2/0.84 [$\chi^2$]\\
\hline 
MCG+04-48-002 & 72.7$^{+14.8}_{-8.0}$ & \nodata & 4.4$^{+8.6}_{-3.0}$ & 22.4$^{+18.8}_{-4.8}$ & 60$^{\rm (f)}$ & 1.66$^{+0.09}_{-0.10}$ & 15.4$^{+6.1}_{-4.4}$ & 3.00$_{-0.59}$ & 1.5$^{+0.9}_{-0.6}$ & 60.1$^{\rm +4.9[X/B]}_{-4.7}$\\
 & 0.49$^{+0.22}_{-0.18}$ & \nodata & \nodata & \nodata & 1.14$^{+0.06}_{-0.10}$ & \nodata & \nodata & \nodata & 1$^{\rm (f)}$ & 3.55$^{\rm +0.67[X]}_{-0.56}$/5.46$^{\rm +0.65[B]}_{-0.58}$\\
 & $-$13.02 & $-$12.16 & $-$11.07 & 40.24 & 40.37 & 40.62 & 41.46 & 42.57 & 385 & 142.9/1.13 [$\chi^2$]\\
\hline 
NGC 7469 & $<$0.0015 & \nodata & 2.1$^{+1.2}_{-0.5}$ & 12.4$^{+0.8}_{-1.3}$ & 45$^{\rm (f)}$ & 1.86$^{+0.02}_{-0.02}$ & 101$^{+6}_{-5}$ & \nodata & \nodata & \nodata\\
 & \nodata & \nodata & \nodata & \nodata & 1.13$^{+0.01}_{-0.01}$ & \nodata & \nodata & \nodata & \nodata & \nodata\\
 & $-$10.54 & $-$10.54 & $-$10.41 & \nodata & \nodata & 43.24 & 43.24 & 43.37 & 80 & 1696.2/1.09 [$\chi^2$]\\
\hline 
NGC 7674 & 29.6$^{+2.5}_{-1.9}$ & \nodata & 3.6$^{+3.8}_{-1.6}$ & 37.9$^{+5.2}_{-4.1}$ & 30$^{\rm (f)}$ & 1.66$^{+0.23}_{-0.16}$ & 5.1$^{+5.7}_{-2.1}$ & 2.79$^{+0.21}_{-0.39}$ & 10.4$^{+8.0}_{-5.7}$ & 13.6$^{\rm +5.0[X]}_{-3.5}$/10.3$^{\rm +3.3[S]}_{-2.4}$\\
 & 0.77$^{+0.05}_{-0.05}$ & 0.11$^{+0.03}_{-0.02}$ & \nodata & \nodata & 0.93$^{+0.07}_{-0.06}$ & 1.20$_{-0.06}$ & 1.20$_{-0.03}$ & \nodata & 1.20$_{-0.32}$ & 0.44$^{\rm +0.11[X/S]}_{-0.11}$\\
 & $-$12.67 & $-$12.14 & $-$11.36 & 41.29 & 41.35 & 41.62 & 42.12 & 42.92 & 248 & 202.2/1.10 [$\chi^2$]\\
\hline 
NGC 7679 & $<$3e-7 & \nodata & 0.1$^{+0.1}$ & 10.0$^{+5.4}$ & 30$^{\rm (f)}$ & 1.77$^{+0.07}_{-0.04}$ & 9.5$^{+1.3}_{-0.8}$ & (=$\Gamma_{\rm AGN}$) & \nodata & \nodata\\
 & 0.38$^{+0.33}_{-0.07}$ & \nodata & \nodata & \nodata & 1.04$^{+0.05}_{-0.05}$ & \nodata & \nodata & 1.05$^{+0.15}_{-0.14}$ & 1.20$_{-0.10}$ & 0.16$^{\rm +0.01[X]}_{-0.01}$\\
 & $-$11.68 & $-$11.47 & $-$11.34 & 40.52 & \nodata & 42.14 & 42.34 & 42.48 & 11 & 423.8/1.10 [$\chi^2$]\\
\tablebreak
NGC 7682 & 38.2$^{+6.5}_{-7.0}$ & \nodata & 0.5$^{+1.0}_{-0.1}$ & 60.2$^{+9.8}_{-35.2}$ & 60$^{\rm (f)}$ & 1.8$^{\rm (f)}$ & 4.0$^{+0.6}_{-0.5}$ & 1.56$^{+0.47}_{-0.06}$ & 2.3$^{+0.8}_{-0.5}$ & \nodata\\
 & 0.26$^{+0.35}_{-0.06}$ & \nodata & \nodata & \nodata & 1.09$^{+0.11}_{-0.16}$ & \nodata & \nodata & \nodata & \nodata & 0.51$^{\rm +0.21[X]}_{-0.17}$\\
 & $-$13.52 & $-$12.42 & $-$11.79 & 39.82 & 40.12 & 40.30 & 41.38 & 42.02 & 231 & 58.8/1.20 [$\chi^2$]\\
\hline
\multicolumn{11}{c}{Stage-B}\\
\hline
NGC 235$^{\dagger}$ & 51.4$^{+8.3}_{-6.1}$ & \nodata & 6.2$^{+6.0}_{-3.7}$ & 19.0$^{+4.9}_{-2.0}$ & 60$^{\rm (f)}$ & 1.68$^{+0.11}_{-0.10}$ & 43$^{+23}_{-13}$ & 3.00$_{-0.10}$ & 0.5$^{+0.2}_{-0.2}$ & 29.3$^{\rm +2.4[C]}_{-2.3}$\\
 & 0.92$^{+0.12}_{-0.14}$ & \nodata & \nodata & 0.94$^{+0.15}_{-0.14}$ & \nodata & 1.17$^{+0.03}_{-0.29}$ & (=$C_{\rm FI}$) & \nodata & \nodata & 0.52$^{\rm +0.17[S]}_{-0.11}$\\
 & $-$13.10 & $-$11.56 & $-$10.61 & 40.49 & 40.73 & 40.95 & 42.47 & 43.43 & 216 & 119.8/1.17 [$\chi^2$]\\
\hline 
CGCG 468-002W & 1.5$^{+0.1}_{-0.1}$ & \nodata & 1.0$^{+1.4}_{-0.4}$ & 14.7$^{+0.9}_{-1.3}$ & 60$^{\rm (f)}$ & 1.63$^{+0.05}_{-0.03}$ & 23.2$^{+2.2}_{-1.3}$ & (=$\Gamma_{\rm AGN}$) & 3.7$^{+1.4}_{-1.4}$ & \nodata\\
 & \nodata & \nodata & \nodata & \nodata & \nodata & \nodata & \nodata & 0.80$^{+0.01}$ & 0.80$^{+0.05}$ & \nodata\\
 & $-$12.04 & $-$11.01 & $-$10.71 & \nodata & 41.11 & 41.78 & 42.82 & 43.12 & 71 & 442.4/1.08 [$\chi^2$]\\
\hline 
ESO 060-IG016 East & 8.4$^{+4.0}_{-2.9}$ & \nodata & 0.8$^{+0.4}_{-0.3}$ & 20$^{\rm (f)}$ & 60$^{\rm (f)}$ & 1.8$^{\rm (f)}$ & 0.6$^{+0.1}_{-0.1}$ & 1.8$^{\rm (f)}$ & 9.6$^{+5.4}_{-3.6}$ & \nodata\\
 & \nodata & \nodata & \nodata & 1.20$_{-0.29}$ & \nodata & \nodata & \nodata & \nodata & \nodata & \nodata\\
 & $-$13.93 & $-$12.88 & $-$12.56 & \nodata & 40.74 & 40.75 & 41.78 & 42.11 & 83 & 18.4/1.15 [$\chi^2$]\\
\hline 
IC 4518A & 17.1$^{+0.5}_{-0.9}$ & \nodata & 3.2$^{+1.3}_{-0.8}$ & 17.5$^{+0.8}_{-0.9}$ & 60$^{\rm (f)}$ & 1.81$^{+0.06}_{-0.07}$ & 35.0$^{+5.4}_{-5.7}$ & 1.66$^{+0.34}_{-0.16}$ & 0.6$^{+0.2}_{-0.1}$ & 21.6$^{\rm +1.7[X]}_{-1.6}$\\
 & 0.90$^{+0.06}_{-0.10}$ & 0.26$^{+0.06}_{-0.08}$ & \nodata & \nodata & 1.07$^{+0.04}_{-0.04}$ & 1.20$_{-0.04}$ & 1.20$_{-0.03}$ & \nodata & 1.14$^{+0.06}_{-0.17}$ & 0.34$^{\rm +0.03[X]}_{-0.03}$/0.82$^{\rm +0.05[S]}_{-0.04}$\\
 & $-$12.87 & $-$11.32 & $-$10.80 & 40.71 & 40.42 & 40.90 & 42.45 & 42.98 & 105 & 772.6/1.12 [$\chi^2$]\\
\hline 
NGC 6286 & 140.2$^{+26.7}_{-34.6}$ & \nodata & 1.8$^{+3.1}_{-0.7}$ & 20$^{\rm (f)}$ & 80$^{\rm (f)}$ & 1.50$^{+0.77}$ & 2.5$^{+38.3}_{-1.0}$ & 1.90$^{+0.27}_{-0.27}$ & 5.6$^{+4.3}_{-5.3}$ & \nodata\\
 & 0.79$^{+0.04}_{-0.04}$ & \nodata & \nodata & 0.80$^{+0.05}$ & 0.99$^{+0.13}_{-0.11}$ & \nodata & \nodata & \nodata & \nodata & \nodata\\
 & $-$12.95 & $-$13.03 & $-$11.88 & 40.79 & 40.36 & 40.93 & 40.84 & 42.00 & 559 & 893.4/1.01 [C]\\
\hline 
\multicolumn{11}{c}{Stage-C}\\
\hline 
MCG+12-02-001 & 707.5$^{+268.4}_{-277.1}$ & \nodata & 67.1$^{+32.9}_{-38.8}$ & 20$^{\rm (f)}$ & 60$^{\rm (f)}$ & 1.8$^{\rm (f)}$ & 3.1$^{+3.5}_{-0.5}$ & 2.20$^{+0.20}_{-0.18}$ & 20.0$_{-10.7}$ & \nodata\\
 & \nodata & \nodata & \nodata & 1$^{\rm (f)}$ & \nodata & \nodata & \nodata & \nodata & \nodata & \nodata\\
 & $-$12.88 & $-$12.91 & $-$12.54 & \nodata & 40.86 & 40.86 & 40.83 & 41.20 & 222 & 262.8/0.90 [C]\\
\hline 
IRAS F06076--2139 & 42.2$^{+24.0}_{-12.0}$ & \nodata & 4.0$^{+1.9}_{-1.3}$ & 20$^{\rm (f)}$ & 60$^{\rm (f)}$ & 1.8$^{\rm (f)}$ & 1.5$^{+0.6}_{-0.4}$ & 1.8$^{\rm (f)}$ & 3.3$^{+1.9}_{-1.2}$ & \nodata\\
 & \nodata & \nodata & \nodata & 1$^{\rm (f)}$ & \nodata & \nodata & \nodata & \nodata & \nodata & \nodata\\
 & $-$13.99 & $-$12.96 & $-$12.22 & \nodata & 40.51 & 40.52 & 41.53 & 42.28 & 177 & 249.8/0.92 [C]\\
\hline 
NGC 3690 West & 302.6$^{+60.6}_{-49.7}$ & 0.45$^{+0.05}_{-0.05}$ & 18.0$^{+28.5}_{-4.3}$ & 44.9$^{+8.5}_{-9.7}$ & 30$^{\rm (f)}$ & 2.23$^{+0.21}_{-0.21}$ & 111$^{+132}_{-58}$ & 1.84$^{+0.08}_{-0.06}$ & 2.3$^{+2.5}_{-1.5}$ & \nodata\\
 & 0.95$^{+0.03}_{-0.03}$ & 0.26$^{+0.02}_{-0.01}$ & 0.04$^{+0.01}_{-0.01}$ & 1.13$^{+0.03}_{-0.03}$ & 0.92$^{+0.02}_{-0.02}$ & \nodata & \nodata & \nodata & \nodata & \nodata\\
 & $-$12.07 & $-$12.00 & $-$11.23 & 42.02 & 41.10 & 41.29 & 41.36 & 42.14 & 391 & 612.1/1.18 [$\chi^2$]\\
\hline 
NGC 4922N & 75.9$^{+285.1}_{-21.6}$ & \nodata & 7.2$^{+4.0}_{-2.3}$ & 20$^{\rm (f)}$ & 60$^{\rm (f)}$ & 1.8$^{\rm (f)}$ & 2.4$^{+1.4}_{-0.8}$ & 1.79$^{+0.32}_{-0.29}$ & 4.7$^{+3.0}_{-2.0}$ & \nodata\\
 & 1.00$^{+0.29}_{-0.18}$ & \nodata & \nodata & 1$^{\rm (f)}$ & \nodata & \nodata & \nodata & \nodata & \nodata & \nodata\\
 & $-$13.41 & $-$12.96 & $-$12.07 & 40.29 & 40.48 & 40.69 & 41.12 & 42.03 & 261 & 323.3/0.83 [C]\\
\hline 
ESO 148-2 & 158.5$^{+45.0}_{-47.6}$ & \nodata & 2.0$^{+0.9}_{-1.0}$ & 20$^{\rm (f)}$ & 80$^{\rm (f)}$ & 1.8$^{\rm (f)}$ & 1.6$^{+1.2}_{-0.8}$ & 1.70$^{+0.19}_{-0.15}$ & 7.2$^{+6.5}_{-3.2}$ & 2.7$^{\rm +0.9[C/X]}_{-0.6}$/24.6$^{\rm +975.4[XRT]}_{-18.1}$\\
 & 0.79$^{+0.04}_{-0.04}$ & \nodata & \nodata & 0.95$^{+0.09}_{-0.08}$ & 1.01$^{+0.12}_{-0.11}$ & \nodata & \nodata & 1$^{\rm (f)}$ & \nodata & \nodata\\
 & $-$13.31 & $-$13.22 & $-$12.47 & 41.05 & 41.06 & 41.35 & 41.43 & 42.19 & 277 & 2323.2/0.97 [C]\\
\hline 
\multicolumn{11}{c}{Stage-D}\\
\hline 
NGC 34 & 50.3$^{+10.3}_{-9.9}$ & \nodata & 4.8$^{+1.1}_{-0.9}$ & 20$^{\rm (f)}$ & 60$^{\rm (f)}$ & 1.8$^{\rm (f)}$ & 2.8$^{+0.7}_{-0.6}$ & 1.65$^{+0.12}_{-0.13}$ & 7.9$^{+2.3}_{-1.7}$ & \nodata\\
 & 0.77$^{+0.08}_{-0.10}$ & \nodata & \nodata & 1.16$^{+0.04}_{-0.14}$ & 1.11$^{+0.09}_{-0.11}$ & \nodata & \nodata & \nodata & \nodata & \nodata\\
 & $-$13.18 & $-$12.62 & $-$11.91 & 40.20 & 40.61 & 40.75 & 41.31 & 42.02 & 163 & 103.4/1.04 [$\chi^2$]\\
\hline 
IRAS F05189--2524 & 7.5$^{+0.1}_{-0.1}$ & \nodata & 49.0$^{+48.8}_{-23.6}$ & 11.8$^{+0.7}_{-0.6}$ & 60$^{\rm (f)}$ & 2.19$^{+0.02}_{-0.02}$ & 35.3$^{+1.1}_{-1.7}$ & 2.92$^{+0.08}_{-0.08}$ & 1.1$^{+0.1}_{-0.1}$ & 299.3$^{\rm +700.7[S]}_{-60.7}$\\
 & 0.96$^{+0.08}_{-0.09}$ & 0.11$^{+0.01}_{-0.01}$ & \nodata & 0.80$^{+0.02}$ & 1.04$^{+0.01}_{-0.01}$ & 0.88$^{+0.07}_{-0.07}$ & 0.93$^{+0.09}_{-0.09}$ & \nodata & 1.20$_{-0.23}$ & 0.87$^{\rm +0.03[C]}_{-0.03}$/0.90$^{\rm +0.02[X]}_{-0.01}$\\
 & $-$12.89 & $-$11.46 & $-$11.35 & 41.03 & 41.54 & 41.73 & 43.15 & 43.28 & 25 & 3522.7/1.17 [$\chi^2$]\\
\hline 
NGC 2623 & 6.0$^{+4.5}_{-2.1}$ & \nodata & 0.6$^{+0.4}_{-0.2}$ & 20$^{\rm (f)}$ & 60$^{\rm (f)}$ & 1.8$^{\rm (f)}$ & 0.3$^{+0.1}_{-0.1}$ & 1.50$^{+0.49}$ & 15.4$^{+4.6}_{-7.7}$ & \nodata\\
 & 1.27$^{+0.39}_{-0.29}$ & \nodata & \nodata & 0.88$^{+0.21}_{-0.08}$ & 1$^{\rm (f)}$ & \nodata & \nodata & \nodata & \nodata & \nodata\\
 & $-$13.79 & $-$13.01 & $-$12.71 & 39.60 & 39.91 & 40.10 & 40.87 & 41.17 & 61 & 597.9/0.90 [C]\\
\hline 
IRAS F08572+3915 & 84.6$^{+129.3}_{-28.3}$ & \nodata & 8.0$^{+5.3}_{-2.7}$ & 20$^{\rm (f)}$ & 60$^{\rm (f)}$ & 1.8$^{\rm (f)}$ & 0.24$^{+0.04}_{-0.03}$ & 1.50$^{+0.10}$ & 10$^{\rm (f)}$ & \nodata\\
 & \nodata & \nodata & \nodata & \nodata & \nodata & 1$^{\rm (f)}$ & \nodata & \nodata & \nodata & \nodata\\
 & $-$14.31 & $-$13.75 & $-$13.01 & \nodata & 40.58 & 40.59 & 41.13 & 41.88 & 170 & 1042.0/0.91 [C]\\
\hline 
UGC 5101 & 96.3$^{+4.3}_{-1.6}$ & 0.11$^{+0.06}_{-0.05}$ & 100.0$_{-84.4}$ & 11.2$^{+6.7}_{-1.2}$ & 65.9$^{+21.1}_{-15.4}$ & 1.50$^{+0.17}$ & 7.6$^{+6.0}_{-2.2}$ & 1.66$^{+0.26}_{-0.16}$ & 1.6$^{+0.8}_{-0.7}$ & 147.9$^{\rm +27.1[C/X/S]}_{-19.4}$\\
 & 0.89$^{+0.13}_{-0.13}$ & \nodata & \nodata & 0.87$^{+0.10}_{-0.07}$ & 1.02$^{+0.11}_{-0.11}$ & 1.12$^{+0.08}_{-0.13}$ & 1.19$^{+0.01}_{-0.21}$ & \nodata & 1$^{\rm (f)}$ & \nodata\\
 & $-$13.58 & $-$12.56 & $-$11.33 & 40.44 & 40.95 & 40.96 & 41.96 & 43.21 & 249 & 172.6/1.02 [$\chi^2$]\\
\hline 
Mrk 231 & 8.5$^{+0.2}_{-0.2}$ & \nodata & 0.5$^{+2.4}_{-0.2}$ & 22.6$^{+5.9}_{-6.5}$ & 60$^{\rm (f)}$ & 1.50$^{+0.06}$ & 2.1$^{+0.1}_{-0.1}$ & 1.50$^{+0.02}$ & 16.6$^{+0.9}_{-0.8}$ & \nodata\\
 & 0.81$^{+0.03}_{-0.03}$ & \nodata & \nodata & 1.18$^{+0.02}_{-0.06}$ & 1.00$^{+0.03}_{-0.03}$ & 1.06$^{+0.03}_{-0.03}$ & 1.09$^{+0.05}_{-0.05}$ & \nodata & \nodata & 0.70$^{\rm +0.05[C]}_{-0.04}$/0.63$^{\rm +0.03[X]}_{-0.03}$\\
 & $-$12.97 & $-$12.04 & $-$11.56 & 41.11 & 41.49 & 41.64 & 42.56 & 43.05 & 81 & 1301.7/1.19 [$\chi^2$]\\
\hline 
IRAS 13120--5453 & 164.8$^{+21.6}_{-20.5}$ & 0.43$^{+0.15}_{-0.18}$ & 2.1$^{+8.3}_{-0.7}$ & 20$^{\rm (f)}$ & 80$^{\rm (f)}$ & 1.8$^{\rm (f)}$ & 2.1$^{+6.1}_{-0.5}$ & 1.50$^{+0.10}$ & 16.3$^{+3.7}_{-12.2}$ & \nodata\\
 & 1.26$^{+0.10}_{-0.12}$ & 0.18$^{+0.04}_{-0.04}$ & \nodata & 1.11$^{+0.09}_{-0.11}$ & 1.01$^{+0.05}_{-0.05}$ & \nodata & \nodata & \nodata & \nodata & \nodata\\
 & $-$13.00 & $-$12.68 & $-$12.15 & 42.09 & 41.20 & 41.33 & 41.65 & 42.18 & 107 & 327.6/1.15 [$\chi^2$]\\
\hline 
Mrk 273 & 49.6$^{+5.5}_{-2.8}$ & \nodata & 10.0$^{+15.5}_{-4.8}$ & 17.3$^{+1.9}_{-2.0}$ & 60$^{\rm (f)}$ & 1.71$^{+0.13}_{-0.14}$ & 9.7$^{+6.2}_{-3.7}$ & 1.51$^{+0.12}_{-0.01}$ & 2.3$^{+1.5}_{-0.9}$ & 92.5$^{\rm +11.3[C]}_{-9.9}$/172.8$^{\rm +54.0[S]}_{-27.9}$\\
 & 1.00$^{+0.05}_{-0.05}$ & 0.31$^{+0.49}_{-0.07}$ & \nodata & 0.92$^{+0.07}_{-0.06}$ & 1.00$^{+0.07}_{-0.07}$ & 1.02$^{+0.09}_{-0.08}$ & 1.10$^{+0.10}_{-0.12}$ & \nodata & 0.80$^{+0.15}$ & \nodata\\
 & $-$13.04 & $-$12.15 & $-$11.29 & 41.16 & 41.20 & 41.48 & 42.34 & 43.22 & 148 & 349.5/1.12 [$\chi^2$]\\
\hline 
IRAS~F14348--1447 & 128.3$^{+95.0}_{-55.4}$ & \nodata & 1.6$^{+19.5}_{-1.0}$ & 20$^{\rm (f)}$ & 80$^{\rm (f)}$ & 1.8$^{\rm (f)}$ & 1.0$^{+7.6}_{-0.6}$ & 1.74$^{+0.24}_{-0.24}$ & 9.7$^{+10.3}_{-8.6}$ & \nodata\\
 & 0.77$^{+0.25}_{-0.23}$ & \nodata & \nodata & 0.80$^{+0.15}$ & 1.20$_{-0.16}$ & \nodata & \nodata & \nodata & \nodata & \nodata\\
 & $-$13.59 & $-$13.34 & $-$12.64 & 41.08 & 41.50 & 41.64 & 41.85 & 42.57 & 232 & 776.9/0.91 [C]\\
\hline 
Arp~220W & 1000.0$_{-22.1}$ & \nodata & 100.0$_{-61.3}$ & 20$^{\rm (f)}$ & 60$^{\rm (f)}$ & 1.8$^{\rm (f)}$ & 1.6$^{+0.1}_{-0.1}$ & 1.50$^{+0.02}$ & 10$^{\rm (f)}$ & \nodata\\
 & 0.91$^{+0.04}_{-0.05}$ & \nodata & \nodata & 0.90$^{+0.05}_{-0.05}$ & 0.97$^{+0.06}_{-0.05}$ & 1.20$_{-0.06}$ & 1.20$_{-0.10}$ & \nodata & \nodata & \nodata\\
 & $-$13.22 & $-$12.97 & $-$12.57 & 40.24 & 40.41 & 40.64 & 40.89 & 41.29 & 120 & 350.6/1.20 [$\chi^2$]\\
\hline 
IRAS F17138--1017 & 1000.0$_{-643.5}$ & \nodata & 100.0$_{-69.5}$ & 20$^{\rm (f)}$ & 60$^{\rm (f)}$ & 1.8$^{\rm (f)}$ & 2.1$^{+0.5}_{-0.3}$ & 1.50$^{+0.17}$ & 10$^{\rm (f)}$ & \nodata\\
 & \nodata & \nodata & \nodata & 1$^{\rm (f)}$ & \nodata & \nodata & \nodata & \nodata & \nodata & \nodata\\
 & $-$13.33 & $-$12.92 & $-$12.45 & \nodata & 40.49 & 40.50 & 40.91 & 41.38 & 120 & 283.9/0.81 [C]\\
\hline
\multicolumn{11}{c}{Stage-N}\\
\hline
UGC 2608 & 540$^{+700}_{-310}$ & \nodata & 15$^{+19}_{-9}$ & 10.0$^{+8.2}$ & 80$^{\rm (f)}$ & 1.8$^{\rm (f)}$ & 96$^{+54}_{-41}$ & 2.87$^{+0.13}_{-0.59}$ & 0.6$^{+0.4}_{-0.3}$ & \nodata\\
 & 0.82$^{+0.14}_{-0.06}$ & \nodata & \nodata & \nodata & 0.97$^{+0.15}_{-0.15}$ & 1.20$_{-0.13}$ & 1.20$_{-0.11}$ & \nodata & \nodata & \nodata\\
 & $-$12.60 & $-$12.67 & $-$11.48 & 41.20 & 41.18 & 41.49 & 41.41 & 42.60 & 1430 & 75.1/1.04 [$\chi^2$]\\
\hline 
MCG--03-34-064 & 98.4$^{+2.8}_{-2.4}$ & \nodata & 9.2$^{+1.3}_{-1.4}$ & 20.1$^{+0.7}_{-0.5}$ & 60$^{\rm (f)}$ & 2.01$^{+0.04}_{-0.05}$ & 127$^{+20}_{-19}$ & 3.00$_{-0.12}$ & 0.6$^{+0.1}_{-0.1}$ & 56.1$^{\rm +1.7[C/X]}_{-2.0}$\\
 & 1.93$^{+0.08}_{-0.15}$ & 0.82$^{+0.01}_{-0.01}$ & 0.12$^{+0.01}_{-0.01}$ & 0.80$^{+0.03}$ & 0.98$^{+0.01}_{-0.01}$ & \nodata & \nodata & \nodata & 1.01$^{+0.12}_{-0.11}$ & 1.46$^{\rm +0.18[C]}_{-0.18}$\\
 & $-$12.22 & $-$11.77 & $-$10.73 & 41.40 & 41.02 & 41.57 & 42.00 & 43.06 & 425 & 2410.3/1.20 [$\chi^2$]\\
\tablebreak
NGC 5135 & 670$^{+1660}_{-280}$ & \nodata & 9.5$^{+24.0}_{-3.9}$ & 10.0$^{+4.5}$ & 84.1$^{+2.9}_{-6.1}$ & 1.71$^{+0.19}_{-0.15}$ & 126$^{+203}_{-57}$ & 2.64$^{+0.15}_{-0.11}$ & 0.9$^{+0.8}_{-0.6}$ & \nodata\\
 & 0.88$^{+0.03}_{-0.05}$ & \nodata & \nodata & 0.94$^{+0.13}_{-0.12}$ & \nodata & 1.03$^{+0.14}_{-0.13}$ & 1.08$^{+0.12}_{-0.14}$ & \nodata & \nodata & \nodata\\
 & $-$12.37 & $-$12.43 & $-$11.35 & 40.84 & 41.03 & 41.25 & 41.19 & 42.26 & 2520 & 164.1/1.04 [$\chi^2$]\\
\hline 
NGC 7130 & 413.3$^{+13.1}_{-7.4}$ & \nodata & 4.9$^{+2.9}_{-1.3}$ & 24.7$^{+9.2}_{-11.5}$ & 80$^{\rm (f)}$ & 1.65$^{+0.22}_{-0.15}$ & 30$^{+31}_{-13}$ & 2.39$^{+0.10}_{-0.11}$ & 2.0$^{+1.7}_{-1.0}$ & \nodata\\
 & 1.09$^{+0.42}_{-0.17}$ & 0.68$^{+0.11}_{-0.12}$ & \nodata & 0.96$^{+0.06}_{-0.08}$ & \nodata & 1.20$_{-0.11}$ & 1.20$_{-0.09}$ & \nodata & 1.20$_{-0.19}$ & \nodata\\
 & $-$12.62 & $-$12.67 & $-$11.33 & 40.79 & 40.89 & 41.16 & 41.10 & 42.43 & 1426 & 186.7/1.13 [$\chi^2$]\\\enddata
\tablecomments{Columns: 
(1) object name; 
(2) line-of-sight hydrogen column density measured with NuSTAR 
in [10$^{22}$ cm$^{-2}$]. 
The error is estimated by linking the torus parameters and using a Markov Chain Monte Carlo (MCMC) chain with a length of 40,000;
(3) hydrogen column density of the host galaxy of the target in [10$^{22}$ cm$^{-2}$];
(4) equatorial hydrogen column density in [10$^{24}$ cm$^{-2}$];
(5--6) torus angular width and inclination in [degree].
When these values are not well constrained, 
we fix the parameters as described in Section~\ref{subsub4-2-1_AGNmodel};
(7--8) power-law photon index of the AGN-transmitted component, and its normalization at 1~keV in [10$^{-4}$~keV$^{-1}$ cm$^{-2}$ s$^{-1}$];
(9--10) power-law photon index of the XRB and scattered
components, and the fraction relative to the transmitted component at 
1~keV in [\%];
(11) time variable line-of-sight hydrogen column densities in the
 epochs of the Chandra (C), XMM-Newton (X), Suzaku (S), Swift/XRT (XRT)
 and BAT (B) observations in [10$^{22}$ cm$^{-2}$];
(12--14) temperatures of the \textsf{apec} components in [keV];
(15--20) instrumental cross normalizations of Chandra/ACIS, XMM-Newton/pn, Suzaku/XIS-FI, XIS-BI, Swift/XRT, and BAT, respectively;
(21) time variability constant of the Chandra, XMM-Newton, Suzaku, and Swift/BAT observations;
(22--24) logarithmic observed fluxes in the 0.5--2~keV, 2--10~keV, and 10--50~keV bands in [erg s$^{-1}$ cm$^{-2}$];
(25--26) logarithmic luminosities of the components of \textsf{apec} and
 the unabsorbed X-ray emission in the 0.5--2~keV band in [erg s$^{-1}$];
(27--29) logarithmic observed luminosities in the 0.5--2~keV, 2--10~keV, and 10--50~keV bands in [erg s$^{-1}$].
The fluxes and luminosities are corrected for Galactic absorption;
(30) EW of the Fe-K$\alpha$ line in [eV];
(31) value of Cash [C] or $\chi^2$ statistics, and the reduced one divided by degrees of freedom.
The mark (f) means a fixed value.
The results of UGC~2608, NGC~5135 \citep{Yamada2020}, and NGC~7469
 \citep{Ogawa2021} are obtained by analyzing the spectra with XCLUMPY 
in the same manner.
}
\tablenotetext{\dagger}{For NGC 235, the cross normalization is set to unity
 for the Swift/BAT as the calibration reference.}
\end{deluxetable*}
\end{longrotatetable}

\startlongtable
\begin{longrotatetable}
\begin{deluxetable*}{lccccccccc}
\label{T6_bestpara-SB}
\tablecaption{Summary of the X-ray Results for Starburst-dominant or Hard X-ray Undetected Sources}
\tabletypesize{\small}
\tablehead{
\colhead{Object} &
\colhead{$N^{\rm Host}_{\rm H}$} &
\colhead{$\Gamma_{\rm bin}$} &
\colhead{$A_{\rm bin}$} &
\colhead{$E_{\rm cut}$} &
\colhead{$kT$} &
\colhead{$C_{\rm ACIS}$} &
\colhead{$C_{\rm MOS}$} &
\colhead{$C_{\rm pn}$} &
\colhead{$C_{\rm FI}$} \\
\ \ \ \ (1) & (2) & (3) & (4) & (5) & (6) & (7) & (8) & (9) & (10)\\
 & $C_{\rm BI}$ & $C_{\rm XRT}$ & log$F^{\rm obs}_{\rm 0.5-2}$ & log$F^{\rm obs}_{2-10}$ & log$L^{\rm apec}_{\rm 0.5-2}$ & log$L^{\rm bin}_{\rm 0.5-2}$ & log$L^{\rm obs}_{\rm 0.5-2}$ & log$L^{\rm obs}_{2-10}$ & $\chi^2$/$\chi^2_{\nu}$ (C/C$_{\nu}$)\\
 & (11) & (12) & (13) & (14) & (15) & (16) & (17) & (18) & (19)
}
\startdata
\multicolumn{10}{c}{Stage-A}\\
\hline
NGC~838 & \nodata & 1.50$^{+0.02}$ & 41.6$^{+3.1}_{-3.0}$ & 4.7$^{+0.7}_{-0.6}$ & 0.79$^{+0.02}_{-0.02}$ & 0.83$^{+0.04}_{-0.03}$ & 1$^{\rm (f)}$ & 0.96$^{+0.04}_{-0.04}$ & \nodata\\
& \nodata & \nodata & $-$12.81 & $-$13.09 & 40.48 & 40.42 & 40.75 & 40.48 & 2206.1/1.07 [C]\\
\hline
NGC~839 & \nodata & 1.50$^{+0.01}$ & 19.0$^{+1.0}_{-0.9}$ & \nodata & 0.84$^{+0.08}_{-0.05}$ & 1$^{\rm (ref)}$ & 1.18$^{+0.02}_{-0.08}$ & 1.12$^{+0.07}_{-0.07}$ & \nodata\\
& \nodata & \nodata & $-$13.27 & $-$12.99 & 39.64 & 40.19 & 40.30 & 40.58 & 1820.6/0.92 [C]\\
\hline
MCG+08-18-013 & \nodata & 1.50$^{+0.34}$ & 9.8$^{+2.5}_{-2.0}$ & \nodata & \nodata & 1$^{\rm (ref)}$ & \nodata & \nodata & \nodata\\
& \nodata & \nodata & $-$13.67 & $-$13.29 & \nodata & 40.51 & 40.51 & 40.89 & 63.1/0.73 [C]\\
\hline
MCG--01-26-013 & \nodata & 1.8$^{\rm (f)}$ & 2.3$^{+1.6}_{-1.4}$ & \nodata & 0.38$^{+0.37}_{-0.12}$ & 1$^{\rm (ref)}$ & 1$^{\rm (f)}$ & 1$^{\rm (f)}$ & \nodata\\
& \nodata & \nodata & $-$13.98 & $-$14.11 & 39.51 & 39.47 & 39.79 & 39.66 & 397.4/0.94 [C]\\
\hline
NGC~3110 & \nodata & 1.77$^{+0.10}_{-0.11}$ & 35.0$^{+2.9}_{-3.7}$ & \nodata & 0.74$^{+0.04}_{-0.04}$ & 1$^{\rm (ref)}$ & 0.98$^{+0.11}_{-0.07}$ & 0.80$^{+0.09}$ & \nodata\\
& \nodata & \nodata & $-$12.90 & $-$12.90 & 40.51 & 40.69 & 40.91 & 40.90 & 1102.3/0.96 [C]\\
\hline
NGC~4418 & \nodata & 1.50$^{+0.15}$ & 3.7$^{+0.9}_{-0.8}$ & \nodata & 0.68$^{+0.37}_{-0.38}$ & 1$^{\rm (ref)}$ & \nodata & \nodata & 1.20$_{-0.11}$\\
& (=$C_{\rm FI}$) & \nodata & $-$13.98 & $-$13.70 & 38.47 & 39.06 & 39.16 & 39.44 & 852.6/1.02 [C]\\
\hline
\multicolumn{10}{c}{Stage-B}\\
\hline
MCG--02-01-052 & \nodata & 1.52$^{+0.81}_{-0.02}$ & 2.3$^{+1.7}_{-1.0}$ & \nodata & 0.56$^{+0.39}_{-0.20}$ & 1$^{\rm (ref)}$ & \nodata & \nodata & \nodata\\
& \nodata & \nodata & $-$14.10 & $-$13.93 & 39.72 & 39.92 & 40.13 & 40.29 & 68.9/0.78 [C]\\
\hline
MCG--02-01-051 & \nodata & 1.66$^{+0.19}_{-0.16}$ & 18.7$^{+3.7}_{-3.5}$ & \nodata & 0.83$^{+0.19}_{-0.13}$ & 1$^{\rm (ref)}$ & \nodata & \nodata & \nodata\\
& \nodata & \nodata & $-$13.25 & $-$13.11 & 40.44 & 40.82 & 40.97 & 41.11 & 135.2/0.78 [C]\\
\hline
NGC~232 & \nodata & 1.63$^{+0.31}_{-0.13}$ & 11.6$^{+4.0}_{-3.5}$ & \nodata & 0.97$^{+0.13}_{-0.17}$ & 1$^{\rm (ref)}$ & \nodata & \nodata & \nodata\\
& \nodata & \nodata & $-$13.34 & $-$13.29 & 40.37 & 40.46 & 40.72 & 40.77 & 131.1/0.80 [C]\\
\hline
ESO~440-58 & \nodata & 1.70$^{+0.47}_{-0.20}$ & 6.0$^{+2.2}_{-1.7}$ & \nodata & \nodata & 1$^{\rm (ref)}$ & \nodata & \nodata & \nodata\\
& \nodata & \nodata & $-$13.89 & $-$13.63 & \nodata & 40.20 & 40.20 & 40.45 & 36.9/0.84 [C]\\
\hline
MCG--05-29-017 & \nodata & 1.8$^{\rm (f)}$ & 4.4$^{+3.0}_{-2.6}$ & \nodata & 0.97$^{+0.11}_{-0.14}$ & 1$^{\rm (ref)}$ & \nodata & \nodata & \nodata\\
& \nodata & \nodata & $-$13.45 & $-$13.79 & 40.48 & 40.04 & 40.62 & 40.29 & 56.6/0.81 [C]\\
\tablebreak
IC~4518B & \nodata & 1.50$^{+0.75}$ & 2.2$^{+1.5}_{-0.9}$ & \nodata & 0.28$^{+0.42}_{-0.11}$ & \nodata & \nodata & 1$^{\rm (ref)}$ & \nodata\\
& \nodata & \nodata & $-$14.06 & $-$13.92 & 39.33 & 39.42 & 39.68 & 39.81 & 226.9/0.85 [C]\\
\hline
NGC~6285 & \nodata & 2.14$^{+0.36}_{-0.33}$ & 8.9$^{+1.7}_{-1.6}$ & \nodata & \nodata & 1$^{\rm (ref)}$ & 1$^{\rm (f)}$ & 1$^{\rm (f)}$ & \nodata\\
& \nodata & \nodata & $-$13.72 & $-$13.76 & \nodata & 40.19 & 40.19 & 40.16 & 534.1/1.02 [C]\\
\hline
NGC~6907 & \nodata & 1.8$^{\rm (f)}$ & 21.2$^{+7.5}_{-6.6}$ & \nodata & 0.55$^{+0.22}_{-0.17}$ & \nodata & \nodata & \nodata & \nodata\\
& \nodata & 1$^{\rm (f)}$ & $-$12.84 & $-$13.14 & 40.39 & 40.06 & 40.56 & 40.26 & 104.4/1.03 [C]\\
\hline
\multicolumn{10}{c}{Stage-C}\\
\hline
ESO~350-38 & \nodata & 1.50$^{+0.12}$ & 39.2$^{+8.3}_{-4.9}$ & 6.6$^{+3.4}_{-1.5}$ & 0.94$^{+0.10}_{-0.12}$ & 1.20$_{-0.16}$ & \nodata & \nodata & \nodata\\
& \nodata & \nodata & $-$13.03 & $-$13.01 & 40.30 & 40.84 & 40.95 & 40.98 & 446.9/1.12 [C]\\
\hline
IC~1623A & \nodata & 1.50$^{+0.01}$ & 70.5$^{+3.3}_{-3.2}$ & 8.1$^{+0.8}_{-0.7}$ & 0.79$^{+0.02}_{-0.02}$ & 1.16$^{+0.04}_{-0.04}$ & 1$^{\rm (f)}$ & 0.93$^{+0.04}_{-0.03}$ & \nodata\\
& \nodata & \nodata & $-$12.65 & $-$12.69 & 40.91 & 41.09 & 41.31 & 41.28 & 526.6/1.13 [$\chi^2$]\\
\hline
NGC~4922S & \nodata & 1.8$^{\rm (f)}$ & 1.5$^{+1.3}_{-1.3}$ & \nodata & 0.82$^{+0.14}_{-0.19}$ & 1$^{\rm (ref)}$ & \nodata & \nodata & \nodata\\
& \nodata & \nodata & $-$13.91 & $-$14.28 & 40.08 & 39.61 & 40.21 & 39.84 & 63.0/0.90 [C]\\
\hline
II~Zw~096 & \nodata & 1.50$^{+0.52}$ & 2.4$^{+1.6}_{-1.0}$ & \nodata & 0.27$^{+0.22}_{-0.09}$ & 1$^{\rm (ref)}$ & \nodata & \nodata & \nodata\\
& \nodata & \nodata & $-$13.97 & $-$13.91 & 40.24 & 40.16 & 40.50 & 40.55 & 23.0/0.77 [C]\\
\hline
IRAS~F20550+1655~SE & \nodata & 1.50$^{+0.03}$ & 26.9$^{+2.1}_{-2.0}$ & 12.2$^{+4.5}_{-2.7}$ & 0.80$^{+0.03}_{-0.03}$ & 0.80$^{+0.04}$ & 1$^{\rm (f)}$ & 1.03$^{+0.05}_{-0.05}$ & \nodata\\
& \nodata & \nodata & $-$13.09 & $-$13.03 & 40.93 & 41.17 & 41.36 & 41.42 & 1674.3/0.90 [C]\\
\hline
\multicolumn{10}{c}{Stage-D}\\
\hline
ESO 374-IG032 & \nodata & 1.8$^{\rm (f)}$ & 5.1$^{+2.3}_{-2.0}$ & \nodata & 0.33$^{+0.63}_{-0.11}$ & 1$^{\rm (ref)}$ & \nodata & \nodata & \nodata\\
& \nodata & \nodata & $-$13.70 & $-$13.78 & 40.41 & 40.45 & 40.73 & 40.64 & 52.6/0.75 [C]\\
\hline
NGC~3256 & 0.18$^{+0.04}_{-0.03}$ & 2.06$^{+0.12}_{-0.12}$ & 235.8$^{+19.7}_{-13.1}$ & 16.1$^{+11.1}_{-4.9}$ & 0.37$^{+0.12}_{-0.05}$/0.98$^{+0.03}_{-0.03}$ & 1.19$^{+0.01}_{-0.05}$ & 1$^{\rm (f)}$ & 0.98$^{+0.03}_{-0.04}$ & \nodata\\
& \nodata & \nodata & $-$12.07 & $-$12.35 & 41.27 & 40.98 & 41.22 & 40.95 & 763.5/1.20 [$\chi^2$]\\
\hline
IRAS~F10565+2448 & \nodata & 1.50$^{+0.05}$ & 10.1$^{+1.1}_{-1.1}$ & \nodata & 0.94$^{+0.06}_{-0.07}$ & 1$^{\rm (ref)}$ & 0.95$^{+0.12}_{-0.11}$ & 1.03$^{+0.11}_{-0.10}$ & 1.20$_{-0.08}$\\
& 1.20$_{-0.14}$ & \nodata & $-$13.47 & $-$13.28 & 40.74 & 40.96 & 41.17 & 41.36 & 1627.5/0.93 [C]\\
\hline
IRAS~F12112+0305 & \nodata & 1.75$^{+0.22}_{-0.22}$ & 4.9$^{+1.0}_{-1.4}$ & \nodata & 0.94$^{+0.18}_{-0.19}$ & 1$^{\rm (ref)}$ & 1.03$^{+0.17}_{-0.19}$ & 0.82$^{+0.28}_{-0.02}$ & \nodata\\
& \nodata & \nodata & $-$13.85 & $-$13.79 & 40.76 & 41.10 & 41.26 & 41.33 & 418.8/0.89 [C]\\
\tablebreak
IRAS~F14378--3651 & \nodata & 1.50$^{+0.33}$ & 4.2$^{+1.6}_{-1.4}$ & \nodata & \nodata & 1$^{\rm (ref)}$ & \nodata & \nodata & \nodata\\
& \nodata & \nodata & $-$14.06 & $-$13.68 & \nodata & 40.97 & 40.97 & 41.36 & 41.3/0.78 [C]\\
\hline
IRAS~F15250+3608 & \nodata & 2.07$^{+0.23}_{-0.25}$ & 4.4$^{+1.0}_{-0.9}$ & \nodata & 0.58$^{+0.10}_{-0.23}$ & 1$^{\rm (ref)}$ & 1.20$_{-0.10}$ & 1.00$^{+0.19}_{-0.16}$ & \nodata\\
& \nodata & \nodata & $-$13.73 & $-$14.04 & 40.85 & 40.81 & 41.13 & 40.83 & 399.3/0.90 [C]\\
\hline
IRAS~F19297--0406 & \nodata & 1.8$^{\rm (f)}$ & 4.6$^{+3.2}_{-2.8}$ & \nodata & 1.01$^{+0.33}_{-0.23}$ & 1$^{\rm (ref)}$ & \nodata & \nodata & \nodata\\
& \nodata & \nodata & $-$13.52 & $-$13.82 & 41.58 & 41.20 & 41.74 & 41.45 & 47.5/0.70 [C]\\
\hline
ESO~286-19 & \nodata & 1.64$^{+0.11}_{-0.12}$ & 13.6$^{+2.0}_{-1.9}$ & \nodata & 0.94$^{+0.04}_{-0.05}$ & 0.92$^{+0.10}_{-0.09}$ & 1$^{\rm (f)}$ & 0.95$^{+0.12}_{-0.11}$ & \nodata\\
& \nodata & \nodata & $-$13.20 & $-$13.21 & 41.14 & 41.08 & 41.44 & 41.43 & 1183.5/0.88 [C]\\
\hline
\multicolumn{10}{c}{Stage-N}\\ 
\hline
NGC~5104 & \nodata & 2.67$^{+0.17}_{-0.52}$ & 32.5$^{+11.5}_{-9.7}$ & \nodata & \nodata & \nodata & \nodata & \nodata & \nodata\\
& \nodata & 1$^{\rm (ref)}$ & $-$13.15 & $-$13.52 & \nodata & 40.75 & 40.75 & 40.38 & 39.4/0.90 [C]\\
\hline
IRAS~F18293--3413 & \nodata & 1.50$^{+0.06}$ & 53.2$^{+6.1}_{-5.7}$ & 7.6$^{+2.4}_{-1.4}$ & 1.03$^{+0.10}_{-0.11}$ & 1.04$^{+0.08}_{-0.08}$ & 1$^{\rm (f)}$ & 1.02$^{+0.09}_{-0.09}$ & \nodata\\
& \nodata & \nodata & $-$12.92 & $-$12.83 & 40.15 & 40.87 & 40.95 & 41.04 & 1626.9/1.02 [C]\\
\enddata
\tablecomments{Columns: 
(1) object name; 
(2) hydrogen column density of the host galaxy of the target in [10$^{22}$ cm$^{-2}$];
(3--4) power-law photon index of the XRB and scattered components, and its normalization at 1~keV in [10$^{-6}$~keV$^{-1}$ cm$^{-2}$ s$^{-1}$];
(5) cutoff energy in [keV];
(6) temperatures of the \textsf{apec} components in [keV];
(7--12) instrumental cross normalizations of Chandra/ACIS, XMM-Newton/MOS, pn, Suzaku/XIS-FI, XIS-BI, Swift/XRT, and BAT, respectively.
The mark 1$^{\rm (ref)}$ means that it is set to unity 
as the calibration reference when the NuSTAR spectra have too poor statistics;
(13--14) logarithmic observed fluxes in the 0.5--2~keV and 2--10~keV bands in [erg s$^{-1}$ cm$^{-2}$];
(15--16) logarithmic luminosities of the components of \textsf{apec} and unabsorbed X-ray emission in the 0.5--2~keV band in [erg s$^{-1}$];
(17--18) logarithmic observed luminosities in the 0.5--2~keV and 2--10~keV bands in [erg s$^{-1}$].
The fluxes and luminosities are corrected for Galactic absorption;
(19) value of Cash [C] or $\chi^2$ statistics, and the reduced one divided by degrees of freedom.
The mark (f) means a fixed value.
}
\end{deluxetable*}
\end{longrotatetable}

\clearpage
\begin{deluxetable}{lccc}[t]
\label{T7_bestpara-lines}
\tablecaption{Additional Emission Lines in X-ray Spectra}
\tablewidth{\columnwidth}
\tabletypesize{\scriptsize}
\tablehead{
\colhead{Object} &
\colhead{\ \ \ \ $A_{\rm Si(1.86\, keV)}$\ \ \ \ } &
\colhead{\ \ \ \ $E_{\rm Fe}$\ \ \ \ } &
\colhead{\ \ \ \ $A_{\rm Fe}$\ \ \ \ }\\
& (10$^{-6}$) & (keV) & (10$^{-6}$) 
}
\decimalcolnumbers
\startdata
\multicolumn{4}{c}{Stage-A,\,N (AGNs)}\\
\hline
MCG--03-34-064$^{ab}$ & 1.67$^{+0.27}_{-0.27}$ & \nodata & \nodata \\
NGC 7130 & 1.55$^{+0.66}_{-0.65}$ & \nodata & \nodata \\
NGC 7674 & \nodata & 6.97$^{\rm (f)}$ & 1.96$^{+0.83}_{-0.83}$ \\
\hline
\multicolumn{4}{c}{Stage-C,\,D (AGNs)}\\
\hline
UGC 5101 & 0.38$^{+0.23}_{-0.23}$ & \nodata & \nodata \\
NGC 3690 West & 2.44$^{+1.17}_{-1.18}$ & \nodata & \nodata \\
IRAS 13120--5453 & 1.02$^{+0.36}_{-0.35}$ & 6.67$^{+0.02}_{-0.03}$ & 1.04$^{+0.27}_{-0.26}$ \\
Arp 220W & 0.34$^{+0.17}_{-0.17}$ & 6.67$^{+0.02}_{-0.02}$ & 1.31$^{+0.24}_{-0.24}$ \\
\hline
\multicolumn{4}{c}{Stage-D (starburst-dominant or hard X-ray undetected sources)}\\
\hline
NGC 3256$^{b}$ & 9.08$^{+0.99}_{-0.91}$ & 6.66$^{+0.05}_{-0.04}$ & 1.07$^{+0.28}_{-0.27}$ \\
ESO 286-19$^{c}$ & \nodata & 6.56$^{+0.12}_{-0.32}$ & 0.45$^{+0.41}_{-0.31}$ \\
\enddata
\tablecomments{Columns:
(1) object name;
(2) normalization of the Si K$\alpha$ line at 1.86~keV;
(3--4) rest-frame central energy and normalization of the highly ionized Fe K$\alpha$ line.}
\tablenotetext{a}{The Ni K$\alpha$ line at 7.48~keV ($A_{\rm Ni}$ = 3.06$^{+1.25}_{-1.23}$ $\times 10^{-6}$) is added.}
\tablenotetext{b}{The \ion{Mg}{11} K$\alpha$ line at 1.342~keV ($A_{\rm Mg}$ = 2.07 $\pm$ 0.39 $\times 10^{-6}$ for MGC-03-34-064 and 7.39$^{+2.19}_{-2.11}$ $\times 10^{-6}$ for NGC 3256) is added.}
\tablenotetext{c}{The \ion{O}{8} K$\alpha$ line at 0.654~keV ($A_{\rm O}$ = 3.29$^{+1.66}_{-1.52}$ $\times 10^{-6}$) is added.}
\end{deluxetable}

\section{X-ray and multiwavelength Results}
\label{S5_results}

Out of our sample (57 U/LIRG systems consisting of 84 individual
galaxies observed with NuSTAR and/or Swift/BAT), we have carried out a
systematic spectral analysis for 59 galaxies by adopting the AGN model
or the starburst-dominant model. For the other 10 galaxies mentioned in 
Section~\ref{sub2-5_Xray-ref}, we refer to the literature 
for the spectral analysis results. Among these 69 galaxies, 
there are 40 hard X-ray detected AGNs
(Section~\ref{subsub4-2-1_AGNmodel}). 
According to their hydrogen column densities, we
categorize them as two unobscured AGNs ($N_{\rm H}^{\rm LoS} <
10^{22}$~cm$^{-2}$), 21 obscured AGNs ($N_{\rm H}^{\rm LoS}$ =
10$^{22}$--10$^{24}$~cm$^{-2}$), 16 CT AGNs ($N_{\rm H}^{\rm LoS} \geq
10^{24}$~cm$^{-2}$), and one jet-dominant AGN (NGC~1275).

In this section, we discuss the X-ray and multiwavelength properties of
our sample as follows: Section~\ref{sub5-1_Xcolor} describes the X-ray color 
and AGN classification, Section~\ref{sub5-2_CTfraction} the fractions of 
CT AGNs for different merger stages, Section~\ref{sub5-3_properties} 
the properties of AGNs and host galaxies, in particular the bolometric 
AGN luminosities (Section~\ref{subsub5-3-1_bolometric}), the
features of multiphase outflows (Section~\ref{subsub5-3-2_outflows}), 
the black hole masses (Section~\ref{subsub5-3-3_BHmass}), 
and the Eddington ratios and bolometric correction
factors (Section~\ref{subsub5-3-4_Edd}).

\subsection{X-ray Color and AGN Classification}
\label{sub5-1_Xcolor}

The NuSTAR band ratio (BR$^{\rm Nu}$) is useful to roughly
constrain the presence of an AGN and its classification even for hard
X-ray undetected sources. In fact, the two starburst-dominant sources have much
smaller band ratios than those of the AGNs. 
\citet{Lansbury2017} report that CT AGNs can be 
efficiently selected 
by applying a cut at BR$^{\rm Nu} > 1.7$ in the 40-month NuSTAR 
serendipitous survey.   

\begin{figure*}
    \epsscale{0.75}
    \plotone{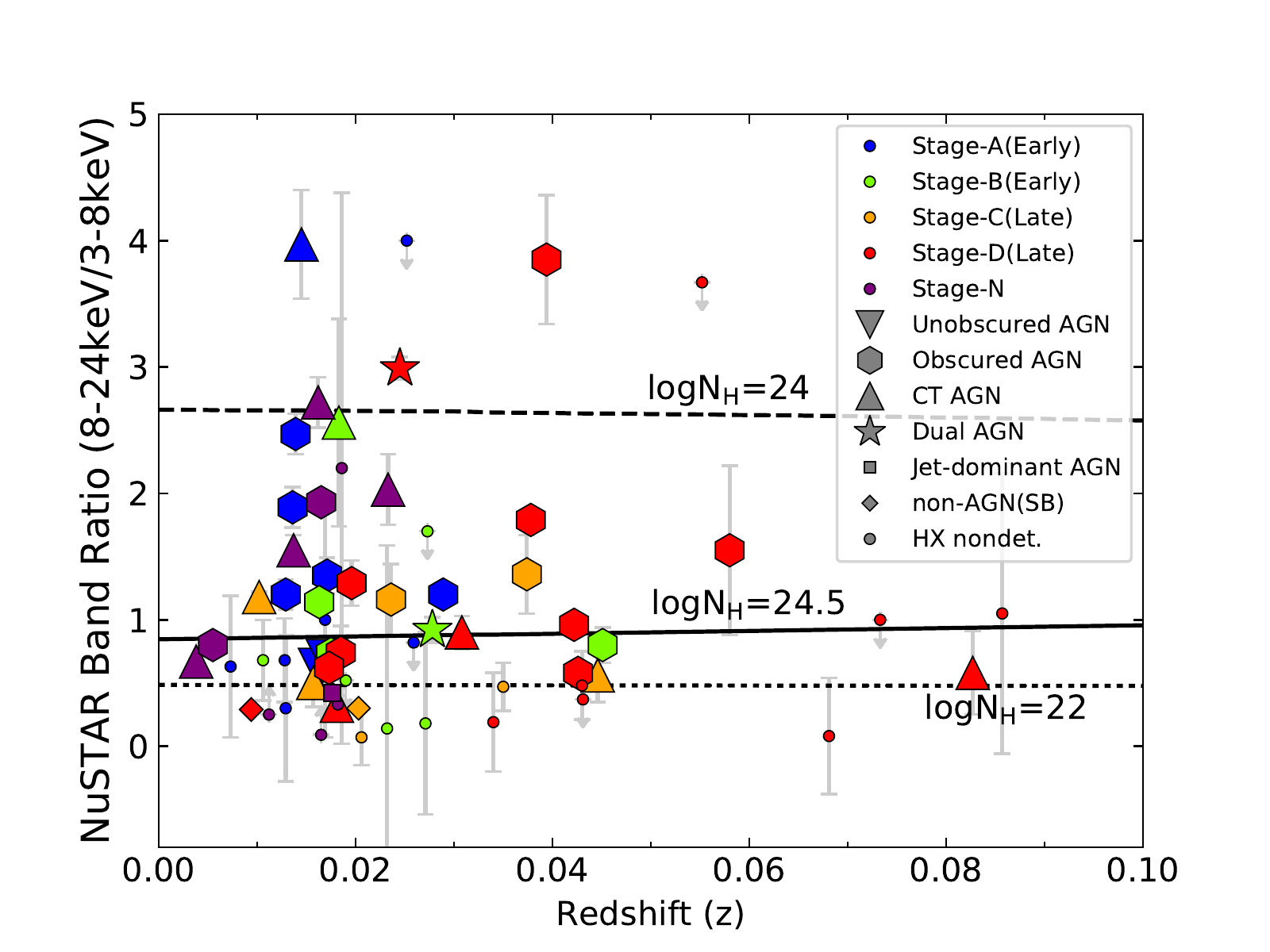}
        \caption{NuSTAR band ratio vs. redshift for the unobscured AGNs, obscured AGNs, CT AGNs,
        dual-AGNs (Mrk 266 and NGC 6240), jet-dominant AGN (NGC 1275),
 8--24~keV detected but starburst-dominant sources, and 8--24~keV undetected sources.
        These marks are color coded by merger stage. The black dotted, dashed and solid curves show the band ratio 
        assuming an unobserved soft X-ray fraction of $f_{\rm bin}$ = 1\% and $N_{\rm H}^{\rm LoS}$ = 10$^{22}$, 10$^{24}$, and 10$^{24.5}$~cm$^{-2}$, respectively.\\
        \label{F2_BR-NuSTAR}}
\end{figure*}

Figure~\ref{F2_BR-NuSTAR} plots the NuSTAR band ratio against redshift 
for all objects. 
The curves are calculated based on the AGN model by assuming 
a photon index of $\Gamma_{\rm AGN}$ (= $\Gamma_{\rm bin}$) = 1.8, an
unabsorbed X-ray emission fraction of $f_{\rm bin} = 1\%$, a torus
angular width of $\sigma = 20\degr$, and an inclination of $i =60\degr$
(for $N_{\rm H}^{\rm LoS}$ = 10$^{22}$~cm$^{-2}$) or $i =80\degr$ (for
$N_{\rm H}^{\rm LoS} \geq$ 10$^{24}$~cm$^{-2}$).
As noticed, BR$^{\rm Nu} \gtrsim 0.5$, corresponding to 
$N_{\rm H}^{\rm LoS} > 10^{22}$~cm$^{-2}$ (black
dotted curve), may be used as a selection criterion 
to identify AGNs in local U/LIRGs, mostly obscured ones
(37/40 AGNs), by 
discriminating from starburst-dominant galaxies.
This implies that there may be a few more AGN candidates even in the hard
X-ray undetected (S/N $<$ 3$\sigma$) sources, which show BR$^{\rm Nu}
\sim$0.0--1.0 within the uncertainties.

We note that the relation of the NuSTAR band ratio to AGN type (CT or
not) differs with merger stage. Most of CT AGNs in stage-A, B, and N
systems (5/7 CT AGNs, including the dual AGN in Mrk~266)
have the band ratios of $\gtrsim$1.6, well consistent with the
threshold value of 1.7 for normal AGNs \citep{Lansbury2017}. By
contrast, the CT AGNs in stage-C and stage-D mergers have smaller band
ratios of $\approx$0.5--1.2. This is because (1)
some of them are heavily CT ($N_{\rm H} \gtrsim 10^{24.5}$~cm$^{-2}$)
where the transmitted AGN components become fainter, and (2) the
unabsorbed emission fractions are large ($f_{\rm bin} \sim$ 10\%) most
probably due to the enhanced contribution from the XRBs
associated with the intense starbursts \citep[e.g.,][]{Mineo2012a}.


The hard X-ray undetected sources show small BR$^{\rm Nu}$
values ($\sim$0.0--1.0). This suggests that many of them may be starburst-dominant 
(see also Section~\ref{sub6-2_upperlim}), although
we cannot exclude the presence of AGNs in these objects.
The possible AGNs in stage-A, B, and N systems 
are not likely to be CT by considering the
threshold value of 1.7. Whereas, those in stage-C and D mergers 
may well be CT AGNs within the uncertainties in BR$^{\rm Nu}$. 
These implications are consistent with 
a larger fraction of CT AGNs in late mergers than in early
ones for the hard X-ray detected AGNs as reported by \citet{Ricci2017bMNRAS}
and shown in the next section.

It is also noticeable that the fractions of 
merger stages among the hard X-ray detected AGNs are different between
the low and high redshift ranges. At $z \lesssim 0.03$, AGNs are
detected in all merger stages, whereas
most of AGNs at $z \gtrsim 0.03$ are obscured ones in 
stage-C and stage-D mergers (consisting of 
six obscured and three CT AGNs). This may be subject to a 
selection bias that only luminous (but not CT) AGNs can be
detectable with NuSTAR at higher redshifts.
We have to bear it in mind when discussing statistical properties
of our sample.

\subsection{Fraction of CT AGNs}
\label{sub5-2_CTfraction}

It is theoretically predicted that merger-driven dynamics may trigger
heavy obscuration from the galactic scales to nuclear scales around the
SMBH \citep[e.g.,][]{Barnes1991,Mihos1996,Springel2003,Di-Matteo2005,Springel2005,Hopkins2006,Angles-Alcazar2017}.
If outflows caused by the AGN do not significantly affect the gas
distribution, the nucleus should be surrounded roughly spherically by
gas and dust falling from random directions. The numerical simulations of
mergers on a scale of $\gtrsim$48~pc \citep{Blecha2018} suggest that the
peak column densities reach the CT regime, and even the average ones
become $\sim$3~$\times$~10$^{23}$~cm$^{-2}$ during mergers. Recent
higher resolution simulations by \citet{Kawaguchi2020} also show the most
luminous phase of the AGN is obscured by gas with $N_{\rm H} \gtrsim
10^{24}$~cm$^{-2}$ for an edge-on view and $N_{\rm H} \gtrsim
10^{23}$~cm$^{-2}$ for a face-on view. 
It is expected that the central SMBHs rapidly grow in these 
heavily obscured environments.

Several observational approaches to prove heavy obscuration in
galaxy mergers have been attempted. Recent observations in the submillimeter
band report the evidence for compact obscuring material with $N_{\rm
H} > 10^{24}$~cm$^{-2}$ around the nuclei on the tens-pc scales in
ULIRGs, known as compact obscuring nuclei (CON;
\citealt{Sakamoto2010,Costagliola2013,Aalto2015,Aalto2019,Scoville2015,Falstad2021}).
In the mid-IR band, \citet{Yamada2019} suggest that the AGNs in
late mergers are deeply buried in geometrically thick tori
\citep[e.g.,][]{Ueda2007,Ueda2015,Imanishi2008}, by using the luminosity
ratio between the nuclear 12~$\mu$m \citep{Asmus2014} and [\ion{O}{4}]
25.89~$\mu$m emission, which originate from dusty tori and narrow
line regions (NLRs), respectively. In addition, by analyzing the X-ray
spectra of 30 interacting U/LIRGs including 25 AGNs,
\citet{Ricci2017bMNRAS} study the relation between AGN obscuration and
galaxy mergers. They find that all AGNs in late mergers have
large column densities of $N_{\rm H} > 10^{23}$~cm$^{-2}$ toward the
nuclei. The estimated fraction of CT AGNs in late mergers is
higher ($f_{\rm CT}$ = 65$^{+12}_{-13}$\%) than in early-merger
stage ($f_{\rm CT}$ = 35$^{+13}_{-12}$\%) and in local hard X-ray
selected AGN ($f_{\rm CT}$ = 27 $\pm$ 4\%;
\citealt{Burlon2011,Ricci2015}). This is in line with 
the presence of compact CT material in late mergers.

\begin{figure}
\epsscale{1.20}
    \plotone{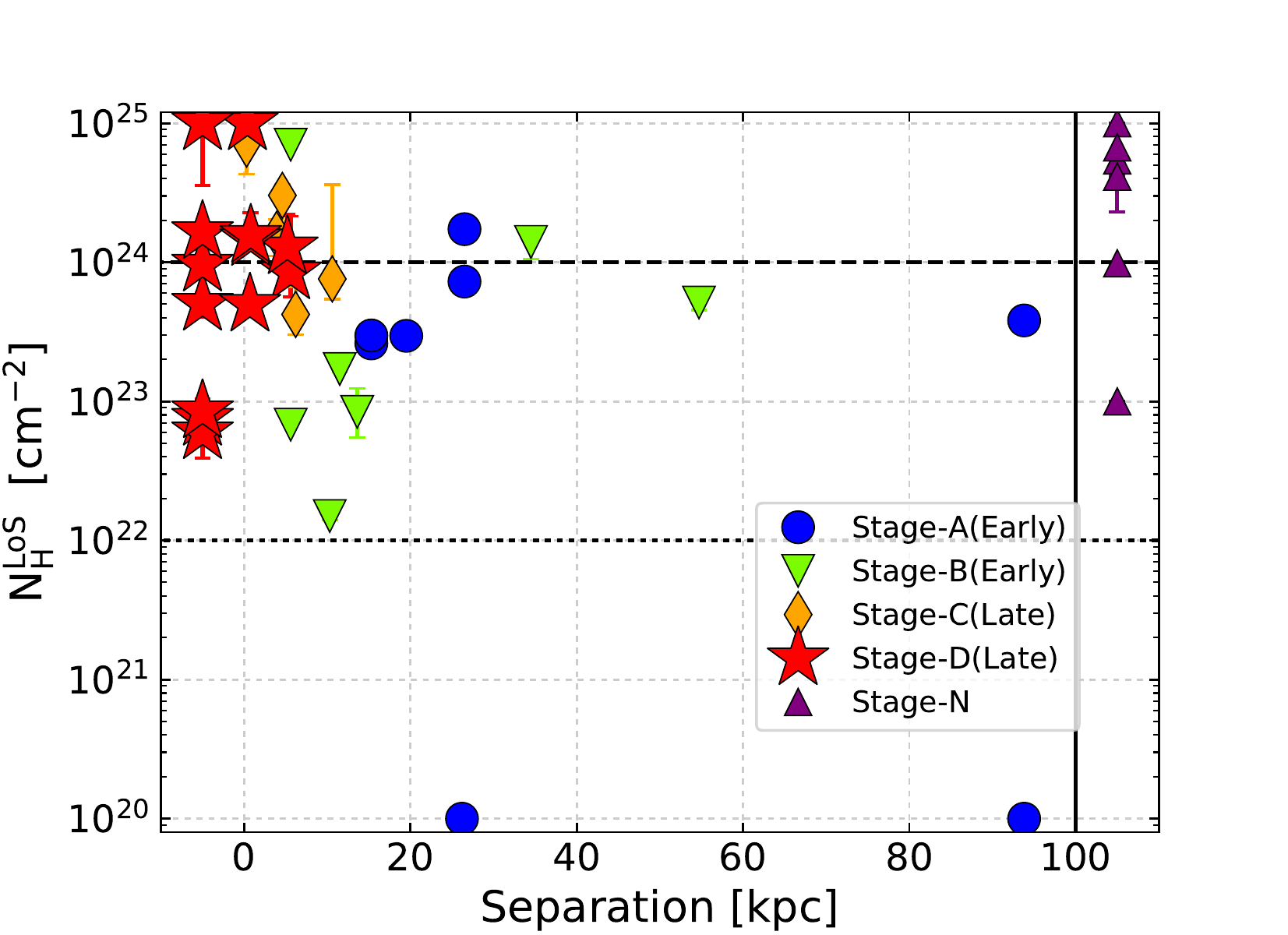}
        \caption{Line-of-sight hydrogen column density vs. galaxy separation.
        The blue circles, green triangles, orange diamonds, red stars, and purple triangles denote the AGNs 
        in stage-A, B (early mergers), C, D (late mergers), and stage-N (nonmergers) U/LIRGs, respectively.
        Single nuclei in merging (stage-D) and nonmerging (stage-N)
 sources are plotted on the left and right ends, respectively.
        The black dotted and dashed lines denote $N_{\rm H}^{\rm LoS} = 10^{22}$~cm$^{-2}$ and $10^{24}$~cm$^{-2}$, respectively.
        The unobscured AGNs (NGC 7469 and NGC 7679) are assigned upper limits of $N_{\rm H}^{\rm LoS}$ = 10$^{20}$~cm$^{-2}$.
        \label{F3_NH-kpc}}
\end{figure}

In this work, we obtain thus-far the largest sample consisting of 39
hard X-ray detected AGNs selected from the GOALS sample (except for the
jet-dominant AGN in NGC~1275). Figure~\ref{F3_NH-kpc} represents 
the relation between
galaxy separation and line-of-sight column density, color-coded by the
merger stages. In our sample, the fractions of CT AGNs are found to be
$f_{\rm CT}$ = 50~$\pm$~11\% (9/18 sources)
in the late mergers and $f_{\rm CT}$ = 22$^{+12}_{-9}$\% 
(3/15 sources) in the early mergers. Here, the values and
uncertainties represent the 50th and 16th--84th quantiles of a binomial
distribution, respectively, calculated with the beta function
\citep{Cameron2011}. The obtained values are a little smaller than the
results of \citet{Ricci2017bMNRAS}.\footnote{The fraction of the CT AGNs
in late mergers with a separation of 0.4--10.8~kpc is
53$^{+13}_{-12}$\% (7/13 sources), 
which is also smaller than the peaked value
(77$^{+13}_{-17}$\%) reported by \citet{Ricci2017bMNRAS}.} 

We find that the absorption column densities of NGC 7674 
and NGC 7682 are smaller than those reported in previous works,
mainly because of the difference of the X-ray spectral data utilized 
(see Appendix~\ref{Appendix-B}).
The slightly smaller column density of UGC 5101 
($N_{\rm H} \sim 9.6 \times 10^{23}$~cm$^{-2}$) than 
the one in the previous NuSTAR study 
($\sim$1.3 $\times 10^{24}$~cm$^{-2}$; \citealt{Oda2017}) 
also works to decrease the fraction of CT AGNs.
Nevertheless, our results support the
earlier result that most AGNs in late mergers are obscured with $N_{\rm
H} \gtrsim 10^{23}$~cm$^{-2}$, which is consistent with the numerical
simulations.

To avoid the possible selection bias mentioned in Section~\ref{sub5-1_Xcolor}, 
we also calculate the CT AGN fraction by using only AGNs at $z < 0.03$, 
which would give a more likely estimate of the intrinsic value. 
In this
case, the fraction of AGNs with $N_{\rm H} > 10^{24}$~cm$^{-2}$ in
late-merger stage is $f_{\rm CT}$ = 64$^{+14}_{-15}$\% 
(6/9 sources), where the torus
opening angle of the CT materials is calculated to be 50 $\pm$
11~degrees assuming the uniform-density spherical geometry with bipolar
cone holes.  
We obtain a similar value, $f_{\rm CT}$ = 80$^{+11}_{-16}$\% 
(6/7 sources), by using
sources at projected separations of 0.3--5.6~kpc.\footnote{If we
exclude the AGNs for which we assume $f_{\rm scat}$ = 10\%, these
fractions are calculated to be 68$^{+14}_{-17}$\% 
(5/7 sources) in late mergers and
77$^{+13}_{-17}$\% (5/6 sources)
among sources with 0.3--5.6~kpc separations.}
The fraction of CT AGNs in early mergers ($f_{\rm CT}$ =
24$^{+12}_{-10}$\%; 3/14 sources) coincides with that obtained from Swift/BAT AGNs
(27 $\pm$ 4\%; \citealt{Ricci2015}) within the uncertainty. 
Among the stage-N
sources, four out of six AGNs are CT AGNs;  the two CT AGNs in NGC~1068 and
NGC~7130 may be affected by past mergers
\citep{Davies2014,Tanaka2017}. By analyzing the X-ray spectra with
XCLUMPY, \citet{Yamada2020} indicate that the other two CT AGNs in
UGC~2608 and NGC~5135 are a normal AGN population seen close to
edge-on through a large line-of-sight column density.
We note that this CT-AGN fraction in stage-N sources ($f_{\rm CT}$ =64$^{+15}_{-18}$\%; 4/6 sources)
is highly uncertain under the small statistics and may be subject to the
sample selection bias as mentioned in Section~\ref{sub2-4_bias}.
Therefore, these results for AGNs at $z < 0.03$ support the scenario
that AGNs in late mergers are heavily obscured by surrounding CT
material, while those in early mergers and nonmergers are not deeply
buried, as reported by recent works
\citep[e.g.,][]{Ricci2017bMNRAS,Yamada2019}.

\subsection{Properties of AGNs and Host Galaxies from the X-ray and Multiwavelength Observations}
\label{sub5-3_properties}

\subsubsection{Bolometric AGN Luminosities}
\label{subsub5-3-1_bolometric}

The bolometric luminosity is one of the most important quantity to characterize
the AGN activity. It can be calculated from the 
absorption-corrected 2--10~keV luminosity with a bolometric correction
factor, $\kappa_{\rm bol,X} = L_{\rm bol,AGN}/L_{\rm
2-10}$. In many works, $\kappa_{\rm bol,X}$ = 20 or 30 is adopted as a
typical value \citep[e.g.,][]{Ricci2017cNature}.
Analyzing the optical, UV, to X-ray spectral energy distributions (SEDs) of
nearby AGNs, \citet{Vasudevan2007} show that $\kappa_{\rm bol,X}$
actually depends on the Eddington ratio ($\lambda_{\rm Edd} \equiv
L_{\rm bol,AGN}/L_{\rm Edd}$, where $L_{\rm Edd}= 1.26 \times 10^{38}$
$M_{\rm BH}$/$M_{\odot}$ is the Eddington limit for a black-hole mass of
$M_{\rm BH}$): $\kappa_{\rm bol,X} \approx$ 15--30 for $\lambda_{\rm
Edd} \lesssim 0.1$, $\kappa_{\rm bol,X} \approx$ 20--70 for
$\lambda_{\rm Edd} \sim$ 0.1--0.2, and $\kappa_{\rm bol,X} \approx$
70--150 for $\lambda_{\rm Edd} \gtrsim 0.2$.
However, \citet{Teng2015} report that AGNs in some ULIRGs have much
higher bolometric corrections than these values obtained for
normal AGNs, such as Mrk~231 ($\kappa_{\rm bol,X} \sim 3 \times 10^3$)
and IRAS~F08572+3915 ($\kappa_{\rm bol,X} \gtrsim 10^4$).  Thus, it is
important to estimate the bolometric luminosities from other wavelength
observations, in particular, those in the IR band
\citep[e.g.,][]{Veilleux2009a,Diaz-Santos2017,Shangguan2019,Toba2021a},
independently of X-ray measurements.

We summarize four major methodologies to estimate the bolometric AGN
luminosity of an U/LIRG from the IR spectroscopic or photometric
data. The first one is to use the [\ion{O}{4}] 25.89~$\mu$m luminosity. Its
ratio to the bolometric AGN luminosity is obtained for local
12~$\mu$m-selected Seyfert galaxies (12MGS; \citealt{Rush1993}) by
performing broadband SED decomposition
\citep{Gruppioni2016}. This can be widely applied for the GOALS objects
because their [\ion{O}{4}] fluxes are systematically measured from the
Spitzer/IRS spectra \citep[e.g.,][]{Inami2013}. The line emission comes
from AGN-excited ions in the NLR, whose luminosity should be
proportional to the AGN luminosity for a given geometry of the
NLR. Since the [\ion{O}{4}] line is far more robust to extinction and is less
contaminated by star formation activity than the optical [\ion{O}{3}]$\lambda
5007$ line, it gives a more reliable estimate of the bolometric AGN
luminosity for heavily obscured objects with high star formation rates
like U/LIRGs.  We note that the [\ion{O}{4}] 25.89~$\mu$m-to-bolometric
luminosity ratio becomes smaller for deeply buried AGNs in ULIRGs where
the NLR is less developed compared with normal AGNs
\citep[e.g.,][]{Imanishi2008,Yamada2019}.

The second one is to use the fractional AGN contribution to the 
bolometric ``total'' luminosity that includes the AGN and starburst
components. The total bolometric luminosity of an U/LIRG can be
calculated as 1.15 times the IR luminosity in the 8--1000~$\mu$m
band ($L_{\rm IR}$) \citep[e.g.,][]{Kim1998,Veilleux2009a,Teng2015}.
Table~\ref{T1_properties} lists $L_{\rm IR}$ for each U/LIRG system; 
for the interacting
systems, we also show identification by \citet{Inami2013} which galaxy
(or both) in the system mainly contributes to the observed IR
luminosity on the basis of the detection with Spitzer/IRS.
\citet{Diaz-Santos2017} calculate the fractional AGN contribution to the
bolometric total luminosity for each system in the GOALS sample with the
IRS observations (see also \citealt{Veilleux2009a,Petric2011}). They
employ up to five methods: the [\ion{Ne}{5}]$_{14.3}$/[\ion{Ne}{2}]$_{12.8}$ and
[\ion{O}{4}]$_{25.9}$/[\ion{Ne}{2}]$_{12.8}$ line flux ratios, the EW of the
6.2~$\mu$m polycyclic aromatic hydrocarbons (PAH; see also
\citealt{Armus2007}), the $S_{30}$/$S_{15}$ dust continuum ratio, and
the Laurent diagram \citep{Laurent2000}. We adopt the averaged
bolometric AGN fractions based on all
diagnostics.\footnote{\citet{Yamada2019} investigate the difference
between the mean AGN fractions calculated by including and excluding the
[\ion{Ne}{5}]$_{14.3}$/[\ion{Ne}{2}]$_{12.8}$ and [\ion{O}{4}]$_{25.9}$/[\ion{Ne}{2}]$_{12.8}$
line flux ratios because these may not good indicators in buried AGNs.
The averaged value of the three diagnostics is slightly larger than that
from the five methods by a factor of 1.1--1.7.}

The third and fourth methodologies are to directly estimate 
the bolometric AGN luminosities from
broadband SED decomposition and Spitzer/IRS spectral fitting,
respectively. \citet{Shangguan2019} present a detailed study of the
IR $\sim$1--500~$\mu$m SEDs of 193 U/LIRGs in the GOALS sample. The
entire sample has been uniformly observed with the Two Micron All Sky
Survey (2MASS; \citealt{Skrutskie2006}), Wide-field Infrared Survey
Explorer (WISE; \citealt{Wright2010}), and Herschel Space Observatory
\citep{Pilbratt2010}. They fit the SEDs by considering the torus
component for which they adopt a new version of the CAT3D
\citep{Honig2017} torus model. This model takes into account the different
sublimation temperatures for silicate and graphite dust 
(producing more emission from hot dust at the inner edge of the torus
than in the case of a single sublimation temperature)
and a wind component, which allows
greater flexibility to accommodate the diversity of IR SEDs of
quasars \citep{Zhuang2018}. The SED fits are conducted without the wind
component, whereas the Spitzer/IRS spectral fits 
for 61 objects are performed with an additional wind component. 
Assuming the bolometric-to-IR luminosity correction of the AGN
emission to be 3.0 \citep[e.g.,][]{Delvecchio2014}, we obtain the
bolometric AGN luminosities from the IR 8--1000~$\mu$m luminosity
of the torus component derived from the SED fitting and Spitzer/IRS
spectral analysis. For some U/LIRGs, we also refer to the bolometric AGN
luminosities obtained from Spitzer/IRS spectral analysis 
by \citet{Alonso-Herrero2012}.

We list the bolometric AGN luminosities for the hard X-ray detected
sources in Table~\ref{T8_AGN-properties} and 
those for starburst-dominant or hard X-ray undetected
sources, most of which should be considered as upper limits, in
Table~\ref{T9_SB-properties}.
By excluding the upper limits from the [\ion{O}{4}] 25.89~$\mu$m
luminosities, the bolometric AGN luminosities obtained from the four
methods are found to be approximately the same for each object typically with a
scatter of $\sim$0.27 dex (see
Appendix~\ref{Appendix-A}). Thus, we adopt these averaged values as the
best-estimates of the bolometric AGN luminosities.
In Tables~\ref{T8_AGN-properties} and \ref{T9_SB-properties}, 
we also list the star formation rate (SFR)\footnote{The SFRs 
are basically calculated from the IR luminosities of the host galaxies 
through the \citet{Kennicutt1998} relation \citep{Gruppioni2016,Shangguan2019}. 
\citet{Pereira-Santaella2015} obtain SFRs using a stellar emission model
that considers several star 
bursts of different ages, and report that their estimates are also in
good agreement with the \citet{Kennicutt1998} relation.}
obtained with SED decomposition and Spitzer/IRS spectral analysis
\citep{Pereira-Santaella2015,Gruppioni2016,Shangguan2019}.\footnote{We also 
refer to the SFRs obtained from the total (AGN and starburst) IR luminosities \citep{Howell2010}.
Thus, we treat the values for the hard X-ray detected AGNs (i.e., NGC~235 and Mrk~266B/Mrk~266A) as upper limits (listed in Table~\ref{T8_AGN-properties}).}

\subsubsection{Features of Multiphase Outflows}
\label{subsub5-3-2_outflows}

Recent multiwavelength observations have detected signals of outflows
on the various scales in U/LIRGs \citep{Fiore2017,Fluetsch2021}.
Ultrafast outflow (UFO), an extremely fast ($\sim$0.1--0.3$c$) and
highly-ionized wind emanating from a close vicinity ($\sim$10$^{-2}$~pc)
of an SMBH, is seen as X-ray blueshifted absorption lines in some
U/LIRGs \citep[e.g.,][]{Tombesi2015,Feruglio2015,Mizumoto2019}.  On kpc
scales, fast (an order of 1000~km~s$^{-1}$) and massive outflow of
ionized gas has been discovered by deep optical/near-IR
spectroscopy mainly from integral field observations (IFU; e.g.,
\citealt{Fischer2013,Rich2015,Cortijo-Ferrero2017b,Toba2017c,Kakkad2018,Venturi2018,Smith2019,U2019,Boettcher2020,Fluetsch2021}).
At the far-IR and millimeter/submillimeter wavelengths, molecular outflow, a
cold gas wind with a size of $\sim$400~pc and a velocity of
$\sim$500~km~s$^{-1}$, is observed in many U/LIRGs (e.g.,
\citealt{Spoon2013,Veilleux2013,Cicone2014,Gonzalez-Alfonso2017,Laha2018}
and the references therein; \citealt{Lutz2020};
but see e.g., \citealt{Toba2017d}).

\citet{Fiore2017} find strong correlations between the ionized and
molecular mass outflow rates and the bolometric AGN luminosity 
by using 94 AGNs with massive winds at sub-pc to kpc spatial scales.
They also find that the maximum outflow velocity correlates with the
bolometric AGN luminosity.
Thus, physical parameters of outflows, such as outflow
velocity ($V_{\rm outflow}$) and mass outflow rate ($\dot{M}_{\rm
outflow}$), provide rich information on mass transportation 
between an SMBH and the host galaxy. We summarize the properties
of outflows collected from the literature for hard X-ray detected
sources in Table~\ref{T8_AGN-properties} and for starburst-dominant or 
hard X-ray undetected sources in Table~\ref{T9_SB-properties}.

\subsubsection{Black Hole Masses in U/LIRGs}
\label{subsub5-3-3_BHmass}

In Table~\ref{T10_BHmass}, we list black hole masses estimated from four methods for
the hard X-ray detected AGNs. Two dynamical estimates of black hole
masses are available, one from the relation with stellar mass for local AGNs
\citep{Reines2015} and the other from that with stellar velocity
dispersion ($M$--$\sigma_{*}$ relation,
\citealt{Gultekin2009}). The stellar masses are mainly obtained from
SED decomposition \citep[e.g.,][]{Pereira-Santaella2015,Shangguan2019}.
The stellar velocity dispersions are taken from the literature
\citep[e.g.,][]{Dasyra2006a,Dasyra2006b}. We also collect the
photometric black-hole mass estimates
\citep[e.g.,][]{Winter2009,Veilleux2009c,Haan2011a} using the relation
with $H$-band bulge luminosity $L_{H,{\rm bulge}}$ \citep{Marconi2003},
given as log$M_{\rm BH}$ = 8.19($\pm$0.07)+1.16($\pm$0.12) $\times$
(log$L_{H,{\rm bulge}} - 10.8$). Finally, we list black-hole mass
estimates from other various methods: the flux density of old
stellar emission at 2~$\mu$m \citep{Caramete2010}; 
scaling relations from \citet{McConnell2013} and the central velocity
dispersion of the Calcium triplet absorption lines \citep{Koss2016a};
the reverberation mapping study \citep{Bentz2015}; a three-dimensional
plane connecting log($M_{\rm BH}$) to a linear combination of
logarithmic velocity dispersion and luminosities of the NLR lines
\citep{Dasyra2011}; 
the velocity dispersion of [\ion{O}{3}] emission \citep{Alonso-Herrero2012}; 
an empirical
relationship in \citet{Kaspi2000} using the H$\beta$ line width and
optical continuum luminosity at rest-frame 5100~\AA\ 
\citep{Kawakatu2007b}; and water masers
(\citealt{Klockner2004,Lodato2003}; see also \citealt{Izumi2016}).  The
last rows of Table~\ref{T10_BHmass} list the averages of the 
black-hole mass estimates derived from these methods.

\begin{figure*}
    \epsscale{1.15}
    \plottwo{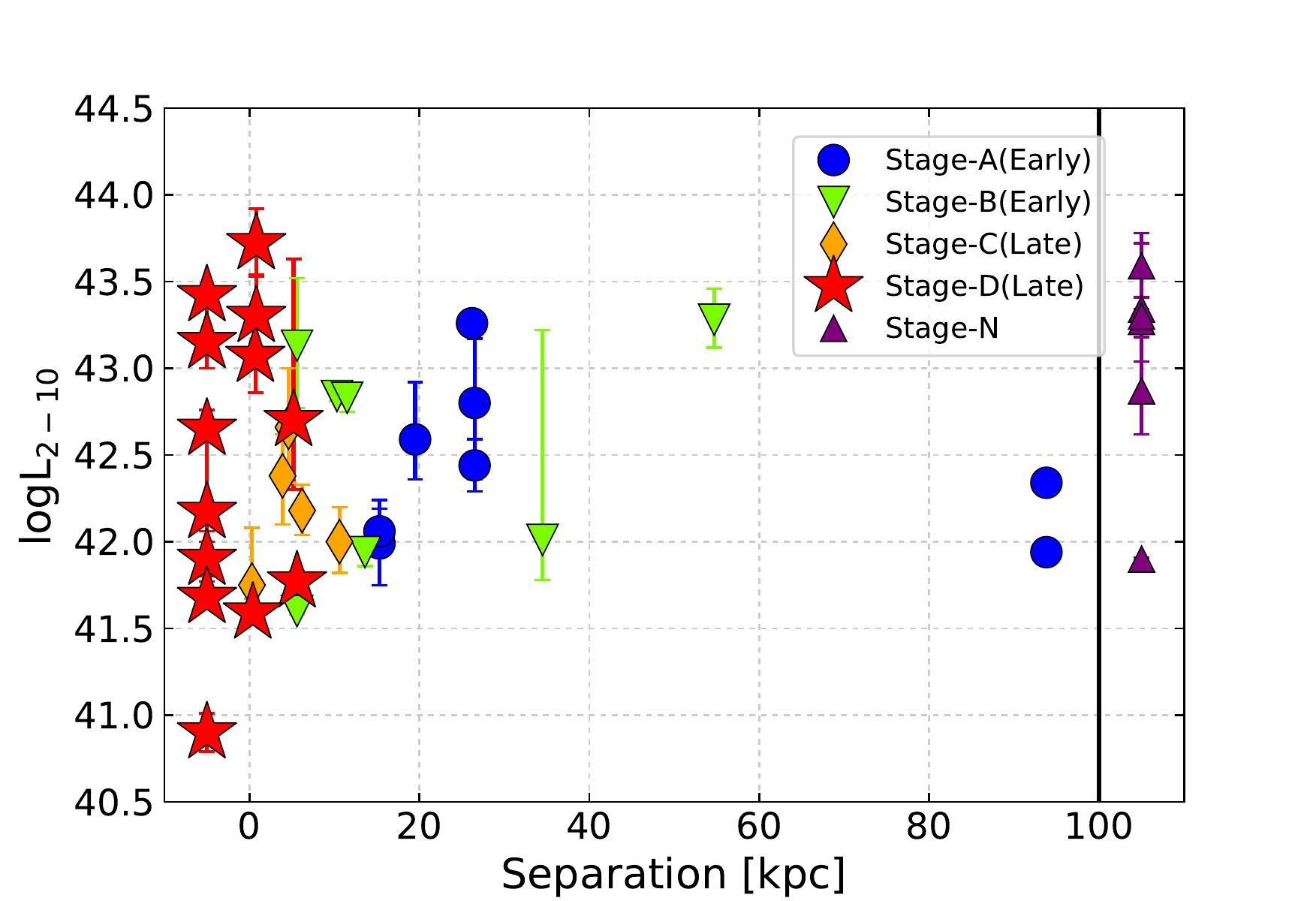}{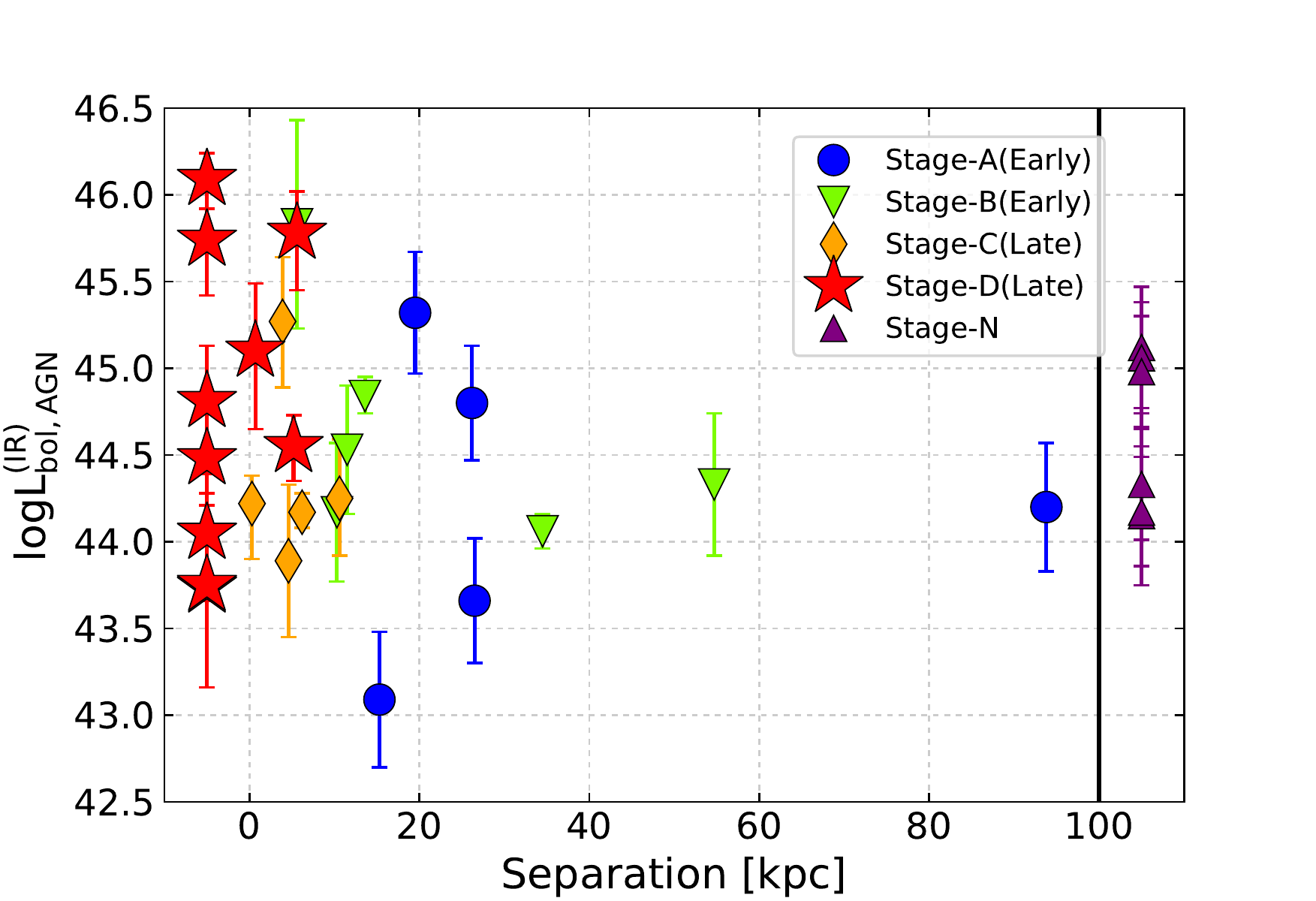}
        \caption{Left panel: absorption-corrected 2--10~keV luminosity vs. galaxy separation. 
        Right panel: bolometric AGN luminosity based on the IR results
 vs. galaxy separation. 
        Symbols and colors are the same as in Figure~\ref{F3_NH-kpc}.
        \label{F4_lumin-kpc}}
\end{figure*}

\begin{figure*}
    \epsscale{1.15}
    \plottwo{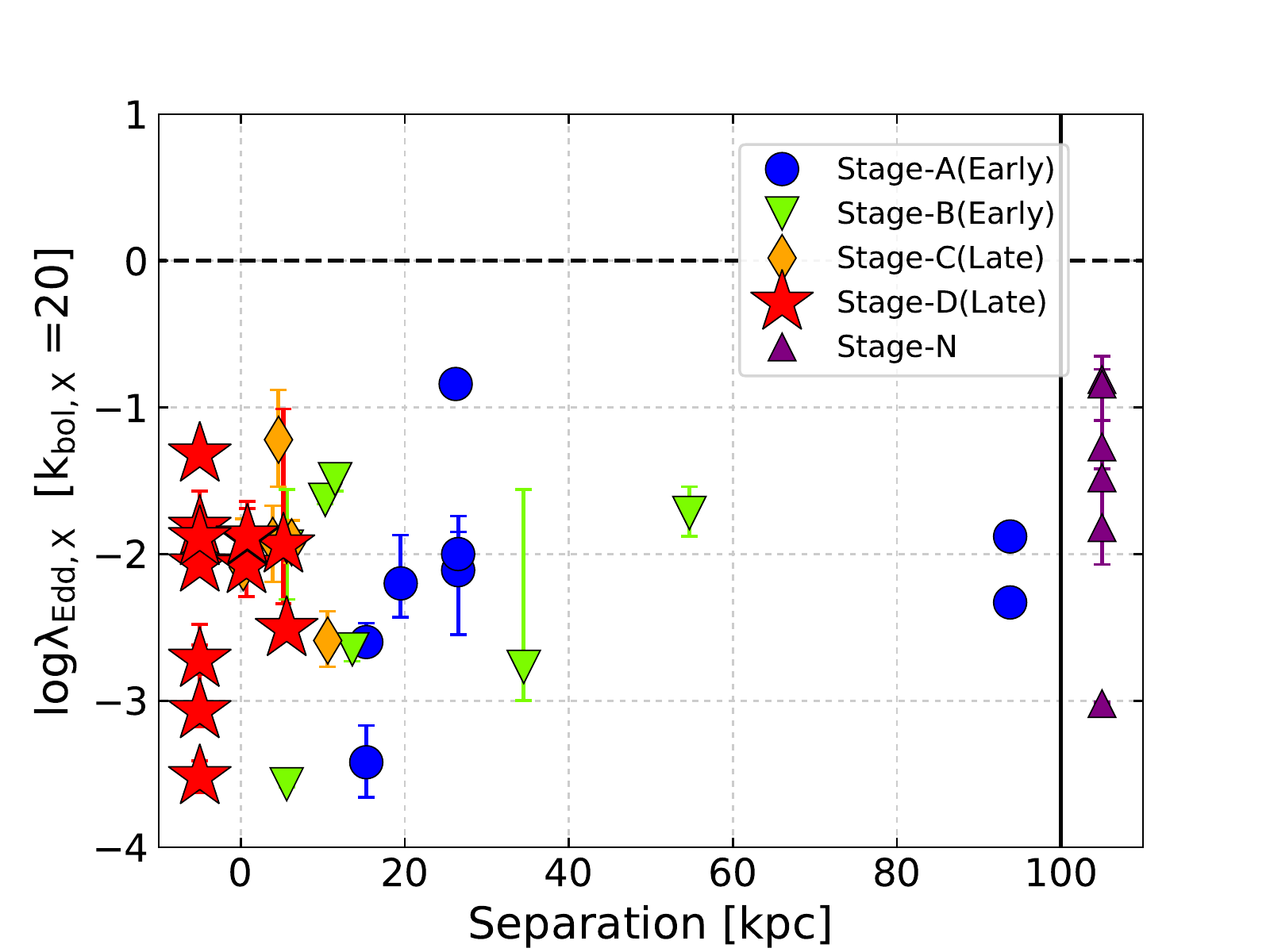}{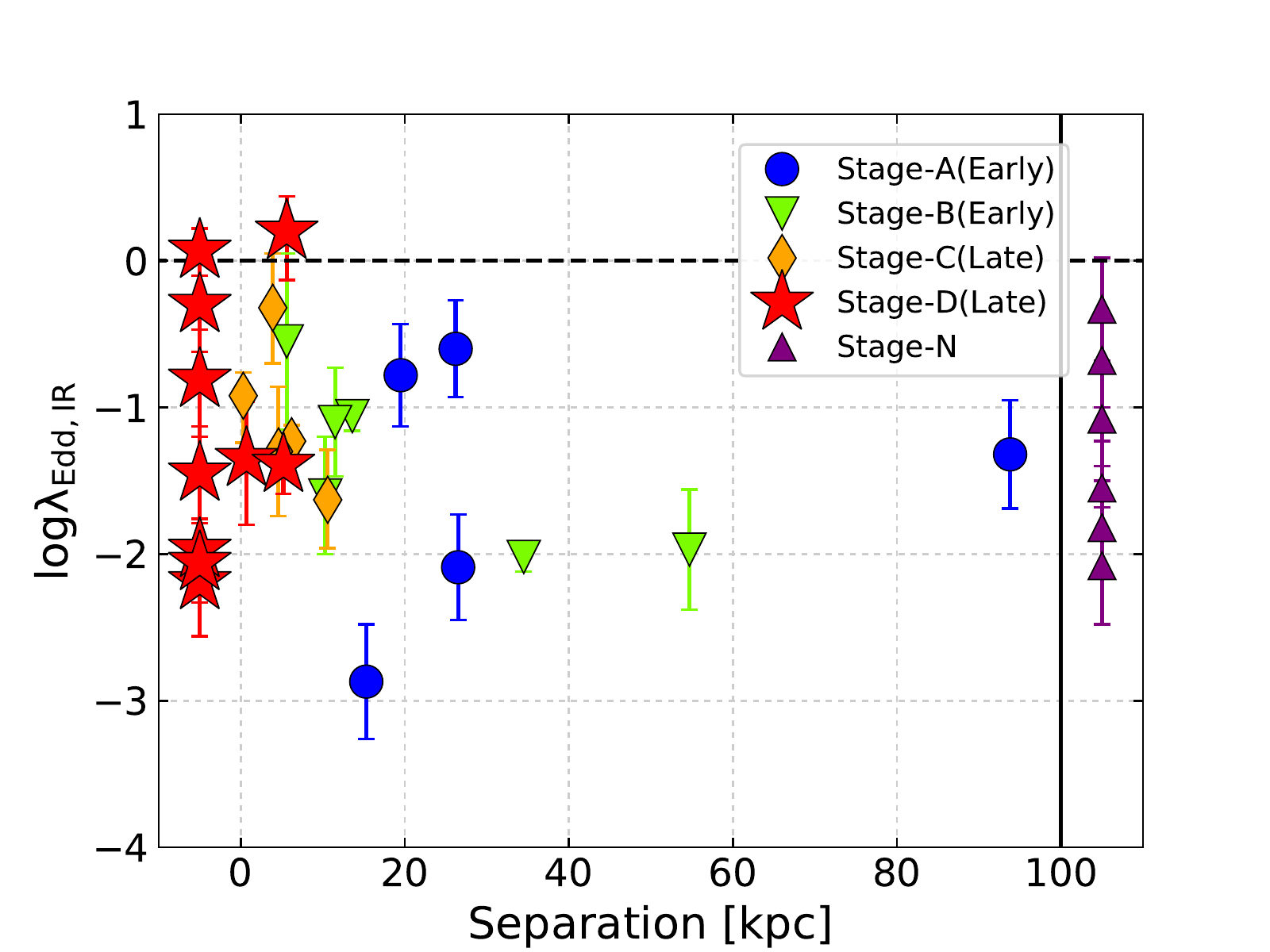}
        \caption{Left panel: X-ray based Eddington ratio vs. galaxy
 separation, where $\kappa_{\rm bol,X}$ (= $L_{\rm bol,AGN}^{\rm
 (IR)}$/$L_{2-10}$) = 20 is assumed.
        Right panel: IR based Eddington ratio vs. galaxy, where 
the bolometric AGN luminosities based on the IR results are adopted.
        The black dashed line shows the Eddington limit ($\lambda_{\rm Edd} = 1$).
        Symbols and colors are the same as in Figure~\ref{F3_NH-kpc}.
        \label{F5_Edd-kpc}}
\end{figure*}

We find that the black-holes mass measurements in U/LIRGs
from dynamical stellar mass and velocity dispersion are
systematically smaller by a factor of $\sim$4--9 ($\sim$0.55--0.95~dex) 
than those from the other methods (see bottom rows in Table~\ref{T10_BHmass}).
\citet{Veilleux2009c} report this tendency in U/LIRGs, who 
conclude that it is caused by a 
systematic difference in the velocity dispersion 
between the optical (\ion{Ca}{2}) and near-IR (CO) measurements. 
The photometric black-hole mass estimates may also have large
uncertainties because of possible effects from non-nuclear star formation,
dust extinction, and point-spread function subtraction.  Particularly,
the photometric estimates would be overestimated in smaller mass hosts
with smaller black-hole masses (see \citealt{Veilleux2009c}). 
These systematic differences 
seem not to be correlated with either the bolometric AGN luminosities or 
Eddington ratios, as noted in Appendix~\ref{Appendix-A}.
To minimize the selection effects of these measurements, we decide 
to adopt the averaged value of the black-hole masses estimated 
from the four methods. 
Hence, the Eddington ratios should be considered to have
systematic uncertainties of $\sim$0.3--0.5~dex originating from those in
the black-hole mass estimates.

\begin{figure*}
    \epsscale{1.15}
    \plottwo{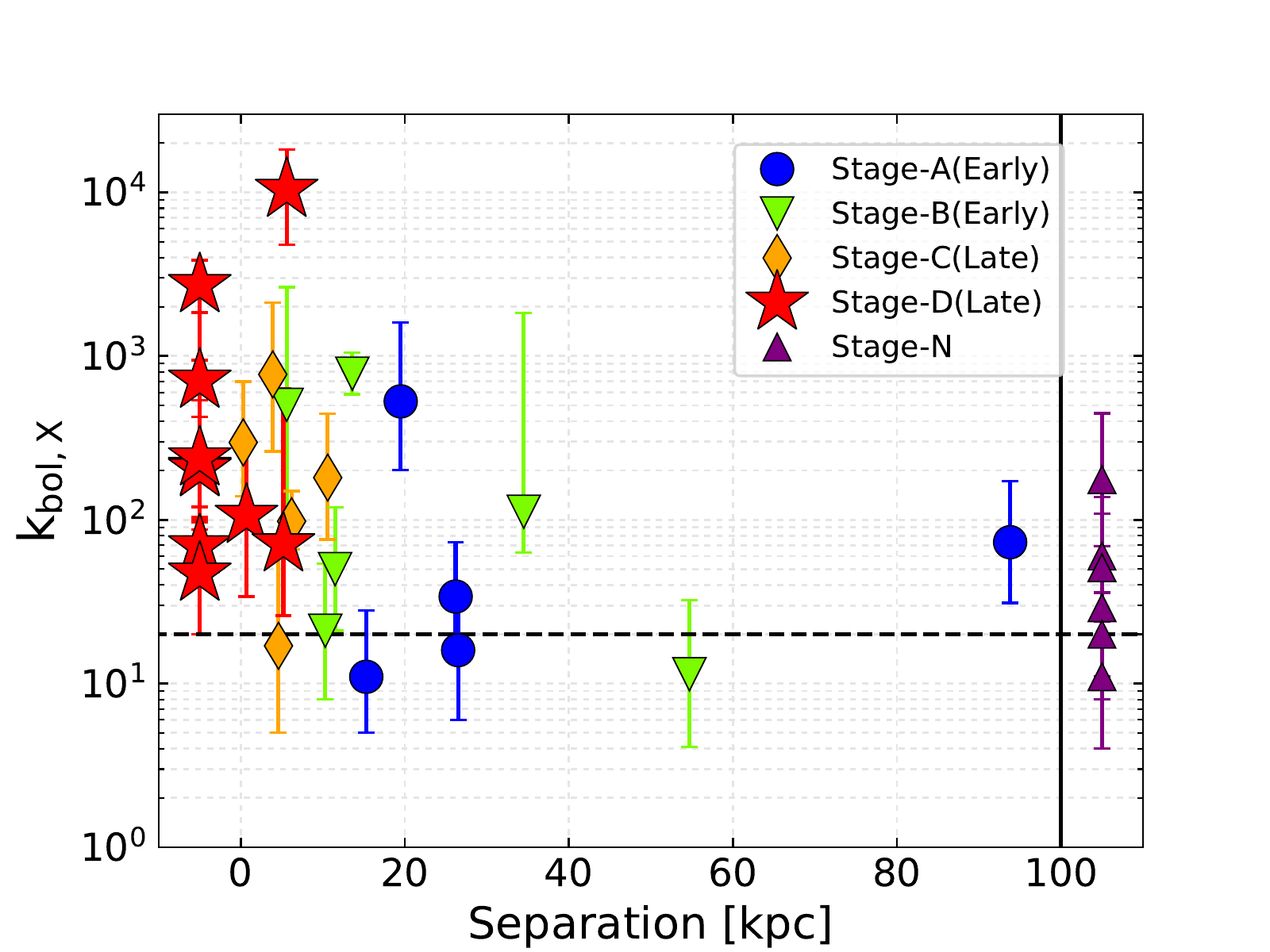}{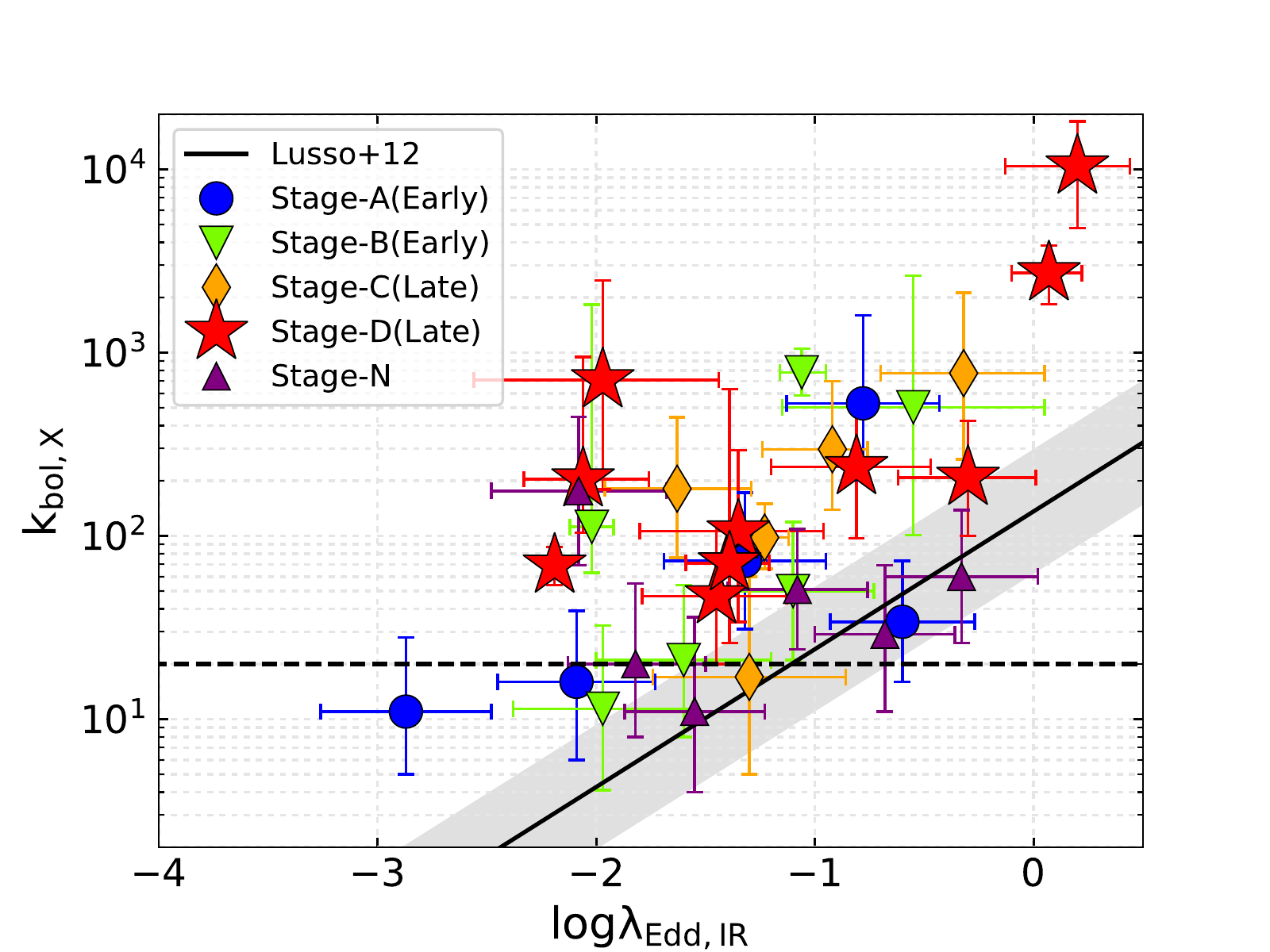}
        \caption{Left panel: bolometric-to-X-ray correction ($\kappa_{\rm
 bol,X}$ = $L^{\rm (IR)}_{\rm bol,AGN}/L_{2-10}$) vs. galaxy separation. 
        Right panel: bolometric-to-X-ray correction vs. IR-based Eddington ratio.
        The black solid curve and gray shaded area represent the
relation and its 1$\sigma$ dispersion found by \citet{Lusso2012} for type-1 AGNs.
        The black dashed line shows a typical value of $\kappa_{\rm
 bol,X}$ = 20. Symbols and colors are the same as in Figure~\ref{F3_NH-kpc}.
        \label{F6_Kx-Edd}}
\end{figure*}

\begin{figure}
    \epsscale{1.20}
    \plotone{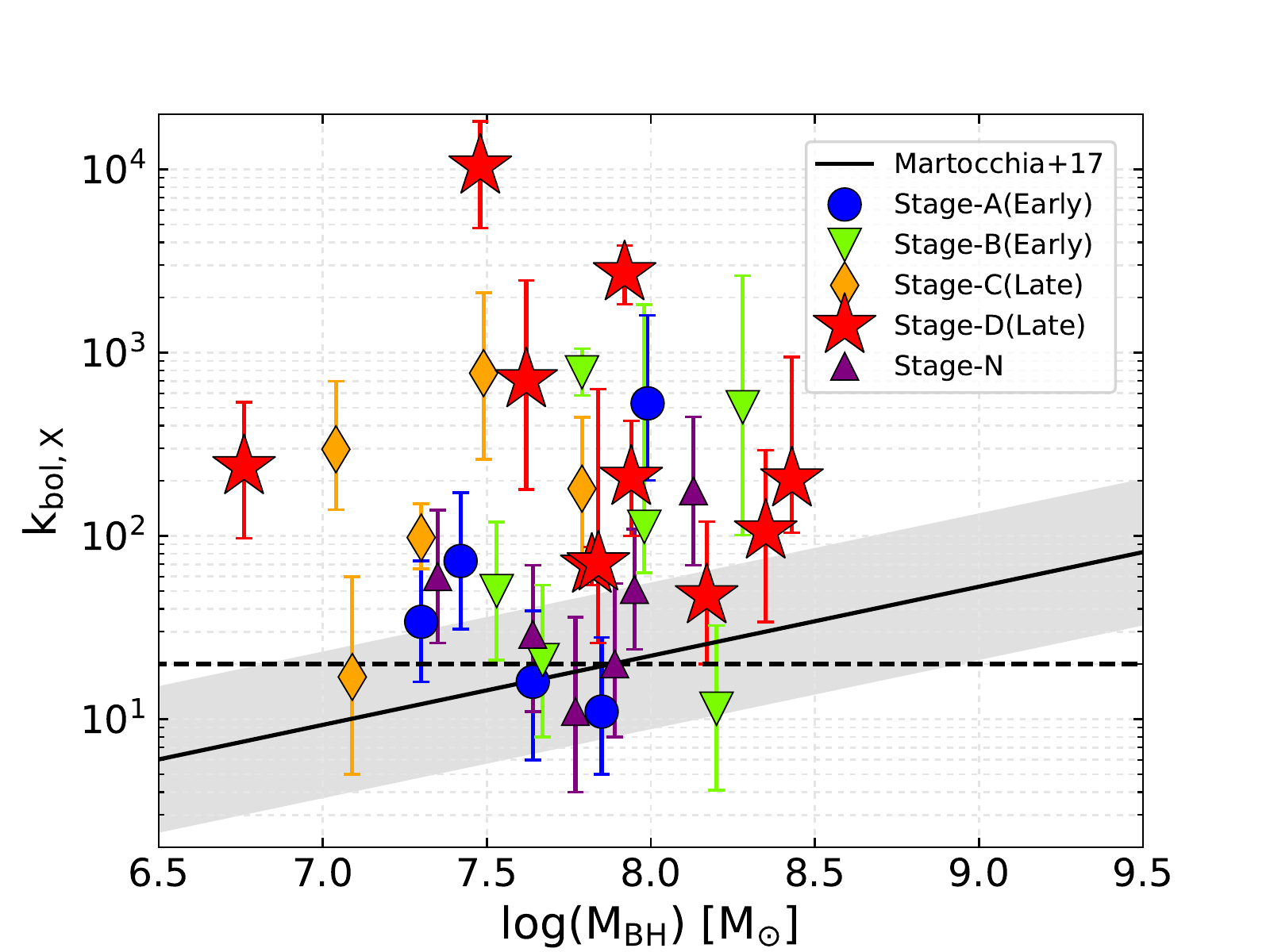}
        \caption{Bolometric-to-X-ray correction vs. black hole mass. 
        The black solid curve and gray shaded area represent the relation and its 1$\sigma$ dispersion found by \citet{Martocchia2017} for WISSH quasars, PG quasars, and XMM-Newton-COSMOS Seyfert 1s.
        The black dashed line shows a typical value of $\kappa_{\rm
 bol,X}$ = 20. Symbols and colors are the same as in Figure~\ref{F3_NH-kpc}.
        \label{F7_Kx-BHmass}}
\end{figure}

\subsubsection{Eddington Ratios and Bolometric Corrections}
\label{subsub5-3-4_Edd}

For the hard X-ray detected AGNs, Table~\ref{T11_Edd-Kx} summarizes 
the absorption-corrected
X-ray luminosities in the 2--10~keV and 10--50~keV bands ($L_{\rm
2-10}$ and $L_{10-50}$), the averaged bolometric
AGN luminosities from the IR results ($L_{\rm bol,AGN}^{\rm
(IR)}$), the averaged black hole mass estimates ($M_{\rm BH}$), the
Eddington ratios from the X-ray (assuming $\kappa_{\rm bol,X} = 20$) and
IR results ($\lambda_{\rm Edd,X}$ and $\lambda_{\rm Edd,IR}$), the
bolometric corrections ($\kappa_{\rm bol,X}$), and the torus covering
factors ($C_{\rm T}$) derived from the torus parameters of XCLUMPY
excluding the case when the torus angular width is fixed to 20\degr\ (see
Section~\ref{subsub6-1-4_CoveringFactor} for detailed discussion of $C_{\rm T}$).

Figures~\ref{F4_lumin-kpc} and~\ref{F5_Edd-kpc} show 
the relation between the $L_{2-10}$, 
$L_{\rm bol,AGN}^{\rm (IR)}$, $\lambda_{\rm Edd,X}$ and
$\lambda_{\rm Edd,IR}$ with the galaxy separation. 
From Figure~\ref{F4_lumin-kpc}, we find that the maximum X-ray 
and bolometric AGN luminosities increase with decreasing galaxy separation. 
A similar tendency of X-ray luminosities is found 
by a study of dual-AGN systems among the Swift/BAT 58-month AGNs
\citep{Koss2012}. From Figure~\ref{F5_Edd-kpc} we find that the maximum Eddington
ratios estimated by the IR results ($\lambda_{\rm Edd,IR}$)
increase with decreasing the separation. The peak values reach the Eddington
limits. This is consistent with recent numerical simulations
\citep[e.g.,][]{Kawaguchi2020}. The peak of the Eddington ratios
assuming $\kappa_{\rm bol,X}$ = 20 ($\lambda_{\rm Edd,X}$) is much
smaller than that from the IR. This may imply the X-ray weakness
in late-stage mergers. 

Whereas, some AGNs in late mergers have moderate or small luminosities
even from the IR estimates. 
The simulations of \citet{Blecha2018}
indicate that the maximum bolometric AGN luminosity depends on the
progenitor galaxy parameters such as total and stellar mass of the
galaxy, initial gas fraction in the galaxy disc, and bulge-to-total mass
ratio. In their simulations, black holes are modeled as gravitational
sink particles that accrete gas via an Eddington-limited, Bondi-Hoyle
like prescription. According to their results, the maximum bolometric
luminosities in gas-rich major mergers can be
$\sim$10$^{45.0}$--10$^{46.5}$~erg~s$^{-1}$, while those in gas-poor
mergers are $\sim$10$^{43.5}$--10$^{45.0}$~erg~s$^{-1}$
\citep[see][]{Blecha2018}. Thus, the bolometric AGN luminosities 
observed in the late mergers of our sample, 
$\sim$10$^{43.5}$--10$^{46.5}$~erg~s$^{-1}$, 
are consistent with these numerical simulations.
Furthermore, more recent numerical simulations of galaxy
mergers on a scale of 8~pc \citep{Yutani2021} show that the
bolometric AGN luminosities are dramatically variable by two orders of
magnitude in a few Myr. This long-term time variability may also explain the
dispersion of AGN luminosity in our sample.

Figure~\ref{F6_Kx-Edd} shows the bolometric correction, $\kappa_{\rm bol,X}$, as a function of
the galaxy separation and the Eddington ratio. From the left panel in
Figure~\ref{F6_Kx-Edd}, we find that the bolometric corrections in the late mergers are
much larger than those in the early mergers and nonmergers. In
fact, the trend of X-ray weakness has been reported for AGNs with
high Eddington ratios in late mergers 
\citep[e.g.,][]{Teng2015,Yamada2018} and higher-$z$ IR-bright dust
obscured galaxies (DOGs; e.g.,
\citealt{Ricci2017aApJ,Vito2018b,Toba2019,Toba2020}). Using the X-ray
selected sample of 929 AGNs from the XMM-Newton survey in the Cosmic
Evolution Survey field (XMM-COSMOS; \citealt{Hasinger2007}),
\citet{Lusso2012} provide the relation between bolometric correction
and Eddington ratio. However, as shown in the right panel
of Figure~\ref{F6_Kx-Edd}, the bolometric corrections in some of the late
(particularly stage-D) mergers are significantly higher than the
expected values from this relation by a factor of $\sim$10--100.

The excess of the bolometric factors in late mergers cannot be explained
by possible uncertainties in the bolometric luminosities because both of
the bolometric correction, $\kappa_{\rm X,bol}$ = $L_{\rm bol,AGN}^{\rm
(IR)}$/$L_{2-10}$, and Eddington ratio, $\lambda_{\rm
Edd,IR}$ = $L_{\rm bol,AGN}^{\rm (IR)}$/(1.26 $\times 10^{38}$ $M_{\rm
BH}$), are proportional to the bolometric luminosity. The uncertainties
in the black hole mass measurements ($\sim$0.3--0.5~dex, see 
Section~\ref{subsub5-3-3_BHmass})
are also too small to account for it. These facts indicate that the AGNs
in late mergers are truly X-ray weak.

\citet{Martocchia2017} perform a survey of the X-ray properties of 41
objects from the WISE/SDSS selected hyper-luminous (WISSH) quasars at $z
\sim$ 2--4.  These IR sources have large bolometric corrections of
$\sim$100--1000, and therefore they may be similar population to our
sample. They point out a possible positive correlation
between bolometric correction and black hole mass, using the AGN sample
of the XMM-COSMOS survey \citep{Lusso2012}, PG quasars, and WISSH
quasars.  Their black hole masses are
$\sim$10$^{7.0}$--10$^{9.5}$~$M_{\odot}$ for the XMM-COSMOS AGNs
and PG quasars, and $\sim$10$^{9.5}$--10$^{11.0}$~$M_{\odot}$ for the WISSH
quasars. Figure~\ref{F7_Kx-BHmass} show the relation of bolometric correction 
and black hole mass of our sample. We find that our AGNs have
$M_{\rm BH} \sim$ 10$^{7.0}$--10$^{8.5}$~$M_{\odot}$ and do not follow
the relation of \citet{Martocchia2017}. We therefore suggest that the extreme X-ray
weakness may not be regulated by the black hole mass, but seems to be a
common feature among AGNs in U/LIRGs, many of which are
gas-rich mergers \citep[e.g.,][]{Veilleux2002,Kartaltepe2010}.
\\

\section{Discussion}
\label{S6_discussion}

\subsection{AGN Structure in U/LIRGs}
\label{sub6-1_AGN-structure}

\subsubsection{X-ray Weakness}
\label{subsub6-1-1_Xweak}

Applying the XCLUMPY model to the highest-quality broadband X-ray
spectra currently available, we have estimated the 
absorption-corrected 2--10~keV
luminosities and absorption column densities of the local U/LIRGs 
with the best accuracy so far. Combining them with the bolometric 
AGN luminosities estimated from the IR data in various methods, 
we have derived their bolometric-to-X-ray correction factors, 
$\kappa_{\rm bol,X}$. Our main finding is that the AGNs in
early mergers and nonmergers tend to follow the relation between 
bolometric correction 
and Eddington ratio obtained for normal AGNs \citep{Lusso2012}, whereas 
those in late mergers often show 
larger $\kappa_{\rm bol,X}$ values than the \citet{Lusso2012}
relation by a factor of $\sim$10--100 (see also \citealt{Teng2015}). As
noted in Section~\ref{subsub5-3-4_Edd}, the discrepancies cannot be attributed to the
uncertainties in the black hole masses and bolometric luminosities. 

We first examine a possibility that the 
absorption-corrected X-ray luminosities
might be underestimated due to time variability.
Since X-ray and IR emission from an AGN arise from the innermost
hot corona and circumnuclear regions ($\sim$1--10~pc scale),
respectively, an object could become apparently X-ray faint 
if it is observed in a fading phase of AGN activity 
\citep[e.g.,][]{Kawamuro2017,Ichikawa2019b,Ichikawa2019c,Cooke2020}.
For example, the AGN in NGC~7674 shows a large bolometric correction of
$\kappa_{\rm bol,X} = 529^{+1071}_{-328}$, although the galaxy is
assigned as a stage-A merger.
The long-term X-ray light curve of NGC~7674 shows an X-ray flux decrease 
by more than one order of magnitude between 1989 and 1996 \citep{Gandhi2017}, 
and hence it could still be in a fading phase
when the X-ray data were taken.\footnote{It is also 
possible that the column density in NGC 7674 is higher than 
our estimate as reported by \citep{Gandhi2017}, if the reflected-dominated AGN 
with no direct detection of transmitted emission is assumed (see also
Appendix~\ref{Appendix-B}).}
However, it is very unlikely that such time variability explains the 
all X-ray weak objects in our sample because 
the timescale of a decline, $10^3$--$10^4$~years 
\citep[e.g.,][]{Schawinski2010}, is much shorter than the lifetime of
ULIRGs ($\sim$30~Myr; e.g., \citealt{Hopkins2008,Inayoshi2018}).

Thus, we suggest that the AGNs in late merger U/LIRGs are 
X-ray weak in nature.
\citet{Teng2015} point out that the X-ray weakness is
also common in broad absorption line (BAL) quasars
\citep[e.g.,][]{Luo2013,Luo2014}, where 
AGN-driven outflows may partially block 
the X-ray emission from the innermost region. 
By Chandra follow-up studies of
BAL quasars at $z$ = 1.6--2.7, \citet{Liu2018} estimate the fraction of
X-ray weak AGNs is $\approx$6\%--23\% among the BAL quasar
population which is much larger than that among non-BAL quasars
($\lesssim$2\%; \citealt{Gibson2008}).
In fact, Mrk 231, a stage-D ULIRG in our sample, is known to be 
an X-ray weak BAL quasar \citep{Teng2014,Veilleux2016}.
In this context, it is important to 
investigate correlation of the X-ray weakness (i.e., bolometric
correction) with the properties of various-scale (sub-pc to
kpc) outflows based on multiwavelength observations.

\begin{figure}
    \epsscale{1.20}
    \plotone{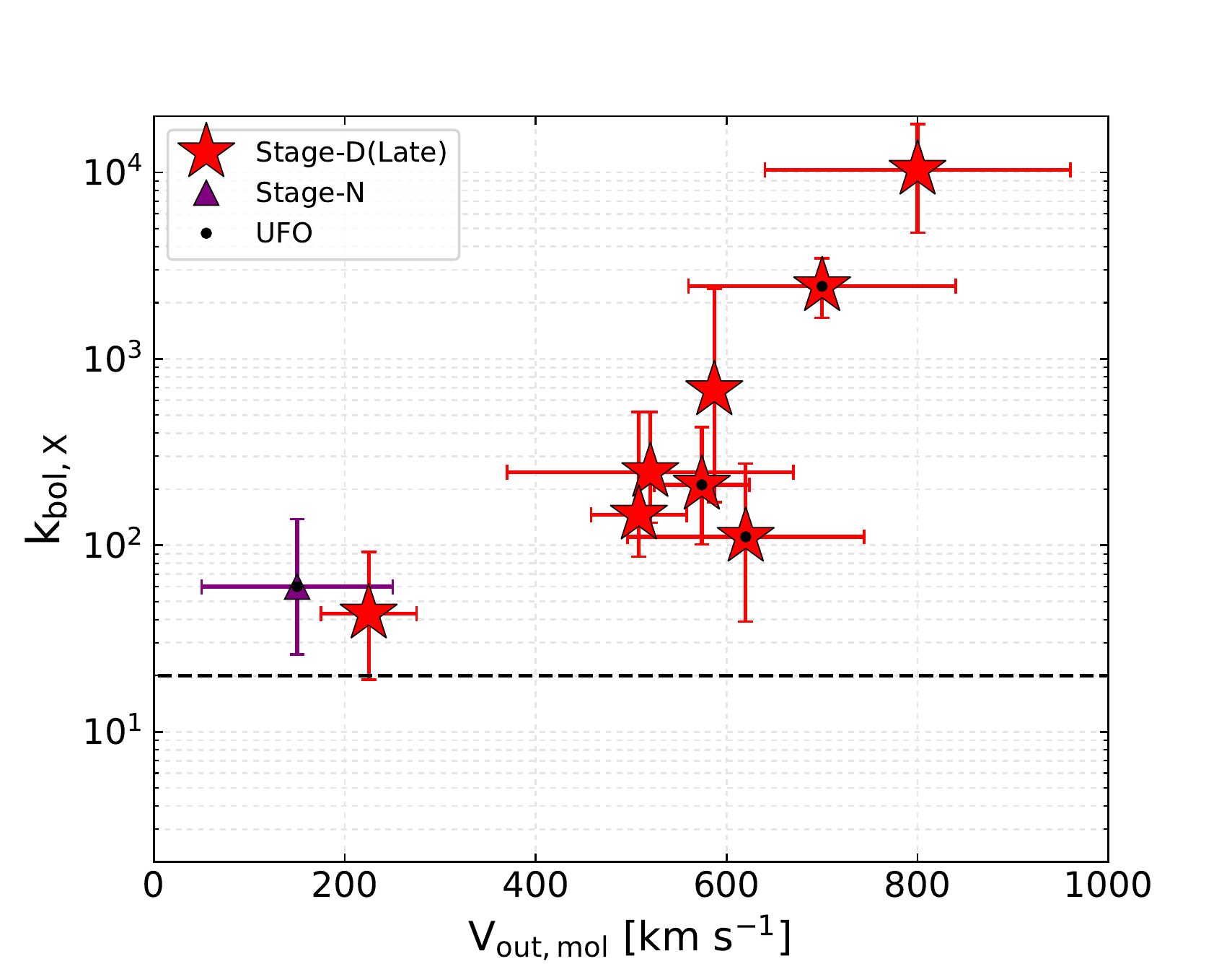}
        \caption{Bolometric-to-X-ray correction vs. molecular outflow velocity.
        Symbols and colors are the same as in Figure~\ref{F3_NH-kpc}.
        The small black circle marks the detection of UFOs \citep{Mizumoto2019,Smith2019}.
        The black dashed line shows a typical value of $\kappa_{\rm bol,X}$ = 20.
        \label{F8_Kx-outflow}}
\end{figure}

\subsubsection{Multiphase Outflows}
\label{subsub6-1-2_outflows}

As mentioned in Section~\ref{subsub5-3-2_outflows},
molecular outflows on large scales ($\gtrsim$0.1~kpc) are often 
discovered in local U/LIRGs (e.g.,
\citealt{Gonzalez-Alfonso2017,Laha2018} and the references therein). It
is suggested that the molecular outflow velocity exhibits correlations
with stellar mass and SFR among star-forming galaxies
\citep{Chisholm2015}. Whereas, on the basis of X-ray observations of
galaxies hosting molecular outflows, \citet{Laha2018} report that
the outflow velocity ($V_{\rm out,mol}$) and mass
transfer rate ($\dot{M}_{\rm out,mol}$) are more strongly correlated with 
AGN luminosity than with starburst luminosity.
These facts imply that, although starbursts can drive massive molecular
outflows, the presence of an AGN always boosts the power of the outflow.

\begin{figure}
    \epsscale{1.20}
    \plotone{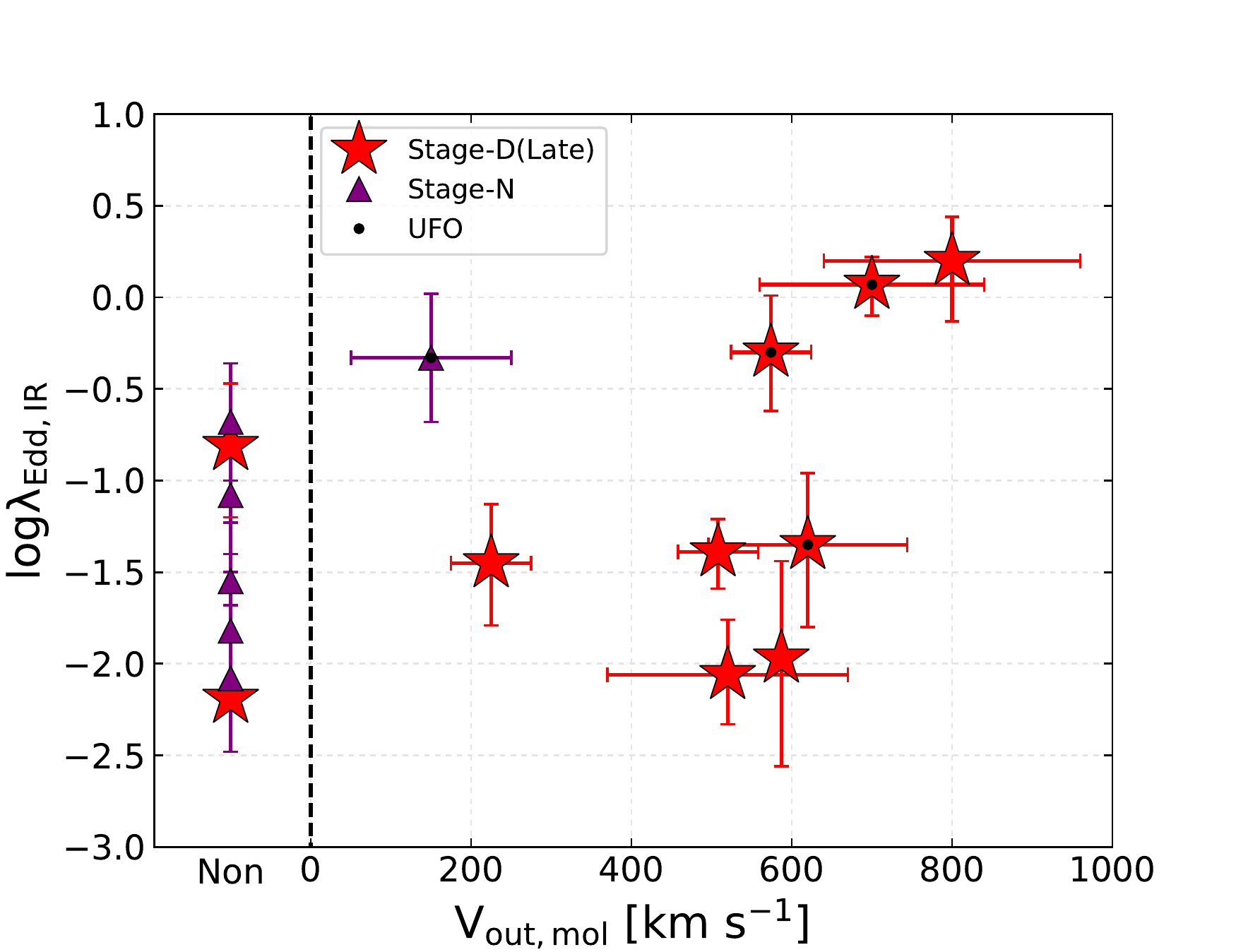}
        \caption{IR-based Eddington ratio vs. molecular outflow velocity
 for the hard X-ray AGNs in the 10 stage-D and 6 stage-N sources.
        Symbols and colors are the same as in Figure \ref{F3_NH-kpc}.
        The small black circle marks the detection of UFOs \citep{Mizumoto2019,Smith2019}.
        The AGNs whose outflow velocities or mass outflow rates are not
 constrained (or the presences of molecular outflows have not yet been
 reported) are plotted on the left side.
        \label{F9_Vout-Edd}}
\end{figure}

In the hard X-ray detected AGNs among our sample, molecular outflows are detected from 10 stage-D 
mergers and 1 nonmerger, although the observations may not be complete. 
Figure~\ref{F8_Kx-outflow} plots the bolometric-to-X-ray correction factors ($\kappa_{\rm bol,X}$) 
against outflow velocities for AGNs with molecular outflow detections (excluding two stage-D mergers, Arp~220W and NGC~6240S/NGC~6240N, since their bolometric AGN luminosities are not constrained).
A strong positive correlation is noticeable, suggesting
a link between the X-ray weakness and outflow activity.
Figure~\ref{F9_Vout-Edd} shows the relation between molecular-outflow velocity and
Eddington ratio; here we include AGNs without molecular outflow
detections. A positive correlation is suggested for stage-D
mergers, supporting the AGN origin for the molecular outflows. It is
noteworthy that the outflow velocity in the stage-N source (NGC~1068)
is smaller than those in state-D mergers at similar Eddington ratios
($\log \lambda_{\rm Edd,IR} \sim -$0.5--0.0). Moreover, molecular
outflows in the other nonmergers have not been detected.
These results imply that powerful outflows can be 
more efficiently launched from late mergers than from nonmergers 
due to environmental effects. 
In fact, \citet{Smethurst2019} show that outflow rates relative to the
mass accretion rates are much larger in merging systems than in
nonmerging systems.

We find that the AGNs with molecular-outflow detections
have ionized outflows detected in the optical band via blueshifted
[\ion{O}{3}] emission lines (see Table~\ref{T8_AGN-properties}). 
Furthermore, as shown in Figures~\ref{T8_AGN-properties}
and~\ref{F9_Vout-Edd}, X-ray spectral features of UFOs are found in 
four stage-D mergers, IRAS F05189--2524, Mrk~231, Mrk~273, NGC~6240S/NGC~6240N
(dual-AGN)\footnote{The results of the dual AGNs in NGC~6240S/NGC~6240N
are not presented in Section~\ref{S6_discussion}, since the bolometric luminosities and
SFRs of individual sources cannot be separately obtained.} and in a 
stage-N LIRG, NGC~1068 \citep{Mizumoto2019,Smith2019}. Their molecular
outflows seem to be driven by 
the AGN activities (see also e.g., \citealt{Tombesi2015,Fiore2017}).
This supports the analogy of late merger U/LIRGs 
with BAL QSOs as mentioned above.

\begin{deluxetable*}{lccccccccc}
\label{T8_AGN-properties}
\tablecaption{Properties of AGN and Starburst Activities for Hard X-ray Detected AGNs}
\tabletypesize{\scriptsize}
\tablehead{
\colhead{Object} &
\colhead{log$L_{\rm [O\,IV]}$} &
\multicolumn{4}{c}{log$L^{\rm (IR)}_{\rm bol,AGN}$} &
\colhead{logSFR} &
\colhead{$V_{\rm out,mol}$} &
\colhead{log$\dot{M}_{\rm out,mol}$} &
\colhead{Ref.} \\
\cline{3-6}
& & [\ion{O}{4}] & $f_{\rm AGN}$ & SED & IRS & & & & \\
& (erg s$^{-1}$) & (erg s$^{-1}$) & (erg s$^{-1}$) & (erg s$^{-1}$) & (erg s$^{-1}$) & ($M_{\odot}$ yr$^{-1}$) & (km s$^{-1}$) & ($M_{\odot}$ yr$^{-1}$) &
}
\decimalcolnumbers
\startdata
\multicolumn{10}{c}{Stage-A}\\
\hline
NGC 833 & $<$39.34 & $<$41.64 & \nodata & \nodata & \nodata & \nodata & \nodata & \nodata & \nodata \\
NGC 835 & 39.98 $\pm$ 0.03 & 42.41 $\pm$ 0.33 & \ \,43.76 $\pm$ 0.07$^{\dagger}$ & \nodata & \nodata & \nodata & \nodata & \nodata & \nodata \\
NGC 6921 & \nodata & \nodata & \nodata & \nodata & \nodata & 0.20$^{+0.10}_{-0.19}$ & \nodata & \nodata & 1 \\
MCG+04-48-002 & 40.72 $\pm$ 0.03 & 43.66 $\pm$ 0.34 & 43.98 $\pm$ 0.06 & \nodata & 43.32$^{+0.15}_{-0.15}$ & 1.17$^{+0.05}_{-0.03}$ & \nodata & \nodata & 2 \\
NGC 7469 & 41.46 $\pm$ 0.02 & 44.92 $\pm$ 0.35 & 44.67 $\pm$ 0.06 & 44.95$^{+0.08}_{-0.07}$ & 44.66$^{+0.15}_{-0.15}$ & 1.63$^{+0.04}_{-0.03}$ & \nodata & \nodata & 2 \\
NGC 7674 & 41.94 $\pm$ 0.01 & 45.73 $\pm$ 0.36 & 45.07 $\pm$ 0.07 & 45.15$^{+0.02}_{-0.02}$ & \nodata & 1.39$^{+0.02}_{-0.01}$ & \nodata & \ \ \ $\cdots^{\rm (i)}$ & 2 \\
NGC 7679 & 41.36 $\pm$ 0.01 & 44.75 $\pm$ 0.35 & 43.98 $\pm$ 0.06 & \nodata & 43.88$^{+0.15}_{-0.15}$ & 1.17$^{+0.01}_{-0.02}$ & \nodata & \nodata & 2 \\
NGC 7682 & \nodata & \nodata & \nodata & \nodata & \nodata & \nodata & \nodata & \nodata & \nodata \\
\hline
\multicolumn{10}{c}{Stage-B}\\
\hline
NGC 235 & 41.28 $\pm$ 0.01 & 44.61 $\pm$ 0.35 & \ \,44.05 $\pm$ 0.06$^{\dagger}$ & [44.29$^{+0.06}_{-0.04}$]: & \nodata & $<$0.96 & \nodata & \nodata & 3\\
CGCG 468-002W & 40.79 $\pm$ 0.01 & 43.78 $\pm$ 0.34 & 44.55 $\pm$ 0.04 & [44.10$^{+0.10}_{-0.16}$]: & \nodata & 0.70$^{+0.34}_{-0.38}$ & \nodata & \nodata & 1\\
ESO 060-IG016 East & $<$41.06 & $<$44.58 & 44.83 $\pm$ 0.06 & 44.85$^{+0.07}_{-0.06}$ & \nodata & 1.83$^{+0.01}_{-0.02}$ & \nodata & \nodata & 2\\
Mrk 266B & 42.00 $\pm$ 0.05 & 45.83 $\pm$ 0.36 & [44.57 $\pm$ 0.07]: & [44.60$^{+0.06}_{-0.09}$]: & \nodata & $<$1.61 & \nodata & \nodata & 3 \\
Mrk 266A & \nodata & \nodata & [44.57 $\pm$ 0.07]: & [44.60$^{+0.06}_{-0.09}$]: & \nodata & $<$1.35 & \nodata & \nodata & 3 \\
IC 4518A & 41.70 & 45.33 $\pm$ 0.36 & \nodata & 44.42$^{+0.02}_{-0.02}$ & 43.85$^{+0.15}_{-0.15}$ & 1.20$^{+0.02}_{-0.02}$ & \nodata & \nodata & 2 \\
NGC 6286 & $<$40.2 & $<$43.11 & 44.06 $\pm$ 0.06 & \nodata & \nodata & 1.44$^{+0.03}_{-0.02}$ & \nodata & \nodata & 2 \\
\hline
\multicolumn{10}{c}{Stage-C}\\
\hline
MCG+12-02-001 & $<$40.55 & $<$43.71 & 43.92 $\pm$ 0.04 & 44.53$^{+0.13}_{-0.27}$ & \nodata & 1.57$^{+0.05}_{-0.03}$ & \nodata & \nodata & 2\\
IRAS F06076--2139 & $<$40.89 & $<$44.29 & 44.34 $\pm$ 0.01 & \nodata & 44.01$^{+0.09}_{-0.08}$ & 1.71$^{+0.03}_{-0.02}$ & \nodata & \nodata & 2\\
NGC 3690 West & 40.82 $\pm$ 0.06 & 43.83 $\pm$ 0.34 & [45.01 $\pm$ 0.07]: & [45.36$^{+0.08}_{-0.11}$]: & 43.94$^{+0.15}_{-0.15}$ & [1.85$^{+0.03}_{-0.03}$]: & \nodata & \nodata & 2 \\
NGC 4922N & 40.74 $\pm$ 0.08 & 43.70 $\pm$ 0.34 & 44.25 $\pm$ 0.05 & 44.81$^{+0.08}_{-0.07}$ & \nodata & 1.31$^{+0.05}_{-0.03}$ & \nodata & \nodata & 2 \\
ESO 148-2 & 41.85 $\pm$ 0.07 & 45.58 $\pm$ 0.36 & 44.98 $\pm$ 0.03 & 45.26$^{+0.14}_{-0.17}$ & \nodata & 1.93$^{+0.04}_{-0.03}$ & \nodata & \ \ \ $\cdots^{\rm (i)}$ & 2 \\
\hline
\multicolumn{10}{c}{Stage-D}\\
\hline
NGC 34 & $<$40.69 & $<$43.95 & 43.74 $\pm$ 0.02 & \nodata & \nodata & 1.47$^{+0.04}_{-0.04}$ & \nodata & \nodata & 2\\
IRAS F05189--2524 & 42.05 $\pm$ 0.03 & 45.92 $\pm$ 0.36 & 45.58 $\pm$ 0.07 & 45.84$^{+0.05}_{-0.11}$ & 45.62$^{+0.01}_{-0.01}$ & 1.89$^{+0.03}_{-0.05}$ & 574 $\pm$ 50 & 2.431$^{\rm (i*)}$ & 2,a,b\\
NGC 2623 & 40.84 $\pm$ 0.07 & 43.87 $\pm$ 0.34 & 44.24 $\pm$ 0.03 & 43.52$^{+0.48}_{-0.57}$ & 43.39$^{+0.26}_{-0.26}$ & 1.63$^{+0.03}_{-0.04}$ & 587 & 1.050$^{\rm (i)}$ & 2,c\\
IRAS F08572+3915 & $<$41.37 & $<$45.12 & 45.48 $\pm$ 0.09 & 46.31$^{+0.23}_{-0.33}$ & 45.57$^{+0.03}_{-0.05}$ & 1.46$^{+0.19}_{-0.38}$ & 800 $\pm$ 160 & 3.082$^{\rm (i)}$ & 2,a\\
UGC 5101 & 41.42 $\pm$ 0.05 & 44.85 $\pm$ 0.35 & 45.05 $\pm$ 0.05 & 44.82$^{+0.15}_{-0.20}$ & 44.54$^{+0.05}_{-0.04}$ & 2.02$^{+0.08}_{-0.06}$ & 225 $\pm$ 50 & \nodata & 2,a\\
Mrk 231 & $<$41.76 & $<$45.78 & 45.94 $\pm$ 0.12 & 46.26$^{+0.10}_{-0.13}$ & 46.07$^{+0.01}_{-0.01}$ & 2.34$^{+0.05}_{-0.05}$ & 700 $\pm$ 140 & 3.02$^{\rm (i*)}$ & 2,a\\
IRAS 13120--5453 & $<$41.17 & $<$44.77 & 44.57 $\pm$ 0.02 & 44.47$^{+0.29}_{-0.22}$ & 44.40$^{+0.12}_{-0.18}$ & 2.40$^{+0.02}_{-0.02}$ & 520 $\pm$ 150 & 2.113$^{\rm (i)}$ & 2,a \\
Mrk 273 & 42.2 $\pm$ 0.01 & 46.17 $\pm$ 0.37 & 45.35 $\pm$ 0.08 & 44.83$^{+0.21}_{-0.39}$ & 44.05$^{+0.19}_{-0.08}$ & 2.15$^{+0.04}_{-0.05}$ & 620 $\pm$ 124 & 2.778$^{\rm (i*)}$ & 2,a\\
IRAS F14348--1447 & $<$41.45 & $<$45.25 & 44.99 $\pm$ 0.05 & \nodata & 44.11$^{+0.15}_{-0.16}$ & 2.37$^{+0.05}_{-0.07}$ & 508 $\pm$ 50 & \nodata & 2,a\\
Arp 220W & [$<$41.06]: & [$<$44.24]: & [44.87 $\pm$ 0.05]: & \nodata & [42.52$^{+0.65}_{-1.61}$]: & [2.35$^{+0.07}_{-0.07}$]: & [153 $\pm$ 50]: & \nodata & 2,a\\
NGC 6240S & [41.48 $\pm$ 0.06]: & [44.95 $\pm$ 0.35]: & [44.61 $\pm$ 0.02]: & [44.35$^{+0.35}_{-1.24}$]: & \nodata & [1.99$^{+0.07}_{-0.04}$]: & [400 $\pm$ 100]: & [2.903]:$^{\rm (i*)}$ & 2,a\\
NGC 6240N & [41.48 $\pm$ 0.06]: & [44.95 $\pm$ 0.35]: & [44.61 $\pm$ 0.02]: & [44.35$^{+0.35}_{-1.24}$]: & \nodata & [1.99$^{+0.07}_{-0.04}$]: & [400 $\pm$ 100]: & [2.903]:$^{\rm (i*)}$ & 2,a\\
IRAS F17138--1017 & $<$40.73 & $<$44.02 & 43.98 $\pm$ 0.04 & 44.13$^{+0.29}_{-0.33}$ & \nodata & 1.58$^{+0.05}_{-0.05}$ & \nodata & \nodata & 2\\
\hline
\multicolumn{10}{c}{Stage-N}\\
\hline
NGC 1068 & 41.81 $\pm$ 0.01 & 45.51 $\pm$ 0.36 & 45.04 $\pm$ 0.01 & 44.82$^{+0.09}_{-0.07}$ & \nodata & 1.28$^{+0.03}_{-0.04}$ & 150 $\pm$ 100 & 1.924$^{\rm (i*)}$ & 2,a\\
UGC 2608 & 42.23 $\pm$ 0.01 & 46.22 $\pm$ 0.37 & 44.75 $\pm$ 0.09 & 44.66$^{+0.05}_{-0.04}$ & 44.61$^{+0.07}_{-0.07}$ & 1.37$^{+0.02}_{-0.02}$ & \nodata & \nodata & 2\\
NGC 1365 & 40.96 $\pm$ 0.01 & 44.07 $\pm$ 0.34 & 44.22 $\pm$ 0.03 & \nodata & \nodata & 1.20$^{+0.02}_{-0.02}$ & \nodata & \ \ \ $\cdots^{\rm (i)}$ & 4\\
MCG--03-34-064 & 41.81 $\pm$ 0.01 & 45.51 $\pm$ 0.36 & 44.87 $\pm$ 0.04 & 45.02$^{+0.01}_{-0.01}$ & 44.52$^{+0.15}_{-0.15}$ & 0.96$^{+0.02}_{-0.04}$ & \nodata & \nodata & 2 \\
NGC 5135 & 41.52 $\pm$ 0.01 & 45.02 $\pm$ 0.35 & 44.32 $\pm$ 0.06 & 44.24$^{+0.03}_{-0.03}$ & 43.72$^{+0.15}_{-0.15}$ & 1.34$^{+0.01}_{-0.01}$ & \nodata & \ \ \ $\cdots^{\rm (i)}$ & 2 \\
NGC 7130 & 41.08 $\pm$ 0.03 & 44.27 $\pm$ 0.35 & 44.41 $\pm$ 0.04 & 44.40$^{+0.07}_{-0.04}$ & 43.60$^{+0.15}_{-0.15}$ & 1.48$^{+0.02}_{-0.03}$ & \nodata & \ \ \ $\cdots^{\rm (i)}$ & 2 \\
\enddata
\tablecomments{Columns:
(1) object name;
(2) logarithmic [\ion{O}{4}] 25.89 $\mu$m luminosity. The values are taken from \citet{Inami2013} except for Mrk 266B \citet{Mazzarella2012} and IC 4518A \citep{LiuTeng2014};
(3) logarithmic bolometric AGN luminosity derived from the relation of [\ion{O}{4}] and AGN luminosities \citep{Gruppioni2016};
(4) logarithmic bolometric AGN luminosity derived from the bolometric AGN fraction \citep{Diaz-Santos2017}.
The bolometric-to-IR correction factor of total luminosity is assumed to
 be 1.15.
(5--6) logarithmic bolometric AGN luminosity derived from the 1--500 $\mu$m SED decomposition and Spitzer/IRS spectral fitting \citep{Alonso-Herrero2012,Shangguan2019}. 
The bolometric-to-IR correction factor of the AGN luminosity is assumed
 to be 3.0 \citep{Delvecchio2014};
(7) logarithmic SFR;
(8--9) molecular outflow velocity and logarithmic mass outflow rate. 
The mark (i) means that ionized outflows are detected \citep{Fischer2013,Rich2015,Cortijo-Ferrero2017b,Kakkad2018,Venturi2018,Smith2019,Boettcher2020,Fluetsch2021},
and * means that UFOs are detected \citep{Mizumoto2019,Smith2019};
(10) References of the columns (7) and (8--9) marked as ``1--4'' and ``a--c'', respectively.
Values in brackets with ``:'' should be considered to be upper limits due to contamination from nearby IR sources.
The error and upper limit are 1$\sigma$ and 3$\sigma$ level.\\
\textbf{References:} (1) \citet{Pereira-Santaella2015}; (2) \citet{Shangguan2019}; (3) \citet{Howell2010}; (4) \citet{Gruppioni2016}; (a) \citet{Laha2018} and the references therein; (b) \citet{Gonzalez-Alfonso2017}; (c) \citet{Lutz2020}.
}
\tablenotetext{\dagger}{The total IR luminosities are calculated by the 40--120 $\mu$m luminosity times 1.9 (see \citealt{Lutz2018}).}
\end{deluxetable*}

\begin{deluxetable*}{lccccccccc}
\label{T9_SB-properties}
\tablecaption{Properties of AGN and Starburst Activities for Starburst-dominant or Hard X-ray Undetected Sources}
\tabletypesize{\scriptsize}
\tablehead{
\colhead{Object} &
\colhead{log$L_{\rm [O\,IV]}$} &
\multicolumn{4}{c}{log$L^{\rm (IR)}_{\rm bol,AGN}$} &
\colhead{logSFR} &
\colhead{$V_{\rm out,mol}$} &
\colhead{log$\dot{M}_{\rm out,mol}$} &
\colhead{Ref.} \\
\cline{3-6}
& & [\ion{O}{4}] & $f_{\rm AGN}$ & SED & IRS & & & & \\
& (erg s$^{-1}$) & (erg s$^{-1}$) & (erg s$^{-1}$) & (erg s$^{-1}$) & (erg s$^{-1}$) & ($M_{\odot}$ yr$^{-1}$) & (km s$^{-1}$) & ($M_{\odot}$ yr$^{-1}$) &
}
\decimalcolnumbers
\startdata
\multicolumn{10}{c}{Stage-A}\\
\hline
NGC 838 & 40.29 $\pm$ 0.03 & 42.93 $\pm$ 0.33 & \ \,43.43 $\pm$ 0.03$^{\dagger}$ & \nodata & \nodata & \nodata & \nodata & \nodata & \nodata\\
NGC 839 & 40.06 $\pm$ 0.23 & 42.54 $\pm$ 0.33 & \ \,43.65 $\pm$ 0.04$^{\dagger}$ & \nodata & \nodata & \nodata & \nodata & \ \ \ $\cdots^{\rm (i)}$ & \nodata\\
MCG+08-18-012 & \nodata & \nodata & \nodata & \nodata & \nodata & \nodata & \nodata & \nodata & \nodata\\
MCG+08-18-013 & $<$39.90 & $<$42.60 & 44.28 $\pm$ 0.01 & 43.77$^{+0.14}_{-0.25}$ & \nodata & 1.43$^{+0.02}_{-0.02}$ & \nodata & \nodata & 1\\
MCG--01-26-013 & \nodata & \nodata & \nodata & \nodata & \nodata & \nodata & \nodata & \nodata & \nodata\\
NGC 3110 & $<$40.30 & $<$43.28 & 44.01 $\pm$ 0.06 & \nodata & \nodata & 1.46$^{+0.03}_{-0.02}$ & \nodata & \nodata & 1\\
NGC 4418 & $<$40.49 & $<$43.61 & 44.51 $\pm$ 0.22 & \nodata & 44.47$^{+0.02}_{-0.02}$ & 1.28$^{+0.02}_{-0.02}$ & 324 & 0.799$^{\rm (i)}$ & 1,a\\
MCG+00-32-013 & \nodata & \nodata & \nodata & \nodata & \nodata & \nodata & \nodata & \ \ \ $\cdots^{\rm (i)}$ & \nodata\\
IC 5283 & \nodata & \nodata & \nodata & \nodata & \nodata & 1.03 & \nodata & \nodata & 2\\
MCG+01-59-081 & \nodata & \nodata & \nodata & \nodata & \nodata & 0.25 & \nodata & \nodata & 2\\
\hline
\multicolumn{10}{c}{Stage-B}\\
\hline
MCG--02-01-052 & \nodata & \nodata & \nodata & \nodata & \nodata & 0.60 & \nodata & \nodata & 2\\
MCG--02-01-051 & 40.60 $\pm$ 0.16 & 43.46 $\pm$ 0.34 & 43.97 $\pm$ 0.05 & 44.76$^{+0.03}_{-0.04}$ & \nodata & 1.56$^{+0.02}_{-0.02}$ & \nodata & \nodata & 1\\
NGC 232 & 40.93 $\pm$ 0.05 & 44.02 $\pm$ 0.34 & \ \,44.13 $\pm$ 0.03$^{\dagger}$ & [44.29$^{+0.06}_{-0.04}$]: & \nodata & 1.59 & \nodata & \nodata & 2\\
ESO 203-1 & $<$41.04 & $<$44.55 & 45.07 $\pm$ 0.26 & \nodata & 44.05$^{+0.20}_{-0.21}$ & 1.76$^{+0.04}_{-0.04}$ & \nodata & \nodata & 1\\
CGCG 468-002E & $<$40.11 & $<$42.96 & \nodata & [44.10$^{+0.10}_{-0.16}$]: & \nodata & 1.18$^{+0.09}_{-0.15}$ & \nodata & \nodata & 3\\
ESO 060-IG016 West & \nodata & \nodata & \nodata & \nodata & \nodata & \nodata & \nodata & \nodata & \nodata\\
ESO 440-58 & $<$39.96 & $<$42.70 & \nodata & \nodata & \nodata & 0.78 & \nodata & \nodata & 2\\
MCG--05-29-017 & 40.17 $\pm$ 0.19 & 42.73 $\pm$ 0.33 & 43.85 $\pm$ 0.03 & \nodata & \nodata & 1.46$^{+0.02}_{-0.02}$ & \nodata & \nodata & 1\\
IC 4518B & \nodata & \nodata & \nodata & \nodata & \nodata & \nodata & \nodata & \nodata & \nodata\\
NGC 6285 & $<$39.77 & $<$42.38 & \ \,43.15 $\pm$ 0.04$^{\dagger}$ & \nodata & \nodata & 0.88 & \nodata & \nodata & 2\\
NGC 6907 & 39.54 $\pm$ 0.12 & 41.66 $\pm$ 0.32 & 43.71 $\pm$ 0.04 & \nodata & \nodata & 1.19$^{+0.02}_{-0.02}$ & \nodata & \nodata & 1\\
NGC 6908 & \nodata & \nodata & \nodata & \nodata & \nodata & \nodata & \nodata & \nodata & \nodata\\
\hline
\multicolumn{10}{c}{Stage-C}\\
\hline
ESO 350-38 & $<$40.77 & $<$44.09 & 44.29 $\pm$ 0.09 & 45.21$^{+0.07}_{-0.11}$ & 44.93$^{+0.02}_{-0.01}$ & 0.77$^{+0.07}_{-0.08}$ & \nodata & \nodata & 1\\
IC 1623A & \nodata & \nodata & \nodata & \nodata & \nodata & 1.32 & \nodata & \nodata & 2\\
IC 1623B & 40.45 $\pm$ 0.38 & 43.21 $\pm$ 0.34 & 44.43 $\pm$ 0.04 & 44.87$^{+0.03}_{-0.03}$ & \nodata & 1.67$^{+0.02}_{-0.02}$ & \nodata & \nodata & 1\\
2MASS 06094601--2140312 & \nodata & \nodata & \nodata & \nodata & \nodata & \nodata & \nodata & \nodata & \nodata\\
NGC 3690 East & 40.59 $\pm$ 0.39 & 43.44 $\pm$ 0.34 & \nodata & [45.36$^{+0.08}_{-0.11}$]: & \nodata & [1.85$^{+0.03}_{-0.03}$]: & \nodata & \nodata & 1\\
NGC 4922S & \nodata & \nodata & \nodata & \nodata & \nodata & $-$0.89 & \nodata & \nodata & 2\\
II Zw 096 & $<$41.28 & $<$44.96 & [44.28 $\pm$ 0.01]: & [45.27$^{+0.21}_{-0.13}$]: & \nodata & [1.98$^{+0.03}_{-0.05}$]: & \nodata & \nodata & 1\\
IRAS F20550+1655 SE & $<$41.21 & $<$44.84 & [44.28 $\pm$ 0.01]: & [45.27$^{+0.21}_{-0.13}$]: & \nodata & [1.98$^{+0.03}_{-0.05}$]: & \nodata & \nodata & 1\\
\hline
\multicolumn{10}{c}{Stage-D}\\
\hline
ESO 374-IG032 & $<$40.96 & $<$44.41 & 44.77 $\pm$ 0.15 & 44.88$^{+0.11}_{-0.07}$ & 44.82$^{+0.06}_{-0.05}$ & 1.70$^{+0.02}_{-0.02}$ & \nodata & \nodata & 1\\
NGC 3256 & 40.69 $\pm$ 0.04 & 43.61 $\pm$ 0.34 & 43.98 $\pm$ 0.02 & 44.76$^{+0.08}_{-0.12}$ & \nodata & 1.69$^{+0.05}_{-0.04}$ & 250 $\pm$ 100 & 1.041--1.204 & 1,b\\
IRAS F10565+2448 & $<$41.02 & $<$44.52 & 44.33 $\pm$ 0.01 & 43.96$^{+0.09}_{-0.09}$ & 44.71$^{+0.09}_{-0.05}$ & 2.14$^{+0.02}_{-0.03}$ & 450 $\pm$ 90 & 2.477$^{\rm (i)}$ & 1,b\\
IRAS F12112+0305 & $<$41.36 & $<$45.10 & 44.78 $\pm$ 0.03 & 44.15$^{+0.32}_{-0.23}$ & 43.75$^{+0.16}_{-0.14}$ & 2.33$^{+0.02}_{-0.03}$ & 237 $\pm$ 50 & \nodata & 1,b\\
IRAS F14378--3651 & $<$41.17 & $<$44.77 & 44.57 $\pm$ 0.02 & 44.01$^{+0.59}_{-0.85}$ & 43.64$^{+0.18}_{-0.11}$ & 2.16$^{+0.11}_{-0.06}$ & 800 $\pm$ 150 & 2.869$^{\rm (i)}$ & 1,b\\
IRAS F15250+3608 & $<$41.48 & $<$45.30 & 44.98 $\pm$ 0.16 & 45.68$^{+0.17}_{-0.32}$ & 44.58$^{+0.14}_{-0.13}$ & 1.79$^{+0.09}_{-0.09}$ & \nodata & \nodata & 1\\
Arp 220E & [$<$41.06]: & [$<$44.24]: & [44.87 $\pm$ 0.05]: & \nodata & [42.52$^{+0.65}_{-1.61}$]: & [2.35$^{+0.07}_{-0.07}$]: & [153 $\pm$ 50]: & \nodata & 1,b\\
IRAS F19297--0406 & $<$41.71 & $<$45.70 & 44.79 $\pm$ 0.03 & \nodata & \nodata & 2.39$^{+0.07}_{-0.04}$ & 532 $\pm$ 50 & \nodata & 1,b\\
ESO 286-19 & $<$41.26 & $<$44.93 & 45.08 $\pm$ 0.05 & 45.50$^{+0.16}_{-0.25}$ & 45.09$^{+0.06}_{-0.08}$ & 1.85$^{+0.05}_{-0.05}$ & 500 & 2.176$^{\rm (i)}$ & 1,c\\
\hline
\multicolumn{10}{c}{Stage-N}\\
\hline
UGC 2612 & \nodata & \nodata & \nodata & \nodata & \nodata & \nodata & \nodata & \nodata & \nodata\\
IC 860 & $<$39.98 & $<$42.74 & 43.56 $\pm$ 0.03 & \nodata & 40.73$^{+1.22}_{-1.32}$ & 1.27$^{+0.02}_{-0.02}$ & \nodata & \nodata & 1\\
NGC 5104 & 40.01 $\pm$ 0.31 & 42.46 $\pm$ 0.33 & 43.91 $\pm$ 0.04 & \nodata & 43.23$^{+0.20}_{-0.20}$ & 1.32$^{+0.02}_{-0.02}$ & \nodata & \nodata & 1\\
IRAS F18293--3413 & 40.74 $\pm$ 0.06 & 43.70 $\pm$ 0.34 & 44.22 $\pm$ 0.03 & \nodata & \nodata & 1.96$^{+0.04}_{-0.04}$ & \nodata & \nodata & 1\\
NGC 7591 & \nodata & \nodata & \nodata & \nodata & \nodata & 1.18$^{+0.02}_{-0.02}$ & \nodata & \nodata & 1\\
\enddata
\tablecomments{Columns: (1--10) same as in Table~\ref{T8_AGN-properties}.\\
\textbf{References:} (1) \citet{Shangguan2019}; (2) \citet{Howell2010}; (3) \citet{Pereira-Santaella2015}; 
(a) \citet{Lutz2020}; (b) \citet{Laha2018}; (c) \citet{Imanishi2017}.
}
\tablenotetext{\dagger}{The total IR luminosities are calculated by the 40--120 $\mu$m luminosity times 1.9 (see \citealt{Lutz2018}).}
\end{deluxetable*}

\begin{deluxetable*}{lcccccccc}
\label{T10_BHmass}
\tablecaption{Stellar and Black Hole Masses of the Hard X-ray Detected AGNs}
\tablewidth{\columnwidth}
\tabletypesize{\scriptsize}
\tablehead{
\colhead{Object} &
\colhead{log$M_{*}$} &
\colhead{Ref. 1} &
\multicolumn{5}{c}{log$M_{\rm BH}$} &
\colhead{Ref. 2}\\
\cline{4-8}
& & & $M_{*}$ & $M$--$\sigma_*$ & $L_{\rm bulge}$ & others & Method &\\
& ($M_{\odot}$) & & ($M_{\odot}$) & ($M_{\odot}$) & ($M_{\odot}$) & ($M_{\odot}$) & & 
}
\decimalcolnumbers
\startdata
\multicolumn{9}{c}{Stage-A}\\
\hline
NGC 833 & \nodata & \nodata & \nodata & 8.97 & \nodata & 8.24 $\pm$ 0.40 & $L_{\rm 2\mu m}$ & 6,7\\
NGC 835 & 10.48 & 1 & 6.90 & 8.16 & \nodata & 8.50 $\pm$ 0.40 & $L_{\rm 2\mu m}$ & 6,7\\
NGC 6921 & 11.16 $\pm$ 0.03 & 2 & 7.62 $\pm$ 0.20 & \nodata & \nodata & 8.60 & $M$--$\sigma$(Ca) & 8\\
MCG+04-48-002 & 11.13 $\pm$ 0.20 & 3 & 7.59 $\pm$ 0.48 & 7.50 $\pm$ 0.40 & \nodata & 7.85 & $M$--$\sigma$(Ca) & 9,8\\
NGC 7469 & 11.18 $\pm$ 0.20 & 3 & 7.64 $\pm$ 0.48 & 7.61 $\pm$ 0.40 & 7.00 $\pm$ 0.14 & 6.97 $\pm$ 0.05 & reverberation & 9,10,11\\
NGC 7674 & 11.25 $\pm$ 0.20 & 3 & 7.71 $\pm$ 0.48 & 7.56 $\pm$ 0.44 & 8.50 $\pm$ 0.09 & 8.20 $\pm$ 0.50 & plane equation & 12,10,13\\
NGC 7679 & 10.79 $\pm$ 0.20 & 3 & 7.23 $\pm$ 0.48 & 6.77 $\pm$ 0.40 & \nodata & 8.27 $\pm$ 0.40 & $L_{\rm 2\mu m}$ & 9,7\\
NGC 7682 & 10.53 & 1 & 6.96 & 7.28 $\pm$ 0.44 & \nodata & 8.20 $\pm$ 0.40 & $L_{\rm 2\mu m}$ & 12,7\\
\hline
\multicolumn{9}{c}{Stage-B}\\
\hline
NGC 235 & 10.93 & 4 & 7.38 & \nodata & 8.76 & 8.47 $\pm$ 0.40 & $L_{\rm 2\mu m}$ & 14,7\\
CGCG 468-002W & 10.83 $\pm$ 0.05 & 2 & 7.27 $\pm$ 0.25 & \nodata & \nodata & 8.06 $\pm$ 0.50 & $M$--$\sigma$([\ion{O}{3}]) & 9\\
ESO 060-IG016 East & 10.90 $\pm$ 0.20 & 3 & 7.35 $\pm$ 0.48 & \nodata & 8.24 $\pm$ 0.09 & \nodata & \nodata & 10\\
Mrk 266B & [11.15 $\pm$ 0.20]: & 3 & [7.61 $\pm$ 0.48]: & \nodata & 8.28 $\pm$ 0.07 & \nodata & \nodata & 10\\
Mrk 266A & [11.15 $\pm$ 0.20]: & 3 & [7.61 $\pm$ 0.48]: & \nodata & 8.37 $\pm$ 0.08 & \nodata & \nodata & 10\\
IC 4518A & 11.12 $\pm$ 0.20 & 3 & 7.58 $\pm$ 0.48 & 7.48 $\pm$ 0.40 & \nodata & \nodata & \nodata & 9\\
NGC 6286 & 11.07 $\pm$ 0.20 & 3 & 7.52 $\pm$ 0.48 & \nodata & \nodata & 8.43 $\pm$ 0.40 & $L_{\rm 2\mu m}$ & 7\\
\hline
\multicolumn{9}{c}{Stage-C}\\
\hline
MCG+12-02-001 & 10.61 $\pm$ 0.20 & 3 & 7.04 $\pm$ 0.48 & \nodata & \nodata & \nodata & \nodata & \nodata \\
IRAS F06076--2139 & 10.86 $\pm$ 0.20 & 3 & 7.30 $\pm$ 0.48 & \nodata & \nodata & \nodata & \nodata & \nodata \\
NGC 3690 West & 10.36 & 5 & 6.77 & 7.40 $\pm$ 0.47 & \nodata & \nodata & \nodata & 15\\
NGC 4922N & 11.24 $\pm$ 0.20 & 3 & 7.70 $\pm$ 0.48 & 7.87 $\pm$ 0.45 & \nodata & \nodata & \nodata & 16$^{\dagger}$\\
ESO 148-2 & 10.97 $\pm$ 0.20 & 3 & 7.42 $\pm$ 0.48 & 7.57 $\pm$ 0.50 & \nodata & \nodata & \nodata & 17$^{\dagger}$\\
\hline
\multicolumn{9}{c}{Stage-D}\\
\hline
NGC 34 & 10.76 $\pm$ 0.20 & 3 & 7.20 $\pm$ 0.48 & 7.71 $\pm$ 0.52 & 8.56 $\pm$ 0.09 & \nodata & \nodata & 18,10\\
IRAS F05189--2524 & 10.93 $\pm$ 0.20 & 3 & 7.38 $\pm$ 0.48 & 7.43 $\pm$ 0.50 & 8.32 & 8.62 & $M$--$\sigma$(Ca) & 13,19,20\\
NGC 2623 & 10.60 $\pm$ 0.20 & 3 & 7.03 $\pm$ 0.48 & 7.61 $\pm$ 0.50 & 7.97 $\pm$ 0.07 & 7.88 $\pm$ 0.50 & plane equation & 13,10\\
IRAS F08572+3915 & 10.49 $\pm$ 0.20 & 3 & 6.91 $\pm$ 0.48 & 8.37 $\pm$ 0.50 & 7.16 $\pm$ 0.13 & \nodata & \nodata & 16$^{\dagger}$,10\\
UGC 5101 & 11.00 $\pm$ 0.20 & 3 & 7.45 $\pm$ 0.48 & 8.00 $\pm$ 0.50 & 8.98 $\pm$ 0.11 & 8.23 $\pm$ 0.50 & plane equation & 13,10\\
Mrk 231 & 11.51 $\pm$ 0.20 & 3 & 7.99 $\pm$ 0.48 & 7.18 $\pm$ 0.48 & 8.58 & 7.94 & H$\beta$ EW & 21$^{\dagger}$,19,22\\
IRAS 13120--5453 & 11.33 $\pm$ 0.20 & 3 & 7.80 $\pm$ 0.48 & \nodata & 9.07 $\pm$ 0.11 & \nodata & \nodata & 10\\
Mrk 273 & 10.99 $\pm$ 0.20 & 3 & 7.44 $\pm$ 0.48 & 8.49 $\pm$ 0.50 & 8.30 & 9.17 $\pm$ 0.05 & maser & 13,19,18\\
IRAS F14348--1447 & 11.14 $\pm$ 0.20 & 3 & 7.60 $\pm$ 0.48 & 7.59 $\pm$ 0.54 & 8.32 $\pm$ 0.07 & \nodata & \nodata & 21$^{\dagger}$,10\\
Arp 220W & [10.97 $\pm$ 0.20]: & 3 & [7.42 $\pm$ 0.48]: & [7.75 $\pm$ 0.46]: & [8.18]: & \nodata & \nodata & 21$^{\dagger}$,19\\
NGC 6240S & [11.50 $\pm$ 0.20]: & 3 & [7.98 $\pm$ 0.48]: & 8.85 $\pm$ 0.05 & 8.76 $\pm$ 0.10 & \nodata & \nodata & 23,10\\
NGC 6240N & [11.50 $\pm$ 0.20]: & 3 & [7.98 $\pm$ 0.48]: & 8.56 $\pm$ 0.10 & 8.20 $\pm$ 0.07 & \nodata & \nodata & 23,10\\
IRAS F17138--1017 & 10.72 $\pm$ 0.20 & 3 & 7.16 $\pm$ 0.48 & 6.37 $\pm$ 0.53 & \nodata & \nodata & \nodata & 24$^{\dagger}$\\
\hline
\multicolumn{9}{c}{Stage-N}\\
\hline
NGC 1068 & 11.06 $\pm$ 0.20 & 3 & 7.51 $\pm$ 0.48 & 7.59 $\pm$ 0.44 & \nodata & 6.96 $\pm$ 0.02 & maser & 12,18\\
UGC 2608 & 10.92 $\pm$ 0.20 & 3 & 7.37 $\pm$ 0.48 & 7.04 $\pm$ 0.46 & \nodata & 8.50 $\pm$ 0.40 & $L_{\rm 2\mu m}$ & 24$^{\dagger}$,7\\
NGC 1365 & 11.16 & 4 & 7.62 & 7.50 $\pm$ 0.51 & 8.88 & 8.50 $\pm$ 0.40 & $L_{\rm 2\mu m}$ & 18,14,7\\
MCG--03-34-064 & 11.09 $\pm$ 0.20 & 3 & 7.54 $\pm$ 0.48 & 7.65 $\pm$ 0.40 & 8.28 & 8.34 $\pm$ 0.50 & plane equation & 9,14,13\\
NGC 5135 & 11.03 $\pm$ 0.20 & 3 & 7.48 $\pm$ 0.48 & 7.29 $\pm$ 0.44 & \nodata & 8.55 $\pm$ 0.40 & $L_{\rm 2\mu m}$ & 12,7\\
NGC 7130 & 11.09 $\pm$ 0.20 & 3 & 7.54 $\pm$ 0.48 & 7.59 $\pm$ 0.44 & \nodata & 8.55 $\pm$ 0.40 & $L_{\rm 2\mu m}$ & 12,7\\
\hline
\multicolumn{9}{c}{Average (Number of Objects)}\\
\hline
Stage-A\ \ (8) & 10.93\ \ (7) & \nodata & 7.38\ \ (7) & 7.69\ \ (7) & 7.75\ \ (2) & 8.10\ \ (8) & \nodata & \nodata \\
Stage-B\ \ (7) & 10.97\ \ (5) & \nodata & 7.42\ \ (5) & 7.48\ \ (1) & 8.41\ \ (4) & 8.32\ \ (3) & \nodata & \nodata \\
Stage-C\ \ (5) & 10.81\ \ (5) & \nodata & 7.25\ \ (5) & 7.61\ \ (3) & \nodata     & \nodata     & \nodata & \nodata \\
Stage-D   (13) & 10.95   (10) & \nodata & 7.39   (10) & 7.83   (11) & 8.38   (11) & 8.37\ \ (5) & \nodata & \nodata \\
Stage-N\ \ (6) & 11.06\ \ (6) & \nodata & 7.51\ \ (6) & 7.44\ \ (6) & 8.58\ \ (2) & 8.23\ \ (6) & \nodata & \nodata \\
All (39)        & 10.95   (33) & \nodata & 7.39   (33) & 7.68   (28) & 8.34   (19) & 8.23   (22) & \nodata & \nodata \\
\enddata
\tablecomments{Columns:
(1) object name;
(2--3) logarithmic stellar mass and their references;
(4) logarithmic black hole mass derived from the relation of stellar mass and black hole mass \citep{Reines2015};
(5) logarithmic black hole mass derived from the $M$--$\sigma_{*}$ relation;
(6) logarithmic black hole mass derived from the photometric bulge luminosity;
(7--8) logarithmic black hole mass and the method for estimation;
(9) references of the results in columns (5--7). 
Values in brackets with ``:'' should be considered to be upper limits
since the components of 
two galaxies are not separated.
The error is 1$\sigma$.\\
\textbf{References:}
(1) \citet{Koss2011b}; (2) \citet{Pereira-Santaella2015}; (3) \citet{Shangguan2019}; (4) \citet{Howell2010};
(5) \citet{Marleau2017}; (6) \citet{Gonzalez-Martin2009}; (7) \citet{Caramete2010}; (8) \citet{Koss2016a};
(9) \citet{Alonso-Herrero2013}; (10) \citet{Haan2011a}; (11) \citet{Bentz2015}; (12) \citet{Marinucci2012};
(13) \citet{Dasyra2011}; (14) \citet{Winter2009}; (15) \citet{Ptak2015}; (16) SDSS Data Release 16;
(17) \citet{Dasyra2006a}; (18) \citet{Izumi2016} and therein references; (19) \citet{Veilleux2009c}; (20) \citet{Xu2017};
(21) \citet{Dasyra2006b}; (22) \citet{Kawakatu2007b}; (23) \citet{Kollatschny2020}; (24) HyperLeda.
The $\dagger$ means that the $M$--$\sigma_{*}$ relation in \citet{Gultekin2009} is adopted.}
\end{deluxetable*}

\begin{deluxetable*}{lccccccll}
\label{T11_Edd-Kx}
\tablecaption{Summary of the AGN Properties Derived from the X-ray/IR Luminosities and Torus Parameters}
\tabletypesize{\scriptsize}
\tablehead{
\colhead{Object} &
\colhead{log$L_{2-10}$} &
\colhead{log$L_{10-50}$} &
\colhead{log$L^{\rm (IR;lit,ave)}_{\rm bol,AGN}$} &
\colhead{log$M^{\rm (lit,ave)}_{\rm BH}$} &
\colhead{log$\lambda_{\rm Edd}$(X/IR)} &
\colhead{$\kappa_{\rm bol,X}$} &
\colhead{$C^{\rm (22)}_{\rm T}$} &
\colhead{$C^{\rm (24)}_{\rm T}$} \\
& (erg s$^{-1}$) & (erg s$^{-1}$) & (erg s$^{-1}$) & ($M_{\odot}$) & & 
}
\decimalcolnumbers
\startdata 
\multicolumn{9}{c}{Stage-A}\\ 
\hline 
NGC 833 & 41.99$^{+0.25}_{-0.24}$ & 41.97$^{+0.25}_{-0.24}$ &  \nodata & 8.61 & $-$3.42$^{+0.25}_{-0.24}$/\ \ \ \ \nodata \ \ \ \ & \nodata & 0.63$^{+0.04}_{-0.03}$ & 0.43$^{+0.01}_{-0.04}$ \\
NGC 835 & 42.06$^{+0.13}_{-0.03}$ & 42.37$^{+0.13}_{-0.03}$ & 43.09$^{+0.39}_{-0.39}$ & 7.85 & $-$2.60$^{+0.13}_{-0.03}$/$-$2.87$^{+0.39}_{-0.39}$ & 11$^{+17}_{-6}$ & 0.61$^{+0.04}_{-0.01}$ & 0.45$^{+0.01}_{-0.02}$ \\
NGC 6921 & 42.80$^{+0.37}_{-0.44}$ & 43.01$^{+0.37}_{-0.44}$ &  \nodata & 8.11 & $-$2.11$^{+0.37}_{-0.44}$/\ \ \ \ \nodata \ \ \ \ & \nodata & 0.73$^{a}$ & 0.31$^{a}$ \\
MCG+04-48-002 & 42.44$^{+0.15}_{-0.15}$ & 42.65$^{+0.15}_{-0.15}$ & 43.66$^{+0.36}_{-0.36}$ & 7.64 & $-$2.00$^{+0.15}_{-0.15}$/$-$2.09$^{+0.36}_{-0.36}$ & 16$^{+23}_{-10}$ & 0.82$^{+0.18a}_{-0.12}$ & 0.46$^{+0.02}_{-0.14}$ \\
NGC 7469 & 43.26$^{+0.03}_{-0.02}$ & 43.33$^{+0.03}_{-0.02}$ & 44.80$^{+0.33}_{-0.33}$ & 7.30 & $-$0.84$^{+0.03}_{-0.02}$/$-$0.60$^{+0.33}_{-0.33}$ & 34$^{+39}_{-18}$ & 0.48$^{+0.04}_{-0.05}$ & 0.18$^{+0.07}_{-0.03}$ \\
NGC 7674 & 42.59$^{+0.33}_{-0.23}$ & 42.80$^{+0.33}_{-0.23}$ & 45.32$^{+0.35}_{-0.35}$ & 7.99 & $-$2.20$^{+0.33}_{-0.23}$/$-$0.78$^{+0.35}_{-0.35}$ & 529$^{+1071}_{-328}$ & 1.00$^{+0.00a}_{-0.01}$ & 0.68$^{+0.07}_{-0.15}$ \\
NGC 7679 & 42.34$^{+0.06}_{-0.04}$ & 42.47$^{+0.06}_{-0.04}$ & 44.20$^{+0.37}_{-0.37}$ & 7.42 & $-$1.88$^{+0.06}_{-0.04}$/$-$1.32$^{+0.37}_{-0.37}$ & 73$^{+99}_{-42}$ & 0.26$^{+0.21a}_{-0.26}$ &  \ \ \nodata \\
NGC 7682 & 41.94$^{+0.06}_{-0.06}$ & 42.05$^{+0.06}_{-0.06}$ &  \nodata & 7.48 & $-$2.33$^{+0.06}_{-0.06}$/\ \ \ \ \nodata \ \ \ \ & \nodata & 1.00$^{+0.00a}_{-0.24}$ &  \ \ \nodata \\
\hline 
\multicolumn{9}{c}{Stage-B}\\ 
\hline 
NGC 235 & 43.28$^{+0.18}_{-0.16}$ & 43.47$^{+0.18}_{-0.16}$ & 44.33$^{+0.41}_{-0.41}$ & 8.20 & $-$1.72$^{+0.18}_{-0.16}$/$-$1.97$^{+0.41}_{-0.41}$ & 11.4$^{+21.0}_{-7.3}$ & 0.75$^{+0.13}_{-0.05}$ & 0.43$^{+0.02}_{-0.08}$ \\
CGCG 468-002W & 42.84$^{+0.04}_{-0.03}$ & 43.07$^{+0.04}_{-0.03}$ & 44.17$^{+0.40}_{-0.40}$ & 7.67 & $-$1.63$^{+0.04}_{-0.03}$/$-$1.60$^{+0.40}_{-0.40}$ & 21$^{+33}_{-13}$ & 0.52$^{+0.01}_{-0.01}$ & 0.01$^{+0.22a}_{-0.01}$ \\
ESO 060-IG016 East & 41.94$^{+0.07}_{-0.08}$ & 42.05$^{+0.07}_{-0.08}$ & 44.84$^{+0.11}_{-0.10}$ & 7.79 & $-$2.65$^{+0.07}_{-0.08}$/$-$1.06$^{+0.11}_{-0.10}$ & 781$^{+271}_{-197}$ &  \ \ \nodata &  \ \ \nodata \\
Mrk 266B & 43.13$^{+0.39}_{-0.36}$ &  \nodata & 45.83$^{+0.60}_{-0.60}$ & 8.28 & $-$1.95$^{+0.39}_{-0.36}$/$-$0.55$^{+0.60}_{-0.60}$ & 504$^{+2125}_{-403}$ &  \ \ \nodata &  \ \ \nodata \\
Mrk 266A & 41.60$^{+0.04}_{-0.02}$ &  \nodata &  \nodata & 8.37 & $-$3.57$^{+0.04}_{-0.02}$/\ \ \ \ \nodata \ \ \ \ & \nodata &  \ \ \nodata &  \ \ \nodata \\
IC 4518A & 42.83$^{+0.06}_{-0.08}$ & 42.94$^{+0.06}_{-0.08}$ & 44.53$^{+0.37}_{-0.37}$ & 7.53 & $-$1.49$^{+0.06}_{-0.08}$/$-$1.10$^{+0.37}_{-0.37}$ & 50$^{+69}_{-29}$ & 0.67$^{+0.02}_{-0.02}$ & 0.32$^{+0.03}_{-0.04}$ \\
NGC 6286 & 42.01$^{+1.21}_{-0.23}$ & 42.32$^{+1.21}_{-0.23}$ & 44.06$^{+0.10}_{-0.10}$ & 7.98 & $-$2.77$^{+1.21}_{-0.23}$/$-$2.02$^{+0.10}_{-0.10}$ & 112$^{+1719}_{-49}$ &  \ \ \nodata &  \ \ \nodata \\
\hline 
\multicolumn{9}{c}{Stage-C}\\ 
\hline 
MCG+12-02-001 & 41.75$^{+0.33}_{-0.08}$ & 41.86$^{+0.33}_{-0.08}$ & 44.22$^{+0.16}_{-0.32}$ & 7.04 & $-$2.09$^{+0.33}_{-0.08}$/$-$0.92$^{+0.16}_{-0.32}$ & 297$^{+400}_{-158}$ &  \ \ \nodata &  \ \ \nodata \\
IRAS F06076--2139 & 42.18$^{+0.15}_{-0.14}$ & 42.29$^{+0.15}_{-0.14}$ & 44.17$^{+0.11}_{-0.09}$ & 7.30 & $-$1.92$^{+0.15}_{-0.14}$/$-$1.23$^{+0.11}_{-0.09}$ & 98$^{+52}_{-32}$ &  \ \ \nodata &  \ \ \nodata \\
NGC 3690 West & 42.66$^{+0.34}_{-0.32}$ & 42.48$^{+0.34}_{-0.32}$ & 43.89$^{+0.44}_{-0.44}$ & 7.09 & $-$1.22$^{+0.34}_{-0.32}$/$-$1.30$^{+0.44}_{-0.44}$ & 17$^{+43}_{-12}$ & 1.00$^{+0.00a}_{-0.01}$ & 0.97$^{+0.03}_{-0.04}$ \\
NGC 4922N & 42.00$^{+0.20}_{-0.18}$ & 42.11$^{+0.20}_{-0.18}$ & 44.25$^{+0.34}_{-0.33}$ & 7.79 & $-$2.59$^{+0.20}_{-0.18}$/$-$1.63$^{+0.34}_{-0.33}$ & 181$^{+264}_{-105}$ &  \ \ \nodata &  \ \ \nodata \\
ESO 148-2 & 42.38$^{+0.24}_{-0.28}$ & 42.49$^{+0.24}_{-0.28}$ & 45.27$^{+0.37}_{-0.38}$ & 7.49 & $-$1.91$^{+0.24}_{-0.28}$/$-$0.32$^{+0.37}_{-0.38}$ & 773$^{+1353}_{-511}$ &  \ \ \nodata &  \ \ \nodata \\
\hline 
\multicolumn{9}{c}{Stage-D}\\ 
\hline 
NGC 34 & 41.90$^{+0.10}_{-0.10}$ & 42.01$^{+0.10}_{-0.10}$ & 43.74$^{+0.03}_{-0.03}$ & 7.82 & $-$2.72$^{+0.10}_{-0.10}$/$-$2.19$^{+0.03}_{-0.03}$ & 69$^{+18}_{-15}$ &  \ \ \nodata &  \ \ \nodata \\
IRAS F05189--2524 & 43.42$^{+0.01}_{-0.02}$ & 43.26$^{+0.01}_{-0.02}$ & 45.74$^{+0.31}_{-0.32}$ & 7.94 & $-$1.32$^{+0.01}_{-0.02}$/$-$0.30$^{+0.31}_{-0.32}$ & 208$^{+216}_{-108}$ & 0.56$^{+0.01}_{-0.01}$ & 0.39$^{+0.01}_{-0.02}$ \\
NGC 2623 & 40.90$^{+0.11}_{-0.11}$ & 41.01$^{+0.11}_{-0.11}$ & 43.75$^{+0.53}_{-0.59}$ & 7.62 & $-$3.52$^{+0.11}_{-0.11}$/$-$1.97$^{+0.53}_{-0.59}$ & 709$^{+1770}_{-530}$ &  \ \ \nodata &  \ \ \nodata \\
IRAS F08572+3915$^{b}$ & 41.77$^{+0.06}_{-0.07}$ & 41.88$^{+0.06}_{-0.07}$ & 45.78$^{+0.24}_{-0.33}$ & 7.48 & $-$2.51$^{+0.06}_{-0.07}$/+0.20$^{+0.24}_{-0.33}$ & 10390$^{+7854}_{-5601}$ &  \ \ \nodata &  \ \ \nodata \\
UGC 5101 & 43.15$^{+0.25}_{-0.15}$ & 43.46$^{+0.25}_{-0.15}$ & 44.81$^{+0.32}_{-0.34}$ & 8.17 & $-$1.82$^{+0.25}_{-0.15}$/$-$1.45$^{+0.32}_{-0.34}$ & 47$^{+73}_{-27}$ & 0.56$^{+0.24a}_{-0.56}$ & 0.41$^{+0.01a}_{-0.41}$ \\
Mrk 231 & 42.65$^{+0.02}_{-0.02}$ & 42.97$^{+0.02}_{-0.02}$ & 46.09$^{+0.15}_{-0.17}$ & 7.92 & $-$2.07$^{+0.02}_{-0.02}$/+0.07$^{+0.15}_{-0.17}$ & 2719$^{+1125}_{-879}$ & 0.70$^{+0.13}_{-0.08}$ &  \ \ \nodata \\
IRAS 13120--5453 & 42.17$^{+0.59}_{-0.11}$ & 42.28$^{+0.59}_{-0.11}$ & 44.48$^{+0.30}_{-0.27}$ & 8.43 & $-$3.07$^{+0.59}_{-0.11}$/$-$2.06$^{+0.30}_{-0.27}$ & 204$^{+741}_{-100}$ &  \ \ \nodata &  \ \ \nodata \\
Mrk 273 & 43.07$^{+0.21}_{-0.21}$ & 43.24$^{+0.21}_{-0.21}$ & 45.10$^{+0.39}_{-0.45}$ & 8.35 & $-$2.08$^{+0.21}_{-0.21}$/$-$1.35$^{+0.39}_{-0.45}$ & 106$^{+188}_{-72}$ & 0.71$^{+0.04}_{-0.04}$ & 0.44$^{+0.02}_{-0.03}$ \\
IRAS F14348--1447 & 42.70$^{+0.93}_{-0.40}$ & 42.81$^{+0.93}_{-0.40}$ & 44.55$^{+0.18}_{-0.20}$ & 7.84 & $-$1.94$^{+0.93}_{-0.40}$/$-$1.39$^{+0.18}_{-0.20}$ & 71$^{+563}_{-45}$ &  \ \ \nodata &  \ \ \nodata \\
Arp 220W$^{b}$ & 41.59$^{+0.02}_{-0.02}$ & 41.70$^{+0.02}_{-0.02}$ &  \nodata &  \nodata  &  \ \ \ \ \nodata \ \ /\ \ \ \ \nodata \ \ \ \  & \nodata &  \ \ \nodata &  \ \ \nodata \\
NGC 6240S & 43.72$^{+0.20}_{-0.19}$ &  \nodata &  \nodata & 8.81 & $-$1.89$^{+0.20}_{-0.19}$/\ \ \ \ \nodata \ \ \ \ & \nodata &  \ \ \nodata &  \ \ \nodata \\
NGC 6240N & 43.30$^{+0.24}_{-0.20}$ &  \nodata &  \nodata & 8.38 & $-$1.88$^{+0.24}_{-0.20}$/\ \ \ \ \nodata \ \ \ \ & \nodata &  \ \ \nodata &  \ \ \nodata \\
IRAS F17138--1017$^{b}$ & 41.68$^{+0.09}_{-0.06}$ & 41.79$^{+0.09}_{-0.06}$ & 44.05$^{+0.34}_{-0.39}$ & 6.76 & $-$1.89$^{+0.09}_{-0.06}$/$-$0.81$^{+0.34}_{-0.39}$ & 238$^{+299}_{-141}$ &  \ \ \nodata &  \ \ \nodata \\
\hline 
\multicolumn{9}{c}{Stage-N}\\ 
\hline 
NGC 1068 & 43.34$^{+0.07}_{-0.08}$ &  \nodata & 45.12$^{+0.35}_{-0.35}$ & 7.35 & $-$0.81$^{+0.07}_{-0.08}$/$-$0.33$^{+0.35}_{-0.35}$ & 60$^{+78}_{-34}$ &  \ \ \nodata &  \ \ \nodata \\
UGC 2608 & 43.59$^{+0.19}_{-0.25}$ & 43.70$^{+0.19}_{-0.25}$ & 45.06$^{+0.32}_{-0.32}$ & 7.64 & $-$0.84$^{+0.19}_{-0.25}$/$-$0.68$^{+0.32}_{-0.32}$ & 29$^{+40}_{-18}$ & 0.45$^{+0.31a}_{-0.45}$ & 0.28$^{+0.23a}_{-0.28}$ \\
NGC 1365 & 41.90$^{+0.01}_{-0.01}$ &  \nodata & 44.15$^{+0.40}_{-0.40}$ & 8.13 & $-$3.02$^{+0.01}_{-0.01}$/$-$2.08$^{+0.40}_{-0.40}$ & 176$^{+270}_{-107}$ &  \ \ \nodata &  \ \ \nodata \\
MCG--03-34-064 & 43.27$^{+0.06}_{-0.07}$ & 43.24$^{+0.06}_{-0.07}$ & 44.98$^{+0.32}_{-0.32}$ & 7.95 & $-$1.48$^{+0.06}_{-0.07}$/$-$1.08$^{+0.32}_{-0.32}$ & 51$^{+58}_{-27}$ & 0.79$^{+0.02}_{-0.02}$ & 0.50$^{+0.01}_{-0.01}$ \\
NGC 5135 & 43.30$^{+0.42}_{-0.26}$ & 43.49$^{+0.41}_{-0.26}$ & 44.33$^{+0.32}_{-0.32}$ & 7.77 & $-$1.27$^{+0.42}_{-0.26}$/$-$1.55$^{+0.32}_{-0.32}$ & 11$^{+25}_{-7}$ & 0.44$^{+0.19a}_{-0.44}$ & 0.26$^{+0.17a}_{-0.26}$ \\
NGC 7130 & 42.87$^{+0.31}_{-0.25}$ & 43.09$^{+0.31}_{-0.25}$ & 44.17$^{+0.32}_{-0.31}$ & 7.89 & $-$1.82$^{+0.31}_{-0.25}$/$-$1.82$^{+0.32}_{-0.31}$ & 20$^{+35}_{-12}$ & 0.88$^{+0.12}_{-0.44}$ & 0.52$^{+0.14}_{-0.26}$ \\
\enddata
\tablecomments{Columns:
(1) object name;
(2--3) logarithmic absorption-corrected X-ray luminosities in the 2--10~keV and 10--50~keV bands.
The errors are estimated by fixing the photon index at the best-fit value;
(4) average value of the logarithmic bolometric AGN luminosities 
(excluding the values of upper limits) derived from the four independent estimates 
based on the IR properties in Table~\ref{T8_AGN-properties}. 
The error is converted to the 90\% confidence level by multiplying a factor of 1.65;
(5) average value of the logarithmic black hole mass derived from the four independent 
estimates in Table~\ref{T10_BHmass};
(6) logarithmic Eddington ratio ($L_{\rm bol}/L_{\rm Edd}$). 
Here, the Eddington luminosity is defined as $1.26 \times 10^{38} M_{\rm BH}/M_{\odot}$.
The bolometric luminosities are obtained from 20$L_{2-10}$ and $L^{\rm (IR;lit,ave)}_{\rm bol,AGN}$, respectively;
(7) bolometric-to-X-ray (2--10~keV) correction factor;
(8--9) torus covering factor derived from the torus parameters for
 Compton-thin ($N_{\rm H} \geq 10^{22}$ cm$^{-2}$) and CT ($N_{\rm H}
 \geq 10^{24}$ cm$^{-2}$) material, respectively, for the AGNs whose
 torus angular widths ($\sigma$) are constrained. 
}
\tablenotetext{a}{The parameter of torus angular width reaches a limit of its allowed range.}
\tablenotetext{b}{X-ray luminosities and bolometric correction factors are uncertain due to the assumption of $f_{\rm bin}$ = 10\% for these three objects.}
\end{deluxetable*}

\begin{figure}
    \epsscale{1.20}
    \plotone{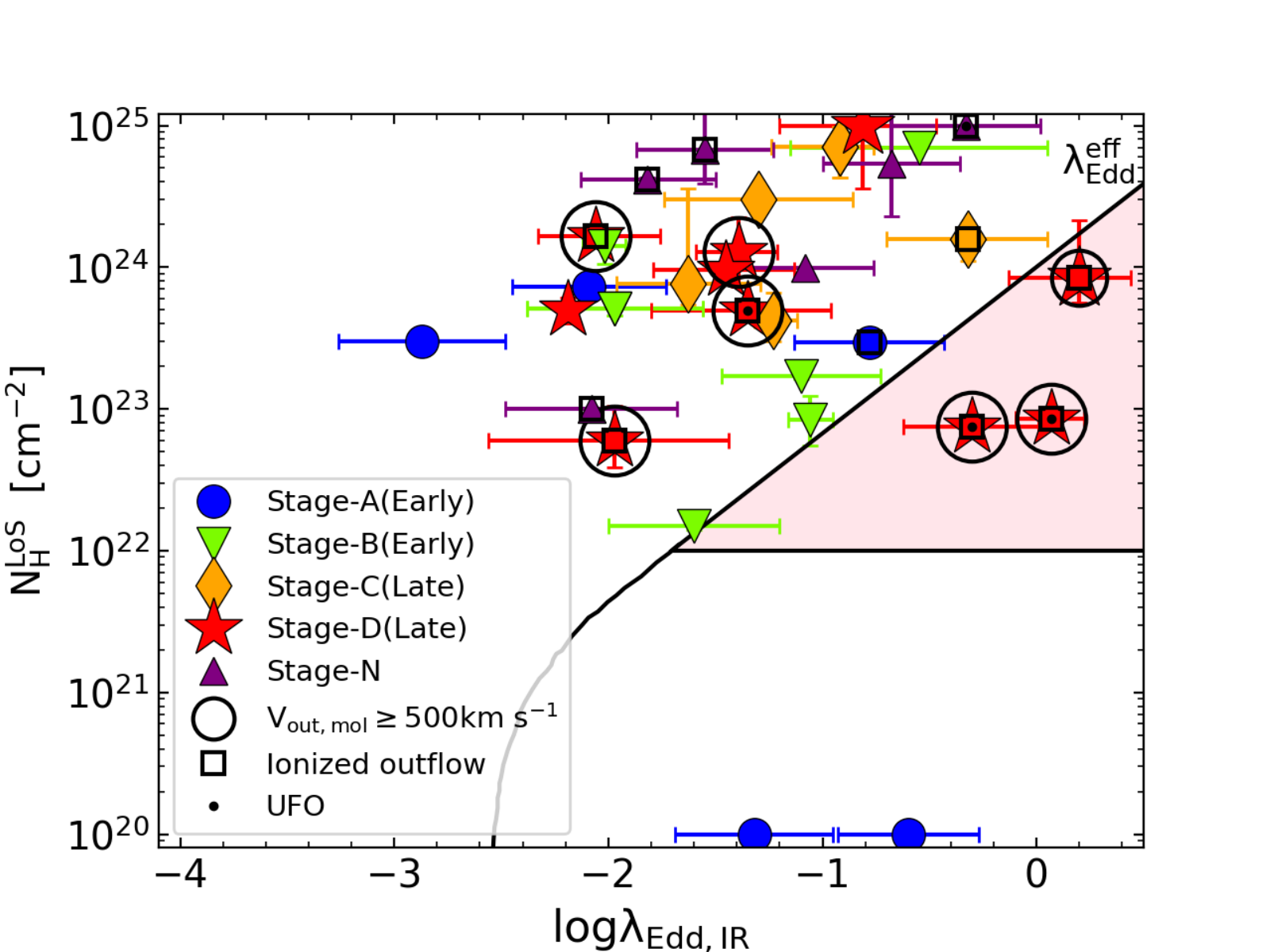}
        \caption{Line-of-sight hydrogen column density vs. IR-based Eddington ratio.
        The black continuous line shows the effective Eddington limit ($\lambda^{\rm eff}_{\rm Edd}$; \citealt{Fabian2008,Fabian2009}).
        The pink triangle area represents the ``forbidden'' region obtained
        by the effective Eddington limit and the $N_{\rm H}^{\rm LoS}$ =
        10$^{22}$ cm$^{-2}$ limit, below which absorption by outer dust lanes may become important.
        Symbols and colors are the same as in Figure~\ref{F3_NH-kpc}.
        The black empty circles, empty squares, and filled points mark the detections of strong molecular outflows with velocities of $V_{\rm out,mol} \geq 500$ km s$^{-1}$ (e.g., \citealt{Gonzalez-Alfonso2017,Laha2018} and the references therein), ionized outflows \citep[e.g.,][]{Rich2015,Kakkad2018,Fluetsch2021}, and UFOs \citep{Mizumoto2019,Smith2019}, respectively.
        \label{F10_NH-Kx}}
\label{F_spectra}
\end{figure}

\subsubsection{$N_{\rm H}$--$\lambda_{\rm Edd}$ Diagram}
\label{subsub6-1-3_NH-Edd}

In Figure~\ref{F10_NH-Kx}, we plot our sample in the $N_{\rm H}$--$\lambda_{\rm Edd}$ diagram,
which is useful to examine the effect of radiation pressure to obscuring
material around an AGN
\citep{Fabian2008,Fabian2009,Kakkad2016,Ricci2017cNature,Ishibashi2018}.
The horizontal line shows the $N_{\rm H} = 10^{22}$~cm$^{-2}$ limit 
below which absorption by outer dust lanes may become important
\citep{Fabian2008,Fabian2009}. 
The pink triangle area above it is
called as ``forbidden'' region where dusty clouds within the 
sphere of influence of the black hole 
are pushed
away by radiation pressure against the gravitational force by the black hole
and hence cannot survive in a stationary state. 
We find that six AGNs in stage-D mergers host both of ionized outflows 
(empty squares) and strong molecular outflows with $V_{\rm out,mol} \geq$
500~km~s$^{-1}$ (empty circles), and three of them with $\lambda_{\rm Edd} \gtrsim 0.5$ (IRAS F05189--2524, IRAS F08572+3915, and Mrk~231)
are located within the forbidden region.
Note that IRAS F05189--2524 and
Mrk~231 show UFOs \citep{Mizumoto2019,Smith2019}, although IRAS
F08572+3915 (with the highest $\lambda_{\rm Edd}$) is too faint to 
examine the presence of an UFO with the current X-ray data.

These results imply that the obscuration 
in these three AGNs is caused 
by in/outflowing materials within the torus scale (a few to tens of pc),
where the gravitational potential of the black hole dominates 
over that of the host galaxy.\footnote{The presence 
of molecular outflows in some objects with lower
Eddington ratios could also be explained if they have column
densities sensitive to radiation pressure outside the line of sight.} 
Alternatively, the absorption could be caused by 
host-galaxy interstellar medium located far away from the torus. 
We infer this possibility unlikely, however, because dramatic
variability of the absorption column density within $\sim$10 years is
detected in IRAS F05189--2524 (Section~\ref{subsub6-1-5_NHvariability}).
The simulations 
\citep[e.g.,][]{Hopkins2006,Angles-Alcazar2017,Kawaguchi2020} 
and observations
\citep[e.g.,][]{Satyapal2014,Kocevski2015,Ricci2017bMNRAS,Yamada2019} 
show that mergers can trigger quasi-spherical inflows of material onto 
the inner pc-scale region of the systems, which may also 
produce powerful outflows (see Section~\ref{subsub6-1-2_outflows}). 
The variability
of $N_{\rm H}$ and the detection of UFOs are consistent with the
scenario that the obscuration in these three AGNs is caused by
the chaotic in/outflows in the nuclear region.

\subsubsection{Torus Covering Factor and Eddington Ratio}
\label{subsub6-1-4_CoveringFactor}

As mentioned in Section~\ref{sub5-2_CTfraction}, 
the large fractions of obscured AGNs and CT
AGNs in late mergers suggest they are deeply buried by surrounding gas
and dust. On the basis of the fraction of obscured AGNs in the Swift/BAT
70-month catalog, \citet{Ricci2017cNature} report that the torus
covering factors by Compton-thin material become smaller in AGNs with
larger Eddington ratios due to radiation pressure from the nuclei. Here
we quantitatively investigate whether the covering fractions of our
AGN sample follow the \citet{Ricci2017cNature} relation obtained from
normal AGNs.

\begin{figure*}
    \epsscale{1.15}
    \plottwo{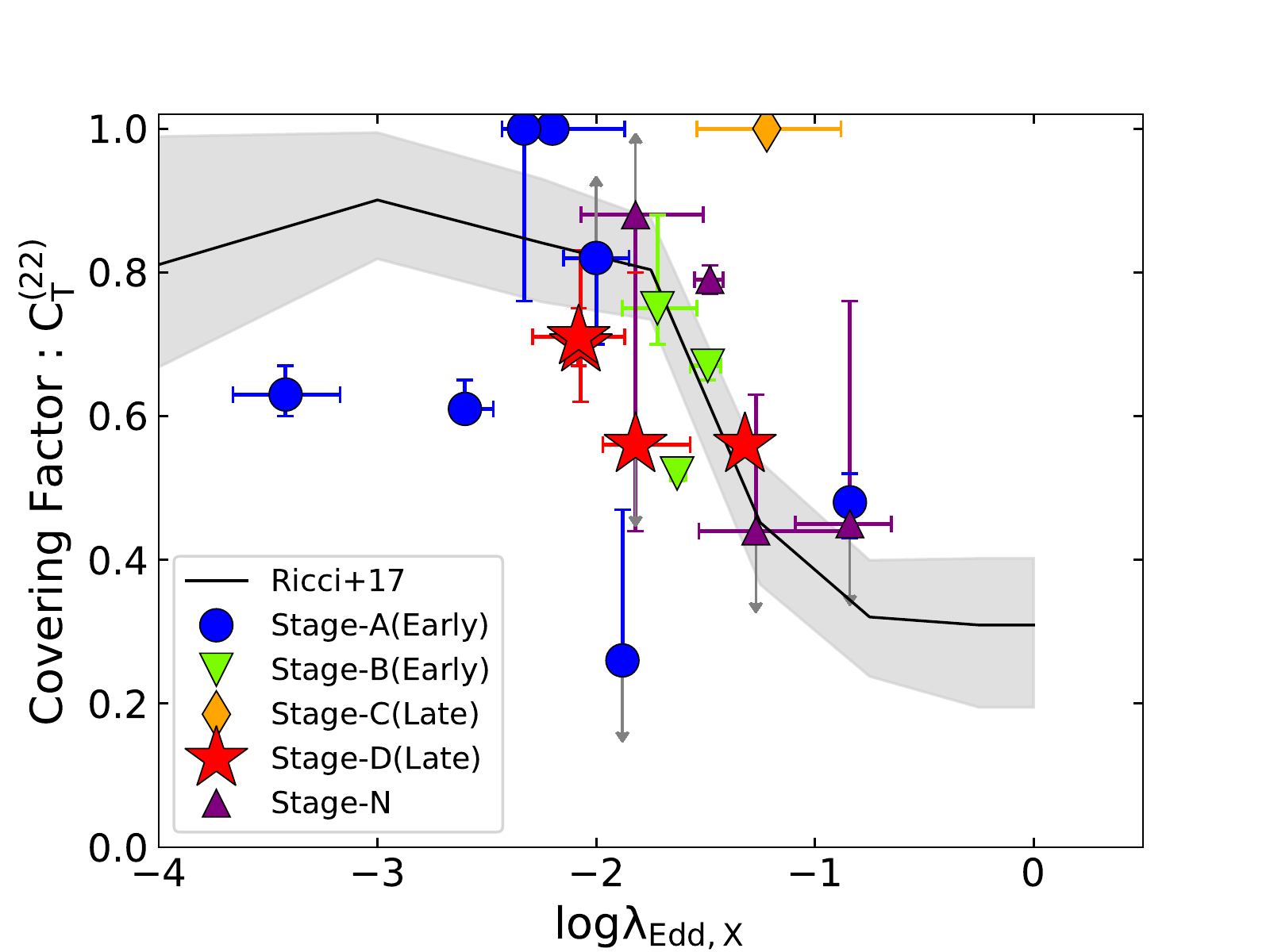}{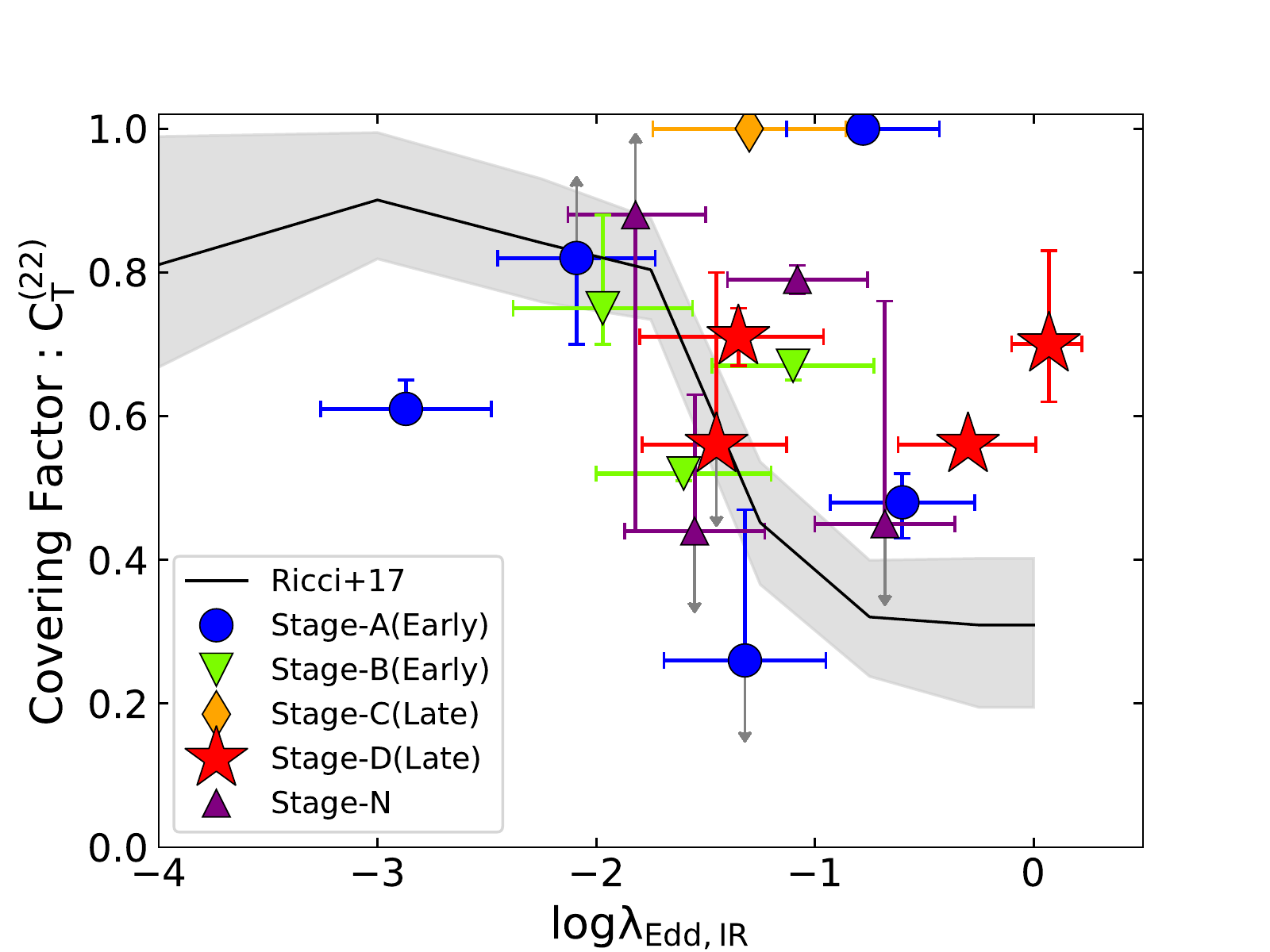}
        \caption{Left panel: torus covering factor vs. X-ray based Eddington ratio assuming $\kappa_{\rm bol,X}$ = 20.
        Right panel: torus covering factor vs. IR-based Eddington ratio.
        The black solid curve and gray shaded area represent the
 relation and its 1$\sigma$ dispersion found by \citet{Ricci2017cNature}
 for AGNs in the Swift/BAT 70-month catalog.
        Symbols and colors are the same as in Figure~\ref{F3_NH-kpc}.\\
        \label{F11_CoveringFactor}}
\end{figure*}

An advantage of the XCLUMPY model is that the torus covering factors
($C_{\rm T}$) of the {\it individual} AGNs can be constrained by using
the torus parameters: the equatorial column density ($N_{\rm H}^{\rm
Equ}$) and torus angular width ($\sigma$).
In the geometry of the XCLUMPY, the torus covering factors for
Compton-thin ($N_{\rm H} \geq 10^{22}$~cm$^{-2}$) material can be 
calculated by Equation~(6) in \citet{Ogawa2021}:
\begin{align}
&C^{\rm (22)}_{\rm T} = {\rm sin}\left(\sigma\sqrt{{\rm In}\left(N_{\rm H}^{\rm Equ}/(10^{22}\ {\rm cm}^{-2})\right)}\right),
\end{align}
and if $N_{\rm H}^{\rm Equ} \geq 10^{24}$~cm$^{-2}$, the value for CT material is
\begin{align}
&C^{\rm (24)}_{\rm T} = {\rm sin}\left(\sigma\sqrt{{\rm In}\left(N_{\rm H}^{\rm Equ}/(10^{24}\ {\rm cm}^{-2})\right)}\right).
\end{align}
The results are listed in Table~\ref{T11_Edd-Kx}.
These correspond to the covering factors defined in \citet{Ricci2017cNature},  
who statistically estimate $C^{\rm (22)}_{\rm T}$ and $C^{\rm (24)}_{\rm
T}$ from the fractions of the obscured and CT AGNs among the
whole Swift/BAT AGNs as a function of Eddington ratio.

Figure~\ref{F11_CoveringFactor} plots the torus covering factors 
against Eddington ratio. 
The Eddington ratios are calculated from the bolometric luminosities
estimated from the X-ray (left) or IR (right) results.
In the case of normal AGNs, 
\citet{Tanimoto2020} and \citet{Ogawa2021} show that the covering 
factors derived from XCLUMPY, are in good agreement with 
the relation of \citet{Ricci2017cNature} by referring to the 
X-ray luminosities.\footnote{\citet{Zhao2020} 
suggest that the covering factors estimated with the homogeneous torus model 
borus02 \citep{Balokovic2018} are also consistent with the 
relation in \citet{Ricci2017cNature}.}
By contrast, the covering factors of our
sample become systematically smaller than the \citet{Ricci2017cNature}
relation when we adopt the X-ray based Eddington ratios by assuming
$\kappa_{\rm bol,X} = 20$ (as done in 
\citealt{Ricci2017cNature}).\footnote{Even if the empirical
relation between bolometric correction and Eddington ratios
\citep[e.g.,][]{Lusso2012} is taken into account for AGNs with
$\lambda_{\rm Edd} \gtrsim 10^{-1}$, e.g., $\kappa_{\rm bol,X} \sim
100$, it has little effect on our discussion.}
Since we have revealed that AGNs in U/LIRGs are X-ray weak
(Section~\ref{subsub6-1-1_Xweak}), we find it more appropriate 
to refer to the IR
based Eddington ratios for this discussion.

The right panel of Figure~\ref{F11_CoveringFactor} shows that the covering factors of U/LIRGs
estimated with the XCLUMPY model are mostly consistent with the
\citet{Ricci2017cNature} relation.\footnote{The trend at $\lambda_{\rm Edd} \lesssim 10^{-1}$ is also consistent
with the IR-based results for $z<0.1$ SDSS quasars by \citet{Toba2021b}.}
However, we find that the AGNs with
the highest Eddington ratios ($\log \lambda_{\rm Edd, IR}\sim 0$) among
our sample, IRAS F05189--2524 and Mrk~231 (both stage-D mergers), have
larger covering factors ($C_{\rm T}^{\rm (22)} \sim 0.6$) than the
expected values for normal AGNs at similar Eddington ratios.
It is noteworthy that both objects are located in the forbidden region 
in the $N_{\rm H}$--$\lambda_{\rm Edd}$ diagram
(Section~\ref{subsub6-1-3_NH-Edd}).
These facts imply that these
AGNs are obscured by dusty in/outflows with large covering fractions
within the torus scale. 
It would be also possible that the nuclei in these gas-rich mergers are
embedded by high amounts of material in the inner regions of the host 
galaxies at larger scales than the tori.
We infer that the presence of such torus-scale in/outflows 
and/or host-galaxy scale material 
enhances the fraction of CT AGNs in late mergers
(Section~\ref{sub5-2_CTfraction}).

\subsubsection{Dramatic Variability of $N_{\rm H}^{\rm LoS}$ in Late Mergers}
\label{subsub6-1-5_NHvariability}

\begin{figure}
    \epsscale{1.15}
    \plotone{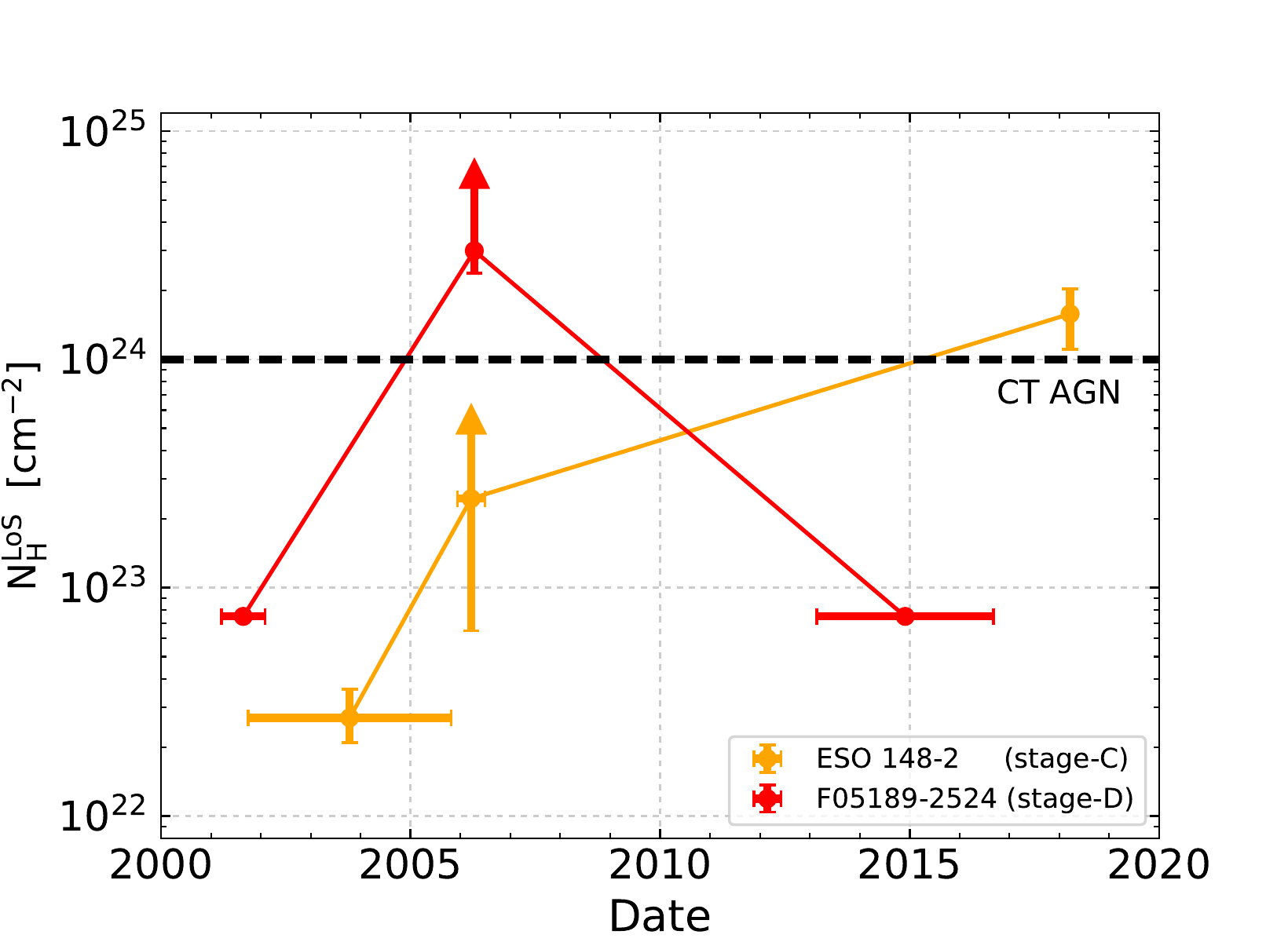}
        \caption{Line-of-sight hydrogen column density against
 observation date for ESO~148-2 (a stage-C merger) and IRAS F05189--2524 (a stage-D merger).\\
        \label{F12_NHvariability}}
\end{figure}

The massive transportation of gas and dust in rapidly accreting AGNs in
late mergers is also supported by variability of the line-of-sight
hydrogen column density. As illustrated in Figure~\ref{F12_NHvariability}, the AGN in
IRAS~F05189--2524, which has large Eddington ratio (log$\lambda_{\rm
Edd} = -0.30^{+0.31}_{-0.32}$) and covering factor ($C_{\rm T}^{\rm
(22)} = 0.56\pm0.01$), shows significant variability in the column
density, from $N_{\rm H}^{\rm LoS} = (7.5 \pm 0.1) \times
10^{22}$~cm$^{-2}$ in the Chandra, XMM-Newton, and NuSTAR observations
(in 2001--2002 and 2013--2016) to $>$2.3 $\times$
$10^{24}$~cm$^{-2}$ in the Suzaku observations (in 2006). We confirm that
the hard X-ray spectra observed with NuSTAR and Suzaku/HXD show little
flux variability, and that this is truly caused 
by variability in the absorption.\footnote{Hence, 
this variability is different from that seen in a changing-look AGN,
which is accompanied by a dramatic change in the absorption-corrected X-ray luminosity
\citep[e.g.,][]{Noda2018,Ruan2019,Ricci2020,Ricci2021a}.}
The stage-C ULIRG ESO~148-2 also shows large variability in
the hydrogen column density. 

By conducting an extensive X-ray spectral
variability study of 20 Compton-thin AGNs, \citet{Laha2020} infer that
the presence of compact-scale (a few pc) inhomogeneous gas,
which show strong variability in full-covering absorption of $N_{\rm H} \sim 10^{20}$--$10^{23}$~cm$^{-2}$, and a long-lived population of clumpy clouds detected as partial-covering absorbers with $N_{\rm H} \sim 10^{21}$--$10^{23.5}$~cm$^{-2}$.
Although there are only two examples showing dramatic variability 
in our sample, 
the observed transitions between Compton-thin and CT absorption 
column densities in late mergers suggest
that the intense in/outflows 
may produce strong turbulence and even much thicker transient
absorbers than in normal AGNs.

\begin{figure*}
    \epsscale{0.82}
    \plotone{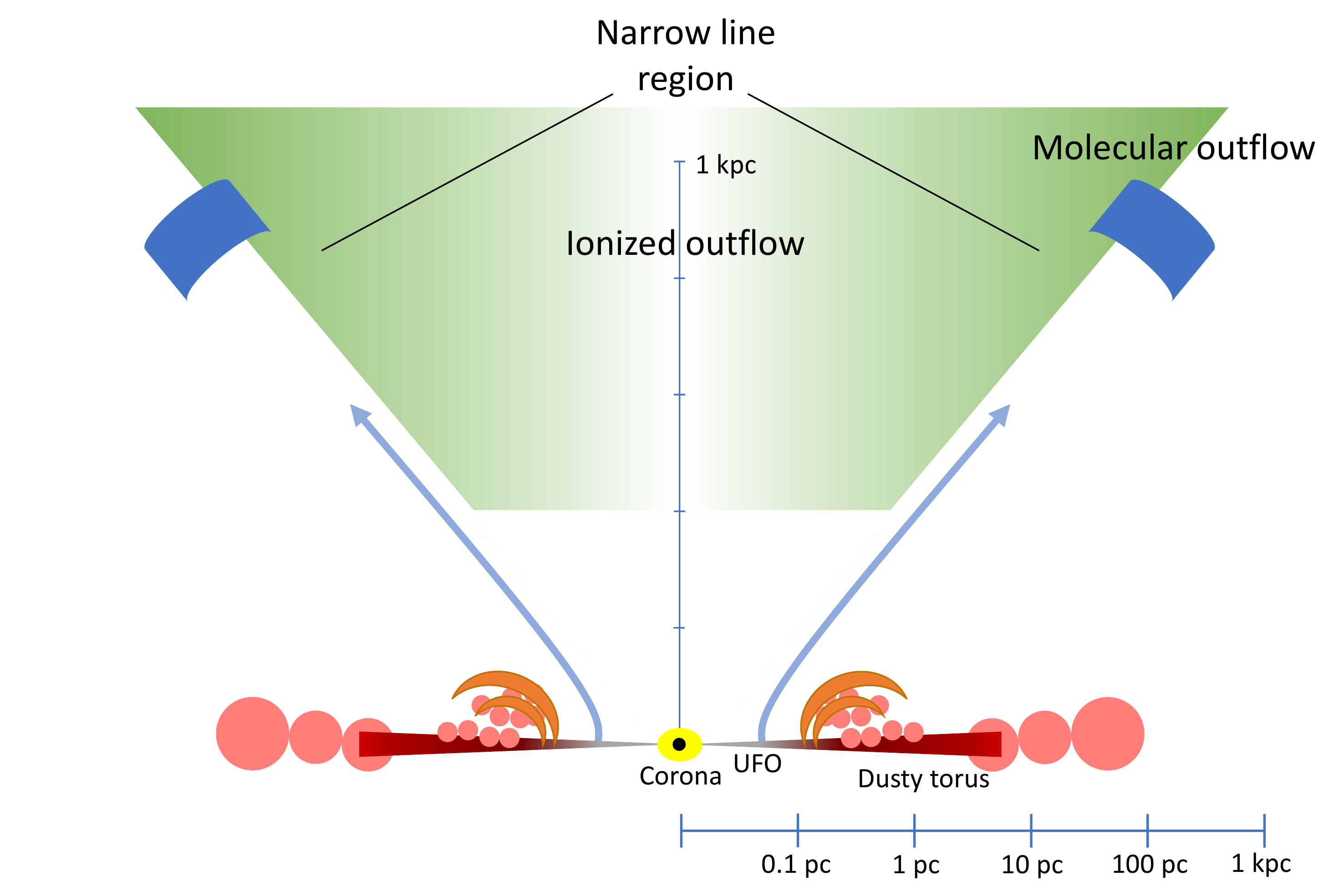}
    \plotone{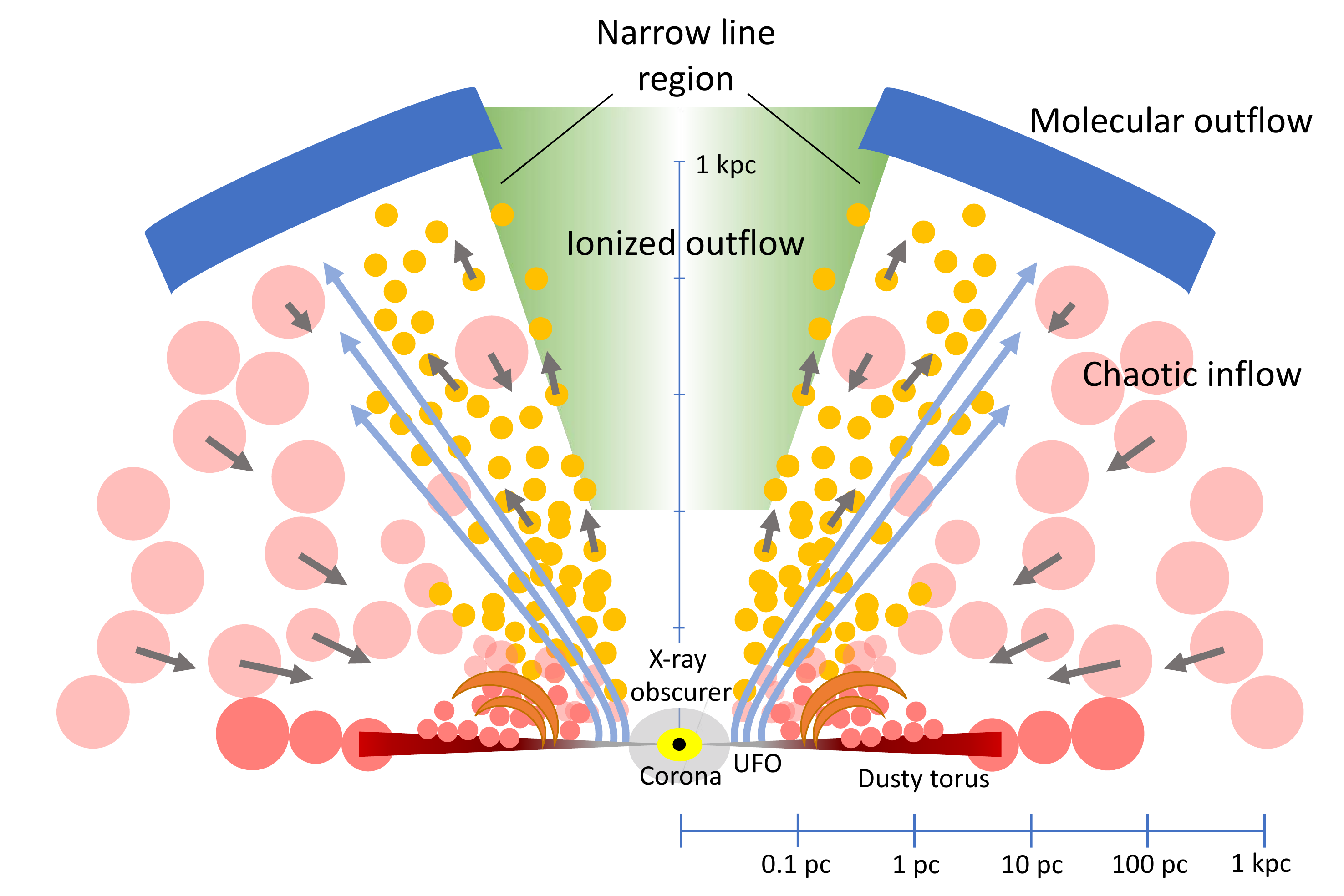}
        \caption{Schematic pictures of AGN structure 
        in nonmergers (top) and late mergers (bottom).
        Light pink, dark pink, and orange circles represent the chaotic
        quasi-spherical inflows, dusty torus region, 
        and outflowing material, respectively.
        In a late merger, X-ray emission from the hot corona is effectively
        blocked by optically-thick failed winds, 
        and stronger UV-driven winds (UFOs) are launched than in a nonmerger 
        (see text). In the high-Eddington ratio AGN in a late merger, 
        a part of the inflowing material is blown away by the radiation pressure. 
        These inflows and/or outflows make the AGN deeply ``buried''.\\
        \label{F13_Schematic-images}}
\end{figure*}

\subsubsection{Unified View of AGNs in Late Merger U/LIRGs}
\label{subsub6-1-6_Unified-view}

To summarize, we have shown that the AGNs in late mergers are 
X-ray weak
(i.e., have large bolometric corrections) compared with normal AGNs at
the same Eddington ratios (Section~\ref{subsub5-3-4_Edd}). 
A large fraction of them show
evidence for multiphase outflows, which are driven by the AGNs 
(Section~\ref{subsub6-1-2_outflows}). 
Even high Eddington-ratio AGNs in late mergers have large
torus covering factors, which are likely contributed by variable
in/outflows (Sections~\ref{subsub6-1-3_NH-Edd}, \ref{subsub6-1-4_CoveringFactor}, 
and \ref{subsub6-1-5_NHvariability}). These differences from normal
AGNs in nonmergers must be attributed to the environmental effects in
late mergers. Simulations show that massive chaotic inflows toward the
SMBHs may be produced in these systems.

We discuss possible physical mechanisms to account for the X-ray
weakness of the AGNs in late mergers (see also \citealt{Laha2018} for
galaxies with molecular outflows). As mentioned in
Section~\ref{subsub6-1-1_Xweak}, it cannot be attributed to the 
current decline in the AGN activities but is likely to be associated 
with AGN-driven strong outflows.
The X-ray luminosity could be suppressed if a part of the hot
corona responsible for Comptonization is destroyed by quasi-spherical
inflows. It is unlikely, however, that inflowing material directly
reaches the hot corona region, which is very compact ($<$10$r_{\rm g}$
where $r_{\rm g}$ is the gravitational radius; e.g., \citealt{Fabian2015}).
As a more plausible scenario, the analogy with BAL quasars suggests the
possibility that the X-rays are attenuated by 
dust-free partial absorbers due to
optically-thick ``failed winds'' in the inner regions 
of the obscuring torus (\citealt{Luo2013,Luo2014}).  
In UV-line driven wind models 
\citep[e.g.,][]{Czerny2017,Nomura2017}\footnote{When the material 
near the disk surface is in a low-ionization state, 
the radiation force on UV line transitions (the bound--bound transition of metals) 
is 10--1000 times larger than the radiation force due to Thomson scattering 
\citep{Stevens1990}, leading to a high velocity disk wind.}, 
a wind once launched from an inner region of the disk (at
$\sim$~10$^2$--10$^3$~$r_{\rm g}$)
is overionized by X-ray irradiation as the height increases,
lose the UV opacity, and eventually falls back into the disk (failed
winds). Under the presence of chaotic quasi-spherical inflows in late
mergers, the densities and covering factors of these failed winds may be
enhanced as material reaching the inner disks could serve as
ingredients of the wind. Then, these enhanced failed 
winds, which are dust free and do not re-emit in the IR band, more
efficiently block X-rays from the central regions, making the AGN SEDs
apparently X-ray weaker compared with normal AGNs.\footnote{We note that
to explain the X-ray weakness of a high-Eddington ratio AGN,
\citet{Veilleux2016} propose strong self-shadowing effects by
geometrically thick disks, so called ``slim disks'', where the photon
trapping and advection are important \citep[e.g.,][]{Abramowicz1988}.}
Thanks to the attenuation of X-rays, UV-driven winds (UFOs) can be more
easily launched at larger radii in AGNs.  Thus, this scenario also
explains the reason why the AGN-driven winds (in the forms of UFOs,
ionized winds, and molecular outflows) are more developed in late
mergers than in nonmerging systems.

In Figure~\ref{F13_Schematic-images}, we compare our schematic views of 
AGNs in late mergers and nonmergers. Mid-IR and X-ray studies of
nonmerger AGNs \citep[e.g.,][]{Yamada2019,Yamada2020} indicate that they
are not deeply buried by gas and dust. Numerical
simulations show that the innermost torus structure is shaped by
radiation-driven fountain-like outflows \citep[e.g.,][]{Wada2018a}, which are confirmed 
by recent ALMA observations \citep[e.g.,][]{Izumi2018c}.
Whereas, the AGNs in late merging ULIRGs (e.g., IRAS F05189--2524 and
Mrk~231) are affected by chaotic quasi-spherical inflows due
to the galaxy collision. The massive inflow produces 
optically-thick failed winds around the 
corona that works as X-ray obscurers 
(see Section~\ref{subsub6-1-4_CoveringFactor}), and hence
powerful UV-line driven winds (UFOs) are launched.
Moreover, a part of inflowing material may be picked up to become 
radiation-driven dusty outflows at high Eddington ratios 
as proposed by \citet{Smethurst2019}.

\startlongtable
\begin{deluxetable*}{lccccc}
\label{T12_upperlim}
\tablecaption{Upper Limits of AGN Luminosities for Starburst-dominant or Hard X-ray Undetected Sources}
\tabletypesize{\footnotesize}
\tablehead{
\colhead{Object\ \ \ } &
\colhead{log$L^{\rm (pre;[O\,IV])}_{2-10}$\ \ \ } &
\multicolumn{3}{c}{log$L^{\rm (lim)}_{2-10}$\ \ \ } &
\colhead{\ \ \ AGN Flag\ \ \ }\\
\cline{3-5}
& & log$N_{\rm H}$=23 \ \ \ \ \ & log$N_{\rm H}$=24 \ \ \ \ \ & log$N_{\rm H}$=24.5 \ \ \ &
}
\decimalcolnumbers
\startdata
\multicolumn{6}{c}{Stage-A}\\
\hline
NGC 838 & 41.92 $\pm$ 0.11 & $<$40.96 & $<$41.55 & $<$42.69 & n\\
NGC 839 & 41.72 $\pm$ 0.23 & $<$40.79 & $<$41.39 & $<$42.53 & n\\
MCG+08-18-012 & \nodata & $<$41.10 & $<$41.69 & $<$42.82 & n\\
MCG+08-18-013 & $<$41.71 & $<$40.99 & $<$41.57 & $<$42.70 & n\\
MCG--01-26-013 & \nodata & $<$40.61 & $<$41.21 & $<$42.35 & n\\
NGC 3110 & $<$42.03 & $<$41.01 & $<$41.60 & $<$42.74 & n\\
NGC 4418 & $<$42.19 & $<$40.09 & $<$40.68 & $<$41.82 & n\\
MCG+00-32-013 & \nodata & \nodata & \nodata & \nodata & n\\
IC 5283 & \nodata & \nodata & \nodata & \nodata & n\\
MCG+01-59-081 & \nodata & \nodata & \nodata & \nodata & n\\
\hline
\multicolumn{6}{c}{Stage-B}\\
\hline
MCG--02-01-052 & \nodata & $<$41.14 & $<$41.73 & $<$42.85 & n\\
MCG--02-01-051 & 42.18 $\pm$ 0.17 & $<$41.19 & $<$41.77 & $<$42.90 & n\\
NGC 232 & 42.47 $\pm$ 0.10 & \nodata & \nodata & \nodata & n\\
ESO 203-1 & $<$42.65 & $<$41.38 & $<$41.95 & $<$43.06 & n?\\
CGCG 468-002E & $<$41.88 & \nodata & \nodata & \nodata & n\\
ESO 060-IG016 West & \nodata & \nodata & \nodata & \nodata & n\\
ESO 440-58 & $<$41.76 & $<$40.97 & $<$41.55 & $<$42.68 & n\\
MCG--05-29-017 & 41.81 $\pm$ 0.20 & \nodata & \nodata & \nodata & n\\
IC 4518B & \nodata & \nodata & \nodata & \nodata & n\\
NGC 6285 & $<$41.60 & $<$40.64 & $<$41.23 & $<$42.36 & n\\
NGC 6907 & 41.27 $\pm$ 0.18 & $<$40.73 & $<$41.32 & $<$42.46 & n\\
NGC 6908 & \nodata & \nodata & \nodata & \nodata & n\\
\hline
\multicolumn{6}{c}{Stage-C}\\
\hline
ESO 350-38 & $<$42.42 & $<$40.96 & $<$41.54 & $<$42.67 & n\\
IC 1623A & \nodata & $<$41.56 & $<$41.88 & $<$42.66 & n(SB)\\
IC 1623B & 42.06 $\pm$ 0.34 & \nodata & \nodata & \nodata & n\\
2MASS 06094601--2140312 & \nodata & \nodata & \nodata & \nodata & n\\
NGC 3690 East & 42.18 $\pm$ 0.35 & \nodata & \nodata & \nodata & n\\
NGC 4922S & \nodata & \nodata & \nodata & \nodata & n\\
II Zw 096 & $<$42.86 & $<$41.72(u) & $<$42.29(u) & $<$43.41(u) & n?(u)\\
IRAS F20550+1655 SE & $<$42.80 & $<$41.72(u) & $<$42.29(u) & $<$43.41(u) & n?(u)\\
\hline
\multicolumn{6}{c}{Stage-D}\\
\hline
ESO 374-IG032 & $<$42.58 & $<$41.24 & $<$41.82 & $<$42.94 & n\\
NGC 3256 & 42.26 $\pm$ 0.10 & $<$40.94 & $<$41.26 & $<$42.05 & n(SB)\\
IRAS F10565+2448 & $<$42.64 & $<$41.18 & $<$41.75 & $<$42.87 & n\\
IRAS F12112+0305 & $<$42.93 & $<$41.81 & $<$42.37 & $<$43.47 & n?\\
IRAS F14378--3651 & $<$42.77 & $<$41.99 & $<$42.54 & $<$43.65 & Y?\\
IRAS F15250+3608 & $<$43.04 & $<$41.73 & $<$42.30 & $<$43.41 & n?\\
Arp 220E & $<$42.67(u) & $<$41.06(u) & $<$41.38(u) & $<$42.16(u) & n(u)\\
IRAS F19297--0406 & $<$43.25 & $<$42.43 & $<$42.98 & $<$44.08 & n?\\
ESO 286-19 & $<$42.84 & $<$41.69 & $<$42.26 & $<$43.38 & n?\\
\hline
\multicolumn{6}{c}{Stage-N}\\
\hline
UGC 2612 & \nodata & $<$41.16 & $<$41.75 & $<$42.89 & n\\
IC 860 & $<$41.77 & $<$40.21 & $<$40.80 & $<$41.94 & n\\
NGC 5104 & 41.68 $\pm$ 0.29 & $<$41.16 & $<$41.75 & $<$42.88 & n\\
IRAS F18293--3413 & 42.31 $\pm$ 0.11 & $<$41.20 & $<$41.79 & $<$42.92 & n\\
NGC 7591 & \nodata & $<$40.79 & $<$41.38 & $<$42.51 & n\\
\enddata
\tablecomments{Columns:
(1) object name;
(2) predicted value of logarithmic absorption-corrected 2--10~keV AGN luminosity derived from the relation of [\ion{O}{4}] 25.89 $\mu$m and X-ray luminosities in \citet{LiuTeng2014}.
The values should be upper limits due to the X-ray weakness and contamination from starburst emission;
(3--5) 3$\sigma$ upper limit of the absorption-corrected 2--10~keV AGN luminosity based on the 8--24~keV net count rates assuming log($N^{\rm LoS}_{\rm H}$/cm$^{-2}$) = 23, 24, and 24.5, respectively.
For the starburst-dominant sources detected with $>$3$\sigma$ level in the 8--24~keV (IC~1623A and NGC~3256),
we utilized the 20--30~keV count rates, which are non-detection but less affected by the confusion of starburst emission;
(6) Flag of a detectable AGN candidate in X-rays. Possible AGNs whose X-ray luminosity could be larger than 10$^{43}$ erg s$^{-1}$ based on its upper limits are described as Y? and n?, both of which are divided on whether the Chandra observations indicate the presence of an AGN or not, respectively \citep{Iwasawa2011,Torres-Alba2018}.
Whereas, n(SB) and n mean that a starburst-dominant source with $>$3$\sigma$ detection or non-detection in the 8--24~keV band, respectively.
The (u) means that the two nuclei are not clearly separated.}
\end{deluxetable*}

\subsection{Upper Limits of AGN Luminosities of Hard X-ray Undetected Sources}
\label{sub6-2_upperlim}

To unveil the X-ray properties of all targets in our sample, here we
evaluate the upper limits of the absorption-corrected X-ray luminosities of
possible AGNs in the hard X-ray undetected sources. Over half of our
targets (44/84 galaxies) in local U/LIRGs are starburst-dominant or not
detected with NuSTAR in the 8--24~keV band. As we mentioned in
Section~\ref{subsub4-2-2_SBmodel}, 
the spectra of the two hard X-ray detected sources and 27
undetected sources are reproduced by adopting the starburst-dominant model,
whereas the spectra of the others are not available due to few statistics
even by combining the soft X-ray observation data.

\begin{figure}
    \epsscale{1.20}
    \plotone{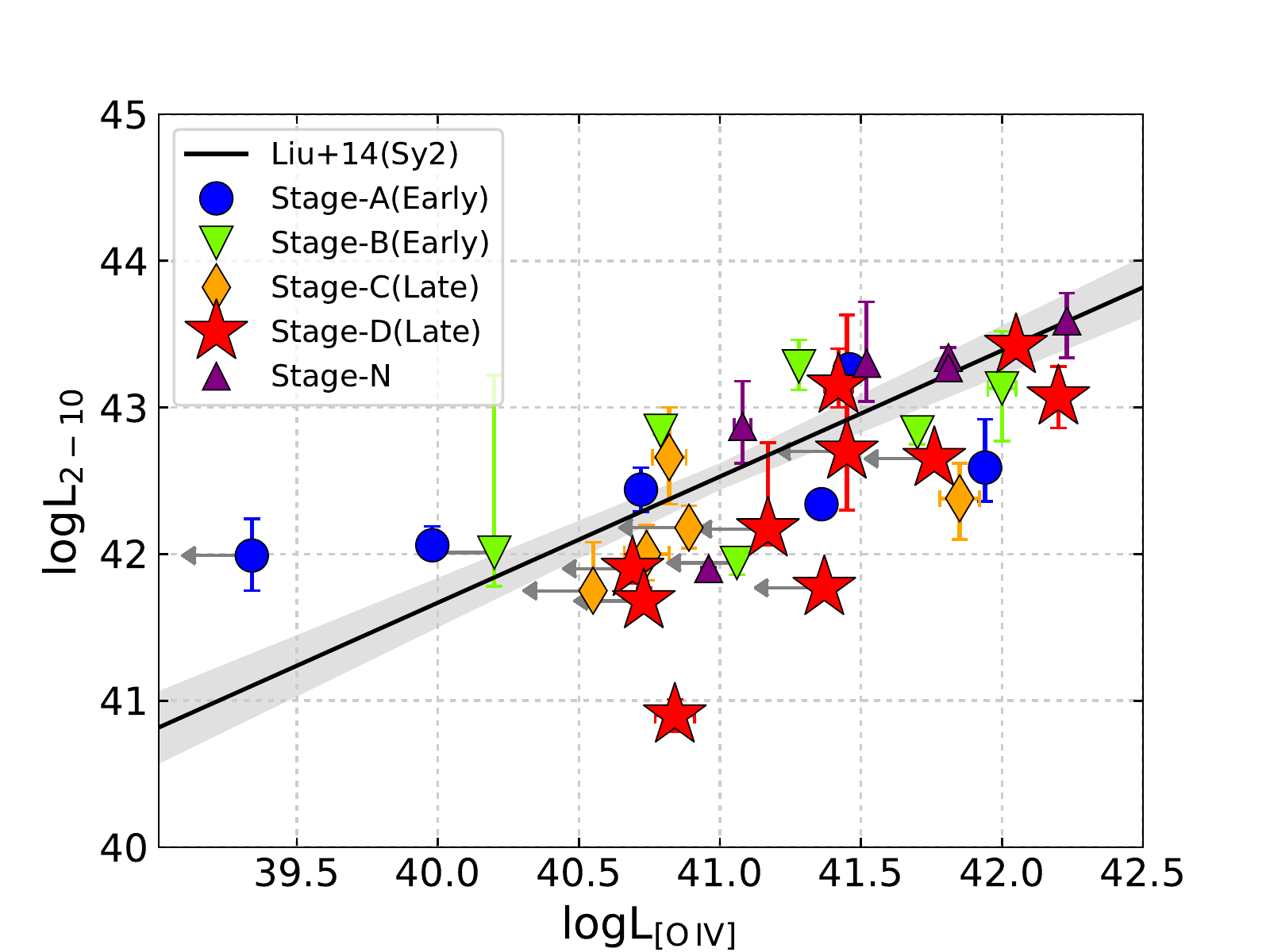}
        \caption{Absorption-corrected 2--10~keV luminosity vs. [\ion{O}{4}] 25.89 $\mu$m luminosity.
        The solid line represents the typical relation for Seyfert 2 galaxies (i.e., Compton-thin obscured AGNs) found by \citet{LiuTeng2014}.
        The gray shaded area shows its 1$\sigma$ dispersion.
        Symbols and colors are the same as in Figure~\ref{F3_NH-kpc}. 
        The arrows show the 3$\sigma$ upper limits when 
        the [\ion{O}{4}] emission lines are not detected.\\
        \label{F14_oiv-x}}
\end{figure}

We first calculate predicted AGN luminosities using the [\ion{O}{4}]
luminosities as an indicator of the AGN emission. \citet{LiuTeng2014}
present the $L_{2-10}$--$L_{\rm [O\,IV]}$ relation using
the sample of total 130 unobscured and Compton-thin obscured AGNs,
including a subsample of Swift/BAT AGNs \citep{Weaver2010}. An 
[\ion{O}{4}] luminosity (mainly radiated from the NLR) relative to
a bolometric AGN luminosity depends on torus covering factor
\citep[e.g.,][]{Yamada2019}. On the other hand, 
this relation should be affected by the X-ray weakness of the AGNs in
U/LIRGs. Figure~\ref{F14_oiv-x} illustrates the relation between [\ion{O}{4}]
and absorption-corrected 2--10~keV luminosities among the hard X-ray
detected AGNs in our sample. In fact, the X-ray luminosities relative to
the [\ion{O}{4}] ones are typically smaller than the relation for
obscured AGNs reported by \citet{LiuTeng2014}. Thus, estimates of
absorption-corrected X-ray luminosities from the [\ion{O}{4}]
luminosities using the \citet{LiuTeng2014} relation should be regarded
as upper limits.

Next, we calculate 3$\sigma$ upper limits of the absorption-corrected X-ray
luminosities from the NuSTAR net count rates in the 8--24~keV band. To
convert the count rates listed in Table~\ref{T4_count-rate} into the 
X-ray luminosities, we utilize the same AGN model as described in
Section~\ref{subsub4-2-1_AGNmodel} by ignoring the scattered and other additional components.
We assume a photon index of 1.8, a torus angular width of 20\degr, and
inclinations of 60\degr\ (for $N_{\rm H}^{\rm LoS} =
10^{23}$~cm$^{-2}$) and 80\degr\ (for $N_{\rm H}^{\rm LoS} = 10^{24}$ and
10$^{24.5}$~cm$^{-2}$). The obtained upper limits are 
listed in Table~\ref{T12_upperlim}.

According to these upper limits, we find that eight sources might have
X-ray luminosities larger than 10$^{43}$~erg~s$^{-1}$. The
multiwavelength results support that most of them show AGN signals
based on the WISE W1--W2 color and EW of the 6.2~$\mu$m polycyclic
aromatic hydrocarbon (PAH) emission (see Appendix~\ref{Appendix-B}). 
Among them, IRAS
F14378--3651 is the only candidate hosting an AGN with Chandra
observations (\citealt{Iwasawa2011,Torres-Alba2018}, and see also
Table~\ref{TA_AGN-signs}). The inferred small fraction of AGNs in the hard X-ray
undetected sources ($\leq$8 out of 45) is consistent with their small
NuSTAR band ratios as reported in Section~\ref{sub5-1_Xcolor}.

Finally, we note that possible existence of heavily obscured AGNs with
$N_{\rm H} \gtrsim 10^{25}$~cm$^{-2}$ cannot be ruled out in some
cases. Among the hard X-ray undetected sources, \citet{Asmus2015}
present predicted hydrogen column densities and absorption-corrected X-ray
luminosities for four sources by utilizing the mid-IR (high
angular-resolution 12~$\mu$m continuum flux) and X-ray correlation. They
classify them as an AGN/starburst composite (NGC~3690E) or uncertain
AGNs (NGC~4418, IRAS 15250+3609, and ESO~286-19), and suggest that the
column densities need to be $N_{\rm H} \gtrsim 10^{25}$~cm$^{-2}$ if
they are AGNs. Moreover, the X-ray weakness makes it difficult to
identify AGNs because the 2--10~keV luminosities might be
$\lesssim$10$^{42}$~erg~s$^{-1}$ (assigned as low-luminosity AGNs in X-rays).

\begin{figure}
    \epsscale{1.15}
    \plotone{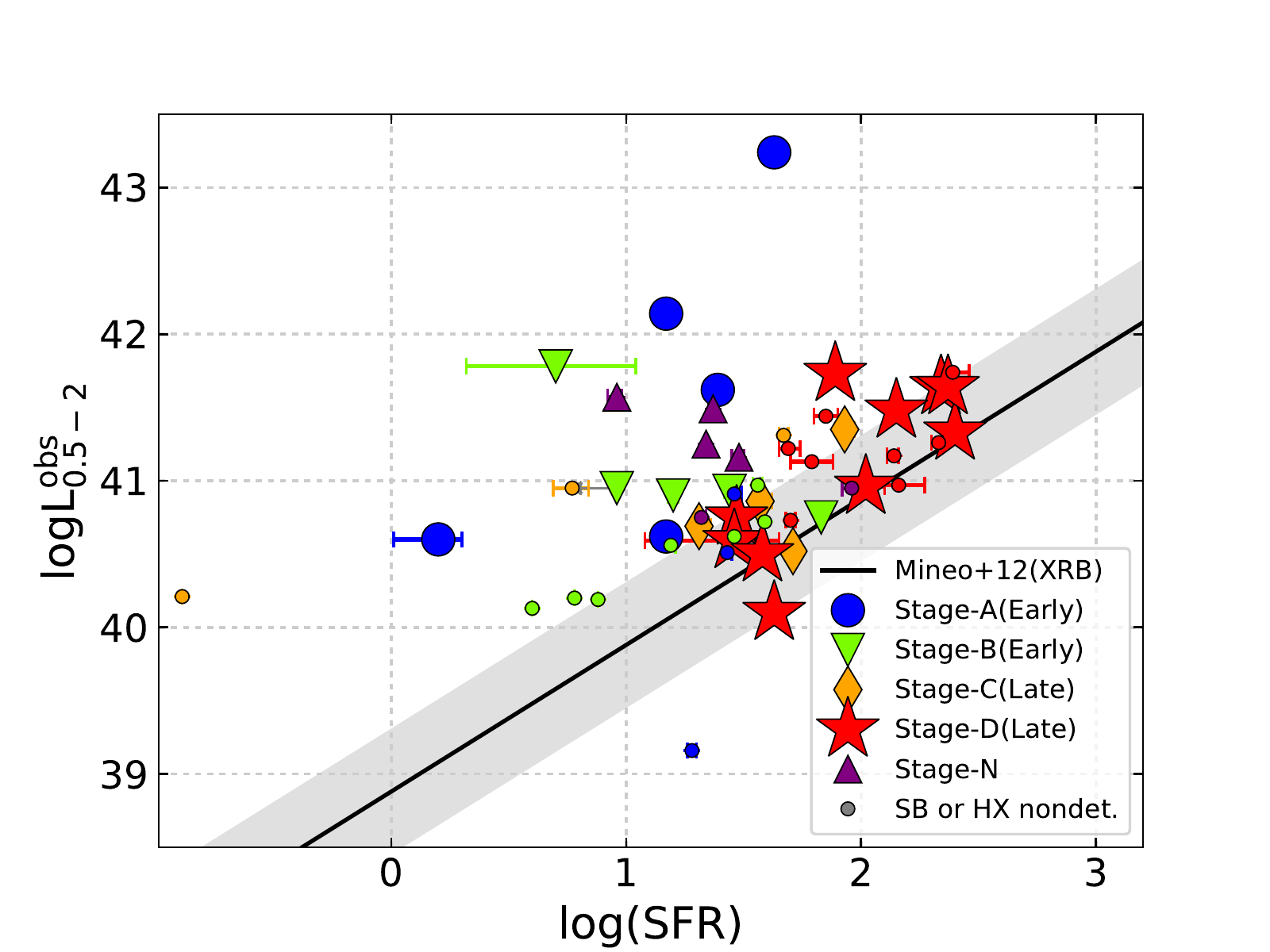}
    \plotone{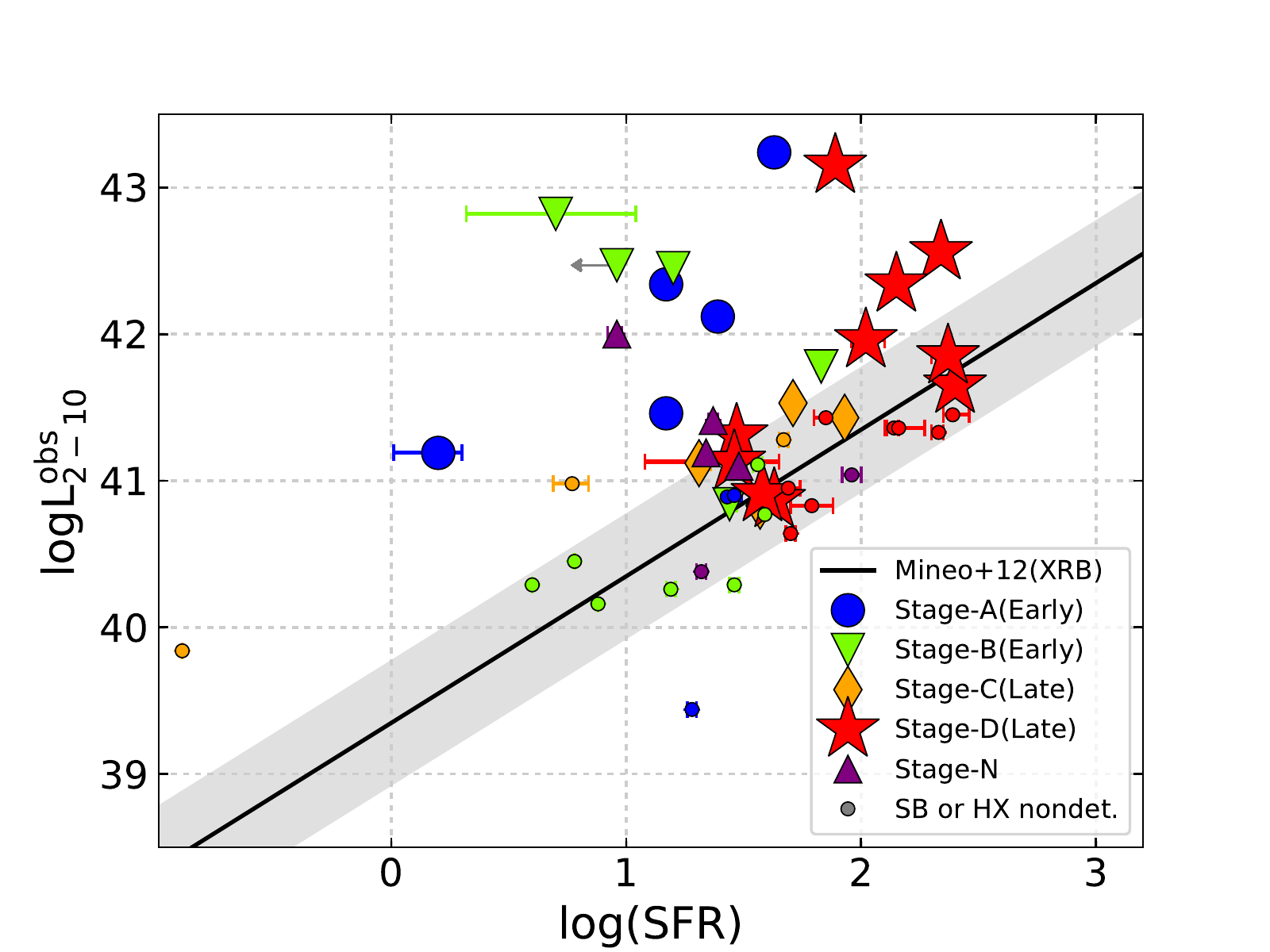}
    \plotone{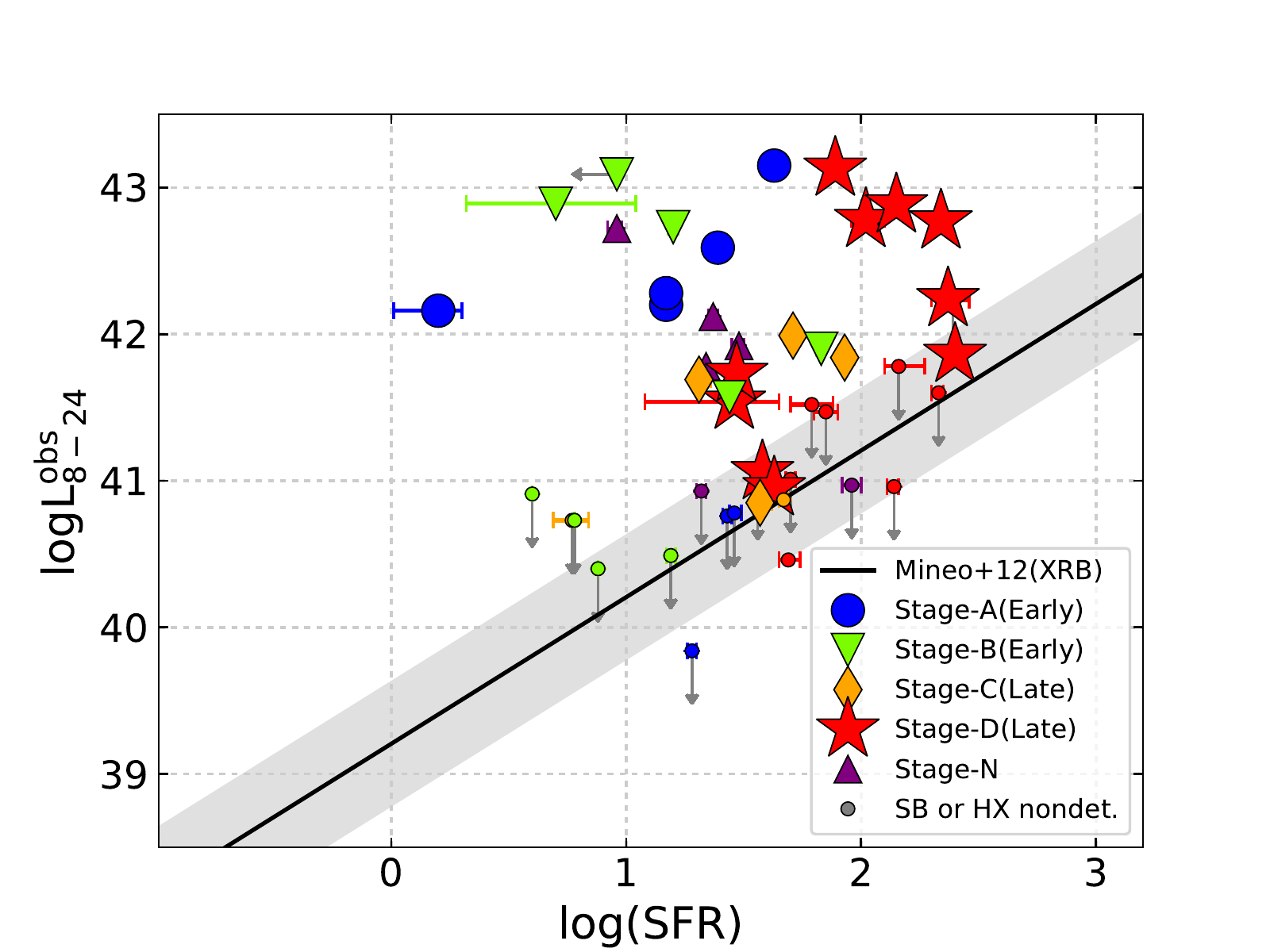}
        \caption{Observed 0.5--2~keV luminosity in the 0.5--2~keV (Top panel), 2--10~keV (Middle panel), and 8--24~keV (Bottom panel) bands vs. SFR.
        The luminosities are corrected for Galactic absorption.
        The black solid curve represents the relation between the X-ray
 luminosity of the XRB emission and SFR found by
 \citet{Mineo2012a}. Here we
        convert the 0.5--8~keV luminosities to the 0.5--2~keV, 2--10~keV, and
 8--24~keV ones by multiplying 0.29, 0.85, and 0.61,
 respectively (based on a power-law with $\Gamma = 1.1$ and $E_{\rm cut}$ = 10~keV).
        The gray shaded area shows its 1$\sigma$ dispersion.
        Symbols and colors are the same as in Figure~\ref{F3_NH-kpc}. The small symbols mark the starburst-dominant or 8--24~keV nondetected sources.\\
        \label{F15_X-SFR}}
\end{figure}

\subsection{X-ray Luminosity to SFR Relation}
\label{sub6-3_X-SFR}

Comparison of {\it observed} X-ray luminosities with host galaxy SFR
could be useful to separate the contributions from star forming activity
and AGNs. Such diagnostics are even applicable to distant U/LIRGs whose
X-ray spectral information are limited. Figure~\ref{F15_X-SFR} plots the observed
0.5--2~keV, 2--10~keV, and 8--24~keV luminosities of the total emission
against SFR for both AGNs and starburst-dominant or hard X-ray undetected
sources. \citet{Mineo2012a} present the relations between the
luminosities of the XRBs and SFRs among star-forming
galaxies. Assuming a cutoff power-law model with $\Gamma$ = 1.1 and
$E_{\rm cut}$ = 10~keV as the spectrum of the XRB emission, we find that
the distribution of most of the starburst-dominant or hard X-ray undetected
sources are well consistent with this relation within a
$\approx$1$\sigma$ dispersion ($\approx$0.43 dex). Here, the 3$\sigma$
upper limits of the 8--24~keV luminosities are converted from the NuSTAR
count rates in the 8--24~keV band.\footnote{To convert a 8--24~keV count
rate to a flux, we assume a cutoff power-law model with $\Gamma$ = 1.8
and $E_{\rm cut}$ = 10~keV for the starburst-dominant and hard X-ray undetected
sources, which is a typical spectral shape based on the results in
Section~\ref{subsub4-2-2_SBmodel}. This model approximately represents the summed emission
from the XRBs and the photoionized plasma by the AGN.}

\begin{figure}
    \epsscale{1.20}
    \plotone{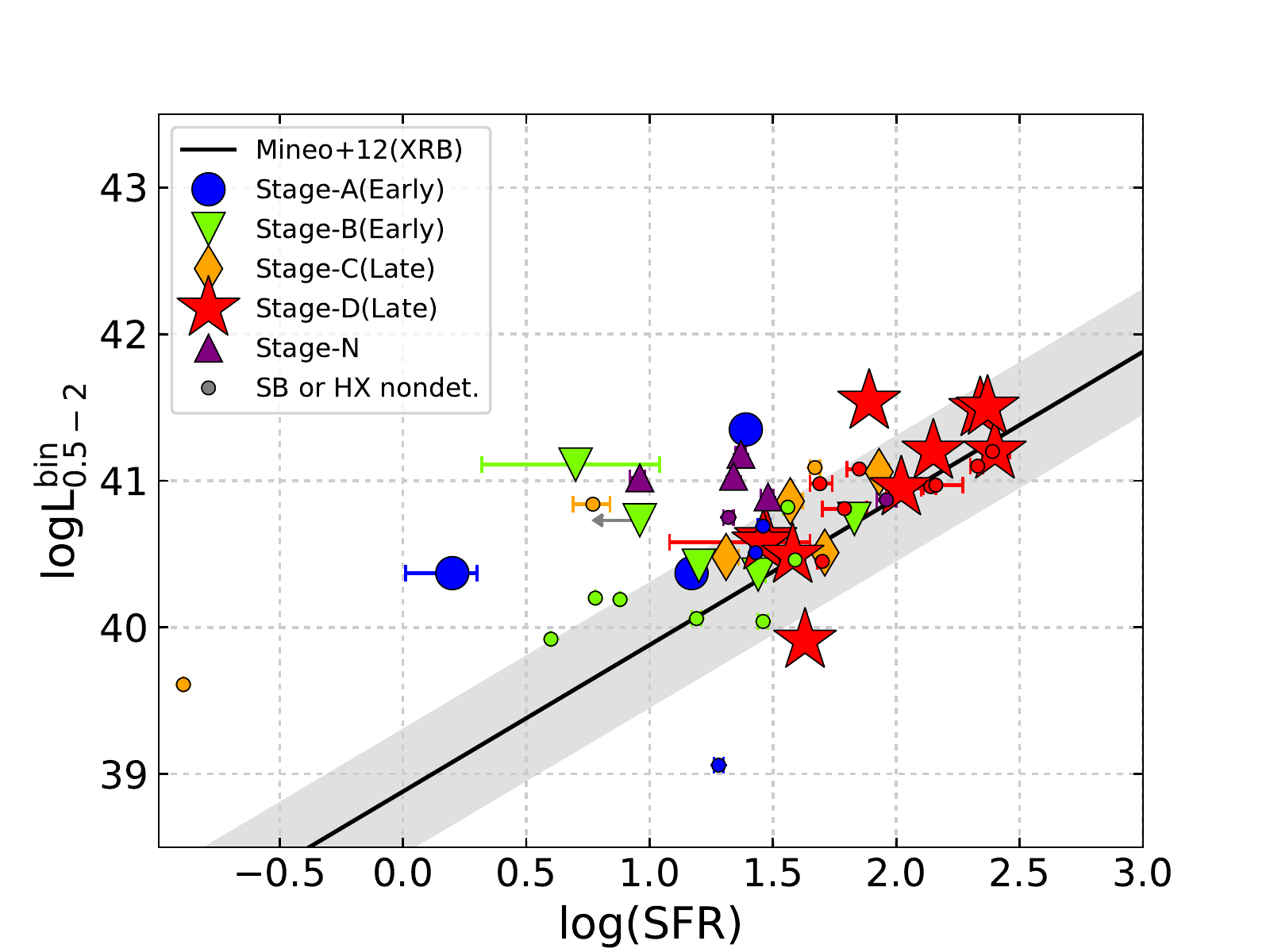}
        \caption{0.5--2~keV luminosity of the unabsorbed X-ray component vs. SFR.
        The luminosities are corrected for Galactic absorption.
        The black solid curve represents the relation between the X-ray
 luminosity of the XRB component and SFR found by \citet{Mineo2012a}. We 
        convert the 0.5--8~keV luminosities to the 0.5--2~keV ones by 
multiplying 0.29 (based on a power-law with $\Gamma = 1.1$ and $E_{\rm cut}$ = 10~keV).
        The gray shaded area shows its 1$\sigma$ dispersion.
        Symbols and colors are the same as in
 Figure~\ref{F3_NH-kpc}. The small symbols mark the starburst-dominant or 
 8--24~keV nondetected sources.\\
        \label{F16_XRB-SFR}}
\end{figure}

As noticed from Figure~\ref{F15_X-SFR}, AGN-dominant sources tend to show larger
X-ray luminosities than those predicted from the $L_{\rm X}$--SFR relation for
star-forming galaxies, indicating the contributions from the AGNs to the
observed luminosities. These excesses are the most significant in the
8--24~keV band, whereas they become less evident in softer bands, 
particularly for late mergers. This is because the fraction of heavily
obscured AGNs, whose contributions to the observed fluxes below 10~keV
are suppressed, increase with merger stage (Section~\ref{sub5-2_CTfraction}).
By using the observed 8--24~keV luminosities, most of the AGNs in
U/LIRGs seems to be separatable from starburst-dominant sources. 
Thus, comparison of the observed X-ray luminosities with the SFRs
enables us to effectively identify the AGNs in early mergers, and also
those in late mergers if hard X-ray ($>10$~keV) fluxes are available.

We also examine the relation of the 0.5--2~keV luminosity of the
unabsorbed X-ray components derived from the spectral analysis with the
SFR (Figure~\ref{F16_XRB-SFR}). The unabsorbed emission consists of the XRB component
and, if an AGN is present, the scattered component.  Excluding the AGNs
in early mergers and nonmergers where the scattering components
significantly affect the results, we find that the estimated
luminosities follow the \citet{Mineo2012a} relation. 


\subsection{View of SMBH-Galaxy Coevolution in Our Sample}
\label{sub6-4_SMBH-Gal}

\begin{figure*}
    \epsscale{0.8}
    \plotone{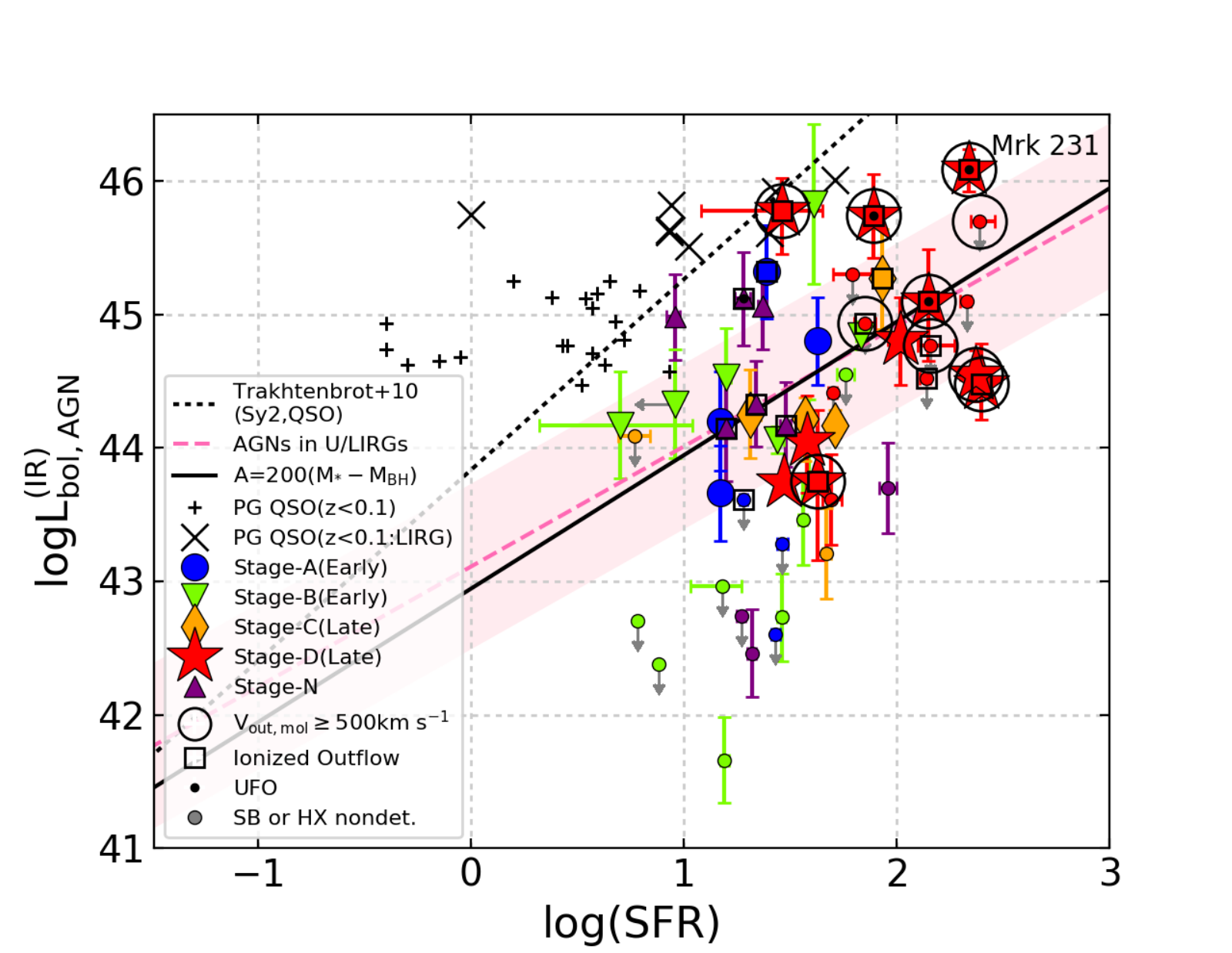}
        \caption{IR-based bolometric AGN luminosity vs. SFR.
        The black dotted line represents the relation found by \citet{Trakhtenbrot2010} for low-$z$ Seyfert 2s, QUEST QSOs, and high-$z$ QSOs.
        The pink dashed line and shaded area show the regression line
        in the log--log space and its 1$\sigma$ dispersion among the hard X-ray detected AGNs in mergers (stages A through D), respectively (see Equation (8)).
        The black solid line is the galaxy-SMBH ``simultaneous evolution'' lines for $A$ = 200 (see also \citealt{Ueda2018}).
        Symbols and colors are the same as in Figure~\ref{F3_NH-kpc}. 
        The small and large crosses represent the PG QSOs at $z < 0.1$
        with $L_{\rm IR} < 10^{11} L_{\odot}$ and $L_{\rm IR}$ =
        10$^{11}$--10$^{12}$ $L_{\odot}$ (LIRGs), respectively, whose $L_{\rm IR}$ 
        values are derived from the UV-to-IR SED decomposition \citep{Lyu2017}. 
        The small symbols mark the starburst-dominant or 8--24~keV nondetected sources, 
        whose $L_{\rm IR}$ values are derived from the [\ion{O}{4}] luminosities.
        The black empty circles, empty squares, and filled points mark the detections of strong molecular outflows with velocities of $V_{\rm out,mol} \geq 500$ km s$^{-1}$ (e.g., \citealt{Gonzalez-Alfonso2017,Laha2018} and the references therein), ionized outflows \citep[e.g.,][]{Rich2015,Kakkad2018,Fluetsch2021}, and UFOs \citep{Mizumoto2019,Smith2019}, respectively.\\
        \label{F17_SFR-AGN}}
\end{figure*}

Finally, we present a view of merger-driven coevolution of SMBHs and
their hosts. Figure~\ref{F17_SFR-AGN} plots the bolometric AGN luminosity $L_{\rm
bol,AGN}^{\rm (IR)}$ as a function of SFR for the hard X-ray detected
AGNs in our sample.  For the hard X-ray undetected sources, we also plot
the nominal bolometric AGN luminosities derived from
the [\ion{O}{4}] luminosities\footnote{This is because the estimates from the
[\ion{O}{4}] emission is contaminated by starburst emission
\citep{Gruppioni2016}, and thus basically present conservative upper
limits of the AGN luminosities for the starburst-dominant sources.} 
in Table~\ref{T9_SB-properties}.
In addition, we mark the detections of multiphase outflows, that is,
strong molecular outflows with velocities of $V_{\rm out,mol} \geq 500$ km
s$^{-1}$ (e.g., \citealt{Gonzalez-Alfonso2017,Laha2018} and the
references therein), ionized outflows
\citep[e.g.,][]{Rich2015,Kakkad2018,Fluetsch2021}, and UFOs
\citep{Mizumoto2019,Smith2019}. Here, the SFRs are mainly obtained from
the broadband SED decomposition \citep[e.g.,][]{Shangguan2019} provided
in Tables~\ref{T8_AGN-properties} and~\ref{T9_SB-properties}. 
Furthermore, we plot the
Palmer--Green (PG) QSOs (i.e., AGN-dominant population) at $z <
0.1$\footnote{We only choose the PG QSOs at $z < 0.1$ to reduce the
selection bias among PG QSOs and our U/LIRGs ($z < 0.088$).} whose
bolometric AGN luminosities and SFRs are estimated by UV-to-IR SED
decomposition \citep{Lyu2017}.

We compare our results with the prediction from the coevolution scenario (see also e.g., \citealt{Ueda2018}).
If the evolution of SMBHs and their hosts is exactly simultaneous, their
AGN and star-forming activities are expected to follow the relation described as
\begin{align}
& A \times \dot{M}_{\rm BH} = {\rm SFR} \times (1-R),
\end{align}
where $A$ represents the ratio of stellar masses to black hole masses in local universe and $R$ is the fraction of stellar masses that are taken back to interstellar medium (return fraction).
As a representative value, we assume $R$ = 0.41 based on a \citet{Chabrier2003} initial mass function.
Considering that the majority of U/LIRGs should be the progenitors of
bulge-dominated galaxies \citep[e.g.,][]{Hopkins2008}, we adopt $A$
$\sim$ 200 for the stellar masses only in bulge components\footnote{We
note that even if we adopt the value for the total stellar mass in both
bulges and disks ($A \sim 400$; e.g., \citealt{Ueda2014,Ueda2018}), the
difference of 0.3 dex does not change our conclusions.} based on the latest calibration by \citet{Kormendy2013}.
The mass accretion rate onto the SMBH can be calculated by using a radiative efficiency ($\eta$) as
\begin{align}
 \dot{M}_{\rm BH} &= L_{\rm bol,AGN}^{\rm (IR)} \times (1-\eta)/(\eta c^2),
\end{align}
where $c$ is the speed of light.
Here, we adopt $\eta$ = 0.05, as estimated by comparison of the local
SMBH mass density and the AGN luminosity fraction \citep{Ueda2014}.
We thus draw the galaxy-SMBH ``simultaneous evolution'' relation in the figure.

To quantitatively evaluate the correlation between 
$L_{\rm bol,AGN}^{\rm (IR)}$ and SFR for AGNs in merging U/LIRGs, 
we perform a linear regression analysis 
in log--log space and a correlation analysis for the hard X-ray
detected AGNs in mergers (stages A through
D).\footnote{We exclude two AGNs in stage-B mergers,
NGC~235 and Mrk~266B, because their SFRs are obtained from the total (AGN
and starburst) IR luminosity \citep{Howell2010}, which are 
treated as upper limits.}  We employ a Bayesian maximum likelihood
method of \citet{Kelly2007} in which the errors in the two parameters
are taken into account (see also e.g.,
\citealt{Toba2019,Setoguchi2021}). The best-fitting linear function is
\begin{align}
 {\rm log}L_{\rm bol,AGN}^{\rm (IR)} = 0.90{\rm \ logSFR} + 43.11,
\end{align}
and the 1$\sigma$ dispersion is 0.62 dex (pink dashed line and shaded area).
The correlation coefficient is $r = 0.51 \pm 0.19$, indicating a moderately
tight correlation.

In general, the AGNs in mergers (stages A through D) roughly follow the
simultaneous evolution line with a scatter of $\sim$1 dex. It is also
notable that later mergers tend to show high SFRs and high bolometric
AGN luminosities. These results support the simplest merger-driven
coevolution picture where the galaxies and SMBHs evolve from
left-bottom to right-top along the simultaneous evolution line after
being triggered by mergers. Note that there is a group of stage-D
mergers with relatively small SFRs
($\sim$10$^{1.5}$~$M_{\odot}$~yr$^{-1}$) and low AGN luminosities
($L_{\rm bol,AGN}^{\rm (IR)} \sim 10^{44}$~erg~s$^{-1}$), which may be
gas poor mergers or minor mergers, as discussed 
in Section~\ref{subsub5-3-4_Edd} \citep{Blecha2018}. The
starburst-dominant or hard X-ray undetected sources are located far below the
simultaneous evolution line. They might be in an earlier phase of
U/LIRGs hosting luminous AGNs before intense mass accretion onto the
SMBHs is triggered.

The black dotted line represents the empirical relation by
\citet{Trakhtenbrot2010} for low-$z$ Seyfert 2 galaxies
\citep{Netzer2009}, QUEST QSOs \citep{Schweitzer2006,Netzer2007}, and
high-$z$ QSOs \citep[e.g.,][]{Lutz2008}. As expected the PG QSOs at $z <
0.1$ follow this relation.  Assuming the evolutionary scenario that star
formation occurs first and an AGN-dominant phase follows later
\citep[e.g.,][]{Sanders1988}, these should be the descendants of the ULIRGs
that are in the peak phase of starburst activities.  The most luminous
AGNs in IRAS F05189--2524 and Mrk~231 ($L_{\rm bol}^{\rm (IR)} \sim
10^{46}$~erg~s$^{-1}$) have relatively small column densities of $N_{\rm
H}^{\rm LoS} \lesssim 10^{23}$~cm$^{-2}$, supporting that they are in an
evolutionary stage between the other stage-D U/LIRGs (hosting CT AGNs)
and PG QSOs (hosting unobscured AGNs; \citealt{Teng2010}).  It is also
noteworthy that (likely) AGN-driven powerful molecular outflows are
ubiquitous in stage-D mergers with high SFRs and high bolometric AGN
luminosities. Such coexistence of massive outflows and intense
starbursts has been reported among 136 outflowing ULIRGs with [\ion{O}{3}]
detections by \citet{Chen2020a}. These results are quantitatively
consistent with a standard AGN feedback scenario where the AGN-driven
outflows quench star formation in the host and eventually luminous
ULIRGs transit to unabsorbed QSOs
\citep[e.g.,][]{Di-Matteo2005,Hopkins2008}. It is, however, still
unclear how efficiently these outflows could affect star formation in
the host galaxy, for which further studies are required.
\\

\section{Conclusions}
\label{S7_conclusion}

To investigate the merger-driven coevolution of SMBHs and their hosts,
we have carried out a systematic X-ray spectral study of 57 local
U/LIRGs (containing 84 individual galaxies) observed with NuSTAR 
and/or Swift/BAT.
Combining available soft X-ray data taken with Chandra,
XMM-Newton, Suzaku, and/or Swift/XRT, 
we have identified 40 hard X-ray detected AGNs, 
for which we have constrained their torus parameters with 
the latest X-ray clumpy torus model (XCLUMPY; \citealt{Tanimoto2019}).
Our sample
consists of two unobscured AGNs, 21 obscured AGNs, 16 CT AGNs, one
jet-dominant AGN, and 44 starburst-dominant or hard X-ray undetected sources.
We list below the main conclusions from our research:

\begin{enumerate}

  \item 
Among the AGNs at $z < 0.03$, for which sample biases are minimized, we
find that the CT AGN fraction in late mergers ($f_{\rm CT}$ =
64$^{+14}_{-15}$\%) is larger than in early mergers ($f_{\rm CT}$ =
$24^{+12}_{-10}$\%), consistent with the trend reported by
\citet{Ricci2017bMNRAS}.
  
  \item 
We have complied the IR-based bolometric AGN luminosities ($L_{\rm
bol,AGN}^{\rm (IR)}$) and black hole masses in the literature derived by
various methods, and adopt their averaged values for each source as the
best estimates. For the hard X-ray detected AGNs, we find that
their bolometric luminosities and Eddington ratios ($\lambda_{\rm Edd}$)
increase with merger stage.
The bolometric AGN luminosities in late mergers cover a wide range of
$L_{\rm bol,AGN}^{\rm (IR)} \sim 10^{43.5}$--$10^{46.5}$~erg~s$^{-1}$,
which is consistent with recent numerical simulations of gas-rich and
gas-poor mergers.

\item 
The bolometric-to-X-ray luminosity correction factors ($\kappa_{\rm
bol,X}$) of the AGNs in late mergers show a systematic excess by a
factor of 10--100 compared with the $\kappa_{\rm bol,X}$--$\lambda_{\rm
Edd}$ relation obtained for normal AGNs, i.e., their SEDs are X-ray
weak.

\item 
Molecular outflows are detected from 10 out of the 12 late (stage-D) mergers 
hosting hard X-ray detected AGNs. The
outflow velocities are correlated with both bolometric correction and
Eddington ratio, supporting the AGN origin for the outflows. The X-ray
weakness is likely associated with AGN-driven strong outflows.

\item 
The AGNs with $\lambda_{\rm Edd} \gtrsim 0.5$ in late mergers (IRAS
F05189--2524, IRAS F08572+3915, and Mrk~231) are located within the
forbidden region of the $N_{\rm H}$--$\lambda_{\rm Edd}$ diagram.
This suggests that the obscuration may be caused by dusty in/outflows. 
Furthermore, IRAS F05189--2524 and Mrk~231 show larger torus
covering factors ($C_{\rm T}^{\rm (22)} \sim 0.6$) as estimated with
XCLUMPY, than the predicted values for normal AGNs at similar Eddington
ratios. We infer that the presence of torus-scale
in/outflows and/or host-galaxy scale material enhances
the fraction of CT AGNs in late mergers.

\item
To comprehensively explain these results, we have presented a unified
view of AGNs in late mergers (Figure~\ref{F13_Schematic-images}).  
Under the presence of chaotic
quasi-spherical inflows in late mergers, the densities and covering
factors of the failed winds launched from the inner disk are
enhanced. Then, these enhanced failed winds more efficiently block
X-rays from the central regions, making the AGN SEDs apparently X-ray
weaker compared with normal AGNs.  Thanks to the attenuation of X-rays,
UV-driven winds (UFOs) can be more easily launched at larger radii in
AGNs.  Thus, this scenario also explains the reason why the AGN-driven
winds (in the forms of UFOs, ionized winds, and molecular outflows) are
more developed in late mergers than in nonmerging systems.  A part of
inflowing material may be picked up to become radiation-driven dusty
outflows at high Eddington ratios.

\item We evaluate the upper limits of AGN luminosities 
for the starburst-dominant and hard X-ray undetected sources 
by utilizing the [\ion{O}{4}] luminosities or
the NuSTAR count rates in the 8--24~keV band. We find few AGN
candidates among the 44 galaxies.

\item 
We have examined the relation of observed X-ray luminosities against SFR
for the AGN-dominant and starburst-dominant U/LIRGs. We find that comparison of
the observed X-ray luminosities with the SFRs enables us to effectively
identify the AGNs in early mergers, and also those in late mergers if
hard X-ray ($>10$~keV) fluxes are available.

\item
Finally, we present a view of merger-driven coevolution of SMBHs and
their hosts in the $L_{\rm bol,AGN}^{\rm (IR)}$--SFR diagram
(Figure~\ref{F17_SFR-AGN}). 
Our AGNs seem to roughly follow the ``simultaneous
evolution line'' from left-bottom to right-top with merger stage,
supporting the merger-driven coevolution picture. The most
luminous AGNs in IRAS F05189--2524 and Mrk~231 have relatively small
column densities of $N_{\rm H}^{\rm LoS} \lesssim 10^{23}$~cm$^{-2}$,
implying that they are in an evolutionary stage between the other
stage-D U/LIRGs (hosting CT AGNs) and PG QSOs (hosting unobscured AGNs;
\citealt{Teng2010}). AGN-driven powerful molecular outflows are
ubiquitous in late mergers with high SFRs and high bolometric AGN
luminosities, consistent with a standard AGN feedback scenario.

\end{enumerate}


The authors first thank the anonymous referee for a careful reading of the manuscript and very helpful comments.
S.Y. as well thanks Misaki Mizumoto for fruitful discussion and comments.
We acknowledge financial support from the Grant-in-Aid for Scientific Research 19J22216 (S.Y.); 17K05384 and 20H01946 (Y.U.); 20J00119 (A.T.); 21K03632 (M.I.); and 18J01050 and 19K14759 (Y.T.).
Y.T. acknowledges support from the Ministry of Science and Technology of Taiwan (MOST 105-2112-M-001-029-MY3).
This work is also supported by the Fondecyt Iniciacion grant 11190831 (C.R.).
\\

This work makes use of data obtained the NuSTAR Data Analysis Software (NuSTARDAS) jointly developed by the ASI Science Data Center (ASDC, Italy) and the California Institute of Technology (USA).
This publication makes use of data obtained with Chandra, supported by the Chandra X-ray Observatory Center at the Smithsonian Astrophysical Observatory, and with XMM-Newton, an ESA science mission with instruments and contributions directly funded by ESA Member States and NASA.
This study makes use of data obtained from the Suzaku satellite, a collaborative mission between the space agencies of Japan (JAXA) and the USA (NASA).
The scientific results reported in this article are based on
observations made by UK Swift Science Data Centre at the University of Leicester.
This research makes use of the SIMBAD database, operated at CDS, Strasbourg, France \citep{Wenger2000}, and the Aladin sky atlas, developed at CDS, Strasbourg Observatory, France \citep{Bonnarel2000,Boch2014}.
Also, this research makes use of data from the NASA/IPAC Extragalactic Database (NED) and NASA/IPAC Infrared Science Archive (IRSA), operated by JPL/California Institute of Technology under contract with the National Aeronautics and Space Administration.


\vspace{5mm}
\facilities{NuSTAR, Chandra, XMM-Newton, Suzaku, Swift.}


\software{XCLUMPY \citep{Tanimoto2019}, HEAsoft (v6.25), XSPEC(v12.10.1; \citealt{Arnaud1996}), NuSTARDAS(v1.8.0), CIAO(v4.11), SAS (v17.0.0; \citealt{Gabriel2004}).}

\appendix
\restartappendixnumbering

\section{Comparison among Different Measurements of $L_{\rm bol,AGN}^{\rm (IR)}$ and $M_{\rm BH}$} \label{Appendix-A}

\begin{figure}
    \epsscale{1.30}
    \plotone{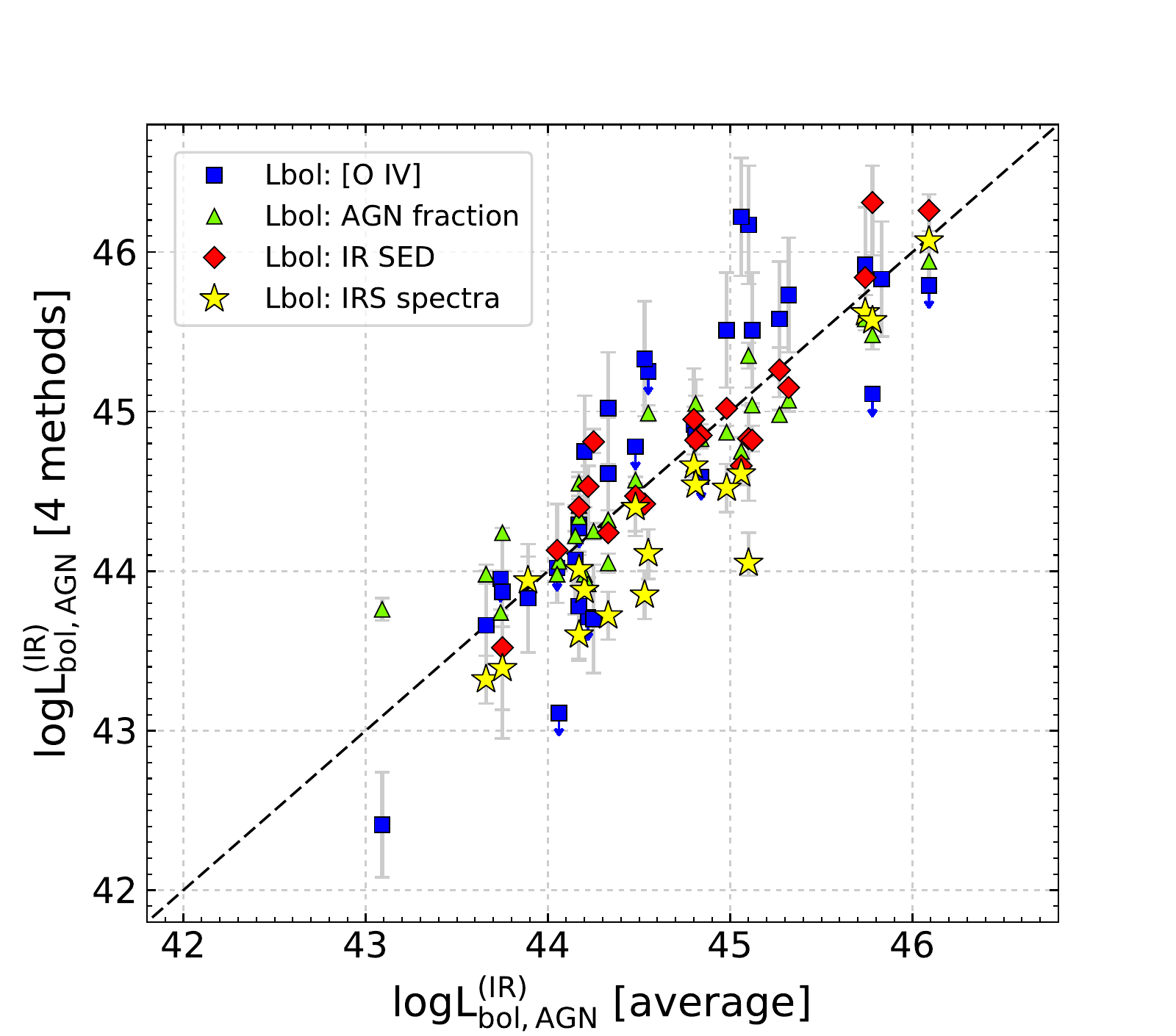}
        \caption{Comparison among the bolometric AGN luminosities derived from different
 methods: the [O IV] luminosity (blue square; e.g., \citealt{Gruppioni2016}), 
        the bolometric AGN fraction (green triangle; e.g.,
 \citealt{Diaz-Santos2017}), 
        the 1--500~$\mu$m SED decomposition 
and Spitzer/IRS spectral fitting  (red diamond and yellow star, respectively;
        \citealt{Alonso-Herrero2012,Shangguan2019}).
        The black dashed line shows the 1:1 relation between the
 individual estimates ({\it y}-axis) and their averages ({\it x}-axis).\\
        \label{A1-F}}
\end{figure}

\begin{figure}
    \epsscale{1.30}
    \plotone{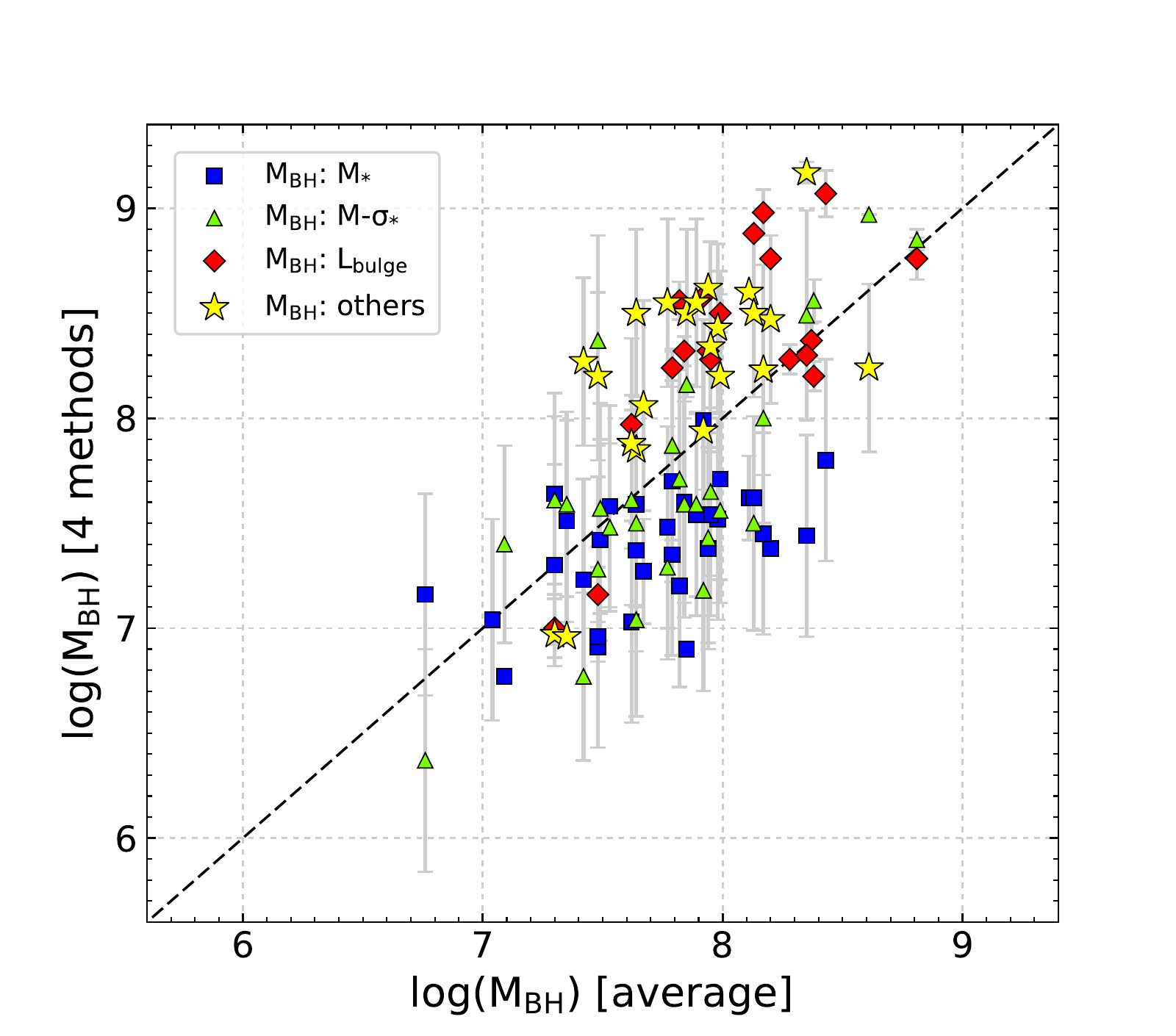}
        \caption{Comparison among the black hole masses derived from different methods: the stellar mass (blue square; e.g., \citealt{Reines2015,Shangguan2019}), 
        the $M$--$\sigma_{*}$ relation (green triangle; e.g., \citealt{Gultekin2009}),
        the photometric bulge luminosity (red diamond; e.g., \citealt{Winter2009,Veilleux2009c,Haan2011a}),
        and the other various methods, such as the flux density of
        old stellar emission at 2~$\mu$m and so on (yellow star; e.g., \citealt{Caramete2010}).
        The black dashed line shows the 1:1 relation between individual estimates ({\it y}-axis) and their averages ({\it x}-axis).\\
        \label{A2-F}}
\end{figure}

\begin{figure}
    \epsscale{1.30}
    \plotone{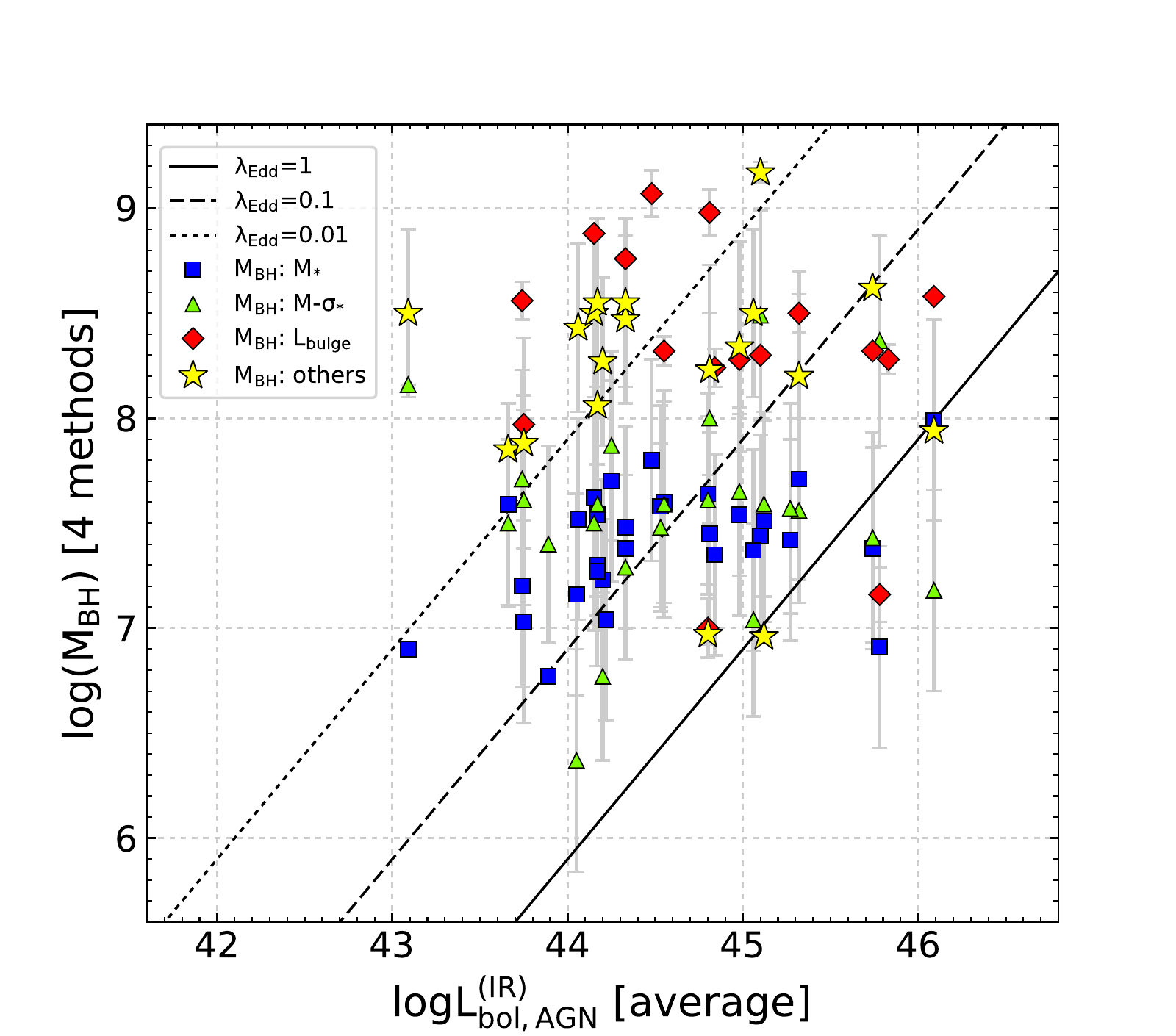}
        \caption{Relation between the averaged bolometric AGN 
        luminosity and the individual black hole masses derived from 
        the four methods. Symbols are as in Figure~\ref{A2-F}.
        The black solid, dashed, and dotted lines represent the relations 
        expected for three different Eddington ratios 
        ($\lambda_{\rm Edd}$ = 1, 0.1, and 0.01, respectively).\\
        \label{A3-F}}
\end{figure}

We investigate if there are systematic trends among different
measurements for bolometric AGN luminosities and black hole masses
(Section~\ref{sub5-3_properties}).
Figure~\ref{A1-F} shows the comparison 
among bolometric AGN luminosities 
derived from different methods: the [O IV] luminosity, bolometric AGN 
fraction, the 1--500~$\mu$m SED decomposition and Spitzer/IRS spectral
fitting. The averaged values obtained with these methods in our sample is 
log$L_{\rm bol,AGN}^{\rm (IR)}$ = 44.80 (22 AGNs), 44.59 (29 AGNs), 
44.86 (20 AGNs), and 44.32 (18 AGNs), respectively.
A typical scatter is about $\pm$0.27 dex when we adopt the 
averaged value of the four measurements for an individual object,  
as noted in Section~\ref{subsub5-3-1_bolometric}.

In Figure~\ref{A2-F}, we 
compare the black hole masses derived from 
the stellar mass,
the $M$--$\sigma_{*}$ relation, the photometric bulge luminosity, and 
the other methods (e.g., the flux density 
of 2~$\mu$m old stellar emission). We find that the measurements 
from dynamical stellar mass and velocity dispersion
(blue squares and green triangles) are smaller
than those from bulge luminosities and other methods 
(red diamonds and yellow stars).
A typical scatter
is $\pm$0.3--0.5~dex when we adopt the averaged value.  
We discuss possible reasons for the systematic differences
in Section~\ref{subsub5-3-3_BHmass} (see also \citealt{Veilleux2009c}).

Figure~\ref{A3-F} represents the relation 
between the averaged bolometric AGN luminosity and 
the four individual black hole measurements for each AGN.
It is notable that the systematic differences seem to depend 
neither the bolometric AGN luminosity nor Eddington ratio.
\\

\ifnum0=1
Thus, our discussions referring to the Eddington ratios
(e.g., Section~\ref{sub5-3_properties} and
Section~\ref{sub6-1_AGN-structure}) are unlikely to be affected
the systematic uncertainties in the black hole mass measurements.
\fi


\section{Notes on Individual Objects} \label{Appendix-B}

In this section, we present the basic characteristics of the individual
objects in our sample (also refer to Appendix A in
\citealt{Ricci2017bMNRAS} for a subsample). We also compare the results
of our X-ray spectral analysis with previous studies. We confirm that
the hard X-ray detected AGNs tend to show AGN signs based on the
following nine diagnostics as summarized in Table~\ref{TA_AGN-signs}:

(i) WISE color of W1$-$W2 $\geq$ 0.8 (i.e., [3.4]$-$[4.6] $\geq$ 0.8
in Vega magnitude) \citep{Stern2012}. We note that this diagnostic is
not applicable for low-luminosity AGNs in hosts with strong starburst
activities \citep[e.g.,][]{Griffith2011,Hainline2016}.

(ii) Near-IR 2.5--5~$\mu$m continuum slope of $\Gamma_{2.5-5} >
1.0$ \citep{Imanishi2010}, which is an indicator of an obscured AGN
\citep{Risaliti2006,Imanishi2008},

(iii) Optical depths of 3.1 $\mu$m ice-covered dust and/or 3.4 $\mu$m
bare carbonaceous dust absorption features ($\tau_{3.1} > 0.3$ and/or
$\tau_{3.4} > 0.2$). These are indicators of an obscured AGN with a
centrally concentrated energy source \citep{Imanishi2010}.

(iv)--(v) EWs of the 3.3~$\mu$m and 6.2~$\mu$m PAH features (EW$_{3.3}
<$~40~nm, \citealt{Imanishi2008,Imanishi2010}; EW$_{6.2} <$~27~nm,
\citealt{Stierwalt2013}).

(vi) Detection of the mid-IR [\ion{Ne}{5}] 14.32 $\mu$m line
\citep[e.g.,][]{Inami2013}.

(vii) Radio spectral index maps ($\alpha$-maps) at 1.49~GHz and 8.44~GHz,
which can be used to find AGN emission from central kpc regions
\citep{Vardoulaki2015}.

(viii) Optical spectroscopic classification on the BPT diagram (i.e., 
[\ion{O}{3}]/H$\beta$ and [\ion{Ne}{3}]/[\ion{O}{2}] ratios,
\citealt{Baldwin1981,Kewley2006}). Sy stands for Seyfert with
subcategories 1, 1.5, 1.8 and 2. L stands for LINERs.  Cp stands for
objects between star-forming and AGN regions in the BPT
\citep[e.g.,][]{Yuan2010}. HII stands for HII regions.  Reference is
given in each case.

(ix) X-ray AGN signatures based on Chandra observations
\citep{Iwasawa2011,Torres-Alba2018}: color between the 0.5--2~keV and
2--7~keV bands, detection of the 6.4~keV line, and continuum absorption
features.

In Table~\ref{TA_AGN-signs}, we also provide X-ray classification based
on our broadband X-ray spectral analysis for the hard X-ray detected
AGNs (unobscured AGN for $N_{\rm H} < 10^{22}$~cm$^{-2}$, obscured AGN
for $N_{\rm H} = 10^{22}$--10$^{24}$~cm$^{-2}$, and CT AGN for $N_{\rm
H} \geq 10^{24}$~cm$^{-2}$). 
\ifnum0=1
For the starburst-dominant or hard X-ray
undetected sources, we examine the possibility of AGNs using the upper
limits of the X-ray luminosities derived in
Section~\ref{sub6-2_upperlim}.
\fi

Finally, we summarize the basic properties of the individual sources 
and discuss the presence or absence of AGNs on the basis of the
broadband X-ray spectral analysis and multiwavelength results.

\begin{enumerate}
\item IRAS F00085--1223 (NGC 34):

This is a LIRG with a single nucleus, assigned as a stage-D merger. 
We identify the source as the hard X-ray detected AGN. The
XMM-Newton studies indicate that NGC 34 hosts an obscured AGN
\citep{Guainazzi2005,Brightman2011a}, and the recent NuSTAR studies
estimate its absorption column density to be $N_{\rm H} \sim 5 \times
10^{23}$~cm$^{-2}$ \citep{Ricci2017bMNRAS,Mingozzi2018}. This is 
consistent with our value of $N_{\rm H}$ = $(5.0 \pm 1.0)$ $\times$
$10^{23}$~cm$^{-2}$ by using the NuSTAR, Chandra, and XMM-Newton data.
The absorption-corrected 2--10~keV luminosity we obtain is log$L_{2-10}$ =
$41.90 \pm 0.10$.

\item IRAS F00163--1039 N/S (MCG--02-01-052 and MCG--02-01-051): 

These two sources are in a stage-B merger, and their separation is
56\farcs1 (a projected distance of 30.7~kpc). While the southern object
MCG--02-01-051 (= Arp 256) is a LIRG, the northern object
MCG--02-01-052 has a lower IR luminosity of log$L_{\rm
IR}$/$L_{\odot}$ = 10.36 \citep{Ricci2017bMNRAS}. Although their hard
X-ray emission is not detected with NuSTAR (see also
\citealt{Ricci2017bMNRAS}), Chandra detects X-ray emission from both
nuclei. The best-fit photon indices obtained with the starburst-dominant model
are $\Gamma$ $\sim$ 1.5 for MCG--02-01-052 and $\Gamma$ $\sim$ 1.7 for
MCG--02-01-051.

\clearpage
\startlongtable
\begin{deluxetable*}{lccccccccccc}
\tablecaption{Summary of the AGN Signatures Obtained by Multiwavelength Observations \label{TA_AGN-signs}}
\tabletypesize{\scriptsize}
\tablehead{
\colhead{Object Name} &
\colhead{W1--W2} &
\colhead{$\Gamma_{2.5-5}$} &
\colhead{$\tau_{\rm 3.1,3.4}$} &
\colhead{PAH$_{3.3}$} &
\colhead{PAH$_{6.2}$} &
\colhead{[\ion{Ne}{5}]} &
\colhead{Radio} &
\colhead{Opt.} &
\colhead{C--GOALS} &
\colhead{This work} &
\colhead{Ref.}
}
\decimalcolnumbers
\startdata
NGC 34 & n & n & n & n & n & n & Y & Sy2 & Y(A) & Obs & 1\\
MCG--02-01-052 & n & \nodata & \nodata & \nodata & \nodata & n & \nodata & HII & \nodata & n & 1\\
MCG--02-01-051 & n & n & \nodata & n & n & n & \nodata & HII & \nodata & n & 1\\
ESO 350-38 & Y(1.336) & \nodata & \nodata & \nodata & Y(0.15) & n & \nodata & HII & n & n & 1\\
NGC 232 & n & n & n & Y & n & n & \nodata & Cp & n & n & 1,2\\
NGC 235 & n & \nodata & \nodata & \nodata & Y(0.16) & Y & \nodata & Sy2 & \nodata & Obs & 3\\
MCG+12-02-001 & n & \nodata & \nodata & \nodata & n & n & \nodata & Cp & \nodata & CT & 4\\
MCG+12-02-001 NED01 & \nodata & \nodata & \nodata & \nodata & \nodata & n & \nodata & \nodata & \nodata & n & \nodata\\
IC 1623A & Y(1.240;u) & n & \nodata & n & n & n & Y(u) & HII & \nodata & n & 1\\
IC 1623B & Y(1.240;u) & Y & Y & Y & \nodata & n & Y(u) & HII & \nodata & n & 1\\
NGC 833 & n & \nodata & \nodata & \nodata & \nodata & n & \nodata & L & n & Obs & 5\\
NGC 835 & n & \nodata & \nodata & \nodata & n & n & \nodata & Sy2 & n & Obs & 5\\
NGC 838 & n & \nodata & \nodata & \nodata & n & n & \nodata & L & n & n & 6\\
NGC 839 & n & \nodata & \nodata & \nodata & n & n & \nodata & L & n & n & 6\\
NGC 1068 & Y(1.125) & \nodata & \nodata & \nodata & \nodata & Y & \nodata & Sy2 & Y(L) & CT & 1,a\\
UGC 2608 & Y(1.282) & \nodata & \nodata & \nodata & Y(0.20) & Y & \nodata & Sy2 & \nodata & CT & 3,b\\
UGC 2612 & n & \nodata & \nodata & \nodata & Y(0.25) & n & \nodata & \nodata & \nodata & n & \nodata\\
NGC 1275 & Y(1.006) & \nodata & \nodata & \nodata & Y(0.02) & n & \nodata & Sy1.5 & \nodata & Jet & 3,c\\
NGC 1365 & n & \nodata & \nodata & \nodata & Y(0.13) & Y & \nodata & Sy1.8 & Y(CLA) & Obs & 7,d\\
ESO 203-1 & Y(1.143) & \nodata & \nodata & \nodata & Y(0.03) & n & \nodata & \nodata & n & n? & \nodata\\
CGCG 468-002W & n & \nodata & \nodata & \nodata & Y(0.12) & Y & \nodata & Sy2 & \nodata & Obs & 8\\
CGCG 468-002E & n & \nodata & \nodata & \nodata & n & n & \nodata & \nodata & \nodata & n & \nodata\\
IRAS F05189--2524 & Y(1.125) & \nodata & n & Y & Y(0.03) & Y & n & Sy2 & Y(CL) & Obs & 1\\
IRAS F06076--2139 & n & \nodata & \nodata & \nodata & n & n & \nodata & Sy? & Y(C) & Obs & 9\\
2MASS 06094601--2140312 & \nodata & \nodata & \nodata & \nodata & \nodata & n & \nodata & HII & n & n & 9\\
NGC 2623 & n & n & Y & Y & Y(0.27) & Y & Y & Cp & Y(C) & Obs & 10\\
ESO 060-IG016 West & \nodata & \nodata & \nodata & \nodata & \nodata & Y(u) & \nodata & \nodata & n & n & \nodata\\
ESO 060-IG016 East & \nodata & \nodata & \nodata & \nodata & Y(0.11) & Y(u) & \nodata & \nodata & Y(C) & Obs & \nodata\\
IRAS F08572+3915 & Y(2.313) & \nodata & Y & Y & Y(0.03) & n & Y & Sy2 & Y(C) & Obs & 1\\
UGC 5101 & Y(1.697) & \nodata & Y & Y & Y(0.13) & Y & Y & Sy2 & Y(CL) & Obs & 1\\
MCG+08-18-012 & n & \nodata & \nodata & \nodata & \nodata & n & \nodata & \nodata & n & n & \nodata\\
MCG+08-18-013 & n & n & \nodata & n & \nodata & n & \nodata & Cp & n & n & 1\\
MCG--01-26-013 & n & \nodata & \nodata & \nodata & \nodata & n & \nodata & \nodata & n & n & \nodata\\
NGC 3110 & n & n & n & n & n & n & \nodata & HII & n & n & 1\\
ESO 374-IG032 & Y(2.534) & \nodata & \nodata & \nodata & Y(0.03) & n & \nodata & HII & n & n & 11\\
NGC 3256 & n & \nodata & \nodata & \nodata & n & n & \nodata & HII & n & n & 12\\
IRAS F10565+2448 & n & \nodata & \nodata & \nodata & n & n & n & HII & n & n & 13\\
NGC 3690 West & \nodata & Y(u) & n & Y(u) & Y(0.12) & n & Y & Sy2 & Y(L) & CT & 14\\
NGC 3690 East & \nodata & Y(u) & n & Y(u) & n & n & Y & L & n & n & 14\\
ESO 440-58 & \nodata & \nodata & \nodata & \nodata & n & n & \nodata & Cp & n & n & 11\\
MCG--05-29-017 & n & \nodata & \nodata & \nodata & n & n & \nodata & Cp & n & n & 11\\
IRAS F12112+0305 & Y(0.802) & n & \nodata & n & n & n & Y & Sy2 & n & n? & 1\\
NGC 4418 & Y(1.258) & n & n & Y & Y(0.02) & n & \nodata & Sy2 & n & n & 5\\
MCG+00-32-013 & n & \nodata & \nodata & \nodata & \nodata & n & \nodata & \nodata & \nodata & n & \nodata\\
Mrk 231 & Y(1.097) & n & \nodata & Y & Y(0.01) & n & Y & Sy1 & Y(CL) & Obs & 1\\
NGC 4922S & \nodata & \nodata & \nodata & \nodata & \nodata & n & \nodata & Sy2(u) & n & n & 1\\
NGC 4922N & \nodata & \nodata & \nodata & \nodata & Y(0.16) & Y & \nodata & Sy2(u) & Y(A) & Obs & 1\\
IC 860 & n & n & \nodata & Y & n & n & \nodata & Sy? & Y(C) & n & 4\\
IRAS 13120--5453 & \nodata & \nodata & \nodata & \nodata & n & n & \nodata & Sy2 & Y(C) & CT & 5\\
NGC 5104 & n & n & \nodata & Y & n & n & \nodata & L & \nodata & n & 1,13\\
MCG--03-34-064 & Y(1.431) & Y & \nodata & Y & Y(0.01) & Y & \nodata & Cp/Sy2 & Y(LA) & Obs &  1,7\\
NGC 5135 & n & n & \nodata & Y & n & Y & \nodata & Sy2 & Y(L) & CT & 1,b\\
Mrk 266B & n & n & n & n & n & Y & Y & Cp/Sy2 & Y(L) & CT & 1,7,e\\
Mrk 266A & n & \nodata & \nodata & \nodata & n & n & n & Sy2 & Y(A) & Obs & 1,e\\
Mrk 273 & Y(1.182) & n & n & n & Y(0.12) & Y & Y & Sy2 & Y(CL) & Obs & 1\\
IRAS F14348--1447 & Y(1.002) & Y & Y & n & Y(0.25) & n & Y & Cp & Y(C) & CT & 1\\
IRAS F14378--3651 & Y(0.871) & \nodata & \nodata & \nodata & n & n & Y & Sy2 & Y(C) & Y? & 15\\
IC 4518A & Y(1.228) & \nodata & \nodata & \nodata & \nodata & Y$^{\dagger}$ & \nodata & Sy2 & \nodata & Obs & 16\\
IC 4518B & \nodata & \nodata & \nodata & \nodata & \nodata & n$^{\dagger}$ & \nodata & \nodata & \nodata & n & \nodata\\
IRAS F15250+3608 & \nodata & Y & n & n & Y(0.03) & n & Y & Cp & n & n? & 1\\
Arp 220W & n & n & n & n & Y(0.17;u) & n & Y(u) & Cp(u) & n & CT & 1\\
Arp 220E & n & n & n & n & Y(0.17;u) & n & Y(u) & Cp(u) & n & n & 1\\
NGC 6240S & n & \nodata & \nodata & \nodata & n & Y(u) & \nodata & Cp(u) & Y(L;u) & CT & 1,f\\
NGC 6240N & n & \nodata & \nodata & \nodata & n & Y(u) & \nodata & Cp(u) & Y(L;u) & CT & 1,f\\
NGC 6285 & n & n & \nodata & n & n & n & \nodata & Cp & n & n & 1\\
NGC 6286 & n & n & n & n & n & n & \nodata & Cp & Y(A) & CT & 1\\
IRAS F17138--1017 & n & \nodata & \nodata & \nodata & n & n & \nodata & Cp & Y(C) & Obs & 1\\
IRAS F18293--3413 & n & \nodata & \nodata & \nodata & n & n & \nodata & HII & n & n & 1\\
IRAS F19297--0406 & n & \nodata & \nodata & \nodata & n & n & \nodata & Cp & n & n? & 1\\
NGC 6907 & n & \nodata & \nodata & \nodata & n & n & \nodata & \nodata & \nodata & n & \nodata\\
NGC 6908 & n & \nodata & \nodata & \nodata & \nodata & n & \nodata & \nodata & \nodata & n & \nodata\\
NGC 6921 & n & \nodata & \nodata & \nodata & \nodata & n & \nodata & L & \nodata & CT & 17\\
MCG+04-48-002 & n & \nodata & \nodata & \nodata & n & Y & \nodata & HII & \nodata & Obs & 17,18\\
II Zw 096 & \nodata & \nodata & \nodata & \nodata & n & n & \nodata & \nodata & n & n? & \nodata\\
IRAS F20550+1655 SE & Y(0.997) & \nodata & \nodata & \nodata & Y(0.27) & n & \nodata & \nodata & n & n? & \nodata\\
ESO 286-19 & Y(1.441) & Y & n & n & Y(0.10) & n & \nodata & HII & n & n? & 1,12\\
NGC 7130 & n & n & \nodata & n & n & Y & \nodata & Sy2 & Y(LA) & CT & 1\\
NGC 7469 & n & n & \nodata & Y & Y(0.23) & Y & \nodata & Sy1 & \nodata & Unobs & 1,g\\
IC 5283 & n & \nodata & \nodata & \nodata & \nodata & n & \nodata & \nodata & \nodata & n & \nodata\\
ESO 148-2 & Y(1.077) & n & \nodata & n & n & Y & \nodata & Cp & Y(C) & CT & 19\\
NGC 7591 & n & \nodata & \nodata & \nodata & n & n & \nodata & Sy2 & n & n & 1\\
NGC 7674 & Y(1.160) & Y & \nodata & Y & Y(0.02) & Y & \nodata & Sy2 & \nodata & Obs & 1\\
MCG+01-59-081 & n & \nodata & \nodata & \nodata & \nodata & n & \nodata & \nodata & \nodata & n & \nodata\\
NGC 7679 & n & \nodata & \nodata & \nodata & n & Y & \nodata & Sy2 & \nodata & Unobs & 1\\
NGC 7682 & n & \nodata & \nodata & \nodata & \nodata & n & \nodata & Sy2 & \nodata & Obs & 20\\
\enddata
\tablecomments{Columns: 
(1) object name;
(2--8) multiwavelength AGN signatures based on WISE color in NASA/IPAC Infrared Science Archive (IRSA), 2.5--5~$\mu$m continuum slope, 3.1~$\mu$m and 3.4~$\mu$m absorption feature \citep{Imanishi2010}, EW of the 3.3 $\mu$m \citep{Imanishi2008,Imanishi2010} and 6.2~$\mu$m PAH emission \citep{Stierwalt2013},
[\ion{Ne}{5}] 14.32 $\mu$m line \citep{Inami2013}, and radio spectral index maps \citep{Vardoulaki2015}.
The criteria for the presence of an AGN is described in the text;
(9) classification based on the optical spectroscopy and the BPT diagram. 
Sy stands for Seyfert with subcategories 1, 1.5, 1.8 and 2.
L stands for LINERs.
Cp stands for AGN/starburst composite \citep[e.g.,][]{Yuan2010}.
HII stands for HII regions;
(10) AGN selection criteria in the C--GOALS sample
 \citep{Iwasawa2011,Torres-Alba2018}; the 0.5--2~keV to 2--7~keV color (C), detection of 6.4~keV line (L), and absorbed AGN feature (A);
(11) X-ray classification based on this work for hard X-ray detected AGNs (Unobs = unobscured AGN, Obs = obscured AGN, CT = CT AGN, Jet = a jet-dominant AGN) and for hard X-ray undetected sources (Y? = good candidate hosting an AGN based on the 8--24~keV upper limits, n? = object with a possible AGN based on the 8--24~keV upper limits, and n = object with no sign of the AGN emission in the hard X-ray band, see also Section~\ref{sub6-2_upperlim});
(12) references of the optical classification in Column (9) marked as ``1--20'', and the previous NuSTAR results for some AGNs in Column (11) marked as ``a--g''.\\
\textbf{References:} (1) \citet{Yuan2010}; (2) \citet{Lopez-Coba2020}; (3) \citet{Veron-Cetty2006}; (4) \citet{Alonso-Herrero2009};
(5) \citet{Veron-Cetty2010}; (6) \citet{Vogt2013}; (7) \citet{Brightman2011b}; (8) \citet{Alonso-Herrero2013};
(9) \citet{Arribas2008}; (10) \citet{Lipari2004}; (11) \citet{Rich2015}; (12) \citet{Lipari2000};
(13) \citet{Veilleux1995}; (14) \citet{Garcia-Martin2006}; (15) \citet{Duc1997}; (16) \citet{Masetti2008};
(17) \citet{Koss2016a}; (18) \citet{Masetti2006}; (19) \citet{Lipari2003}; (20) \citet{Huchra1992};
(a) \citet{Bauer2015}; (b) \citet{Yamada2020}; (c) \citet{Rani2018}; (d) \citet{Rivers2015};
(e) \citet{Iwasawa2020}; (f) \citet{Puccetti2016}; (g) \citet{Ogawa2021}.
}
\tablenotetext{\dagger}{The results of [\ion{Ne}{5}] detection on IC 4518A/IC 4518B are referred from \citet{Pereira-Santaella2010}.}
\end{deluxetable*}

\clearpage

\item IRAS F00344--3349 (ESO 350-38): 

The LIRG ESO 350-38 is a stage-C merger composed of three main
star-forming condensations \citep{Kunth2003,Atek2008}. Among them, the
eastern and western knots whose separation is 5\farcs2
(2.2~kpc) are clearly resolved as X-ray sources with Chandra
\citep[e.g.,][]{Torres-Alba2018}. NuSTAR detects only soft X-ray
emission below $\sim$10~keV. Combining the NuSTAR and Chandra data, we
constrain the photon index and cutoff energy to be $\Gamma \leq$ 1.62
and $E_{\rm cut}$ = 6.6$^{+3.4}_{-1.5}$~keV, respectively, supporting
that this is a starburst-dominant galaxy.

\item IRAS F00402--2349 W/E (NGC 232 and NGC 235): 

This LIRG in merger stage-B is composed of two nuclei (NGC 232 and NGC
235) with a separation
of 120\farcs8 (54.7~kpc). 
We identify NGC 235 as the hard X-ray detected AGN.
The fractions of the Spitzer/MIPS 24~$\mu$m
luminosities are 68\% and 32\% for the western (NGC 232) and eastern
(NGC 235) objects, respectively \citep{Inami2013}. NGC~235 has an X-ray
bright AGN in the Swift/BAT 105-month catalog, and the hydrogen column
density is estimated to be $N_{\rm H} \sim 3 \times 10^{23}$~cm$^{-2}$
by using the Swift/BAT and XRT observations
\citep{Ricci2017dApJS}. Combining the Chandra and Suzaku data with the
Swift/BAT spectrum, we find that the column density varies from
$(2.9~\pm~0.2) \times 10^{23}$~cm$^{-2}$ (Chandra) to $N_{\rm H} =
(5.1^{+0.8}_{-0.6}) \times 10^{23}$~cm$^{-2}$ (Suzaku and Swift/BAT).
The estimated absorption-corrected X-ray luminosity is log$L_{2-10}$ =
43.28$^{+0.18}_{-0.16}$. For NGC 232, only the Chandra spectrum is
obtained, and the starburst-dominant model yields the best-fit photon index of
$\Gamma \sim 1.6$.

\item IRAS F00506+7248 (MCG+12-02-001): 

This system is a stage-C merger composed of four galaxies; the northern
source (MCG+12-02-001a), the western source (MCG+12-02-001 NED01),
and the eastern interacting pair (MCG+12-02-001) with a separation of
0\farcs9 (0.3~kpc; \citealt{Ricci2017bMNRAS}). The eastern pair is
assigned as a LIRG, which is the brightest source of this system in the
IRAC 1--4 (3.6, 4.5, 5.8, and 8.0~$\mu$m) images. 
We identify this pair (MCG+12-02-001) as the hard X-ray detected
AGN.

\citet{Ricci2017bMNRAS} report that the detection significance with
NuSTAR in the 3--10~keV and 10--24~keV bands are $<$3$\sigma$ in this
system.
In our analysis, it is 2.98$\sigma$ 
and 3.37$\sigma$ in the 8--24~keV and 10--40~keV bands, respectively.
By combining the Chandra observations, we analyze the 0.2--40~keV
spectra and suggest the presence of a CT AGN with $N_{\rm H} >
4.3 \times 10^{24}$~cm$^{-2}$. The estimated absorption-corrected
X-ray luminosity is log$L_{2-10}$ = 41.75$^{+0.33}_{-0.08}$,
classifying it as a low-luminosity AGN. The large fraction of
unabsorbed soft X-ray emission ($f_{\rm bin} \gtrsim 9$\%) implies that
the starburst emission seems to be dominant in soft X-rays, in agreement
with the optical classification of the AGN/starburst composite and with no
clear AGN signs from the other multiwavelength diagnostics.

\item IRAS F01053--1746 W/E (IC 1623A and IC 1623B): 

The LIRG IC 1623 (= VV 114) is a stage-C merger consisting of two nuclei with a
separation of 15\farcs7 (6.4~kpc). The eastern object (IC~1623B) is much
brighter than the western source (IC 1623A) in the Spitzer/IRAC
bands. Whereas, the Chandra results by \citet{Grimes2006} show that
IC~1623A is much brighter than IC~1623B in X-rays. 
IC 1623A is also very luminous in the UV and
optical, with many star clusters detected in HST images
\citep{Cortijo-Ferrero2017a} and in soft X-ray images (Chandra and
XMM-Newton). \citet{Grimes2006}
imply the possible presence of a
low-luminosity AGN in IC~1623B ($N_{\rm H} \sim 1.5 \times
10^{22}$~cm$^{-2}$), although its contribution to the total soft X-ray flux 
of the system is small. 
We cannot reproduce the broadband spectra of the system observed with
NuSTAR, Chandra, and XMM-Newton by the AGN model, due to a cutoff
feature at $E_{\rm cut} = 8.1^{+0.8}_{-0.7}$~keV.  Thus, we treat them
as non AGNs in this paper. 
The starburst-dominant model yields $\Gamma \leq$ 1.51.

\item IRAS F02071--1023 W2/W/E/S (NGC 833, NGC 835, NGC 838 and NGC 839): 

The system IRAS F02071--1023 (= HCG 16 or Arp 318) is a stage-A merger
with seven member galaxies, including the central four spiral galaxies
\citep{Hickson1982}. With the Chandra observations,
\citet{OSullivan2014} detect obscured AGNs in the pair of NGC 833 and
NGC 835 separated by 56\farcs4 (15.3~kpc), whereas they conclude that
NGC 838 and NGC 839, separated by 148\farcs8 (39.1~kpc), are
starburst-dominant sources.  \citet{Bitsakis2014} estimate their total
IR luminosities by the UV-to-submillimeter SED decomposition, and
find that only NGC 838 is a LIRG.
We identify both NGC 833 and NGC 835 as the hard X-ray detected AGNs.

\citet{Oda2018} carry out a detailed X-ray study of this system using
NuSTAR, Chandra, and XMM-Newton, detecting the obscured AGNs with
$N_{\rm H} \sim 3 \times 10^{23}$~cm$^{-2}$ in both NGC 833 and NGC 835.
They also report an increase of the line-of-sight column density in NGC
835 from 2000 (XMM-Newton) to 2013/2015 (Chandra/NuSTAR; $\sim$5 $\times
10^{23}$~cm$^{-2}$). We find, however, that our AGN model with XCLUMPY
can reproduce the variability by a change of the absorption-corrected
luminosity rather than that of the column density ($C^{\rm Time} =
1.95^{\rm +0.19}_{-0.21}$ for Chandra; and $C^{\rm Time} = 0.11^{\rm
+0.02}_{-0.01}$ for XMM-Newton). We estimate the column densities to be 
$N_{\rm H} = (2.6 \pm 0.1)$ $\times 10^{23}$~cm$^{-2}$ and $N_{\rm
H} = (3.0^{+0.1}_{-0.2}) \times 10^{23}$~cm$^{-2}$, and the
X-ray luminosities to be log$L_{2-10}$ = 41.99$^{+0.25}_{-0.24}$
and log$L_{2-10}$ = 42.06$^{+0.13}_{-0.03}$ for NGC 833 and
NGC 835, respectively.  NuSTAR does not detect hard X-ray emission
from NGC 838 and NGC 839. Their best-fit photon indices derived from
the Chandra spectra with the starburst-dominant models are $\Gamma \leq 1.52$
(with $E_{\rm cut}$ = 4.7$^{+0.7}_{-0.6}$~keV) and $\Gamma \leq 1.51$,
respectively.

\item IRAS F02401--0013 (NGC 1068): 

The LIRG NGC 1068 is a possible past merger \citep{Tanaka2017}
categorized as a stage-N source in this work. The source has been well
studied in X-rays, e.g., with Chandra \citep[e.g.,][]{Torres-Alba2018}
and NuSTAR \citep[e.g.,][]{Marinucci2016}. We refer to the results
reported in \citet{Bauer2015}, who perform a spectral analysis adopting
a multi-component reflector model to the NuSTAR, Chandra, XMM-Newton,
Suzaku, BeppoSAX, and Swift/BAT spectra. They find that the AGN is
heavily obscured ($N_{\rm H} \sim 10^{25}$~cm$^{-2}$) with an absorption-corrected
X-ray luminosity of log$L_{2-10}$ = 43.34$^{+0.07}_{-0.08}$.

\item IRAS F03117+4151 N/S (UGC 2608 and UGC~2612): 

Each galaxy in IRAS F03117+4151 is a stage-N source with no sign of
interaction. The northern object (UGC 2608) is a LIRG, while the
southern object (UGC~2612) is faint in the IR
\citep[e.g.,][]{Inami2013}. 
\citet{Guainazzi2005} report the results of the XMM-Newton observation
of UGC 2608. Adding the NuSTAR and Suzaku data to the XMM-Newton ones,
\citet{Yamada2020} perform a broadband spectral analysis of UGC 2608
with XCLUMPY and identify a heavily obscured AGN ($N_{\rm H} =
5.4^{+7.0}_{-3.1} \times 10^{24}$~cm$^{-2}$) with an absorption-corrected
luminosity of log$L_{2-10}$ = 43.59$^{+0.19}_{-0.25}$. 
For UGC 2612, no X-rays are detected with NuSTAR, Chandra, or
XMM-Newton. The 3$\sigma$ upper limit of the AGN luminosity in UGC 2612
is log$L_{2-10}^{\rm (lim)} < 42.89$, assuming log$N_{\rm H} \leq$
24.5.

\item IRAS F03164+4119 (NGC 1275): 

The radio galaxy NGC 1275 is a nonmerging (stage-N) LIRG in the Perseus
cluster. \citet{Fukazawa2018} find correlated long-term variability
between X-ray (Suzaku/XIS and Swift/XRT) and GeV gamma-ray (Fermi/LAT)
fluxes, indicating a significant jet contribution to the X-ray
emission. The Hitomi observation reveals that the Fe
K$\alpha$ line at 6.4~keV is emitted from a region at $\sim$1.6~kpc from
the core, most likely a low-covering-faction torus or a rotating
molecular disk \citep{Hitomi-Collaboration2018}.
By analyzing the NuSTAR and Swift/BAT spectra, \citet{Rani2018} suggest
that the hard X-ray power-law component dominating the spectrum
above 20~keV originates from the jet. Thus, we find it difficult 
to constrain the column density and 
AGN luminosity of the nucleus from the broadband spectra, and 
discuss only the NuSTAR band ratio in this paper.

\item IRAS F03316--3618 (NGC 1365): 

The LIRG NGC 1365 is a stage-N object. The source has an AGN showing
extreme absorption variability observed with e.g., Chandra
\citep{Torres-Alba2018,Venturi2018}, XMM-Newton
\citep{Risaliti2005,Whewell2016}, and NuSTAR
\citep[e.g.,][]{Walton2014a,Risaliti2016}.  \citet{Rivers2015} study the
variable absorption seen in the four long-look joint observations of
NuSTAR and XMM-Newton with a multi-layer absorber model.
In this work, we refer to their averaged values of $N_{\rm H} \sim 1
\times 10^{23}$~cm$^{-2}$ and log$L_{2-10}$ = 41.90 $\pm$ 0.01.

\item IRAS F04454--4838 (ESO 203-1): 

This LIRG is a stage-B merger showing two nuclei 
(north-east and south-west components) with a separation of 
7\farcs5 (7.7~kpc; \citealt{Haan2011a}). The north-eastern galaxy is
much brighter than south-western source in the Spitzer/IRAC 1--4 bands.
The Chandra (see also \citealt{Iwasawa2011}) and NuSTAR observations 
do not detect any X-ray emission in this system. 
The 3$\sigma$ upper limit of the AGN luminosity 
is log$L_{2-10}^{\rm (lim)} < 43.06$ assuming log$N_{\rm H} \leq$ 24.5.

\item IRAS F05054+1718 W/E (CGCG 468-002W and CGCG 468-002E): 

These two stage-B merging galaxies have a separation of
29\farcs5 (10.3~kpc). The eastern source (CGCG 468-002E) is a LIRG,
while the western source (CGCG 468-002W) is not luminous in the
IR band (log$L_{\rm IR}$/$L_{\odot}$ = 10.74;
\citealt{Ricci2017bMNRAS}). 
We identify CGCG 468-002E as the hard X-ray detected AGN.
The recent study with NuSTAR and
Swift/XRT observations by \citet{Ricci2017bMNRAS} reveals the presence of
an AGN with $N_{\rm H} \sim 1.5 \times 10^{22}$~cm$^{-2}$ in CGCG
468-002E.  This is consistent with our results using
NuSTAR, Swift/BAT and XRT data ($N_{\rm H} = 1.5 \pm 0.1 \times
10^{22}$~cm$^{-2}$ and log$L_{2-10}$ = 42.84$^{+0.04}_{-0.03}$). Due
to a contamination from the brighter eastern source, the upper limit of
the hard X-ray flux of the western source is not constrained.

\item IRAS F05189--2524: 

This ULIRG is a stage-D merger with a single nucleus. The galaxy shows
multiphase outflows detected as UFO \citep{Smith2019}, ionized
outflow \citep[e.g.,][]{Fluetsch2019,Fluetsch2021}, and  
molecular outflow with $>$500~km~s$^{-1}$ \citep[e.g.,][]{Veilleux2013}.
We identify the galaxy as the hard X-ray detected AGN.

The source has been well studied in X-rays by e.g., Chandra
\citep{Ptak2003,Grimes2005,Iwasawa2011}, XMM-Newton
\citep{Imanishi2004,Teng2010}, Suzaku \citep{Teng2009}, and NuSTAR
\citep[e.g.,][]{Teng2015,Xu2017,Smith2019}. NuSTAR detects the hard
X-ray emission from an obscured AGN. \citet{Teng2015} fit the combined
NuSTAR and XMM-Newton spectra with a two absorber model
($N_{\rm H} \sim 5 \times 10^{22}$~cm$^{-2}$ covering 98\% and $N_{\rm
H} \sim 9 \times 10^{22}$~cm$^{-2}$ covering 74\%). With the XCLUMPY
model, we confirm dramatic variability in absorption between
Compton-thin ($N_{\rm H}$ = $(7.5 \pm 0.1) \times 10^{22}$~cm$^{-2}$
for NuSTAR, Chandra, XMM-Newton, and Swift/BAT) and CT obscuration
($>$2.3 $\times 10^{24}$~cm$^{-2}$ for Suzaku), as discussed
in Section~\ref{subsub6-1-5_NHvariability}.  We estimate the absorption-corrected X-ray luminosity to be
log$L_{2-10}$ = 43.42$^{+0.01}_{-0.02}$. 
\\
\\

\item IRAS F06076--2139 N/S (IRAS F06076--2139 and 2MASS 06094601--2140312): 

This system is a stage-C LIRG consisting of two galaxies separated by
8\farcs3 (6.2~kpc). The northern source (IRAS F06076--2139) is much
brighter than southern source (2MASS 06094601--2140312) in the
Spitzer/IRAC 1--4 bands. 
We identify IRAS F06076--2139 as the hard X-ray detected AGN.
Using NuSTAR and Chandra, \citet{Privon2020}
detect X-ray emission from the obscured AGN in IRAS F06076--2139 with
$N_{\rm H} \sim 6 \times 10^{23}$~cm$^{-2}$, consistent with our result
($N_{\rm H}$ = 4.2$^{+2.4}_{-1.2} \times 10^{23}$~cm$^{-2}$). The
absorption-corrected X-ray luminosity we obtain is log$L_{2-10}$ =
42.18$^{+0.15}_{-0.14}$. No significant X-ray emission from 2MASS
06094601--2140312 is detected with NuSTAR and Chandra.

\item IRAS F08354+2555 (NGC 2623): 

The LIRG NGC 2623 is a stage-D merger with a single nucleus. 
We identify the source as the hard X-ray detected AGN. The Chandra
and XMM-Newton studies suggest the presence of an obscured AGN \citep{Maiolino2003,Evans2008,Iwasawa2011},
We detect the 8--24~keV emission at 4.5$\sigma$ level with NuSTAR. 
With the broadband spectral analysis using the NuSTAR, Chandra, and
XMM-Newton data, we estimate that 
$N_{\rm H}$ = $(6.0^{+4.5}_{-2.1}) \times
10^{22}$~cm$^{-2}$ and log$L_{2-10}$ = 40.90 $\pm$ 0.11.

\item IRAS F08520--6850 W/E (ESO 060-IG016 West and ESO 060-IG016 East): 

This LIRG system is a stage-B merger, composed of two galaxies separated
by 15\farcs4 (13.6~kpc).  
We identify ESO 060-IG016 East as the hard X-ray detected AGN.
The Chandra spectrum of ESO 060-IG016 East
reveals the presence of an obscured AGN with $N_{\rm H} \sim 1 \times
10^{23}$~cm$^{-2}$ \citep{Iwasawa2011}, whose 8--24~keV emission is
detected with NuSTAR. Combining these observations, we constrain the
column density and absorption-corrected AGN luminosity to be $N_{\rm H}$ =
(8.4$^{+4.0}_{-2.9}$) $\times 10^{22}$~cm$^{-2}$ and log$L_{2-10}$ 
= 41.94$^{+0.07}_{-0.08}$, respectively. NuSTAR and Chandra detect
no significant X-ray emission from ESO 060-IG016 West.

\item IRAS F08572+3915: 

This is a double-nucleus ULIRG (north-west and south-east components)
classified as a stage-D merger with a separation of 
4\farcs4 (5.6~kpc; \citealt{Ricci2017bMNRAS}). From the north-western
nucleus, a very deep silicate absorption and little PAH emission are
detected in the Spitzer IRS spectrum, indicating the presence of a heavily
obscured AGN \citep[e.g.,][]{Spoon2007}. Ionized outflow
\citep[e.g.,][]{Fluetsch2019,Fluetsch2021} and molecular outflow
with $>$500~km~s$^{-1}$ \citep[e.g.,][]{Sturm2011,Cicone2014} have been
discovered, whereas UFO is not detected under the poor statistics of the
X-ray data \citep[e.g.,][]{Mizumoto2019}.
We identify IRAS F08572+3915 as the hard X-ray detected AGN.

\citet{Iwasawa2011} report the Chandra results that
the northwest nucleus is
detected in the $\sim$2--4~keV band, whereas 
the south-eastern source is much fainter.
The X-ray color suggests an obscured AGN in the northwest nucleus.
Although the earlier NuSTAR observation detects no X-ray emission
\citep{Teng2015}, we detect the hard X-ray emission at 3.9$\sigma$ level
by utilizing the recent follow-up observation in 2019 (ObsID =
50401004002; PI: C. Ricci). The large NuSTAR band ratio
(BR$^{\rm Nu} \sim 1.6$) supports the presence of an AGN (Section~\ref{sub5-1_Xcolor}).
From the 0.1--20~keV spectra of NuSTAR, XMM-Newton, and Suzaku, we
estimate the absorption and absorption-corrected X-ray luminosity to be $N_{\rm H}
= (8.5^{+12.9}_{-2.8}) \times 10^{23}$~cm$^{-2}$ and log$L_{2-10}$ 
= 41.77$^{+0.06}_{-0.07}$, respectively, by assuming
$f_{\rm bin}$ = 10\%, which may be highly uncertain.

\item IRAS F09320+6134 (UGC 5101): 

The ULIRG UGC 5101 is a stage-D merger showing a single nucleus.
We identify the source as the hard X-ray detected AGN.
\citet{Ricci2015} detect a CT AGN by analyzing the XMM-Newton and
Swift/BAT spectra. The NuSTAR study by \citet{Oda2017} finds the
absorption column density to be $N_{\rm H} \sim 1.3 \times
10^{24}$~cm$^{-2}$. By utilizing all the available data of NuSTAR, Chandra, XMM-Newton,
Suzaku, and Swift/BAT, we obtain almost the same results as \citet{Oda2017} but
newly find that the column density is variable from $N_{\rm H} =
(1.5^{+0.3}_{-0.2}) \times 10^{24}$~cm$^{-2}$ (Chandra, XMM-Newton, and
Suzaku) to $(9.6^{+0.4}_{-0.2}) \times 10^{23}$~cm$^{-2}$ (NuSTAR and
Swift/BAT). The absorption-corrected X-ray luminosity we estimate is 
log$L_{2-10}$ = 43.15$^{+0.25}_{-0.15}$.

\item IRAS F09333+4841 W/E (MCG+08-18-012 and MCG+08-18-013): 

The two galaxies are in a stage-A merger with a separation of
65\farcs3 (33.6~kpc). MCG+08-18-013 is a LIRG, whereas
MCG+08-18-012 is less luminous (log$L_{\rm IR}$/$L_{\odot}$ = 9.93;
\citealt{Ricci2017bMNRAS}). Both objects are not detected wuth NuSTAR in
either the 3--8~keV or 8--24~keV bands, as reported by
\citet{Ricci2017bMNRAS}. Chandra detects X-ray emission only 
from MCG+08-18-013. By adopting the starburst-dominant model without an
\textsf{apec} component, we obtain $\Gamma \leq 1.84$.

\item IRAS F10015--0614 S/N (MCG--01-26-013 and NGC 3110): 

IRAS F10015--0614 is a stage-A merger with two galaxies separated by
108\farcs8 (36.5~kpc). The LIRG NGC 3110 (north) is much brighter than
MCG--01-26-013 (south) in the Spitzer/IRAC and Herschel images
\citep{Chu2017}. \citet{Ricci2017bMNRAS} report that neither of them is
detected with NuSTAR. They obtain photon indices of $\Gamma =
1.8^{+0.6}_{-0.7}$ for MCG--01-26-013 and $\Gamma =
2.2^{+0.3}_{-0.2}$ for NGC 3110 by using the Chandra and XMM-Newton
spectra. Although we fix $\Gamma=1.8$ for MCG--01-26-013 in our
analysis, we obtain $\Gamma$ = 1.8 $\pm$ 0.1 for NGC~3110; the
difference from \citet{Ricci2017bMNRAS} is likely comes from the
inclusion of absorption (\textsf{zphabs}) to the soft X-ray component in
\citet{Ricci2017bMNRAS}. 
We confirm the conclusion by \citet{Ricci2017bMNRAS} 
that the X-ray spectra do not require the presence of 
AGNs in these galaxies.

\item IRAS F10038--3338 (ESO 374-IG032): 

This merger stage-D LIRG shows a single nucleus. It was mistakenly
identified with the optical source IC 2524 in the RBGS catalog; the
correct counterpart is ESO 374-IG032 (see the Essential Notes in NED).
Chandra detects a faint X-ray source \citep{Iwasawa2011}, whereas NuSTAR
does not. The Chandra spectrum is reproduced by the starburst-dominant model
with a photon index fixed at 1.8.

\item IRAS F10257--4339 (NGC 3256): 

NGC 3256 is an LIRG in merger stage-D, showing two nuclei with a
separation of 5\farcs1 (1.0~kpc; \citealt{Ricci2017bMNRAS}).
\citet{Lehmer2015} perform an X-ray spectral analysis using the 
NuSTAR and Chandra data, and conclude that the flux at $>$3~keV is
dominated by those from $\sim$5--10 ultraluminous X-ray sources (ULXs). 
We find that the broadband spectra of NuSTAR, Chandra, and XMM-Newton
can be reproduced with the starburst-dominant model. However, the best-fit
values of $\Gamma$ = 2.06 $\pm$ 0.12 and $E_{\rm cut}$ =
16.1$^{+11.1}_{-4.9}$~keV are not typical of ULXs, which exhibit spectral
turnovers between 6 and 8~keV
\citep[e.g.,][]{Bachetti2013,Walton2013,Walton2014b,Delvecchio2014,Rana2015}.
Thus, we suggest that the hard X-ray flux is not dominated by ULXs,
although their contribution may not be negligible.

\item IRAS F10565+2448: 

This ULIRG is a stage-D merger composed of two nuclei separated by
7\farcs4 (6.7~kpc; \citealt{Ricci2017bMNRAS}). The western galaxy should
be the counterpart of the IRAS source as it is the only source detected
in the Spitzer/IRAC images. Soft X-ray emission also comes only from the
western galaxy, which is detected with Chandra and XMM-Newton
\citep{Teng2010,Iwasawa2011}. Neither of them is detected with NuSTAR
\citep{Teng2010,Ricci2017bMNRAS}. Analyzing the spectra of the system
observed with Chandra, XMM-Newton, and Suzaku with the starburst-dominant
model, we obtain a photon index of $\Gamma \leq 1.55$.

\item IRAS F11257+5850 W/E (NGC 3690 West and NGC 3690 East): 

IRAS F11257+5850 (= Arp 299) is a LIRG in merger stage-C with two
galaxies NGC 3690 West (= Arp 299B) and NGC 3690 East (= Arp 299A) 
separated by 22\farcs0 (4.6~kpc). Both galaxies
are strong IR sources, and the IRAS HIRES flux ratio
between the west and east galaxies is 1:3 \citep{Surace2004}. 
We identify NGC 3690 West as the hard X-ray detected AGN.
The detailed Chandra studies
\citep[e.g.,][]{Iwasawa2011,Anastasopoulou2016} verify the presence of
an AGN in the western galaxy by detecting its prominent neutral Fe~K$\alpha$
line, whereas the eastern galaxy only shows
ionized Fe~K$\alpha$ lines, which could arise from hot gas in the
star-forming region. Combining the Chandra data with NuSTAR observation,
\citet{Ptak2015} confirm the presence of the CT AGN in NGC 3690 West
($N_{\rm H} \sim 4 \times 10^{24}$~cm$^{-2}$) while detecting no hard
X-ray emission from NGC 3690 East. With XCLUMPY, we also confirm the CT
obscuration ($N_{\rm H}$ = 3.0$^{+0.6}_{-0.5} \times
10^{24}$~cm$^{-2}$) with an estimate absorption-corrected luminosity of 
log$L_{2-10}$ = 42.66$^{+0.34}_{-0.32}$ for NGC 3690 West.

\item IRAS F12043--3140 N/S (ESO 440-58 and MCG--05-29-017): 

This is a stage-B merger showing two nuclei with a separation
of 11\farcs8 (5.5~kpc). The northern galaxy (ESO 440-58) is a LIRG,
whereas the southern one (MCG--05-29-017) has a lower IR
luminosity of log$L_{\rm IR}$/$L_{\odot}$ = 10.54 (see
\citealt{Ricci2017bMNRAS}). NuSTAR detects no X-ray emission from
either nucleus in the soft or hard bands \citep{Ricci2017bMNRAS}.
We analyze the Chandra spectra with the starburst-dominant model, where 
we obtain $\Gamma \sim 1.7$ for ESO~440-58 (without an \textsf{apec}
component) but fix $\Gamma=1.8$ for MCG--05-29-017.

\item IRAS F12112+0305: 

This ULIRG is a stage-D merger with two nuclei separated by 3\farcs0
(4.1~kpc; \citealt{Imanishi2020}).  The Chandra
\citep{Teng2005,Iwasawa2011} and XMM-Newton studies
\citep{Franceschini2003} suggest that the object is starburst-dominated.
No X-ray emission is detected with NuSTAR in either the soft or 
hard X-ray band. The 3$\sigma$ upper limit of the X-ray AGN 
luminosity is log$L^{\rm (lim)}_{2-10} < 43.47$ assuming log$N_{\rm
H} \leq 24.5$, and hence the possible presence of a CT AGN cannot be
excluded. By adopting the starburst-dominant model, the Chandra and XMM-Newton
data constrain the photon index to be $\Gamma$ = 1.75 $\pm$ 0.22.

\item IRAS F12243--0036 NW/SE (NGC 4418 and MCG+00-32-013): 

The system is a stage-A merger including a LIRG in the north-western
galaxy NGC 4418 (= NGC 4355). The south-eastern source MCG+00-32-013 
(= VV~655) is a dwarf galaxy separated by 179\farcs9 (29.4~kpc), 
which would enhance the starburst activity in NGC~4418
\citep{Varenius2017,Boettcher2020}. 
NGC 4418 hosts a compact obscured nucleus (CON)
\citep[e.g.,][]{Costagliola2013,Sakamoto2013,Ohyama2019}. 
Chandra detects weak X-ray emission from
NGC~4418 but the presence of an AGN has been a matter of controversy
due to the low statistics
\citep{Maiolino2003,Lehmer2010,Torres-Alba2018}.  Hard
X-rays are not detected with NuSTAR and the 3$\sigma$ upper limit of the
absorption-corrected AGN luminosity is log$L^{\rm (lim)}_{2-10}
< 41.82$ assuming log$N_{\rm H} \leq 24.5$.  Thus, if NGC~4418 has an
AGN, it should be a low-luminosity and/or heavily obscured one
(log$N_{\rm H} \gtrsim 25.0$). Analyzing the Chandra and Suzaku spectra
with the starburst-dominant model, we obtain a photon index of $\Gamma \leq
1.65$. X-ray emission is not detected from MCG+00-32-013 with either
NuSTAR or Chandra.
\\
\\

\item IRAS F12540+5708 (Mrk 231): 

Mrk 231 (= UGC 8058) is the most luminous ULIRG (log$L_{\rm
IR}/L_{\odot} = 12.57$) in our sample, classified as a stage-D merger
with a single nucleus. Previous multiwavelength studies reveal the
presence of multiphase outflows, that is, UFO
\citep[e.g.,][]{Mizumoto2019}, ionized outflow
\citep[e.g.,][]{Fluetsch2019,Fluetsch2021}, and strong molecular outflow
with $>$500~km~s$^{-1}$ \citep{Sturm2011,Cicone2014}. Mrk 231 also
shows spatially unresolved BAL systems in the optical and
near-UV spectra
\citep{Boksenberg1977,Smith1995,Gallagher2002,Gallagher2005,Rupke2002}.
We identify Mrk 231 as the hard X-ray detected AGN.

The source has been detected with Chandra
\citep[e.g.,][]{Ptak2003,Grimes2004,Gallagher2002,Gallagher2005,Iwasawa2011},
XMM-Newton \citep[e.g.,][]{Franceschini2003,Turner2003,Braito2004}, and
NuSTAR \citep[e.g.,][]{Teng2014,Feruglio2015}. 
\citet{Teng2014} suggest the presence of an obscured AGN and
variability in the absorption column density from $N_{\rm H} \sim 2
\times 10^{23}$~cm$^{-2}$ in 2003 (Chandra) to $N_{\rm H} \sim 1 \times
10^{23}$~cm$^{-2}$ in 2012 (Chandra and NuSTAR).  Using the NuSTAR,
Chandra, XMM-Newton, and Suzaku observations, we estimate the
time-averaged column density of $N_{\rm H}$ = ($8.5 \pm 0.2$)
$\times 10^{22}$~cm$^{-2}$ and the X-ray luminosity of 
log$L_{2-10}$ = 42.65 $\pm$ 0.02.

\item IRAS F12590+2934 S/N (NGC 4922S and NGC 4922N): 

The system is a stage-C merger with the two galaxies separated by
22\farcs1 (10.6~kpc). The northern source (NGC 4922N) is a LIRG, whereas
the southern (NGC 4922S) is less luminous (log$L_{\rm IR}$/$L_{\odot}$ =
8.87; \citealt{Ricci2017bMNRAS}). 
We identify NGC 4922N as the hard X-ray detected AGN.
The X-ray study with NuSTAR and
Chandra by \citet{Ricci2017bMNRAS} suggests a CT AGN in NGC 4922N
($N_{\rm H} \geq 4.27 \times 10^{24}$~cm$^{-2}$). Our analysis with
XCLUMPY supports the presence of the obscured AGN, with 
an absorption-corrected X-ray luminosity of log$L_{2-10}$ =
42.00$^{+0.20}_{-0.18}$, although the absorption is
somewhat smaller ($N_{\rm H} = 7.6^{+28.5}_{-2.2} \times
10^{23}$~cm$^{-2}$) than that in \citet{Ricci2017bMNRAS}. 
X-ray emission from NGC 4922S is detected only with Chandra, whose
spectrum is reproduced by the starburst-dominant model with $\Gamma$=1.8
(fixed).
\\
\\

\item IRAS F13126+2453 (IC 860): 

The source is a stage-N LIRG. No significant X-ray emission is detected
with Chandra (see also e.g., \citealt{Lehmer2010}) or NuSTAR.
Despite the low count rates, this galaxy is classified as an AGN in view
of its Chandra X-ray color \citep{Torres-Alba2018}. Using the NuSTAR
count rates, we constrain the 3$\sigma$ upper limit of the 
absorption-corrected AGN luminosity as log$L^{\rm (lim)}_{2-10} < 41.94$
assuming log$N_{\rm H} \leq 24.5$. Since IC 860 hosts a CON
\citep[e.g.,][]{Costagliola2011}, heavily CT obscuration 
is possible and thus the classification remains ambiguous.

\item IRAS 13120--5453: 

This ULIRG is a stage-D merger with a single nucleus. 
We identify the galaxy as the hard X-ray detected AGN.
The NuSTAR and
Chandra study of \citet{Teng2015} identifies a CT AGN. By adding the
XMM-Newton data to them, we also confirm the CT obscuration of $N_{\rm
H}$ = (1.6 $\pm$ 0.2) $\times 10^{24}$~cm$^{-2}$, and estimate the
absorption-corrected luminosity to be log$L_{2-10}$ =
42.17$^{+0.59}_{-0.11}$.

\item IRAS F13188+0036 (NGC 5104): 

This is a LIRG classified as a stage-N source. \citet{Privon2020}
report that NuSTAR does not detect the source, while the soft X-ray
emission is detected with Swift/XRT. They find an observed 2--10~keV
luminosity of $\sim$1.5~$\times 10^{40}$~erg~s$^{-1}$, which is smaller
than a predicted value from the SFR with the \citet{Lehmer2016} relation. 
Thus, no evidence for an AGN is provided.
Analyzing the spectrum with the starburst-dominant model, we 
obtain a photon index of $\Gamma$ = 2.67$^{+0.17}_{-0.52}$.

\item IRAS F13197--1627 (MCG--03-34-064): 

This LIRG is a stage-N object. 
We identify MCG--03-34-064 as the hard X-ray detected AGN.
\citet{Ricci2017bMNRAS} carry out a 
broadband spectral analysis using the XMM-Newton and Swift/BAT spectra,
and reveal the presence of an obscured AGN in this object ($N_{\rm H}
\sim 5 \times 10^{23}$~cm$^{-2}$). By combining these data with the
NuSTAR and Chandra data, we obtain similar results and find that the
hydrogen column density varies from $N_{\rm H}$ = (5.6~$\pm$~0.2)
$\times 10^{23}$~cm$^{-2}$ (Chandra and XMM-Newton) to (9.8$^{+0.3}_{-0.2}$)
$\times 10^{23}$~cm$^{-2}$ (NuSTAR and Swift/BAT). The absorption-corrected X-ray
luminosity we estimate is log$L_{2-10}$ = 43.27$^{+0.06}_{-0.07}$.
\\
\\

\item IRAS F13229--2934 (NGC 5135): 

This LIRG is a stage-N source, with no signs of a merger even in the
nuclear region \citep[e.g.,][]{Sabatini2018}.  The Suzaku 0.5--50~keV
spectrum indicates the presence of a CT AGN \citep{Singh2012}.  Using
NuSTAR, Chandra, and Suzaku, \citet{Yamada2020} perform a broadband
X-ray spectral analysis with the XCLUMPY model and identify the galaxy
as the hard X-ray detected AGN. The estimated column
density and absorption-corrected luminosity are $N_{\rm H} = 6.7^{+16.6}_{-2.8}
\times 10^{24}$~cm$^{-2}$ and log$L_{2-10}$ =
43.30$^{+0.42}_{-0.26}$, respectively.

\item IRAS F13362+4831 S/N (Mrk 266B and Mrk 266A): 

Mrk 266 (= NGC 5256) is a system composed of two galaxies in a stage-B
merger with a separation of 10\farcs1 (5.6~kpc). The southern source
(Mrk 266B) is a LIRG (log$L_{\rm IR}$/$L_{\odot}$ = 11.36), whereas the
northern source (Mrk 266A) is less luminous (log$L_{\rm IR}$/$L_{\odot}$
= 10.82; \citealt{Mazzarella2012}). 
Using the Chandra data, \citet{Torres-Alba2018} find that 
Mrk 266B can be reproduced with an obscured AGN model and that 
Mrk 266A shows a 6.4~keV Fe K$\alpha$ line at $\sim$2.1$\sigma$ level
\citep{Torres-Alba2018}. The XMM-Newton observation also suggests the
presence of CT AGN(s) with bright Fe K$\alpha$ line(s)
\citep{Mazzarella2012} in the Mrk 266 system. Using the NuSTAR, Chandra,
and XMM-Newton data, \citet{Iwasawa2020} confirm the presence of dual
AGNs in the system. They estimate the column densities to be  
$N_{\rm H} \sim 7 \times 10^{24}$~cm$^{-2}$ and $N_{\rm H} \sim 6.8
\times 10^{22}$~cm$^{-2}$ and the absorption-corrected X-ray luminosities to be 
log$L_{2-10}$ = 43.13$^{+0.39}_{-0.36}$ and 
log$L_{2-10}$ = 41.60$^{+0.04}_{-0.02}$ for Mrk 266B and Mrk 266A,
respectively.

\item IRAS F13428+5608 (Mrk 273): 

Mrk 273 (= UGC 8696) is an ULIRG in a stage-D merger, with two nuclei
(northern and south-western sources) at a separation of
0\farcs9 (0.7~kpc; \citealt{Ricci2017bMNRAS}). This shows multiphase
outflows such as UFO \citep[e.g.,][]{Mizumoto2019}, ionized outflow
\citep[e.g.,][]{Fluetsch2019,Fluetsch2021}, and molecular outflow with
$>$500~km~s$^{-1}$ \citep{Cicone2014}.
We identify Mrk 273 as the hard X-ray detected AGN.

The system is well studied with Chandra 
\citep[e.g.,][]{Xia2002,Ptak2003,Grimes2005} and is considered to host an
obscured AGN in the south-western nucleus \citep{Iwasawa2011a}. The NuSTAR
and XMM-Newton spectral analysis performed by \citet{Teng2015} report
that the AGN shows Compton-thin obscuration ($\sim$4$ \times
10^{23}$~cm$^{-2}$). Combining these data with the Suzaku and Swift/BAT
data, we find variability in absorption, $N_{\rm H} =
(1.7^{+0.5}_{-0.3}) \times 10^{24}$~cm$^{-2}$ (Suzaku in 2006), $(5.0^{+0.6}_{-0.3})
\times 10^{23}$~cm$^{-2}$ (NuSTAR and XMM-Newton in 2013, and
Swift/BAT), and $(9.3^{+1.1}_{-1.0}) \times 10^{23}$~cm$^{-2}$ (Chandra in
2016 and 2017). The absorption-corrected X-ray luminosity is estimated to be 
log$L_{2-10}$ = 43.07 $\pm$ 0.21.

\citet{Iwasawa2018} discuss a possibility that Mrk 273 has
two AGNs ($N_{\rm H} \sim 3 \times 10^{23}$~cm$^{-2}$ for the south-western
nucleus and $N_{\rm H} \sim 1.4 \times 10^{24}$~cm$^{-2}$ for the northern
nucleus, both of which have $L_{2-10} \sim 1 \times
10^{43}$~erg~s$^{-1}$). 
If this scenario is the case, our results of the column density and
luminosity should be regarded as the average and sum of the dual
AGNs, respectively.

\item IRAS F14348--1447: 

This ULIRG in merger stage-D is composed of two nuclei with a separation
of 3\farcs4 (5.2~kpc; \citealt{Imanishi2014}). The southern object is
brighter in the IRAC images. 
We identify IRAS F14348--1447 as the hard X-ray detected AGN.
\citet{Teng2010} report the results of the
Chandra and XMM-Newton observations. Chandra detects a higher X-ray flux
from the southern nucleus, whose X-ray color indicates the existence of
an obscured AGN \citep{Iwasawa2011}.
We find that the detection significance with NuSTAR 
is $<$3$\sigma$ in the 8--24~keV band but 3.1$\sigma$ in the 10--40~keV
band.
Combining the NuSTAR, Chandra, and XMM-Newton spectra
($\sim$0.1--30~keV), we identify the CT AGN ($N_{\rm H}$ =
1.3$^{+1.0}_{-0.6} \times 10^{24}$~cm$^{-2}$) and estimate the X-ray
luminosity to be log$L_{2-10}$ = 42.70$^{+0.93}_{-0.40}$.

\item IRAS F14378--3651: 

This ULIRG is a stage-D merger with a single nucleus. The Chandra X-ray
color classifies the object as an AGN with an observed 2--10~keV luminosity of
$\sim$3~$\times 10^{41}$~erg~s$^{-1}$. The small X-ray to IR
luminosity ratio implies that it is a CT AGN candidate
\citep{Iwasawa2011}. NuSTAR does not detect any hard X-ray emission (see also
\citealt{Teng2015}), although the 3$\sigma$ upper limit of the absorption-corrected AGN
luminosity is large (log$L^{\rm (lim)}_{2-10} < 43.65$
assuming log$N_{\rm H} \leq 24.5$). 
Analyzing the Chandra spectrum with the starburst-dominant model (without an
\textsf{apec} component), we obtain the best-fit photon index of $\Gamma
\leq 1.83$.

\item IRAS F14544--4255 W/E (IC 4518A and IC 4518B): 

The system is a stage-B merger showing two nuclei with a separation of
35\farcs5 (11.5~kpc).  IC 4518A (west) is a LIRG, whereas IC 4518B (east)
is less luminous in the IR band (log$L_{\rm IR}$/$L_{\odot}$ =
11.15 and 10.28, respectively, derived from MIPS 24~$\mu$m data;
\citealt{Alonso-Herrero2012}).
We identify IC 4518A as the hard X-ray detected AGN.

\citet{Ricci2017bMNRAS} report that both nuclei are detected with
XMM-Newton and that the hard X-ray emission from IC 4518A is also
detected by NuSTAR. Their spectral analysis implies that each nucleus
harbors an obscured AGN. 
Using the XCLUMPY model, we find the
hydrogen column density varies from $N_{\rm H}$ = (1.7~$\pm$~0.1)
$\times 10^{23}$~cm$^{-2}$ (NuSTAR, Suzaku, and Swift/BAT) to (2.2~$\pm$~0.2) $\times 10^{23}$~cm$^{-2}$ (XMM-Newton). 
We estimate the absorption-corrected luminosity 
of the AGN in IC 4518A to be log$L_{2-10}$ = 42.83$^{+0.06}_{-0.08}$. 
We cannot confirm the presence of an AGN in IC~4518B, however, 
whose spectrum is subject to large uncertainties due to the
contamination from IC 4518A. 
Applying the starburst-dominant model to the XMM-Newton spectrum of IC 4518B,
we constrain the photon index to be $\Gamma \leq 2.25$.

\item IRAS F15250+3608: 

This is an ULIRG in merger stage-D showing two nuclei separated by
0\farcs7 (0.8~kpc; \citealt{Scoville2000}). The Chandra
\citep{Teng2005,Iwasawa2011} and XMM-Newton \citep{Franceschini2003}
detect weak soft X-ray emission from the system. No hard X-ray
emission is detected with NuSTAR with the 3$\sigma$ upper limit of the
absorption-corrected luminosity of log$L^{\rm (lim)}_{2-10} < 43.41$
assuming log$N_{\rm H} \leq 24.5$. 
The photon index derived from the Chandra and XMM-Newton spectra with
the starburst-dominant model is $\Gamma$ = 2.07$^{+0.23}_{-0.25}$.


\item IRAS F15327+2340 W/E (Arp 220W and Arp 220E): 

Arp 220 (= UGC 9913) is the nearest ULIRG among the stage-D mergers,
containing two nuclei (west and east components) with a separation of 
1\farcs0 (0.4~kpc). The far-IR and millimeter/submillimeter observations
\citep[e.g.,][]{Downes2007,Gonzalez-Alfonso2012,Scoville2015,Falstad2021} 
suggest that Arp 220W contains a heavily CT AGN 
($N_{\rm H} \gtrsim 5 \times 10^{24}$~cm$^{-2}$),
whereas no clear hints for an AGN in Arp 220E have been obtained 
in previous multiwavelength studies.
We identify Arp 220W as the hard X-ray detected AGN.

The Chandra image finds that the $>$2~keV emission is peaked at
the western nucleus \citep{Clements2002,Iwasawa2005}, which shows an
X-ray color consistent with those of obscured AGNs
\citep{Iwasawa2011}. \citet{Paggi2017} constrain the absorption-corrected
AGN 2--10~keV luminosities to be $<$1 $\times 10^{42}$~erg~s$^{-1}$ (west)
and $<$0.4 $\times 10^{42}$~erg~s$^{-1}$ (east), which are derived from the
3$\sigma$ upper limits of the neutral Fe~K$\alpha$ flux of the nuclear
regions. The spectral study using the NuSTAR and Chandra data by
\citet{Teng2015} cannot unveil the presence of the CT AGN due to the
weak detection in the NuSTAR $>$10~keV band. Whereas, they estimate the
hydrogen column density of the possible AGN in Arp 220 to be $>$5.3
$\times 10^{24}$~cm$^{-2}$ with the BNTorus model \citep{Brightman2011a} and
$>$1.2 $\times 10^{24}$~cm$^{-2}$ with the MYTorus model \citep{Murphy2009}.

We also confirm the NuSTAR detection of Arp 220 in the 8--24~keV band
(4.0$\sigma$). When the starburst-dominant model is applied to the 0.2--20~keV
spectra (NuSTAR, Chandra, XMM-Newton and Suzaku), we constrain the
cutoff energy to be $>$92~keV. Since this value is too high to be
explained by emission from starburst or ULXs, we suggest that Arp 220W
actually hosts a heavily obscured AGN. We note that intense starburst
coexists with the AGN, considering the detection of a strong
highly-ionized Fe~K$\alpha$ line at $E_{\rm Fe} =
6.67 \pm 0.02$~keV. Assuming $f_{\rm bin}$ = 10\% as a typical
value among the hard X-ray detected AGNs in stage-D mergers, our XCLUMPY model
yields a column density of $N_{\rm H} \gtrsim 9.8 \times
10^{24}$~cm$^{-2}$, supporting the extreme obscuration. The obtained
absorption-corrected X-ray luminosity (log$L_{2-10}$ = 41.59 $\pm$ 0.02)
is consistent with the upper limit reported by \citet{Paggi2017}.

\item IRAS F16504+0228 S/N (NGC 6240S and NGC 6240N): 

This LIRG is a stage-D merger composed of two nuclei, NGC 6240S (south)
and NGC 6240N (north), with a separation of 1\farcs7 (0.8~kpc). 
This target has been well studied with 
Chandra \citep{Ptak2003,Grimes2005,Iwasawa2011}, and it is revealed that
both nuclei contain CT AGNs showing reflected-dominated
spectra\citep{Komossa2003}. \citet{Puccetti2016} carry out a detailed
spectral analysis of the NuSTAR, Chandra, XMM-Newton, and BeppoSAX data,
and confirm the presence of the dual CT AGNs (see also
\citealt{Nardini2017}). They estimate the column densities ($N_{\rm H}
= 1.47^{+0.21}_{-0.17} \times 10^{24}$~cm$^{-2}$ and $N_{\rm H} =
1.55^{+0.72}_{-0.23} \times 10^{24}$~cm$^{-2}$) and the absorption-corrected X-ray
luminosities (log$L_{2-10}$ = 43.72$^{+0.20}_{-0.19}$ and
log$L_{2-10}$ = 43.30$^{+0.24}_{-0.20}$) for NGC 6240S and NGC
6240N, respectively.

\item IRAS F16577+5900 N/S (NGC 6285 and NGC 6286): 

The system is a stage-B merger with the two galaxies separated by
91\farcs0 (34.5~kpc). NGC 6286 is categorized as a LIRG, whereas NGC 6285
is not a luminous IR source (log$L_{\rm IR}$/$L_{\odot}$ = 10.72;
\citealt{Ricci2017bMNRAS}). 
We identify NGC 6286 as the hard X-ray detected AGN.
The Chandra spectrum of NGC 6286 shows
excess emission above 5~keV \citep{Torres-Alba2018}, implying an
obscured AGN. \citet{Ricci2016} find evidence for a CT, 
low-luminosity AGN ($N_{\rm H} \sim 1 \times 10^{24}$~cm$^{-2}$) 
in NGC 6286, using
NuSTAR, Chandra and XMM-Newton. This is in good agreement with our measurements
with XCLUMPY ($N_{\rm H}$ = $1.4 \pm 0.3 \times
10^{24}$~cm$^{-2}$). The absorption-corrected luminosity we obtain is 
log$L_{2-10}$ = 42.01$^{+1.21}_{-0.23}$. 
From NGC 6285, weak soft X-ray emission is detected with Chandra and
XMM-Newton. The starburst-dominant model without an \textsf{apec} component
yields a photon index of $\Gamma$ = 2.14$^{+0.36}_{-0.33}$.

\item IRAS F17138--1017: 

This LIRG is a stage-D merger with a single nucleus.
We identify IRAS F17138--1017 as the hard X-ray detected AGN.
The Chandra X-ray
spectrum indicates its classification as an obscured AGN
\citep{Torres-Alba2018}. \citet{Ricci2017bMNRAS} fit the combined NuSTAR
and Chandra spectra with a simple power-law model and obtain a photon
index of $\sim$1.1, which is harder than typical spectra of starbursts.
We confirm that NuSTAR detects the 8--24~keV emission at 4.4$\sigma$ (or
3.9$\sigma$) level using the source counts in a 30\arcsec\ (or
45\arcsec) radius. 
Assuming $f_{\rm bin}$ = 10\%, 
our spectral analysis with XCLUMPY suggests the presence of a CT AGN 
with a column density of $N_{\rm H}$ $>$~3.5 $\times
10^{24}$~cm$^{-2}$ and an absorption-corrected luminosity of 
log$L_{2-10}$ = 41.68$^{+0.09}_{-0.06}$. These values should be 
referred to with caveats due to the assumption of $f_{\rm bin}$.

\item IRAS F18293--3413: 

This LIRG is a stage-N source. The earlier Chandra data are studied by
\citet{Iwasawa2011}, whereas a much longer Chandra observation has been
performed in 2019.
NuSTAR detects the source only at $\lesssim$20~keV.
Analyzing the NuSTAR, Chandra, and XMM-Newton data with the 
starburst-dominant model, we obtain a photon index of $\Gamma \leq
1.56$ and a cutoff energy of $E_{\rm cut}$ = 7.6$^{+2.4}_{-1.4}$~keV, 
consistent with no contribution from an AGN.

\item IRAS F19297--0406: 

This source is an ULIRG in merger stage-D showing a single nucleus.
Chandra detects the faint X-ray emission \citep{Iwasawa2011}. No
emission is detected with NuSTAR, and the 3$\sigma$ upper limit of the
absorption-corrected AGN luminosity is log$L^{\rm (lim)}_{2-10} < 44.08$
assuming log$N_{\rm H} \leq 24.5$. Although the origin of the emission
is ambiguous, the Chandra spectrum can be reproduced with the starburst-dominant
model with a photon index fixed at 1.8.

\item IRAS F20221--2458 SW/NE (NGC 6907 and NGC 6908): 

The LIRG NGC 6907 and the nearby dwarf galaxy NGC 6908 are interacting
systems classified as stage-B with a separation of 43\farcs9 (9.4~kpc).
\citet{Privon2020} report that NGC~6907 is detected with Swift/XRT 
but not with NuSTAR. The observed 2--10~keV luminosity
($\sim$7~$\times 10^{39}$~erg~s$^{-1}$) is even lower than the expected value
($\sim$8~$\times 10^{40}$~erg~s$^{-1}$) from the SFR, suggesting
that there is no significant AGN contribution to the X-ray flux.
We find that the combined NuSTAR and Swift/XRT spectra below 10~keV can
be well fitted with the starburst-dominant model ($\Gamma=1.8$ fixed). No X-ray
emission is detected from NGC 6908 in these observations.

\item IRAS 20264+2533 W/E (NGC 6921 and MCG+04-48-002): 

This system is a stage-A merger composed of NGC 6921 and the LIRG
MCG+04-48-002 with a separation of 91\farcs4 (26.5~kpc). 
We identify both NGC 6921 and MCG+04-48-002 
as the hard X-ray detected AGNs. The NuSTAR
and XMM-Newton studies \citep{Koss2016a,Ricci2017bMNRAS} suggest that
both nuclei have obscured or CT AGNs, well consistent with our results.
We find that NGC 6921 contains a CT AGN with $N_{\rm H}$ =
(1.7 $\pm$ 0.3) $\times 10^{24}$~cm$^{-2}$ and 
log$L_{2-10}$ = 42.80$^{+0.37}_{-0.44}$.
Whereas, MCG+04-48-002 hosts an obscured AGN, whose column density
varies from $N_{\rm H}$ = (6.0~$\pm$~0.5) $\times 10^{23}$~cm$^{-2}$
(XMM-Newton and Swift/BAT) to (7.3$^{+1.5}_{-0.8}$) $\times
10^{23}$~cm$^{-2}$ (NuSTAR). The absorption-corrected luminosity measured by NuSTAR
is log$L_{2-10}$ = 42.44 $\pm$ 0.15, showing
flux variability ($C^{\rm Time}$ = 3.55$^{\rm +0.67}_{-0.56}$ for
XMM-Newton and 5.46$^{\rm +0.65}_{-0.58}$ for Swift/BAT).

\item IRAS F20550+1655 W/E (II Zw 096 and IRAS F20550+1655 SE): 

The LIRG IRAS F20550+1655 (= CGCG 448-020) is a stage-C merger showing
two main galaxies with a separation of 11\farcs6 (8.1~kpc).
\citet{Zenner1993} identify four near-IR peaks, designated as the
galaxy A (= IRAS F20550+1655 SE or CGCG 448-020W), the galaxy B (= II Zw
096), and the sources C and D (= CGCG 448-020E), which are two prominent red
knots, as shown by Figure 2 in \citet{Goldader1997}. Source D radiates
$>$70\% of the 24~$\mu$m emission in the MIPS image
\citep{Inami2010}, which is the closest to the IRAS peak
position. \citet{Iwasawa2011} report that Chandra detects two compact
X-ray emitters from the galaxy A and the sources C+D. 
The sources C+D show a hard X-ray color consistent with an AGN, although
this may be due to absorption within the red knots. No hard X-rays above
8~keV are detected with NuSTAR. Using the soft X-ray spectra of NuSTAR,
Chandra, and XMM-Newton, we obtain $\Gamma \leq 1.53$ and $E_{\rm cut}$
= 12.2$^{+4.5}_{-2.7}$~keV for the galaxy IRAS F20550+1655 SE (including
sources C+D), indicating that the X-ray source is starburst-dominant.
The galaxy in II Zw 096 is detected only with Chandra and its spectrum
yields a photon index of $\Gamma \leq 2.02$.

\item IRAS F20551--4250 (ESO 286-19): 

This ULIRG is a stage-D merger with a single nucleus. Chandra detects
an X-ray source \citep{Ptak2003,Grimes2005,Iwasawa2011}, which 
is classified as an obscured AGN with XMM-Newton
\citep{Franceschini2003}. NuSTAR detects the 8--24~keV emission 
at a low significance level (2.0$\sigma$),  
and also an ionized or neutral Fe~K$\alpha$ line ($E_{\rm Fe} =
6.56^{+0.12}_{-0.32}$~keV and $A_{\rm Fe} = 0.45^{+0.41}_{-0.31} \times
10^{-6}$). The 3$\sigma$ upper limit of the AGN luminosity is log$L^{\rm
int(lim)}_{2-10} < 43.38$ assuming log$N_{\rm H} \leq 24.5$, and thus
the possibility of an obscured AGN remains. Applying the starburst-dominant
model to the NuSTAR, Chandra, and XMM-Newton spectra, we
obtain a photon index of $\Gamma$ = 1.64$^{+0.11}_{-0.12}$.

\item IRAS F21453--3511 (NGC 7130): 

The LIRG NGC 7130 is a possible past merger \citep{Davies2014},
classified as a stage-N source in this paper. 
We identify the galaxy as the hard X-ray detected AGN.
The Chandra spectrum shows
a hard excess and a strong Fe K$\alpha$ line at 6.4~keV, suggesting an
obscured or CT AGN \citep[e.g.,][]{Levenson2005,Torres-Alba2018}. With a
combined analysis of the NuSTAR and Chandra spectra,
\citet{Ricci2017bMNRAS} and \citet{Pozzi2017} confirm the presence of CT
obscuration. Adding the Suzaku and Swift/BAT spectra to the NuSTAR and
Chandra ones, we estimate the column density and absorption-corrected luminosity to
be $N_{\rm H}$ = (4.1~$\pm$~0.1) $\times 10^{24}$~cm$^{-2}$ and
log$L_{2-10}$ = 42.87$^{+0.31}_{-0.25}$, respectively.

\item IRAS F23007+0836 S/N (NGC 7469 and IC 5283): 

The system is a stage-A merger with the two galaxies NGC 7469 and IC
5283 separated by
79\farcs7 (26.2~kpc). NGC 7469 is a LIRG, whereas IC 5283 has a smaller
IR luminosity of log$L_{\rm IR}$/$L_{\odot}$ = 10.79
\citep{Ricci2017bMNRAS}. NGC 7469 hosts a bright unobscured AGN
detected by Chandra/High Energy Transmission Grating Spectrometer
\citep{Scott2005}, XMM-Newton \citep{Blustin2003}, Suzaku
\citep{Walton2013,Iso2016}, and NuSTAR
\citep[e.g.,][]{Ricci2017bMNRAS,Middei2018,Ogawa2019}. Recently,
\citet{Ogawa2021} perform an spectral analysis
with XCLUMPY, using the NuSTAR and XMM-Newton spectra.
Their results are in good agreement
with the previous works. In this paper, we refer to \citet{Ogawa2021} 
for the column density 
($<$1.5 $\times 10^{19}$~cm$^{-2}$) and absorption-corrected luminosity
(log$L_{2-10}$ = 43.26$^{+0.03}_{-0.02}$).
No significant X-ray emission from IC 5283 is detected with XMM-Newton.

\item IRAS F23128--5919 (ESO 148-2): 

This ULIRG is a stage-C merger with two galaxies separated by 4\farcs5
(3.9~kpc; \citealt{Zenner1993}). 
We identify this pair (ESO 148-2) as the hard X-ray detected AGN.
Previous studies using Chandra
\citep{Ptak2003,Grimes2005,Iwasawa2011} and XMM-Newton
\citep{Franceschini2003} demonstrate that the southern nucleus hosts an
obscured AGN. NuSTAR detects the 8--24~keV emission (3.1$\sigma$) and
variable spectral features between the NuSTAR and other soft
X-ray observations, indicating the presence of an AGN. Adopting the
XCLUMPY model, we find that the column density increases dramatically from
a Compton-thin level with $N_{\rm H}$ = $(2.7^{+0.9}_{-0.6}) \times
10^{22}$~cm$^{-2}$ (Chandra and XMM-Newton) 
to a CT level with $N_{\rm H}$ =
$(1.6 \pm 0.5) \times 10^{24}$~cm$^{-2}$ (NuSTAR), as reported in
Section~\ref{subsub6-1-5_NHvariability}. The estimated absorption-corrected AGN luminosity is log$L_{2-10}$ = 42.38$^{+0.24}_{-0.28}$.

\item IRAS F23157+0618 (NGC 7591): 

This LIRG is a stage-N object. The X-ray emission is not detected with 
Chandra \citep{Torres-Alba2018} or NuSTAR \citep{Privon2020}. 
The low soft X-ray luminosity indicates that it is a starburst-dominant source. Even though we cannot rule out the possibility of a CT AGN, the
3$\sigma$ upper limit of the absorption-corrected AGN luminosity is small
(log$L^{\rm (lim)}_{2-10} < 42.51$ assuming log$N_{\rm H} \leq
24.5$).

\item IRAS F23254+0830 W/E (NGC 7674 and MCG+01-59-081): 

IRAS F23254+0830 has two galaxies in a stage-A merger, separated by
33\farcs3 (19.5~kpc).  NGC 7674 is a LIRG, while MCG+01-59-081 (=
NGC 7674A) has a lower IR luminosity (log$L_{\rm IR}$/$L_{\odot}$
= 10.01; \citealt{Ricci2017bMNRAS}). 
We identify NGC 7674 as the hard X-ray detected AGN.
\citet{Bianchi2005} report a decline
of the 2--10~keV flux between the Ginga (in 1989) and BeppoSAX
observations (in 1996), which may be the reason for the large
bolometric-to-X-ray correction factor, as noted in Section~\ref{subsub6-1-1_Xweak}.
\citet{Gandhi2017} analyze the broadband X-ray spectra of NuSTAR,
Suzaku/XIS, and Swift/XRT, and indicate the presence of a
reflection-dominated CT AGN ($N_{\rm H} \gtrsim 3 \times
10^{24}$~cm$^{-2}$) assuming three geometries of the nuclear obscurer
and reflector. 
By contrast, \citet{Tanimoto2020} report that 
the XCLUMPY model can also reproduce 
the spectra with a smaller column density of $N_{\rm H} \sim 2 \times
10^{23}$~cm$^{-2}$. 
Using the NuSTAR, XMM-Newton, Suzaku/XIS and HXD, and Swift/BAT spectra,
we find that the column density is variable, $N_{\rm
H}$ = (1.4$^{+0.5}_{-0.4}$) $\times 10^{23}$~cm$^{-2}$ (XMM-Newton in
2004), (1.0$^{+0.3}_{-0.2}$) $\times 10^{23}$~cm$^{-2}$ (Suzaku in 2013), and
(3.0$^{+0.3}_{-0.2}$) $\times 10^{23}$~cm$^{-2}$ (NuSTAR in 2014 and
Swift/BAT). The estimated absorption-corrected X-ray luminosity is log$L_{2-10}$ = 42.59$^{+0.33}_{-0.23}$.  No X-ray emission is detected
from MCG+01-59-081 in these observations.
\\

\item IRAS 23262+0314 W/E (NGC 7679 and NGC 7682):

The system is a stage-A merger with the two galaxies NGC 7679 and NGC
7682 separated by 269\farcs8 (93.8~kpc). While NGC 7679 is a LIRG, NGC
7682 is less luminous in the IR band. 
We identify NGC 7679 and NGC 7682 as the hard X-ray detected AGNs.
The BeppoSAX observation in 1998
\citep{Risaliti2002,Dadina2007} suggests that NGC 7679 has an unobscured
AGN with a 2--10~keV luminosity of $3.4 \times 10^{42}$~erg~s$^{-1}$.
Analyzing the XMM-Newton data taken in 2005, \citet{Ricci2017bMNRAS}
find a flux decline by a factor of $\sim$10 ($4\times 10^{41}$~erg~s$^{-1}$). 
Furthermore, using the NuSTAR and Swift/XRT data in 2017 and the
time averaged (2005--2013) Swift/BAT data, we detect a re-brightening of the
unobscured AGN ($N_{\rm H}$~$<$~3 $\times 10^{15}$~cm$^{-2}$) with a
larger absorption-corrected luminosity (log$L_{2-10} = 42.34^{+0.06}_{-0.04}$) compared with that in the XMM-Newton
observation ($C^{\rm Time}$ = 0.16 $\pm$ 0.01). Whereas, NGC 7682 hosts
a CT AGN revealed by XMM-Newton \citep{Singh2011,Ricci2017bMNRAS} and
Swift/BAT \citep{Ricci2015}. Analyzing the NuSTAR and XMM-Newton spectra
with XCLUMPY, however, we obtain a smaller column density of $N_{\rm H}
= (3.8 \pm 0.7) \times 10^{23}$~cm$^{-2}$ with an X-ray
luminosity of log$L_{2-10}$ = 41.94 $\pm$ 0.06.
\\

\end{enumerate}

\restartappendixnumbering
\section{X-ray Spectra and Best-fit Models} \label{Appendix-C}

\begin{figure*}
    \epsscale{1.15}
    \plottwo{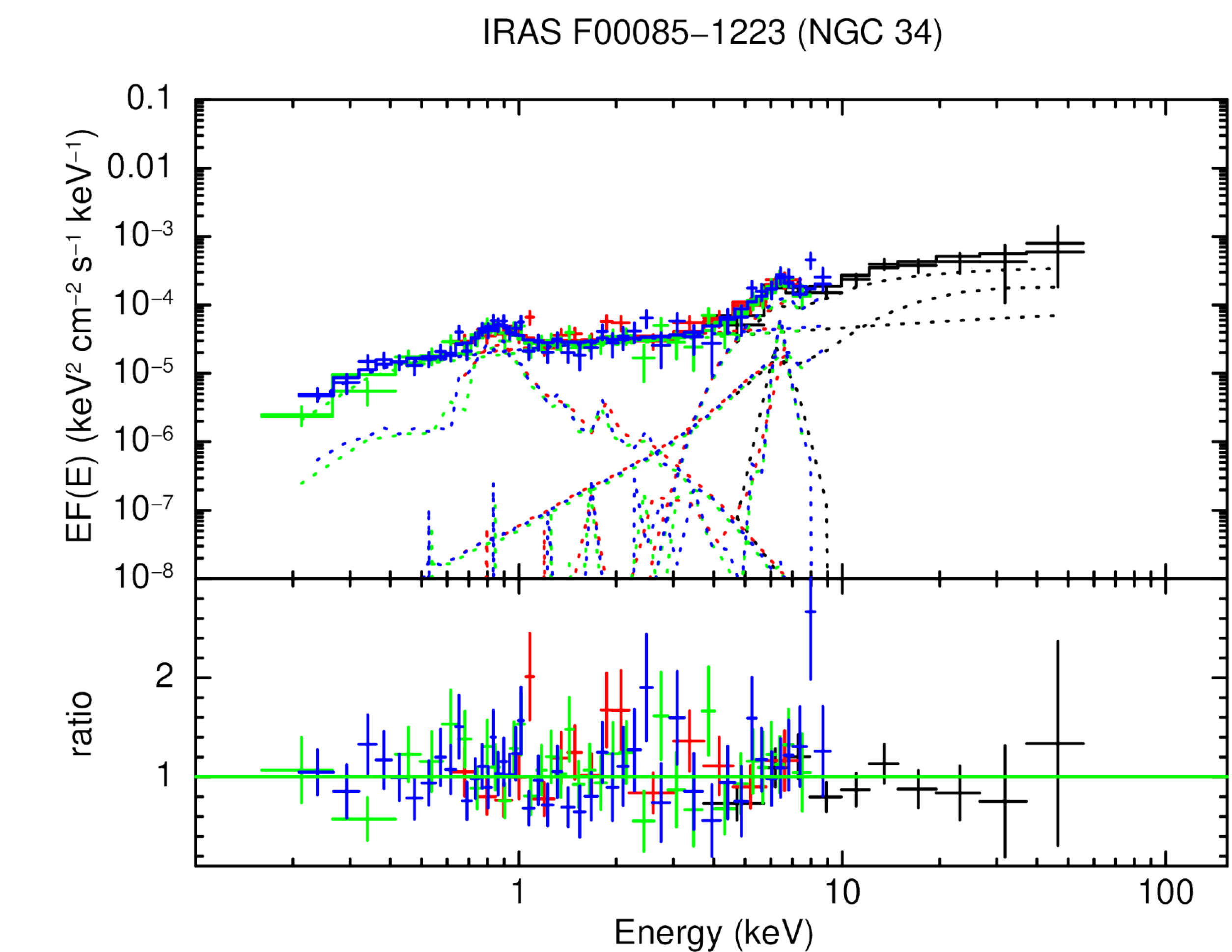}{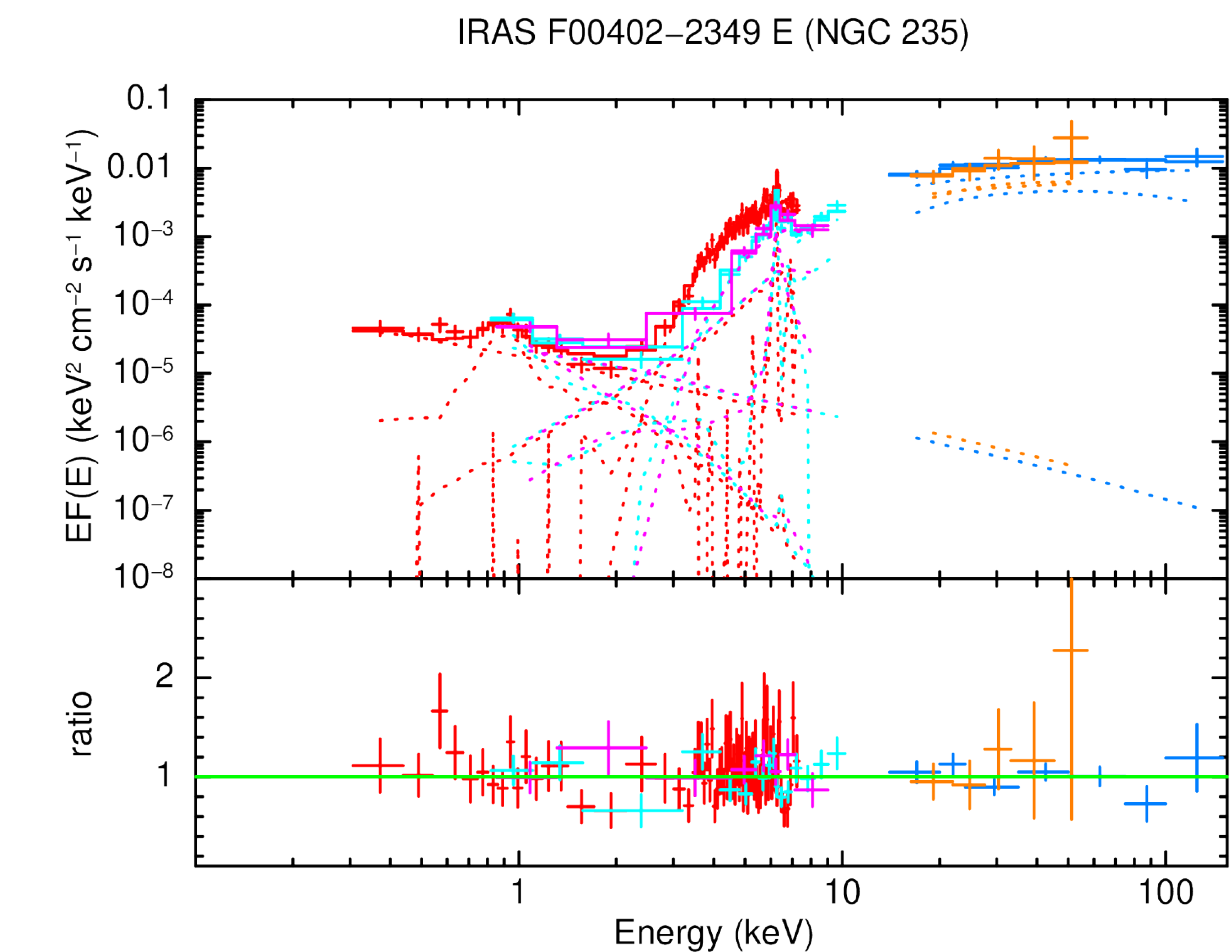}
    \plottwo{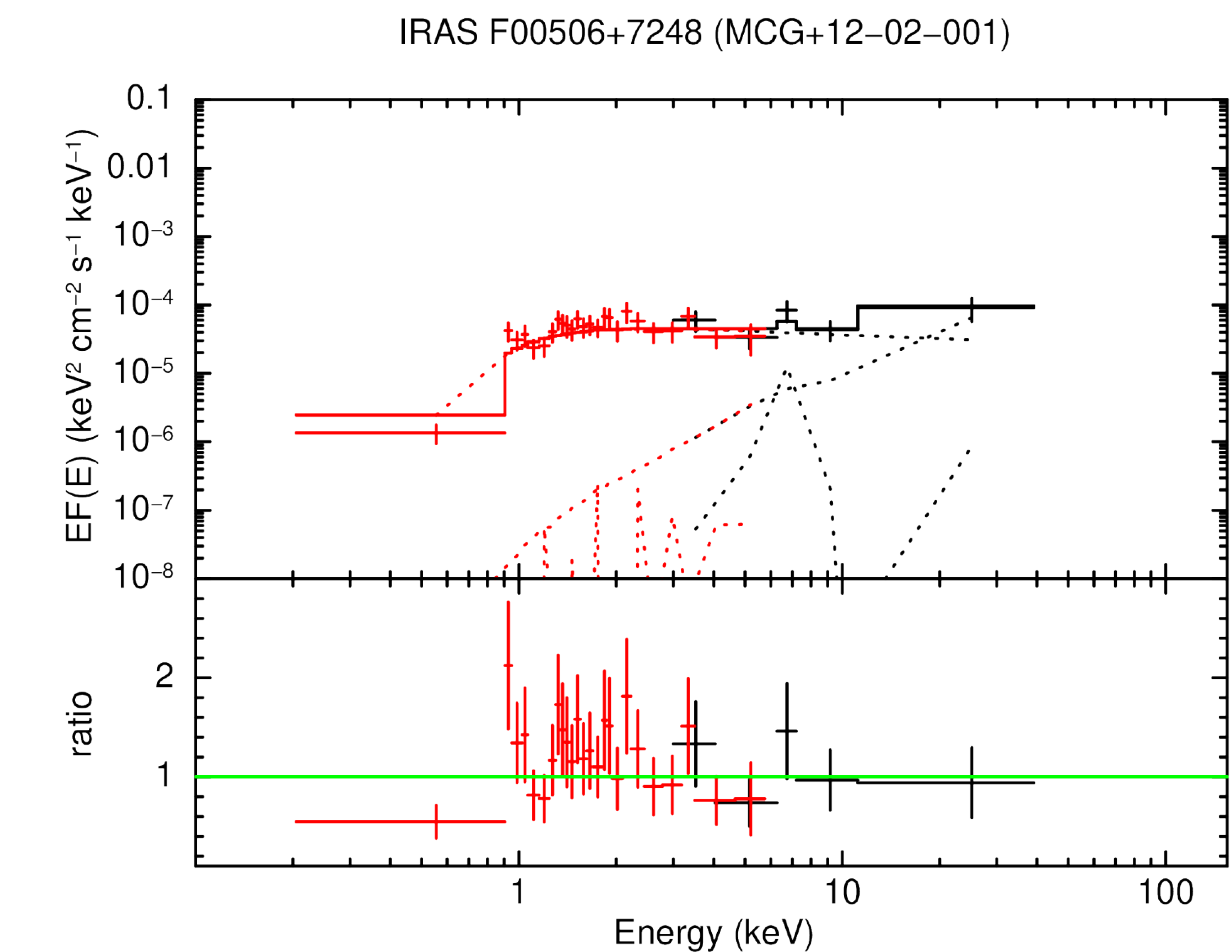}{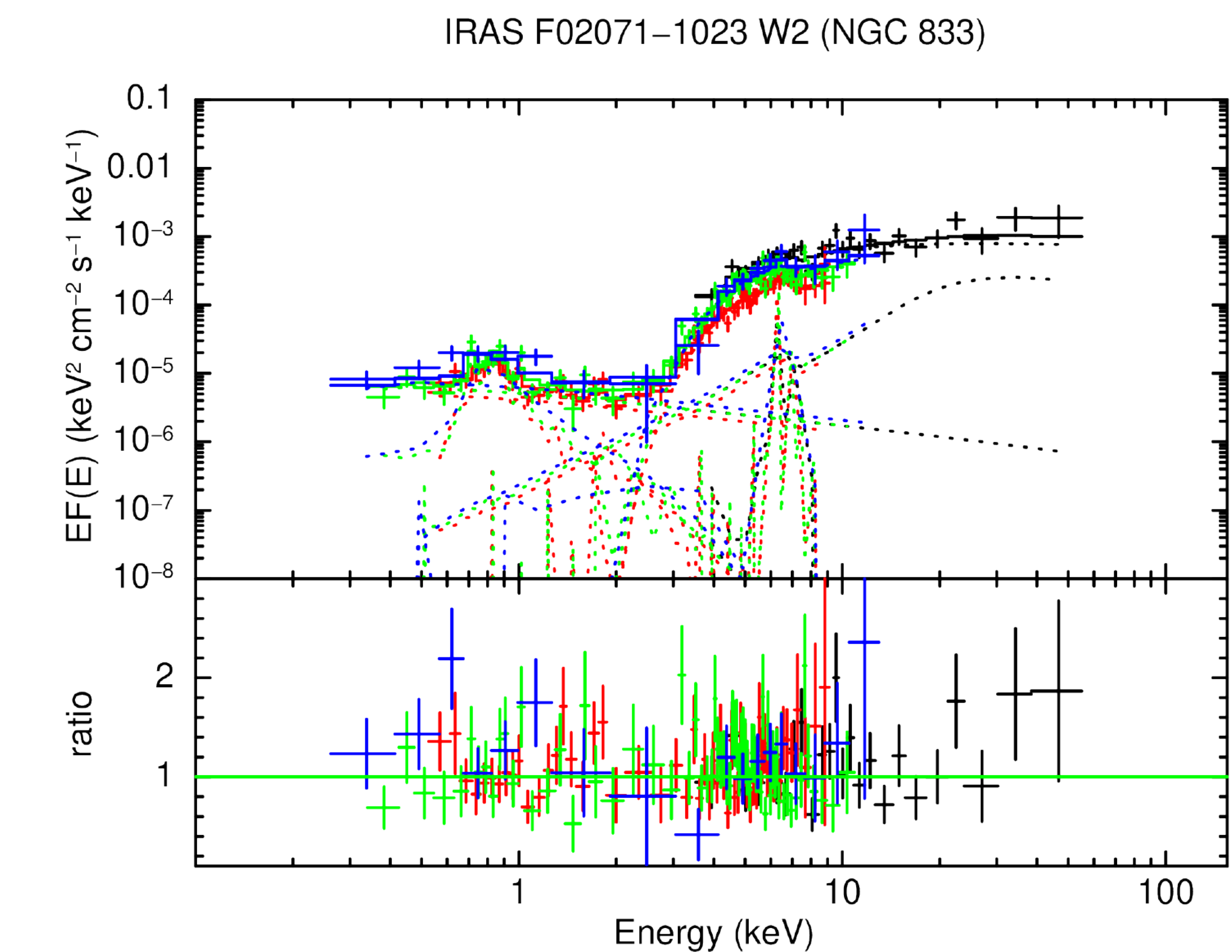}
    \plottwo{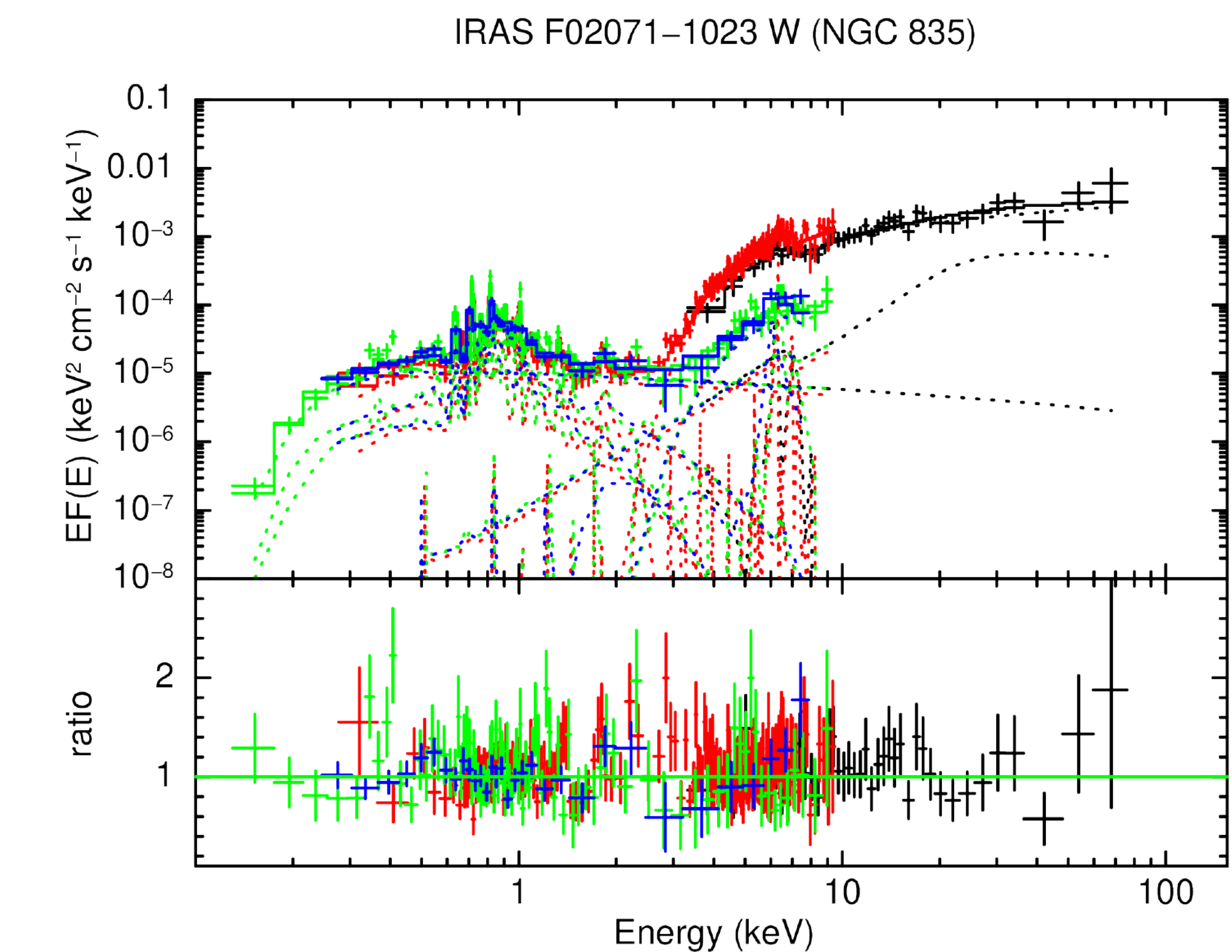}{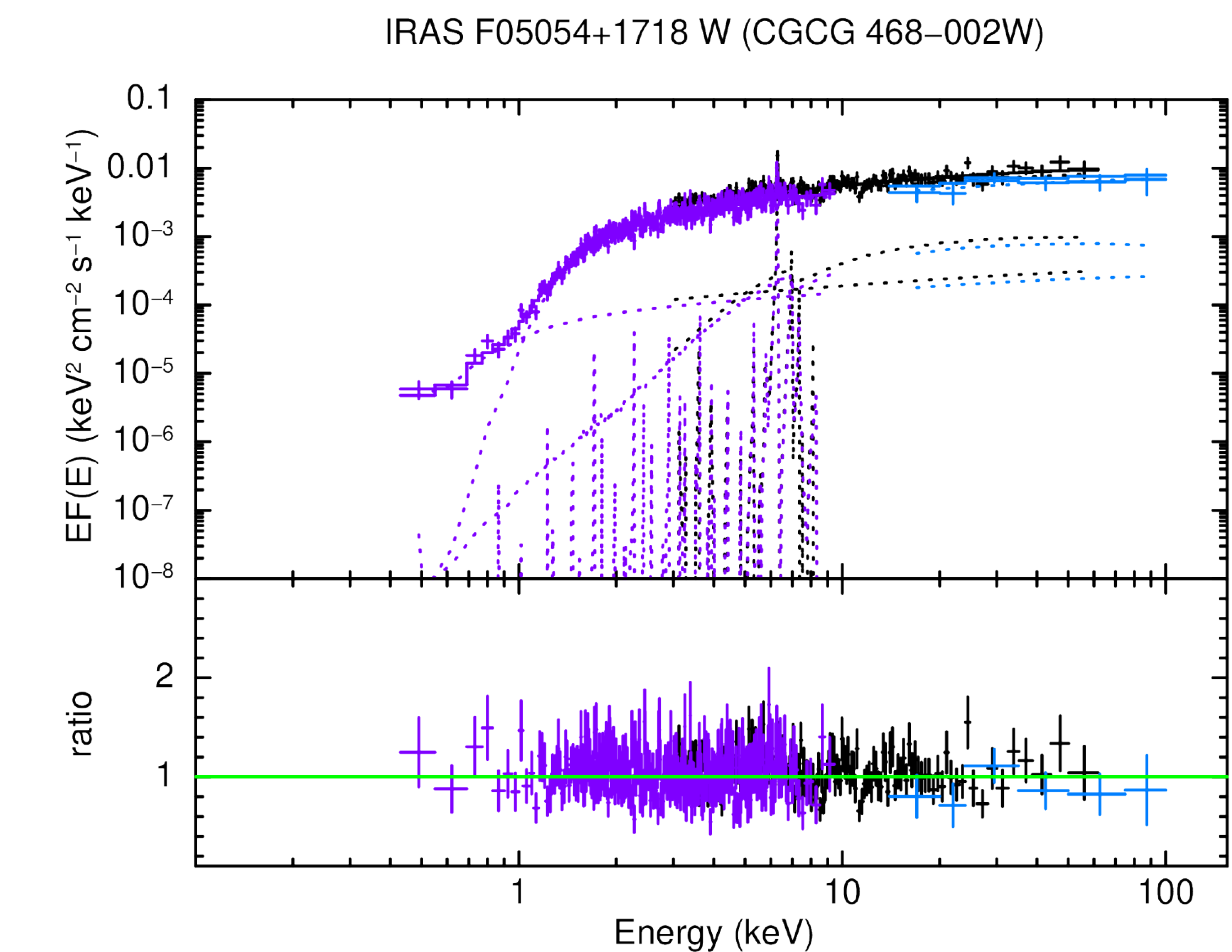}
        \caption{X-ray spectra, best-fit models, and ratios between the data and models of the AGNs in
        NGC~34 
        [$N_{\rm H}$ = (5.0~$\pm$~1.0) $\times 10^{23}$~cm$^{-2}$],
        NGC~235 
        [$N_{\rm H}$ = (5.1$^{+0.8}_{-0.6}$) $\times 10^{23}$~cm$^{-2}$; 
        and (2.9~$\pm$~0.2) $\times 10^{23}$~cm$^{-2}$ for Chandra],
        MCG+12-02-001 
        [$N_{\rm H}$ = ($>$4.3 $\times 10^{24}$~cm$^{-2}$],
        NGC~833 
        [$N_{\rm H}$ = (2.6~$\pm$~0.1) $\times 10^{23}$~cm$^{-2}$],
        NGC~835 
        [$N_{\rm H}$ = (3.0$^{+0.1}_{-0.2}$) $\times 10^{23}$~cm$^{-2}$],
        and CGCG~468-002W 
        [$N_{\rm H}$ = (1.5~$\pm$~0.1) $\times 10^{22}$~cm$^{-2}$].}
\label{C1-F}
\end{figure*}

\begin{figure*}
    \epsscale{1.15}
    \plottwo{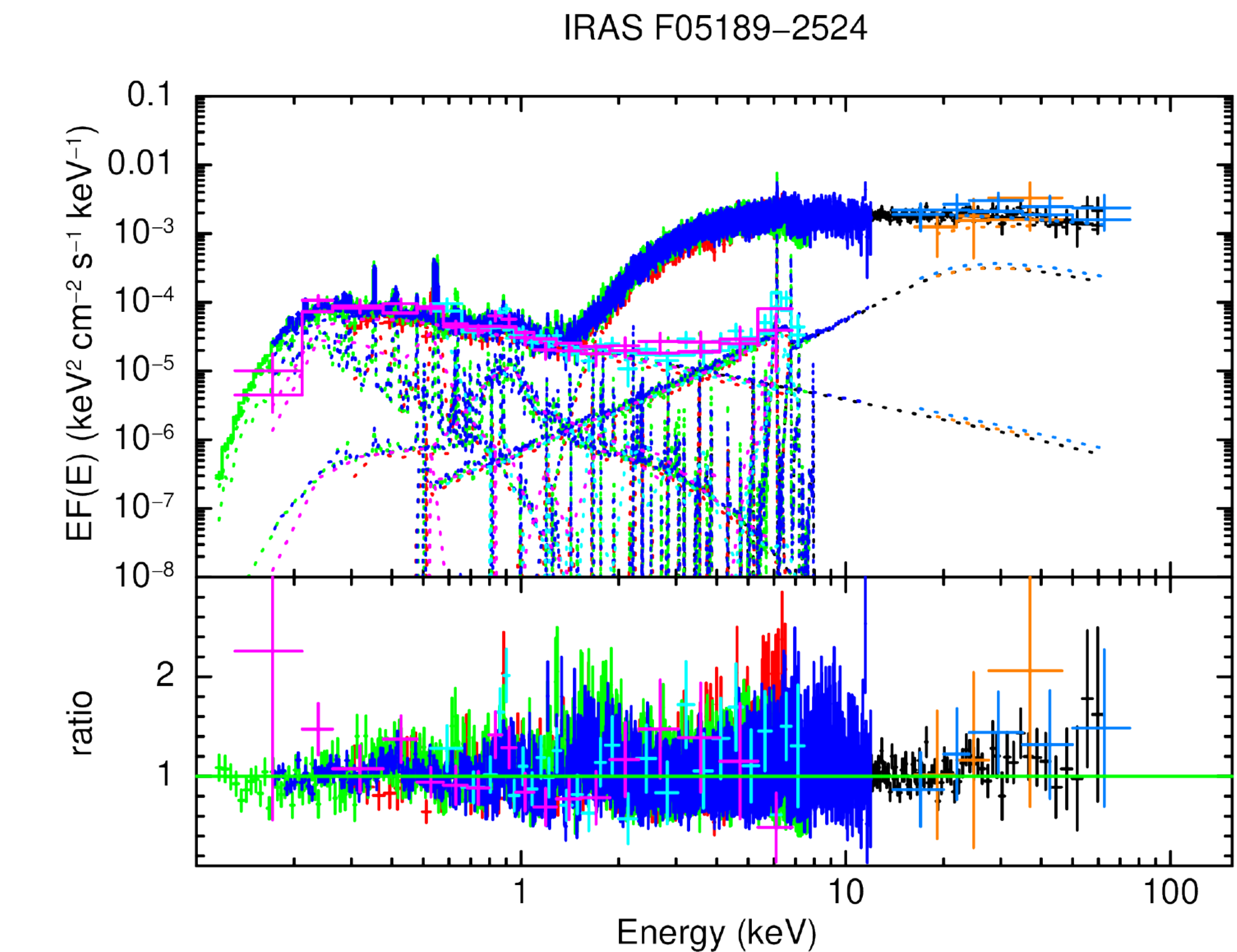}{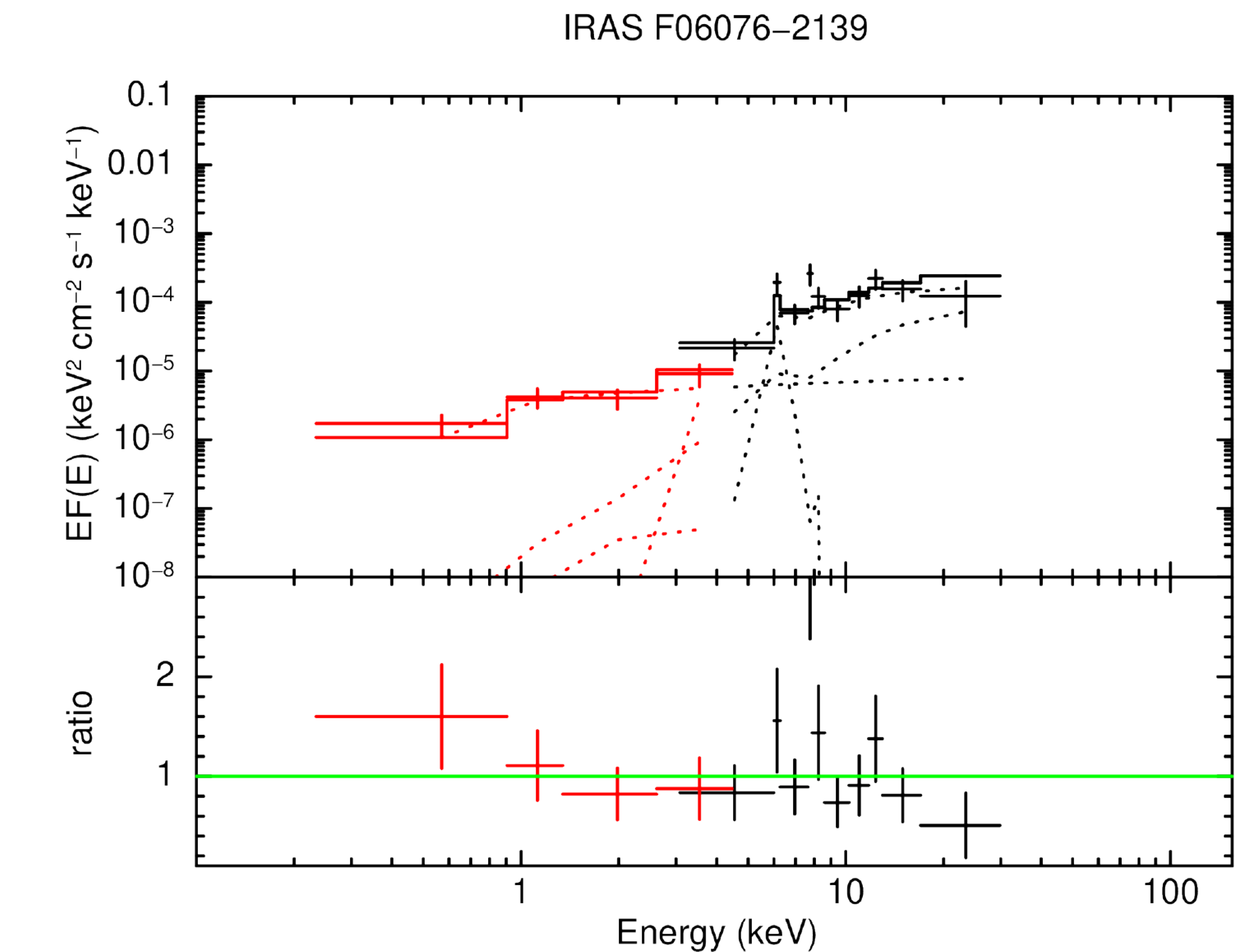}
    \plottwo{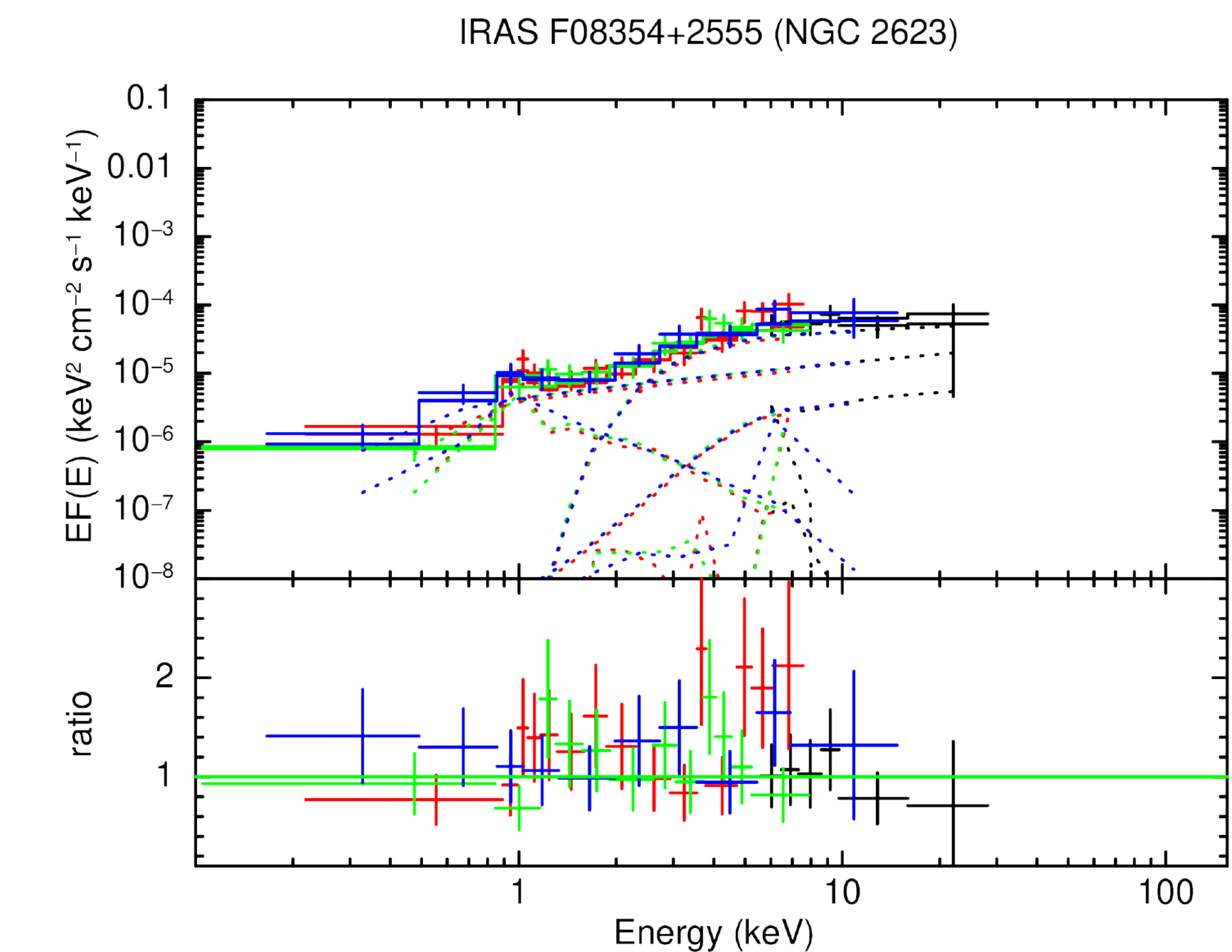}{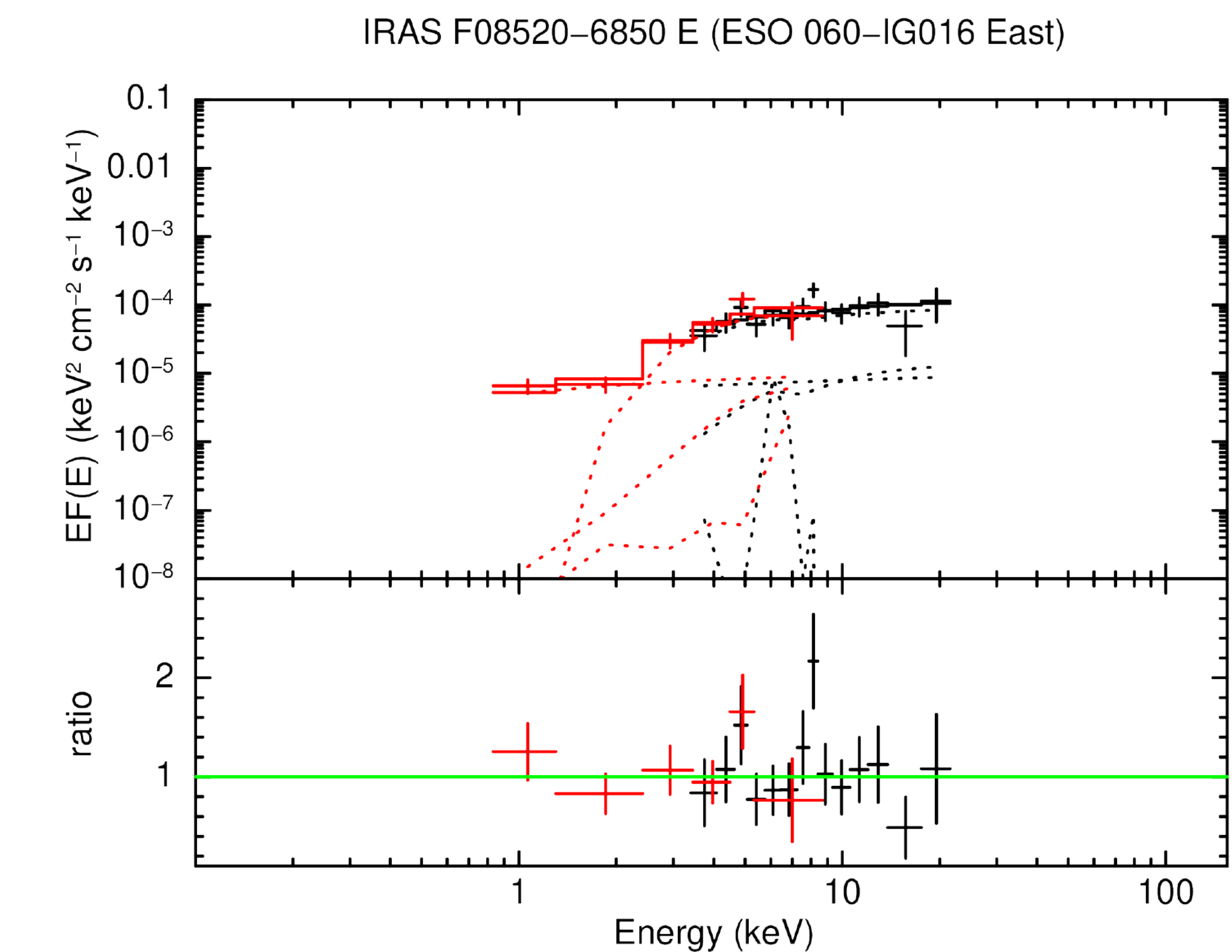}
    \plottwo{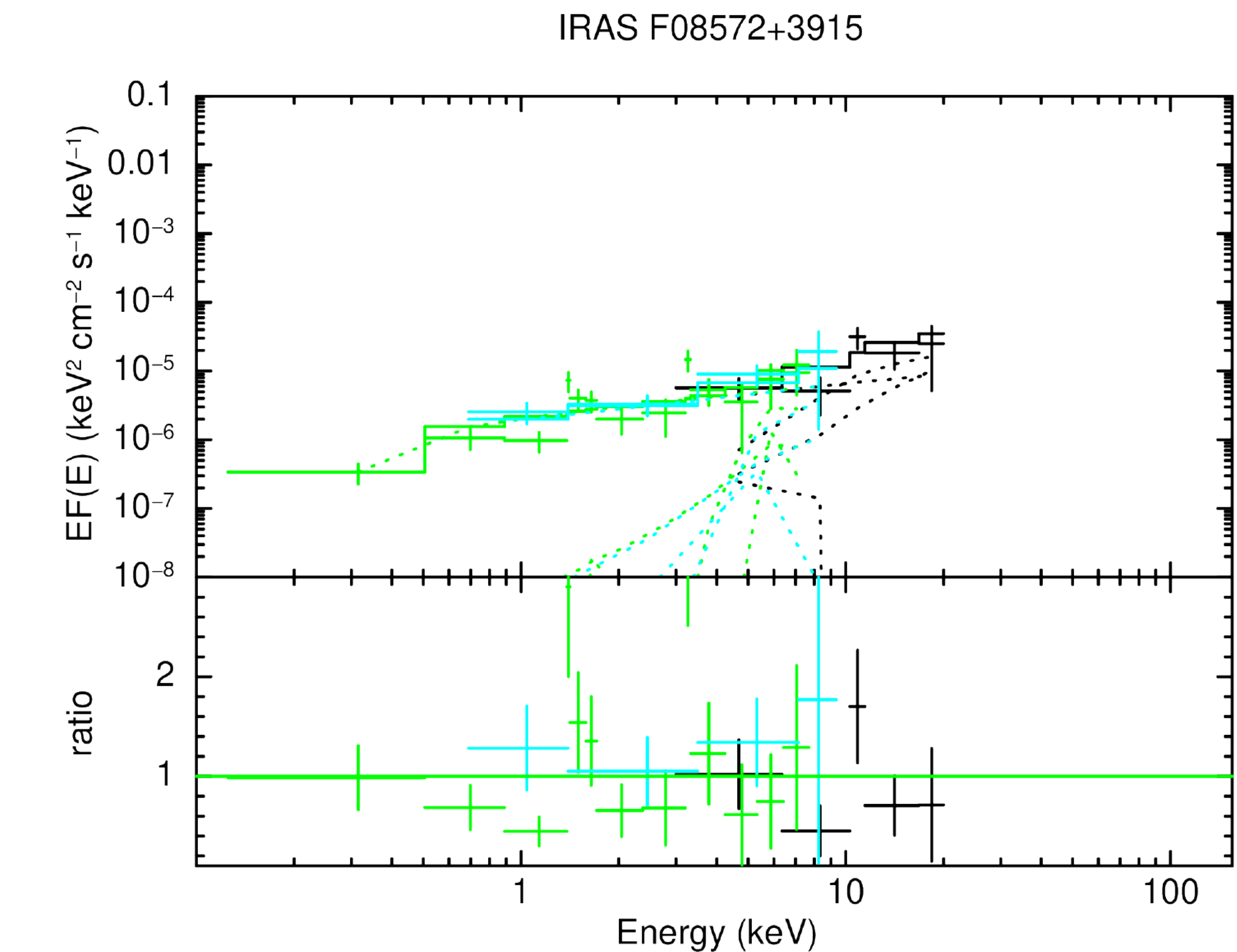}{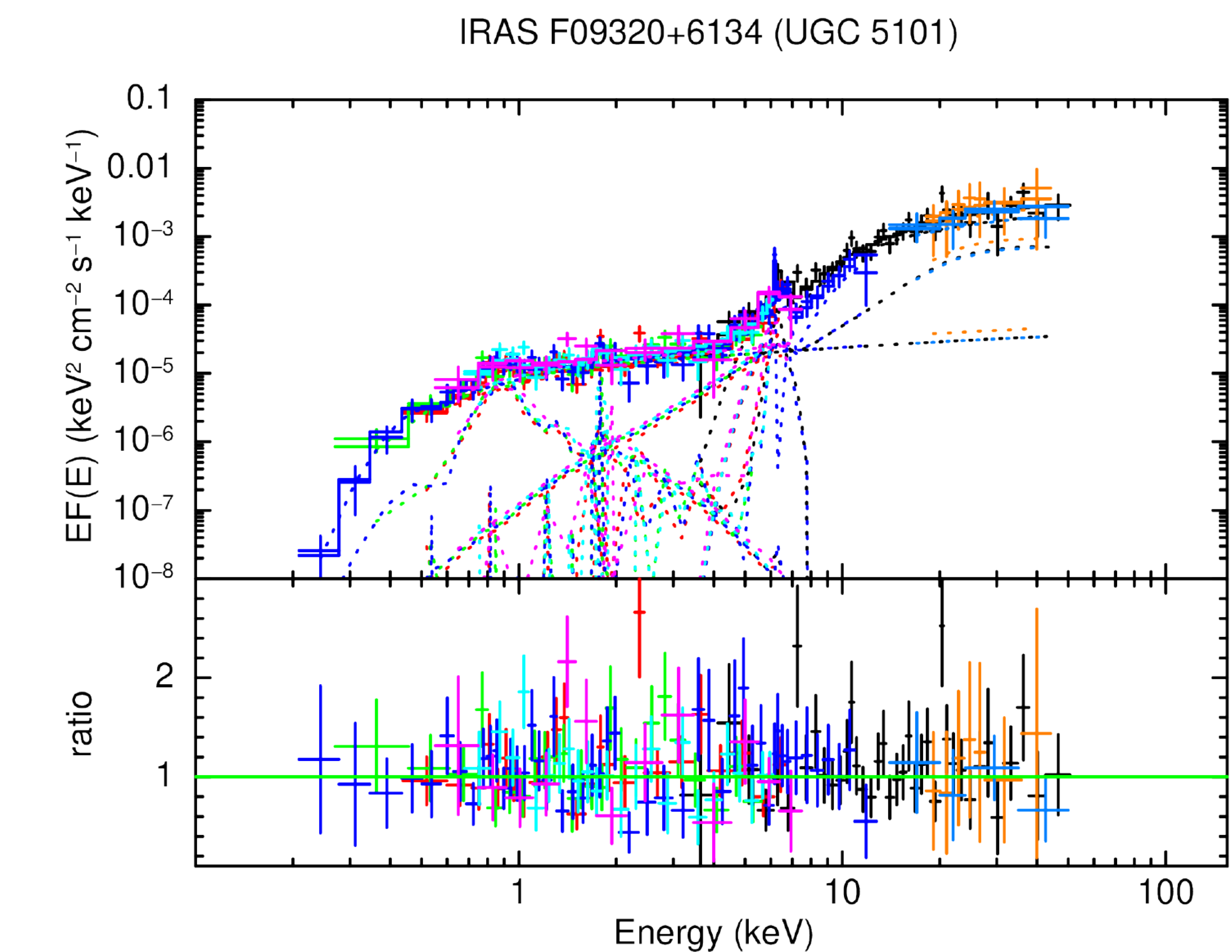}
        \caption{X-ray spectra, best-fit models, and ratios between the data and models of the AGNs in
        IRAS F05189--2524 
        [$N_{\rm H}$ = (7.5~$\pm$~0.1) $\times 10^{22}$~cm$^{-2}$; 
        and $>$2.3 $\times 10^{24}$~cm$^{-2}$ for Suzaku],
        IRAS F06076--2139 
        [$N_{\rm H}$ = (4.2$^{+2.4}_{-1.2}$) $\times 10^{23}$~cm$^{-2}$],
        NGC~2623 
        [$N_{\rm H}$ = (6.0$^{+4.5}_{-2.1}$) $\times 10^{22}$~cm$^{-2}$],
        ESO~060-IG016 East 
        [$N_{\rm H}$ = (8.4$^{+4.0}_{-2.9}$) $\times 10^{22}$~cm$^{-2}$],
        IRAS F08572+3915 
        [$N_{\rm H}$ = (8.5$^{+12.9}_{-2.8}$) $\times 10^{23}$~cm$^{-2}$],
        and UGC~5101
        [$N_{\rm H}$ = (9.6$^{+0.4}_{-0.2}$) $\times 10^{23}$~cm$^{-2}$; 
        and (1.5$^{+0.3}_{-0.2}$) $\times 10^{24}$~cm$^{-2}$ for Chandra/XMM-Newton/Suzaku].}
\label{C2-F}
\end{figure*}

\begin{figure*}
    \epsscale{1.15}
    \plottwo{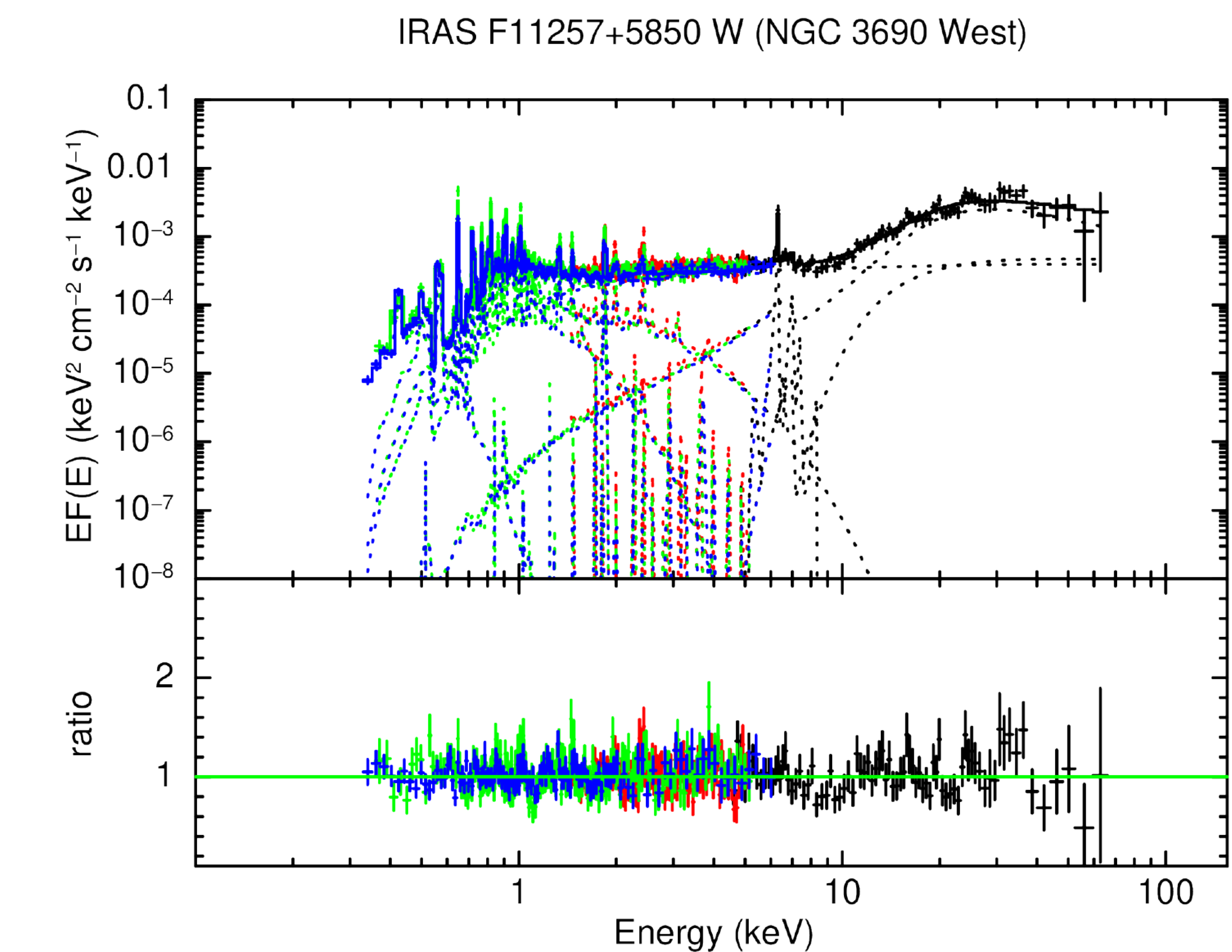}{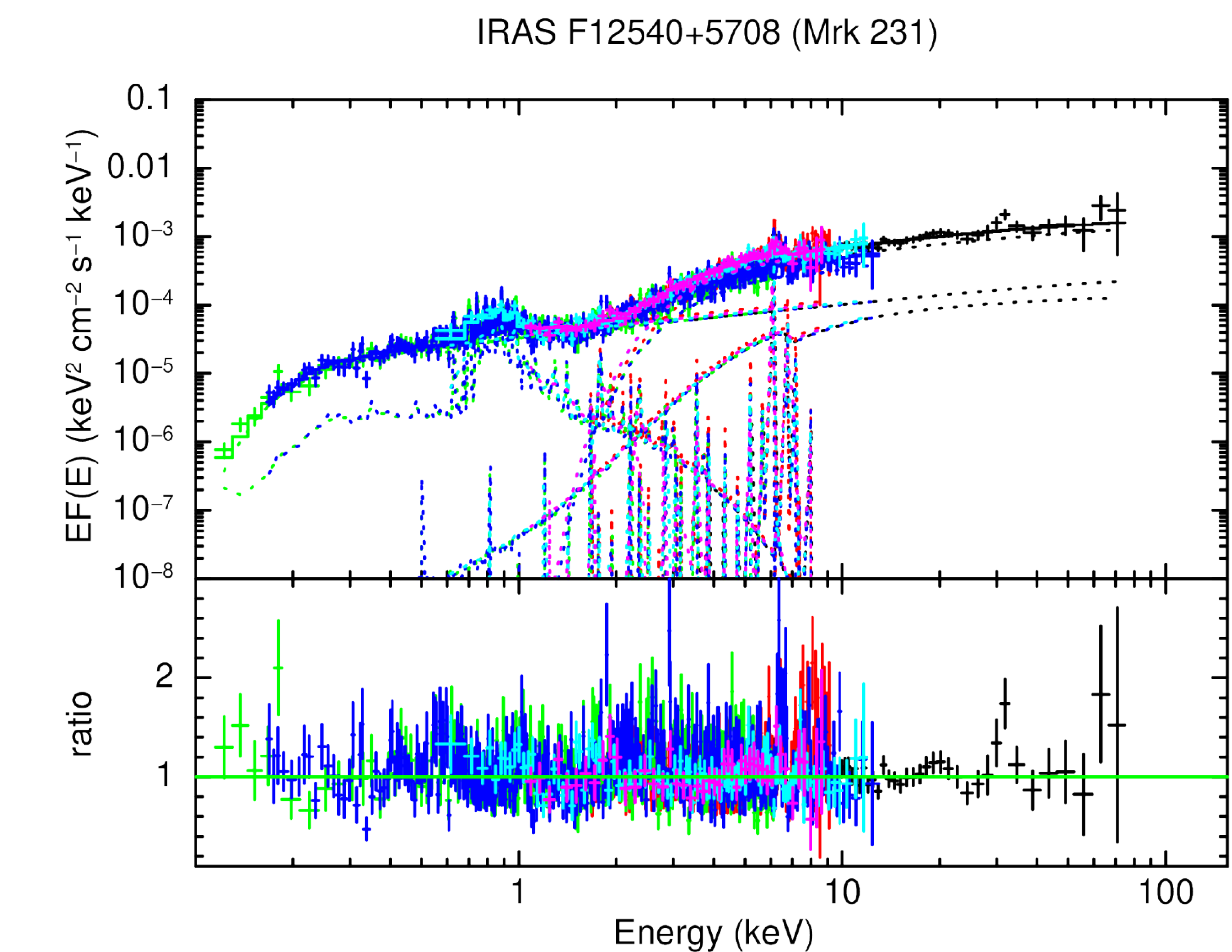}
    \plottwo{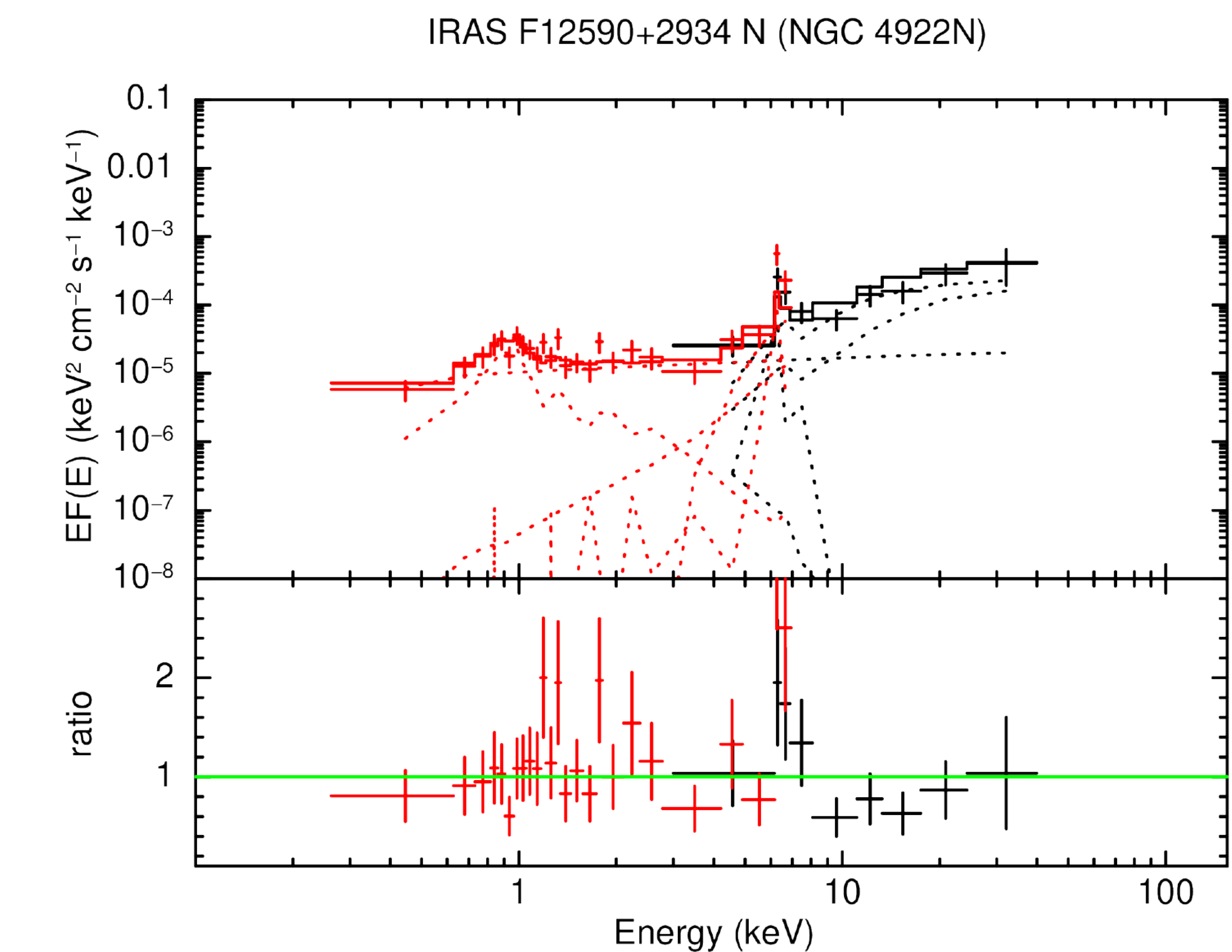}{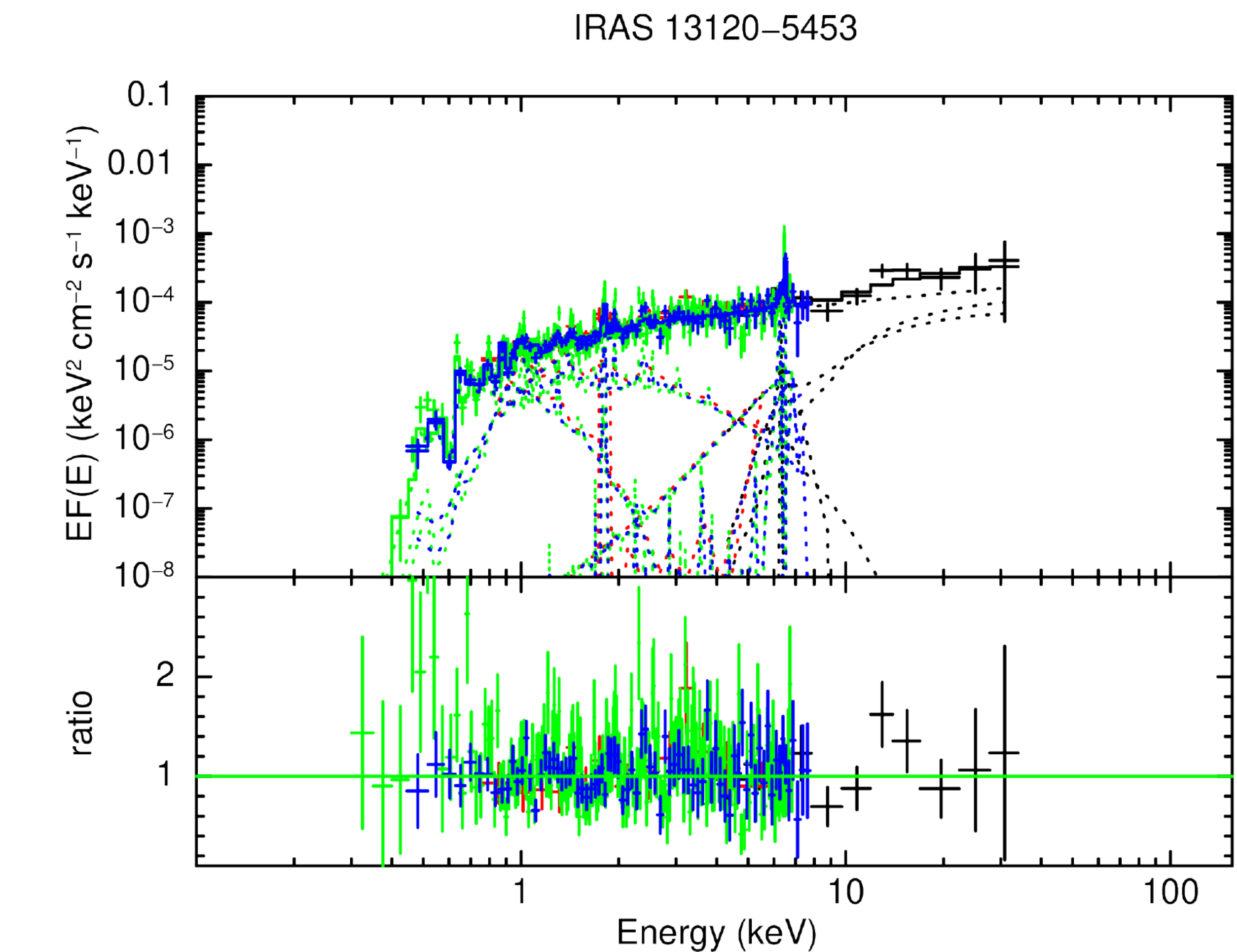}
    \plottwo{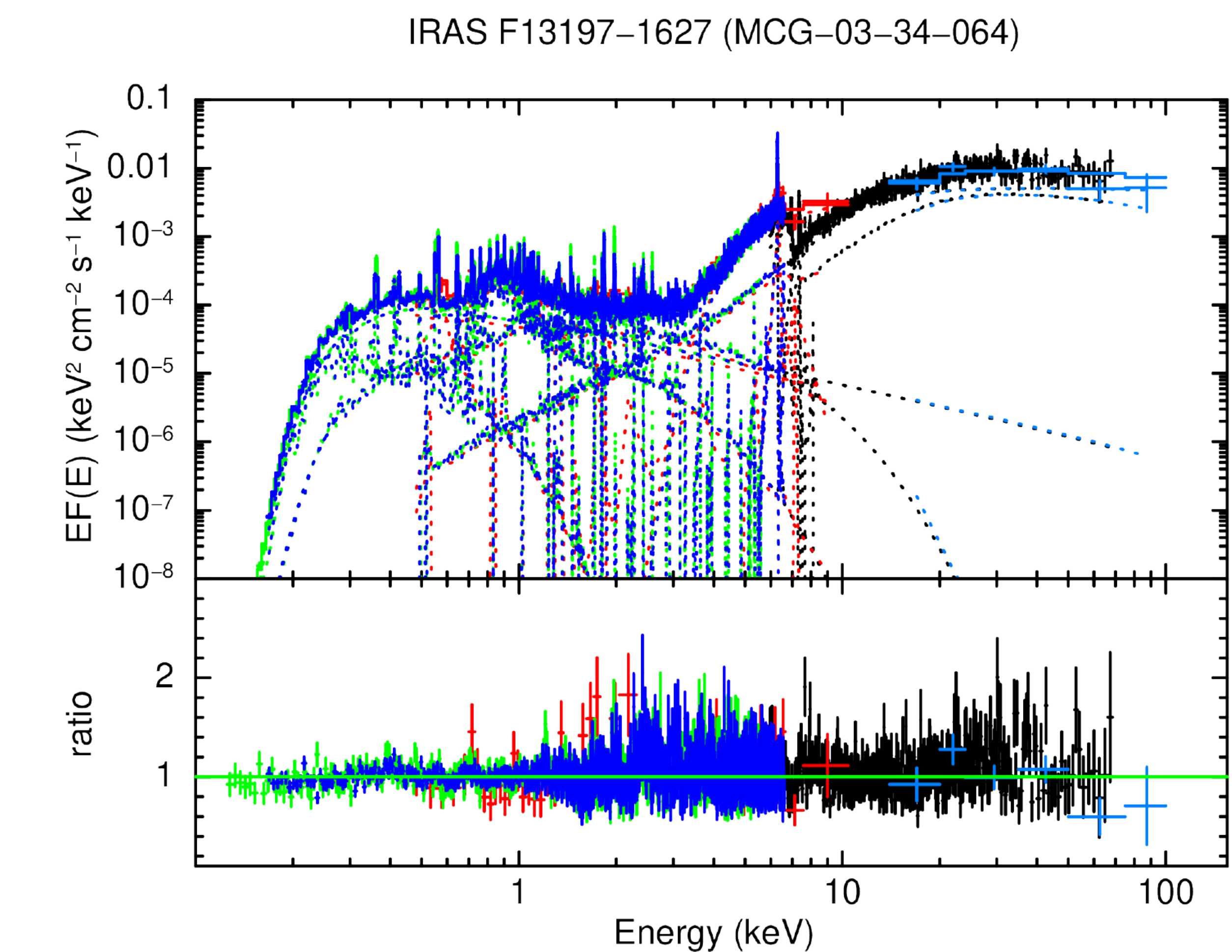}{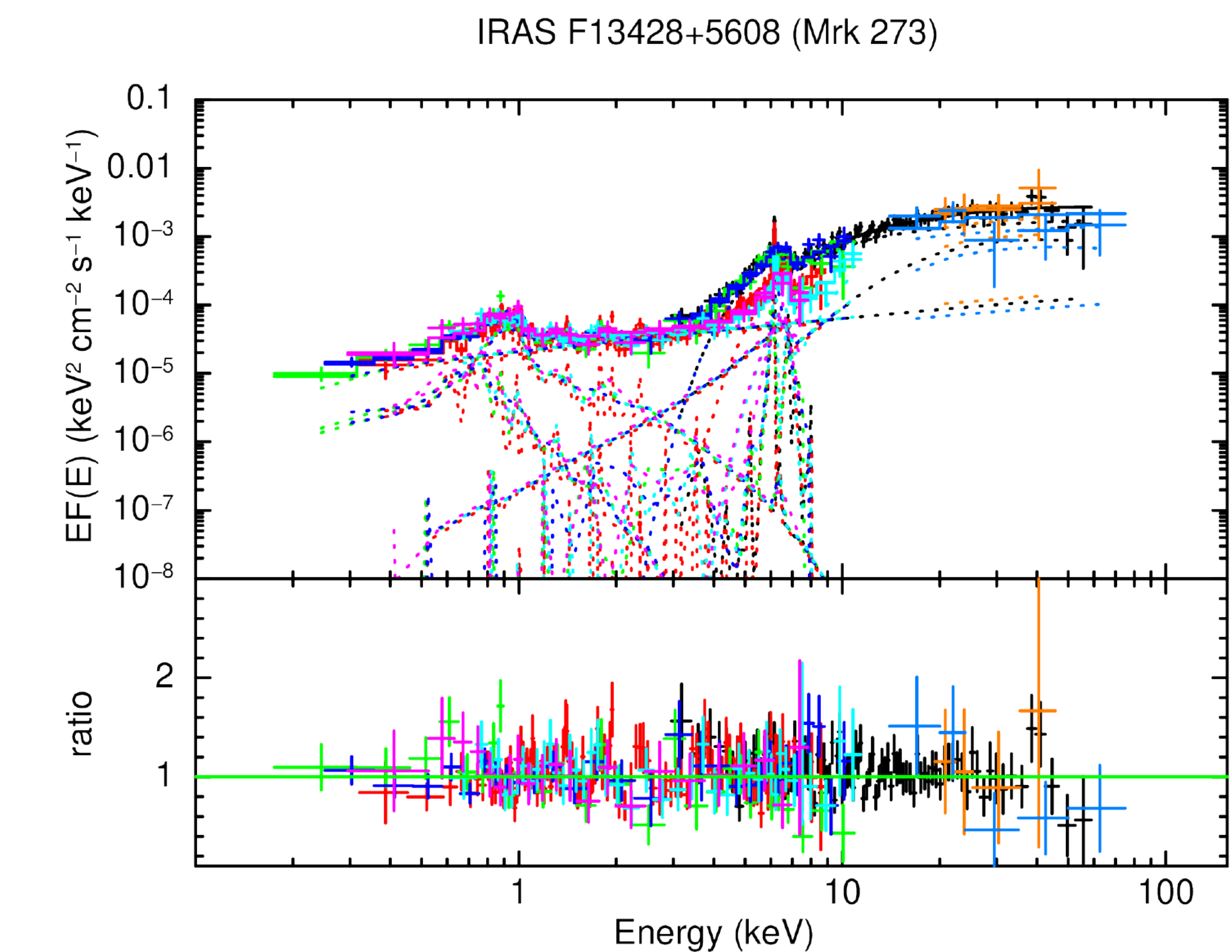}
        \caption{X-ray spectra, best-fit models, and ratios between the data and models of the AGNs in
        NGC~3690 West 
        [$N_{\rm H}$ = (3.0$^{+0.6}_{-0.5}$) $\times 10^{24}$~cm$^{-2}$],
        Mrk~231 
        [$N_{\rm H}$ = (8.5~$\pm$~0.2) $\times 10^{22}$~cm$^{-2}$],
        NGC~4922N 
        [$N_{\rm H}$ = (7.6$^{+28.5}_{-2.2}$) $\times 10^{23}$~cm$^{-2}$],
        IRAS 13120--5453 
        [$N_{\rm H}$ = (1.6~$\pm$~0.2) $\times 10^{24}$~cm$^{-2}$],
        MCG--03-34-064 
        [$N_{\rm H}$ = (9.8$^{+0.3}_{-0.2}$) $\times 10^{23}$~cm$^{-2}$; 
        and (5.6~$\pm$~0.2) $\times 10^{23}$~cm$^{-2}$ for Chandra/XMM-Newton],
        and Mrk~273 
        [$N_{\rm H}$ = (5.0$^{+0.6}_{-0.3}$) $\times 10^{23}$~cm$^{-2}$; 
        (9.3$^{+1.1}_{-1.0}$) $\times 10^{23}$~cm$^{-2}$ for Chandra; 
        and (1.7$^{+0.5}_{-0.3}$) $\times 10^{24}$~cm$^{-2}$ for Suzaku].}
\label{C3-F}
\end{figure*}

\begin{figure*}
    \epsscale{1.15}
    \plottwo{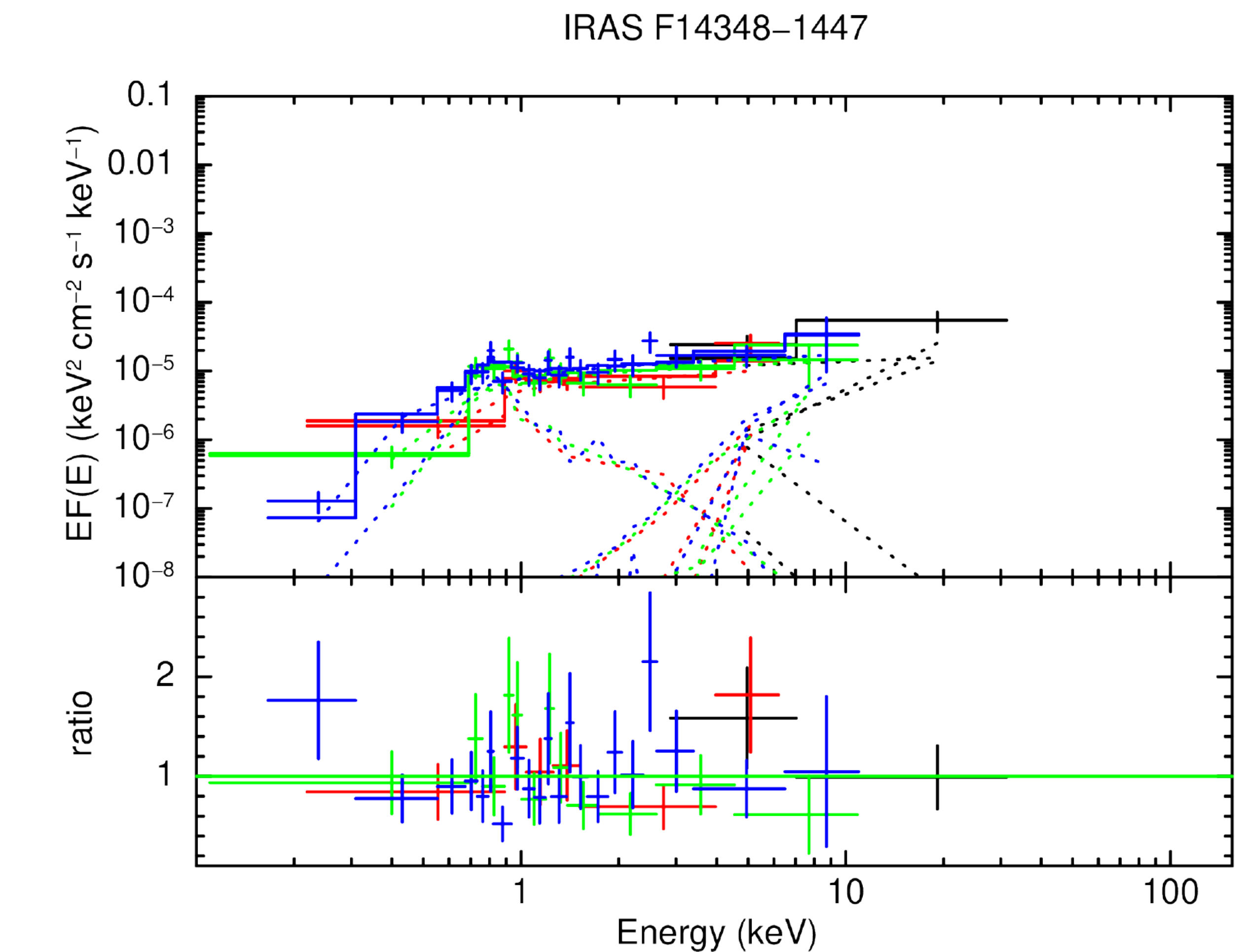}{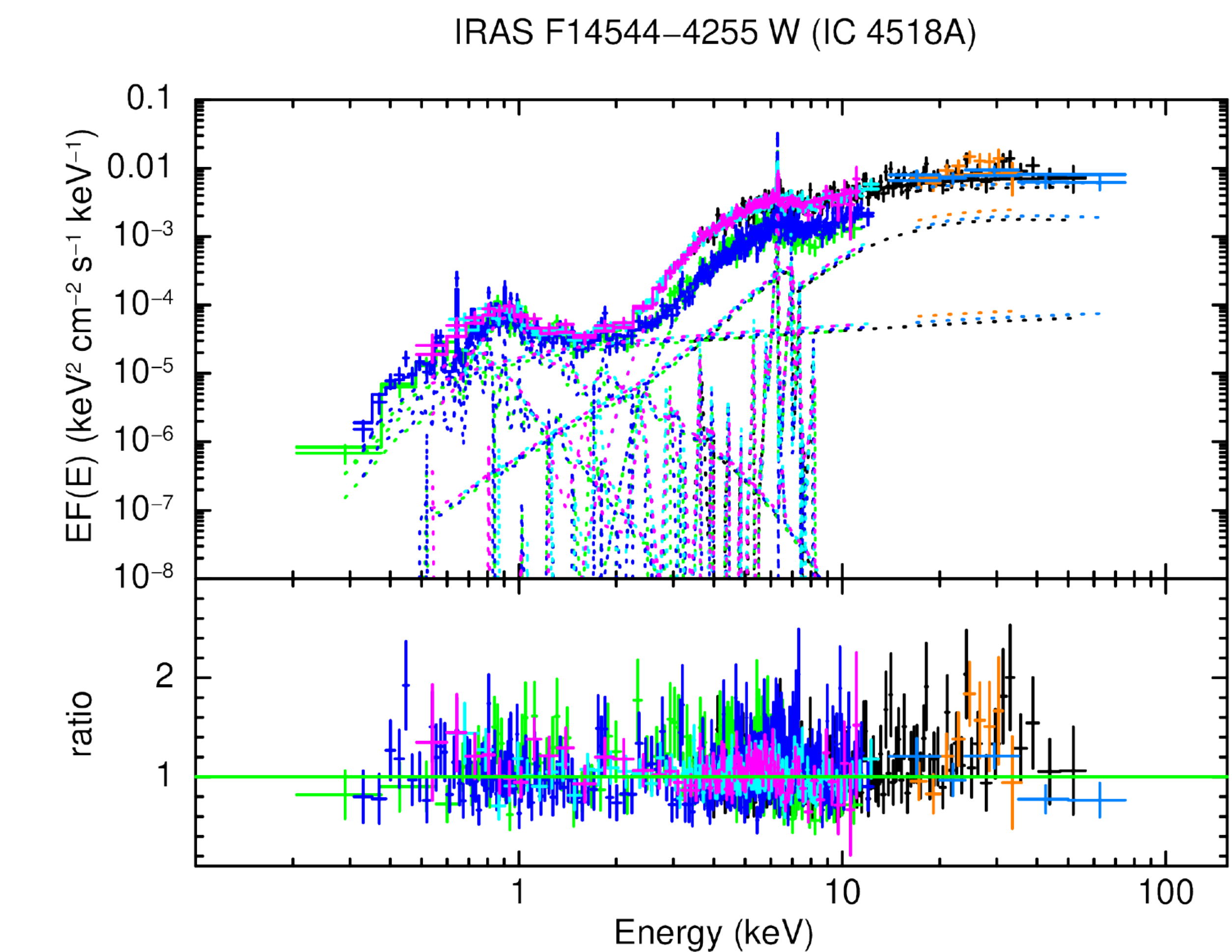}
    \plottwo{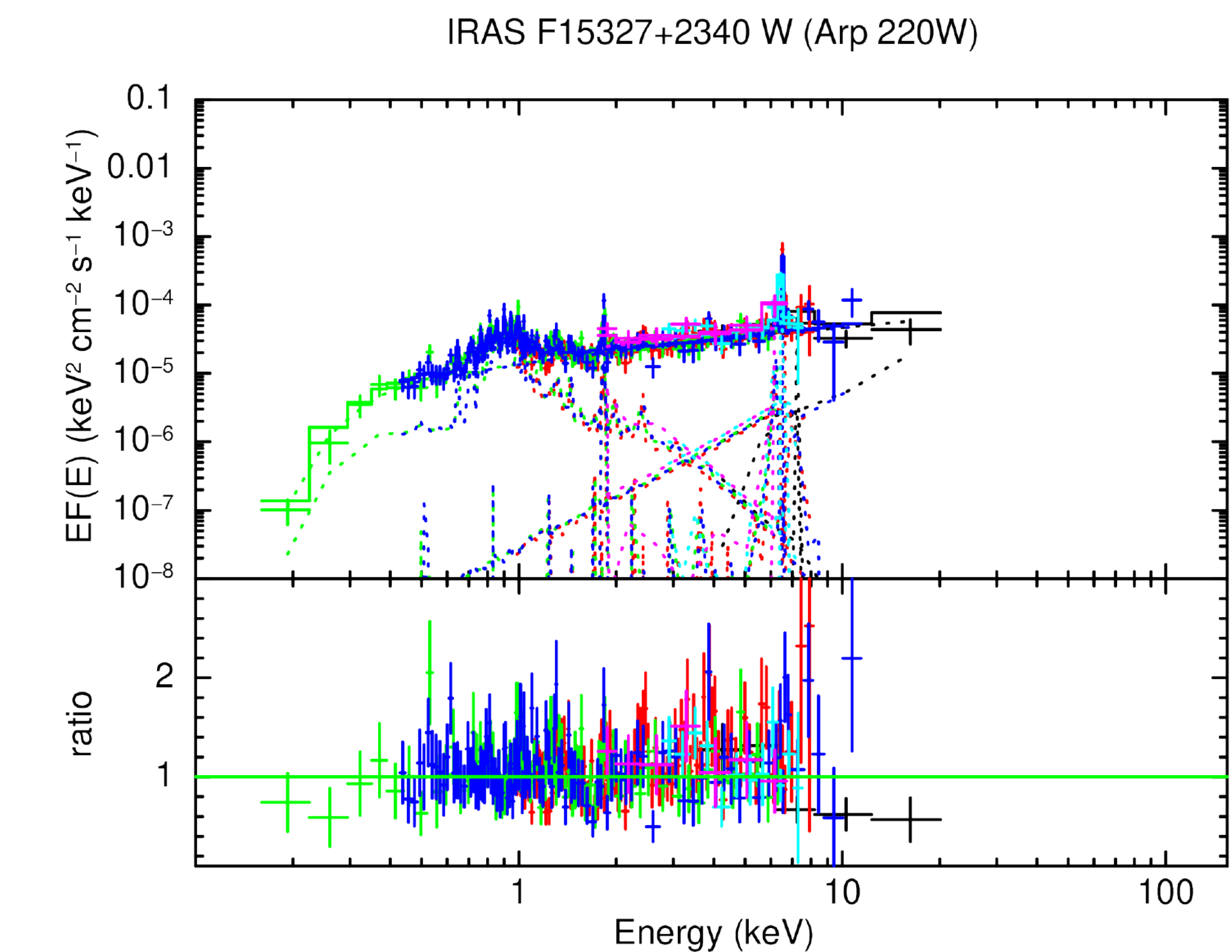}{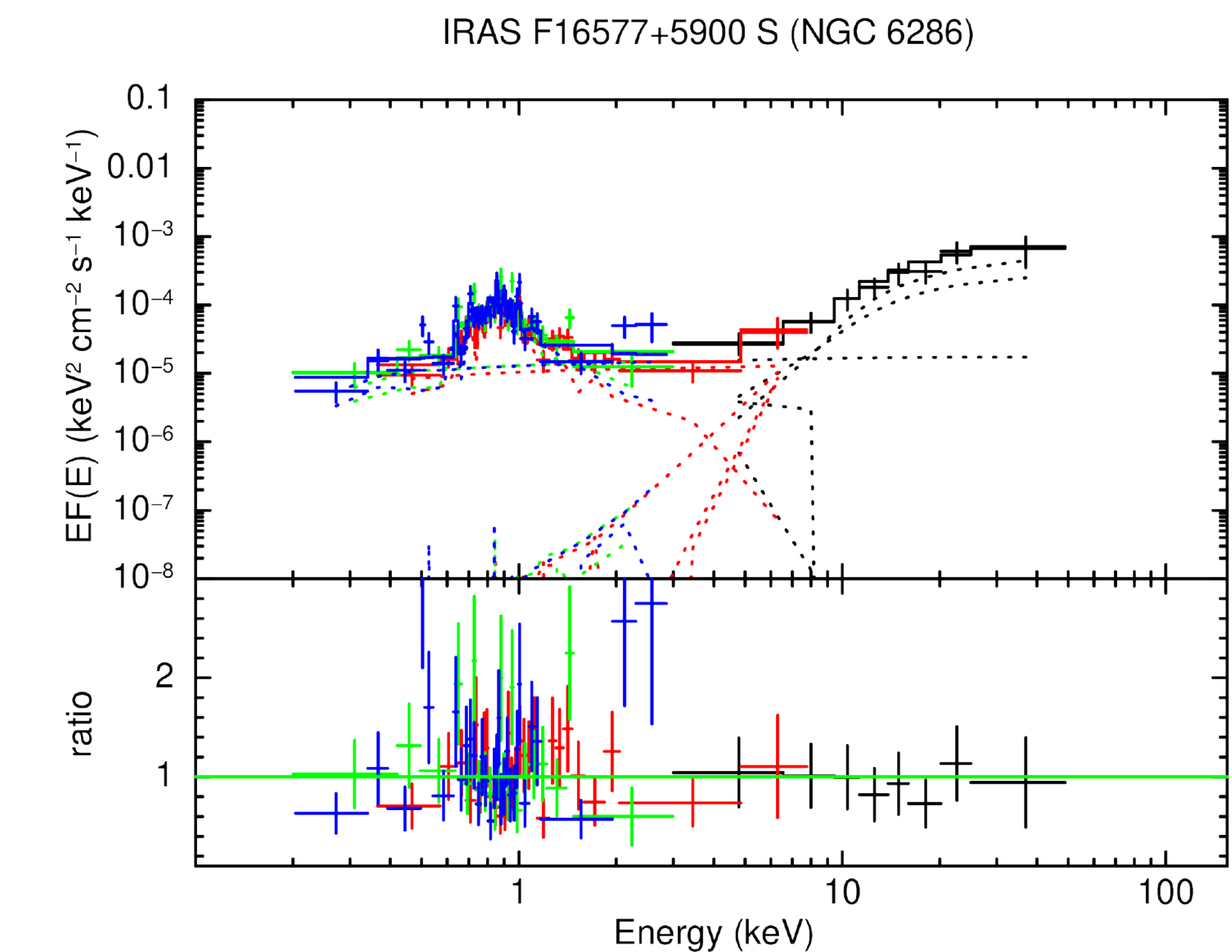}
    \plottwo{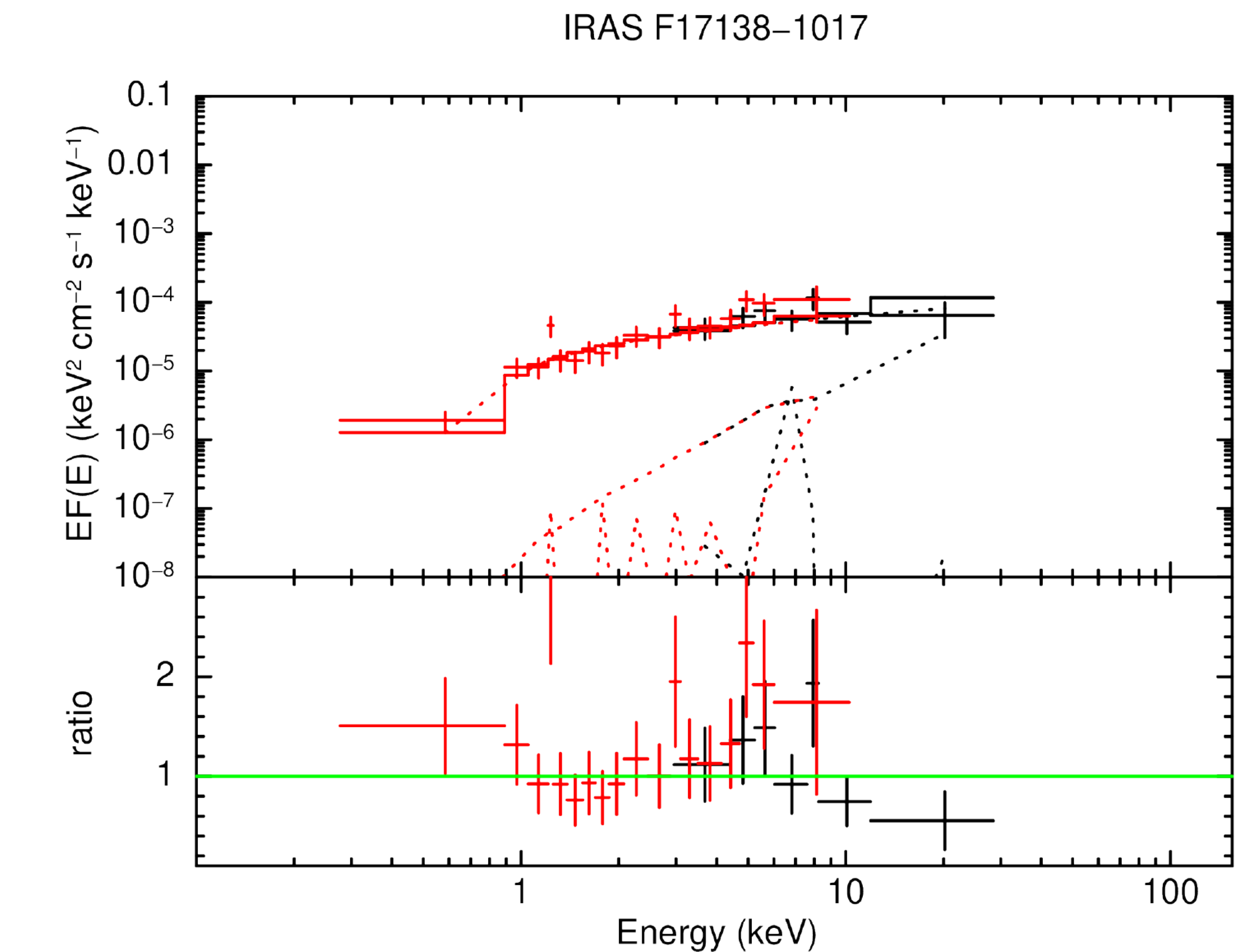}{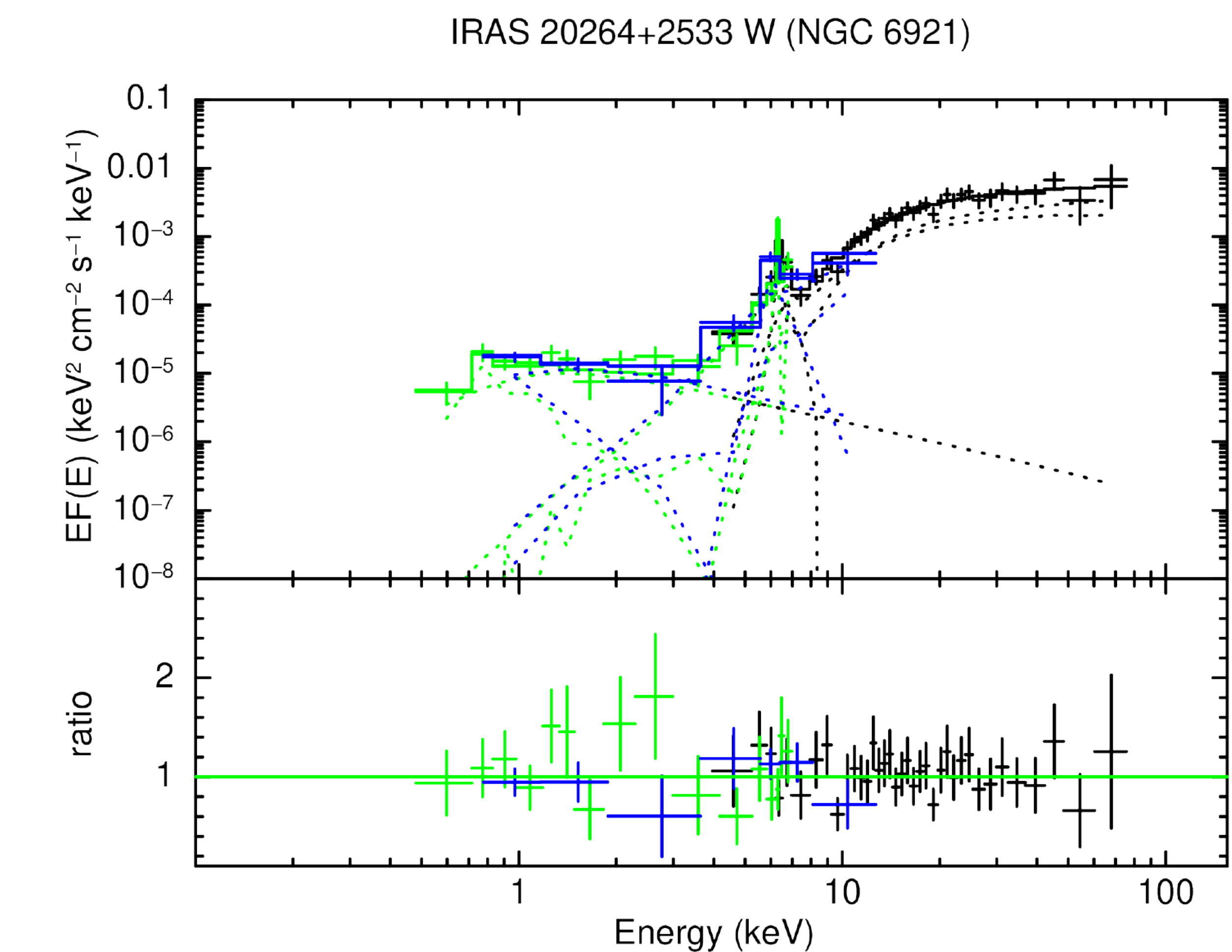}
        \caption{X-ray spectra, best-fit models, and ratios between the data and models of the AGNs in
        IRAS F14348--1447 
        [$N_{\rm H}$ = (1.3$^{+1.0}_{-0.6}$) $\times 10^{24}$~cm$^{-2}$],
        IC~4518A 
        [$N_{\rm H}$ = (1.7~$\pm$~0.1) $\times 10^{23}$~cm$^{-2}$; 
        and (2.2~$\pm$~0.2) $\times 10^{23}$~cm$^{-2}$ for XMM-Newton],
        Arp~220W 
        [$N_{\rm H}$ $\gtrsim$ 9.8 $\times 10^{24}$~cm$^{-2}$],
        NGC~6286 
        [$N_{\rm H}$ = (1.4~$\pm$~0.3) $\times 10^{24}$~cm$^{-2}$],
        IRAS~F17138--1017 
        [$N_{\rm H}$ $>$~3.5 $\times 10^{24}$~cm$^{-2}$],
        and NGC~6921 
        [$N_{\rm H}$ = (1.7~$\pm$~0.3) $\times 10^{24}$~cm$^{-2}$].}
\label{C4-F}
\end{figure*}

\begin{figure*}
    \epsscale{1.15}
    \plottwo{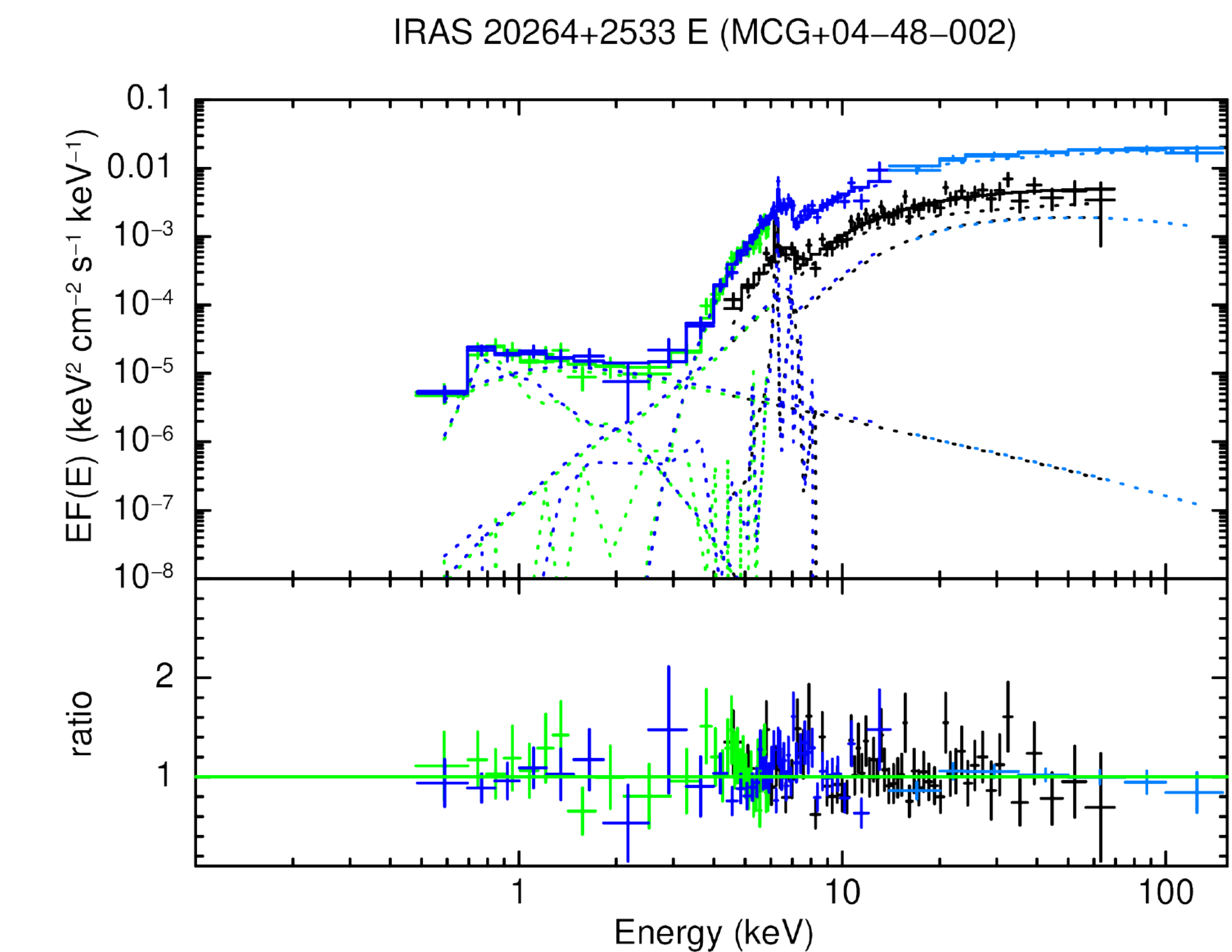}{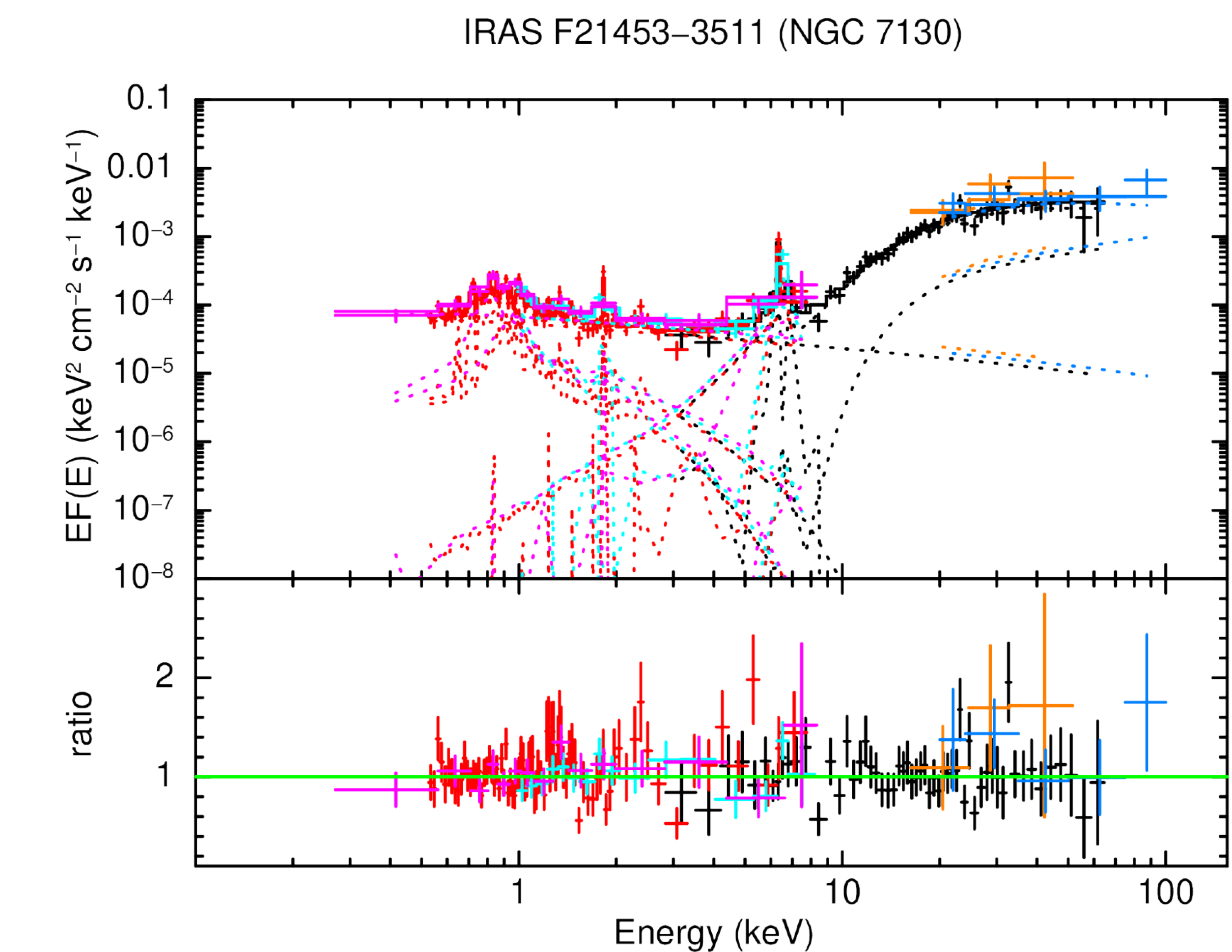}
    \plottwo{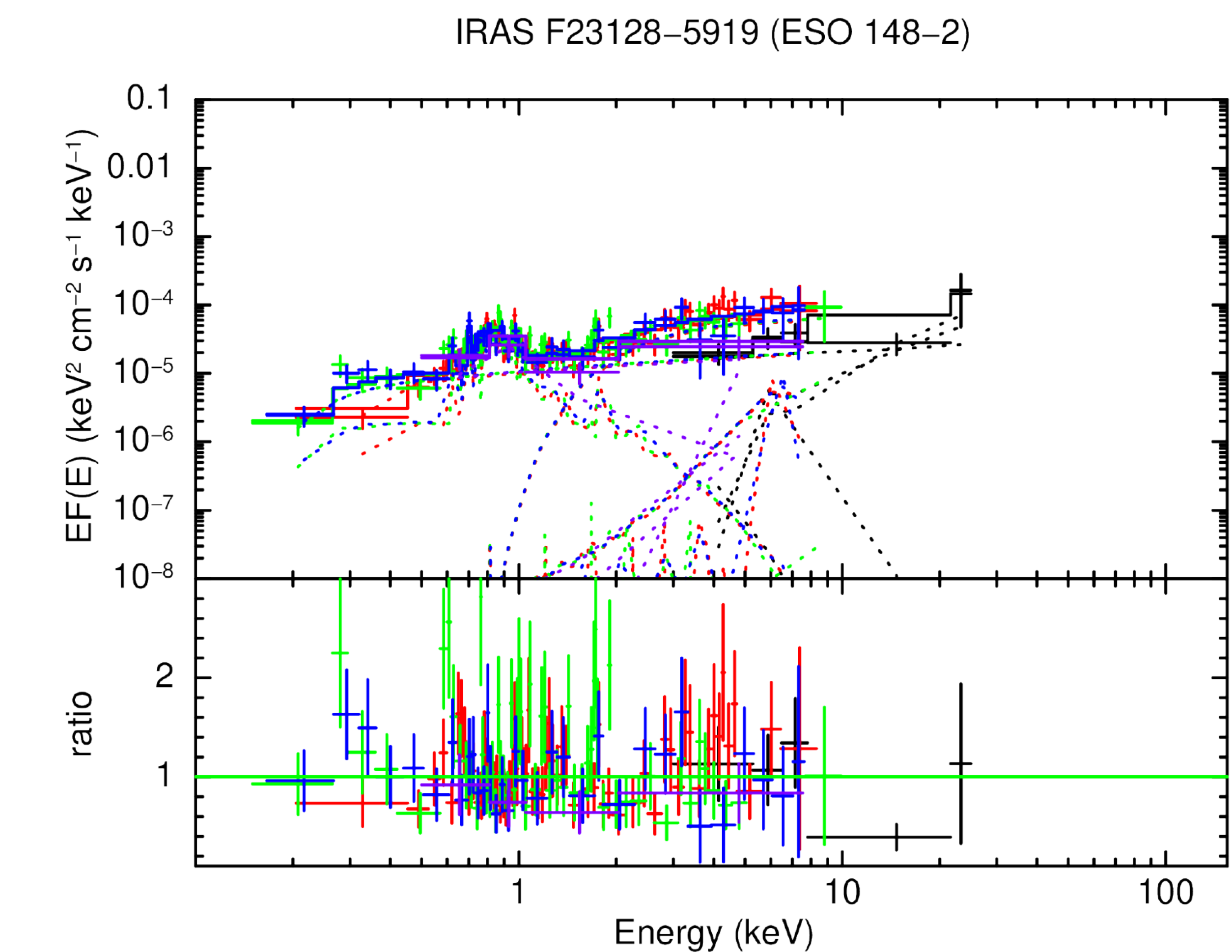}{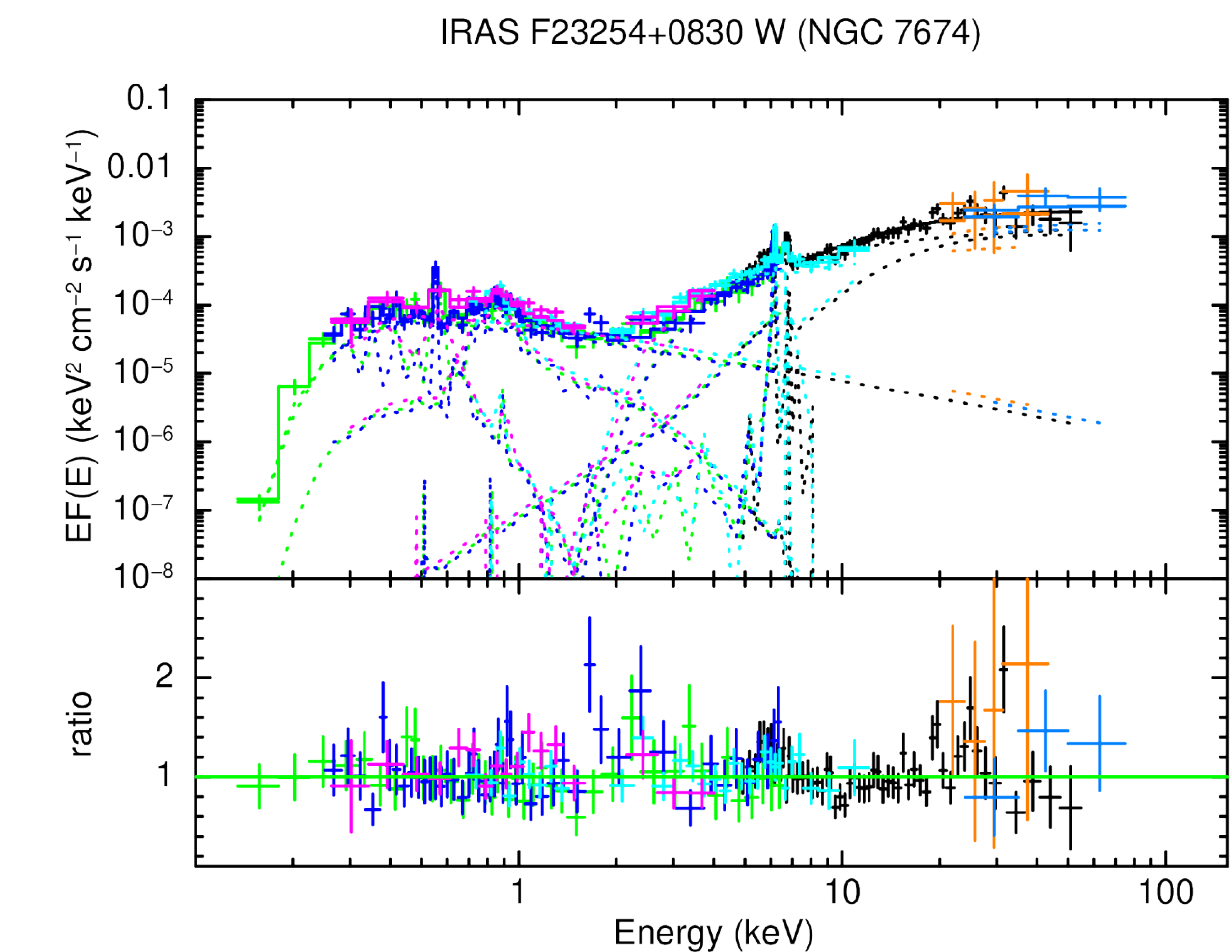}
    \plottwo{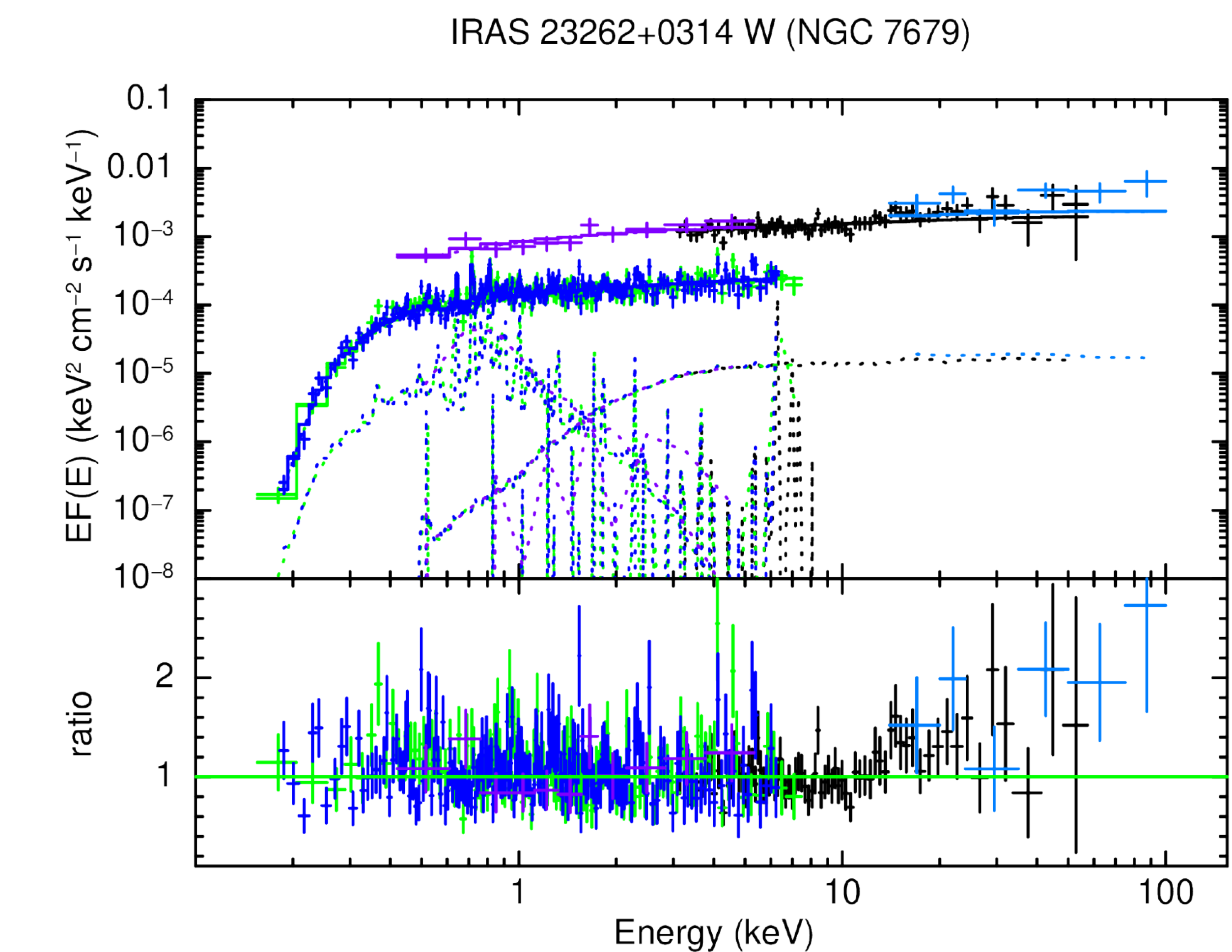}{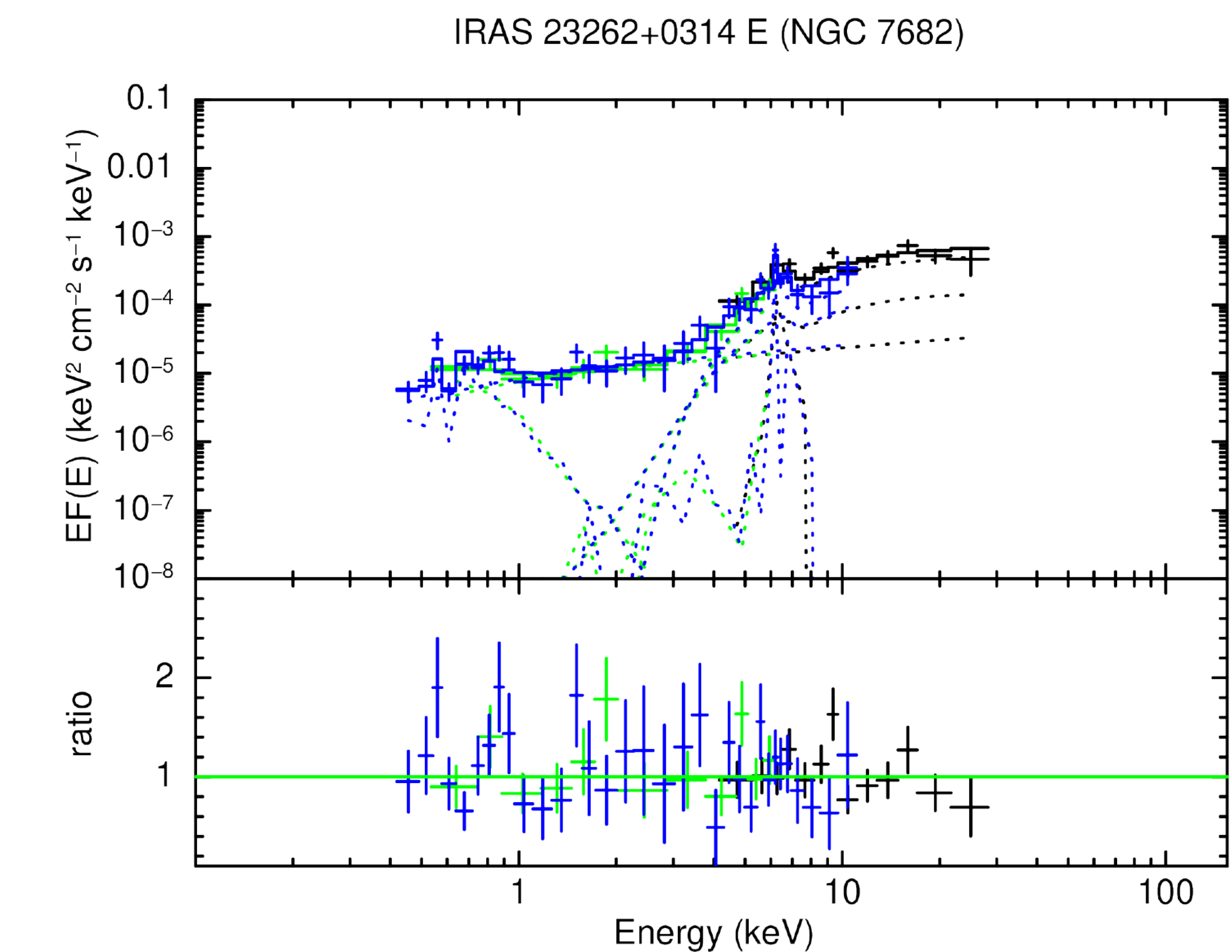}
        \caption{X-ray spectra, best-fit models, and ratios between the data and models of the AGNs in
        MCG+04-48-002 
        [$N_{\rm H}$ = (7.3$^{+1.5}_{-0.8}$) $\times 10^{23}$~cm$^{-2}$; 
        and (6.0~$\pm$~0.5) $\times 10^{23}$~cm$^{-2}$ for XMM-Newton/Swift],
        NGC~7130 
        [$N_{\rm H}$ = (4.1~$\pm$~0.1) $\times 10^{24}$~cm$^{-2}$],
        ESO~148-2 
        [$N_{\rm H}$ = (1.6~$\pm$~0.5) $\times 10^{24}$~cm$^{-2}$; 
        (2.7$^{+0.9}_{-0.6}$) $\times 10^{22}$~cm$^{-2}$ for Chandra/XMM-Newton; 
        and (2.5$^{+97.5}_{-1.8}$) $\times 10^{23}$~cm$^{-2}$ for Swift],
        NGC~7674 
        [$N_{\rm H}$ = (3.0$^{+0.3}_{-0.2}$) $\times 10^{23}$~cm$^{-2}$; 
        (1.4$^{+0.5}_{-0.4}$) $\times 10^{23}$~cm$^{-2}$ for XMM-Newton; 
        and (1.0$^{+0.3}_{-0.2}$) $\times 10^{23}$~cm$^{-2}$ for Suzaku],
        NGC~7679 
        [$N_{\rm H}$~$<$~3 $\times 10^{15}$~cm$^{-2}$],
        and NGC~7682 
        [$N_{\rm H}$ = (3.8~$\pm$~0.7) $\times 10^{23}$~cm$^{-2}$].}
\label{C5-F}
\end{figure*}

\begin{figure*}
    \epsscale{1.15}
    \plottwo{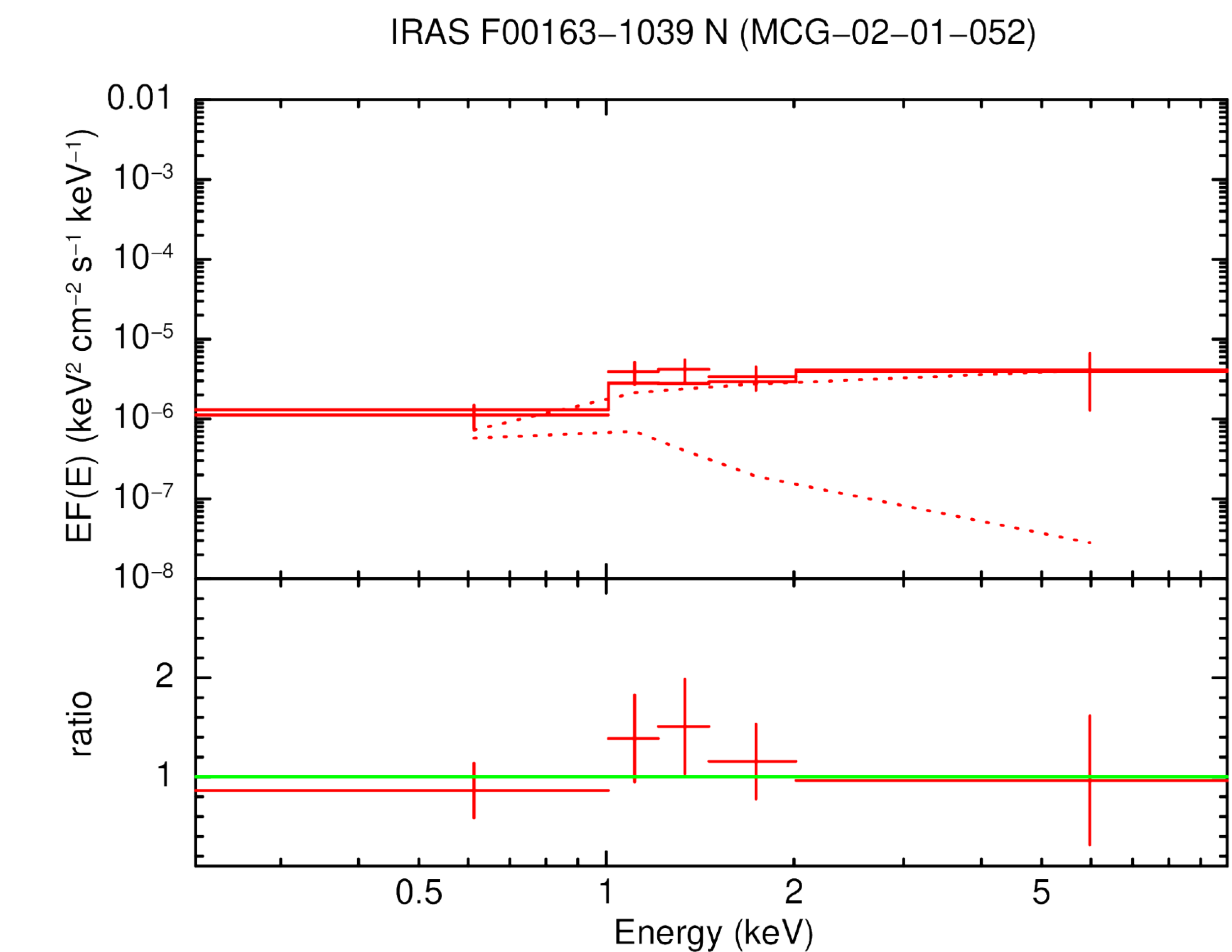}{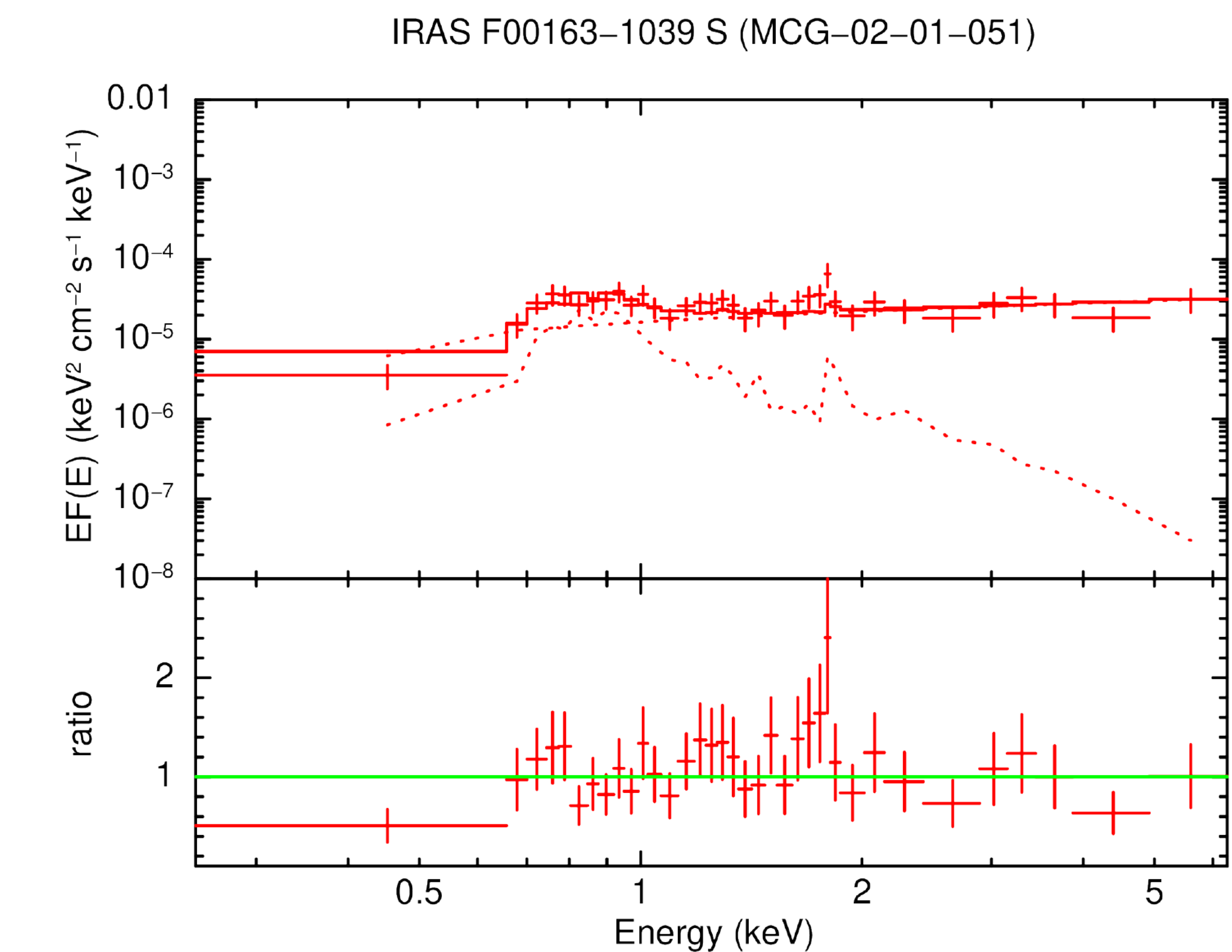}
    \plottwo{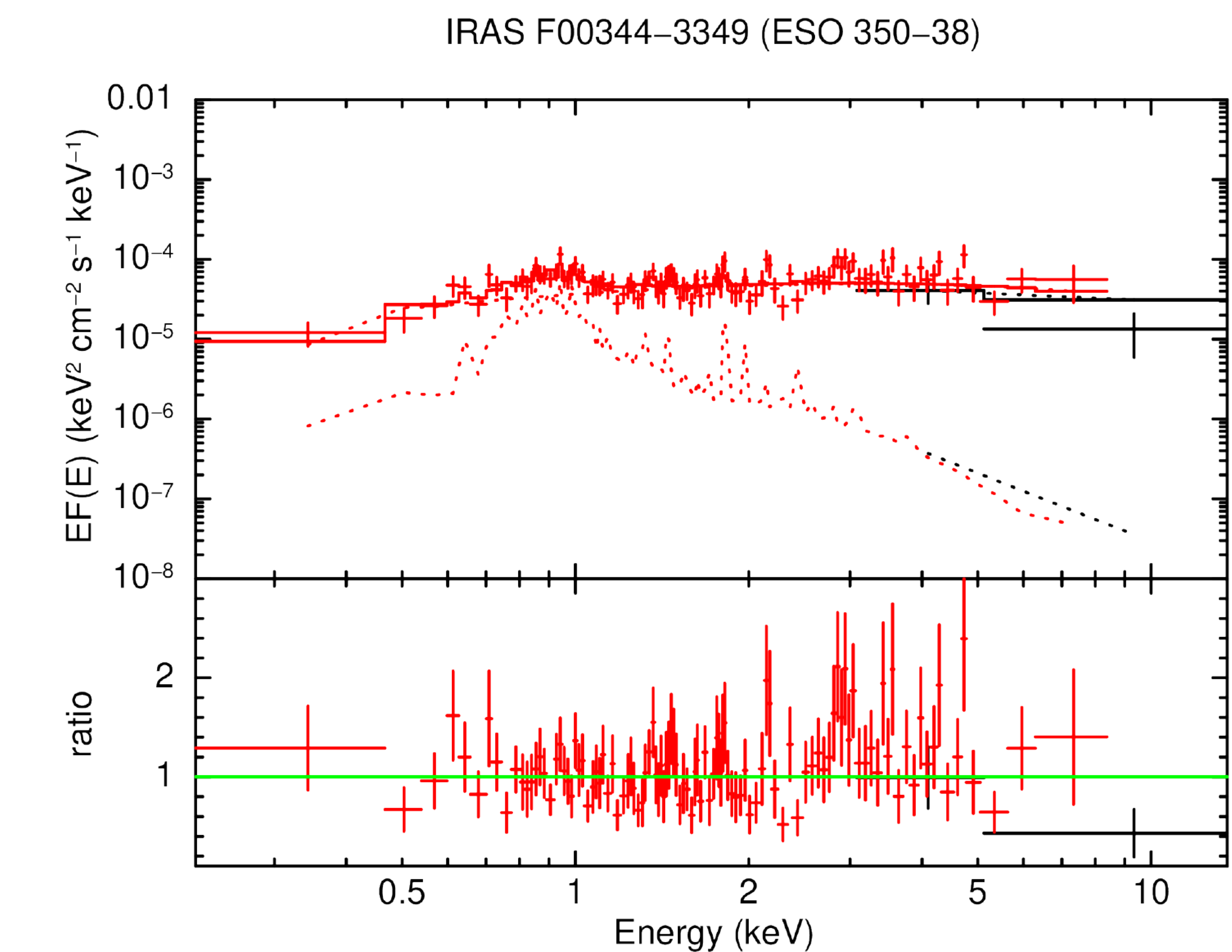}{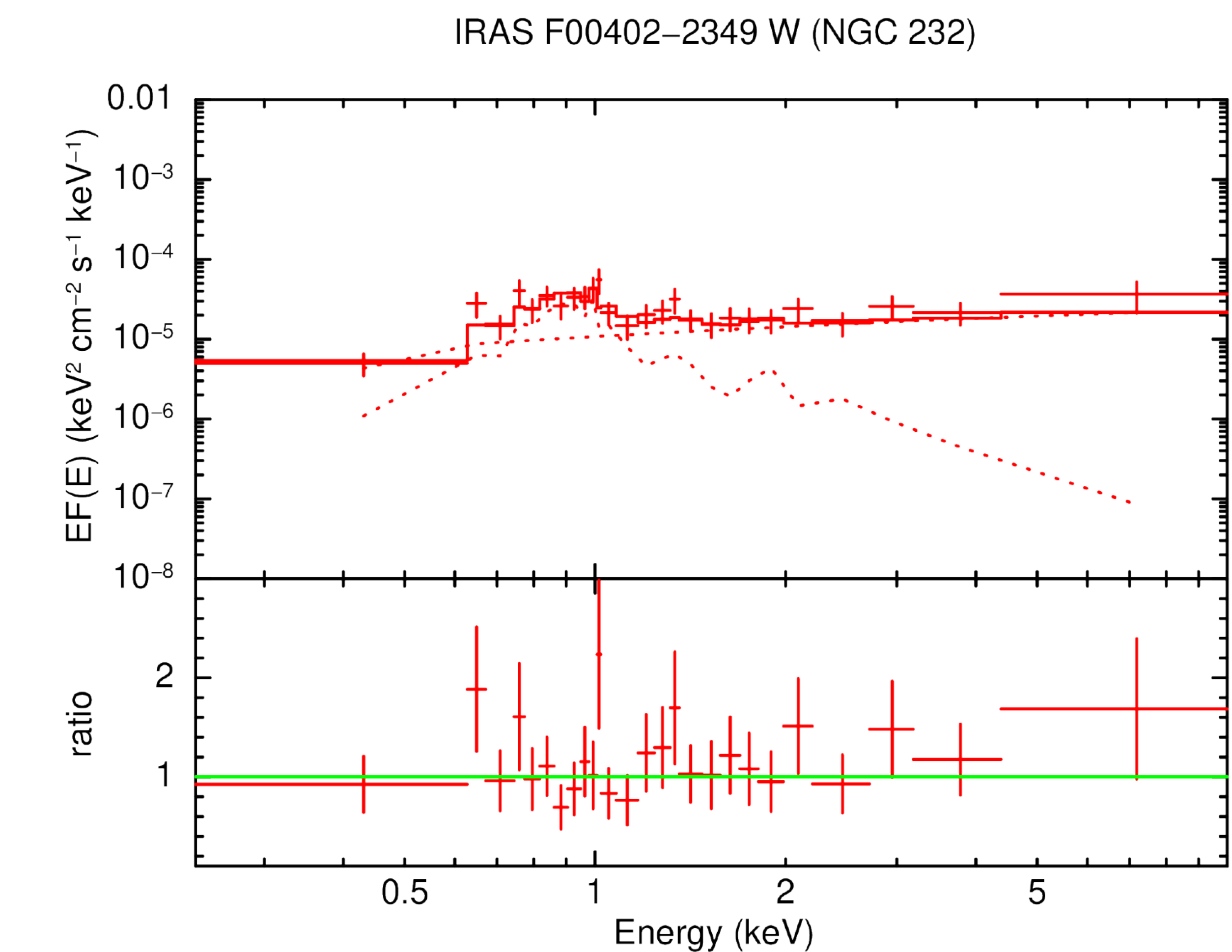}
    \plottwo{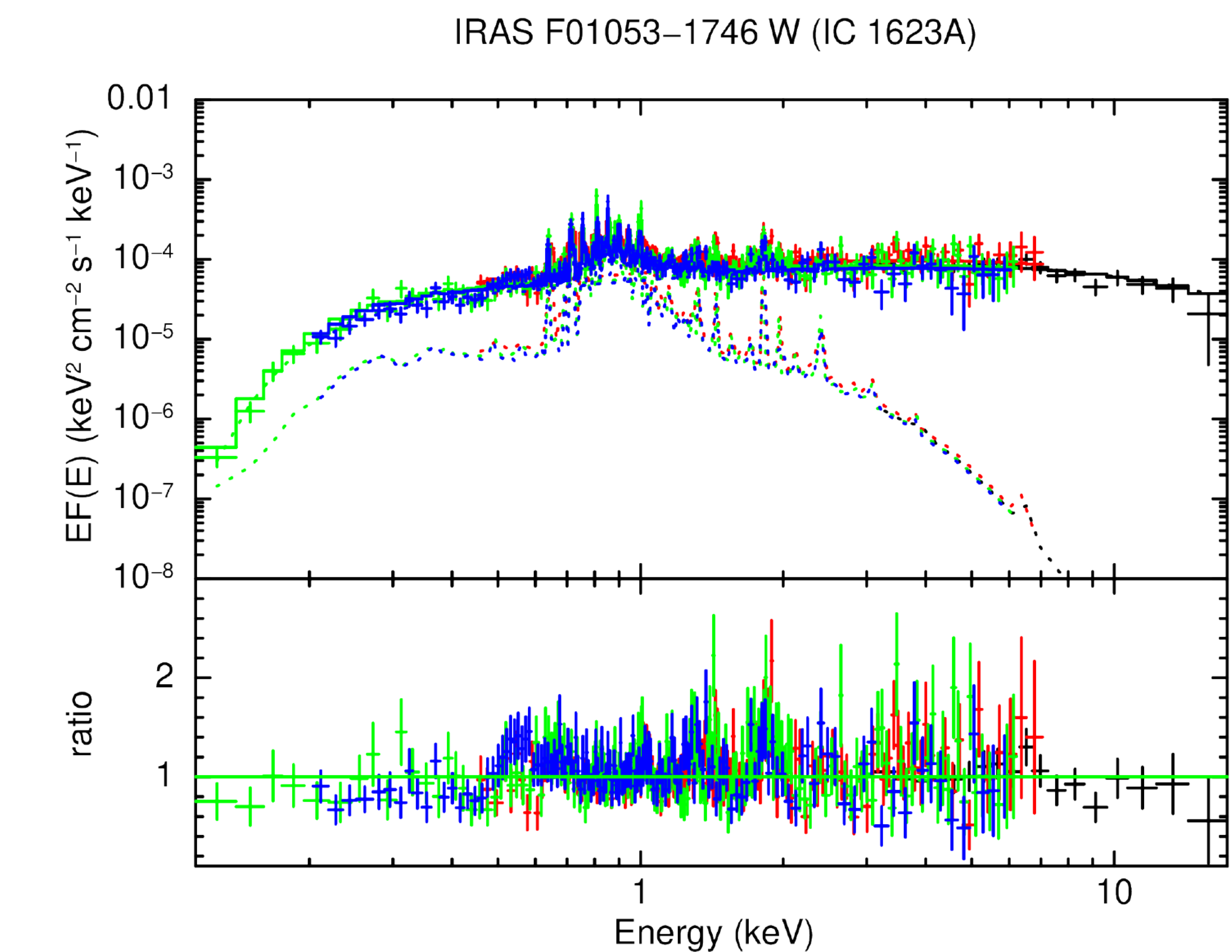}{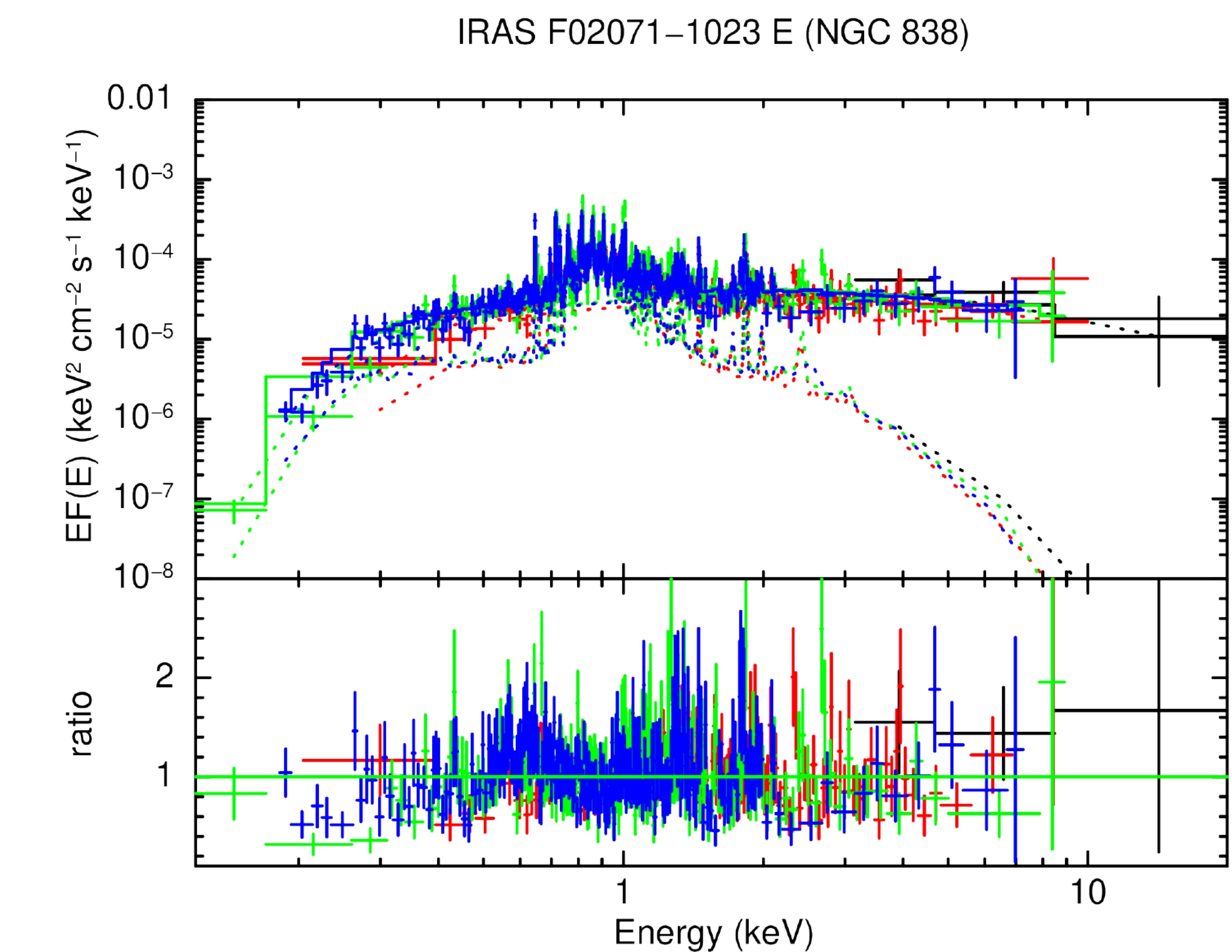}
        \caption{X-ray spectra, best-fit models, and ratios between the data and models of the starburst-dominant or hard X-ray undetected sources in 
        MCG--02-01-052,
        MCG--02-01-051,
        ESO~350-38,
        NGC~232,
        IC~1623B,
        and NGC~838.\\}
\label{C6-F}
\end{figure*}

\begin{figure*}
    \epsscale{1.15}
    \plottwo{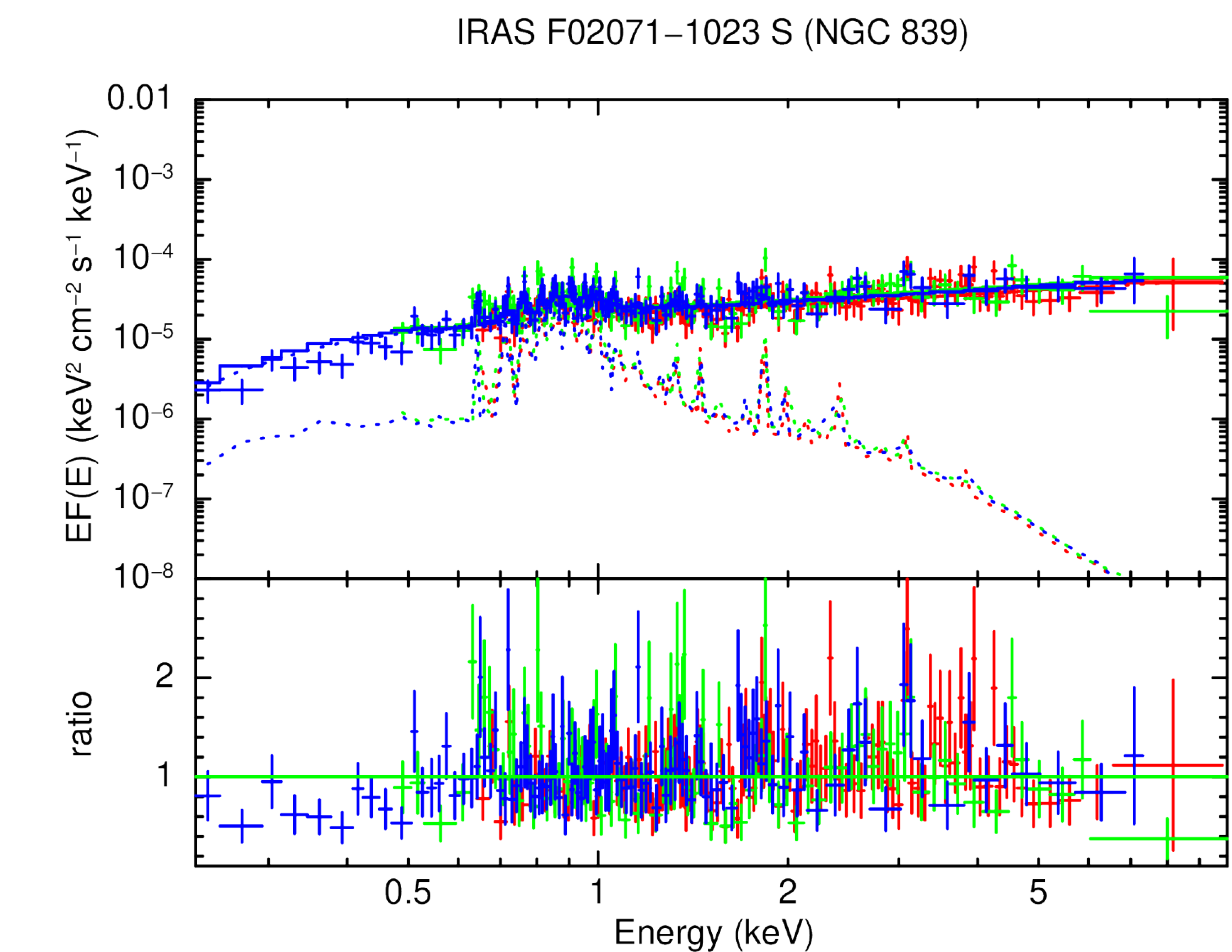}{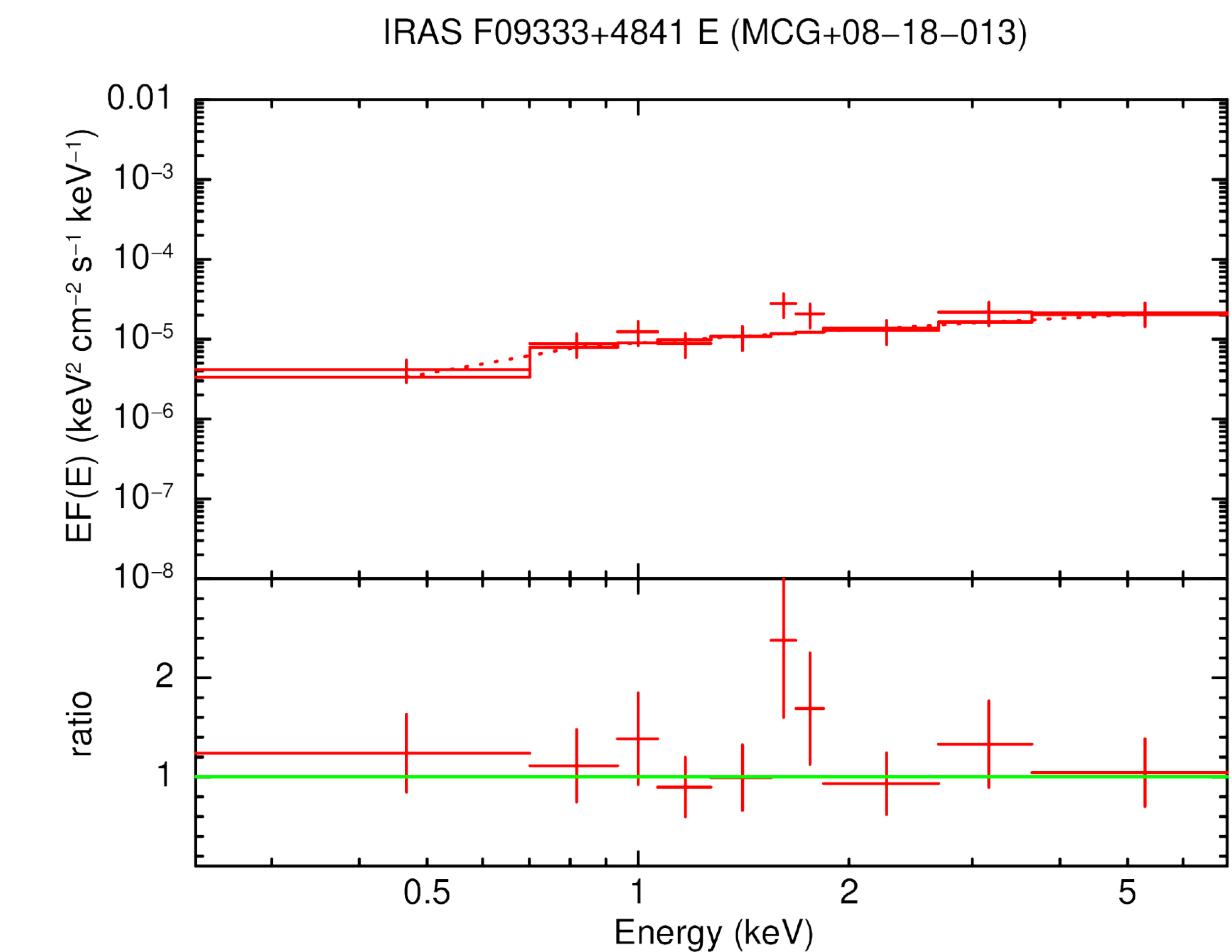}
    \plottwo{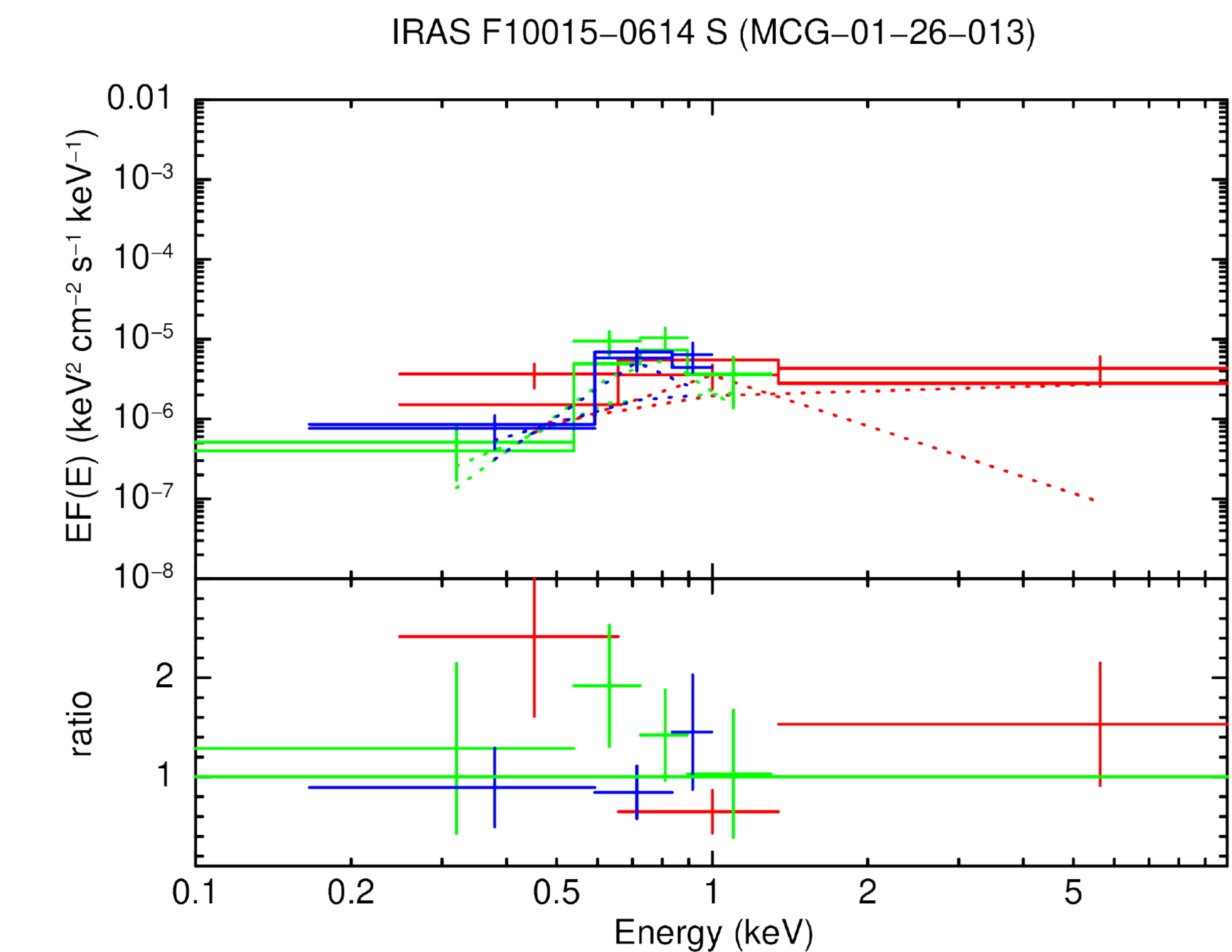}{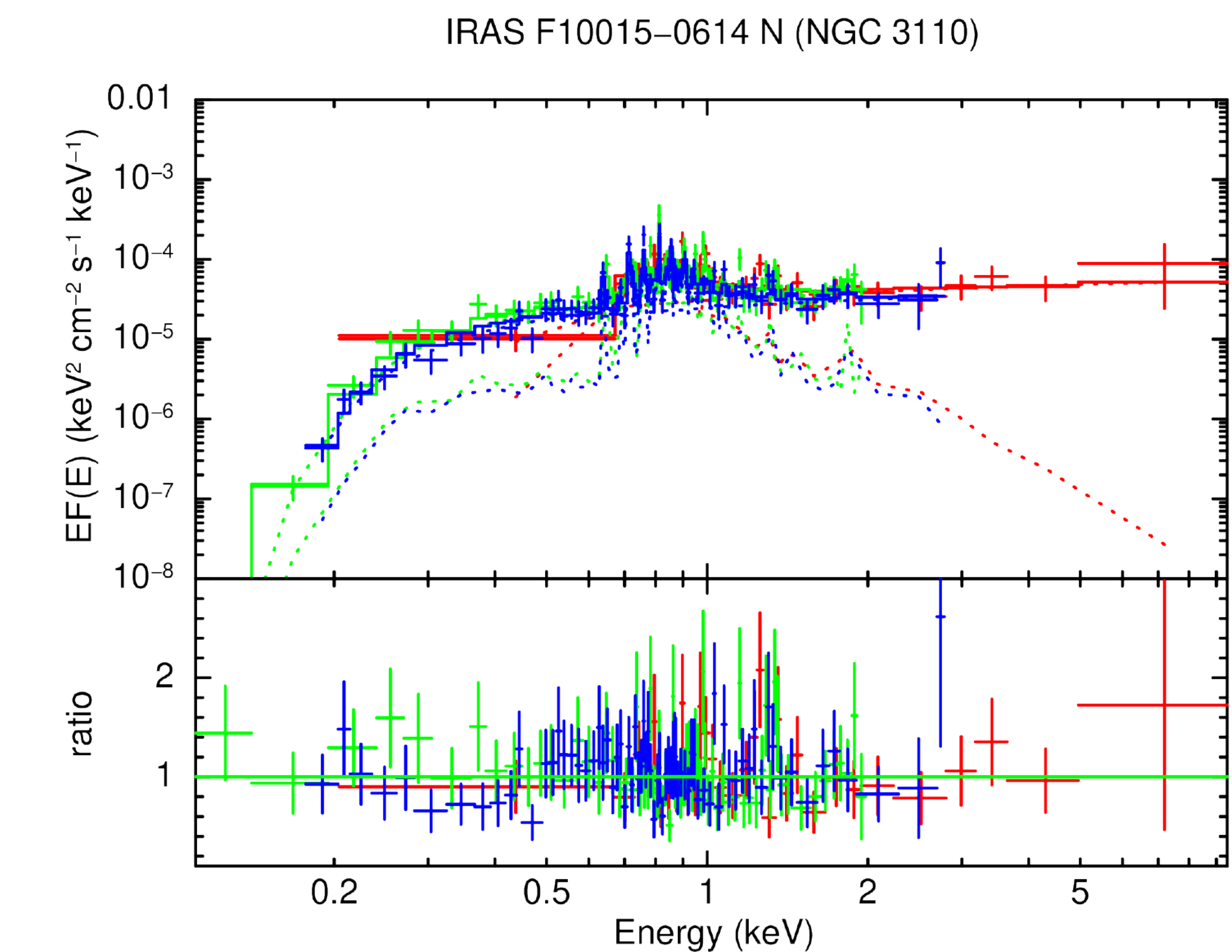}
    \plottwo{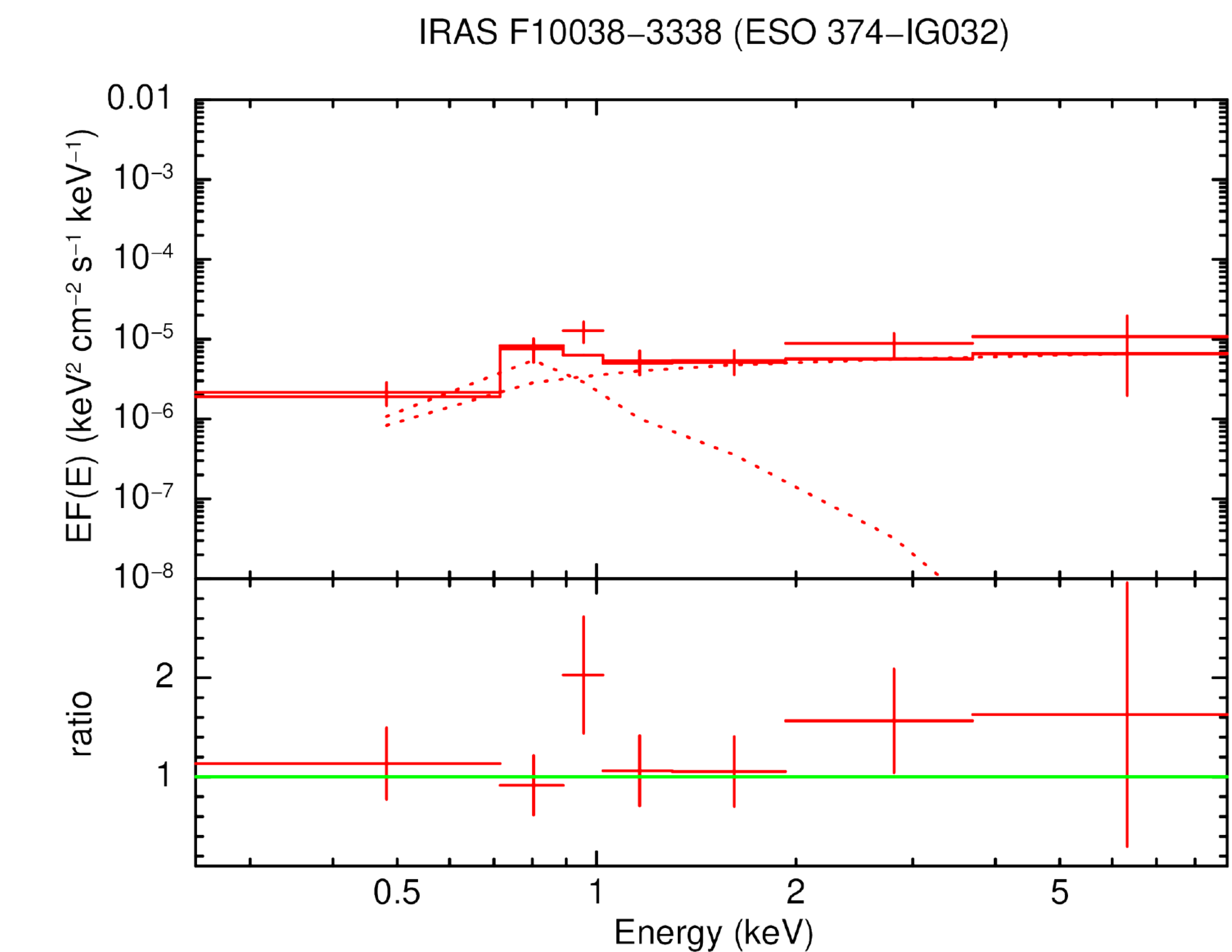}{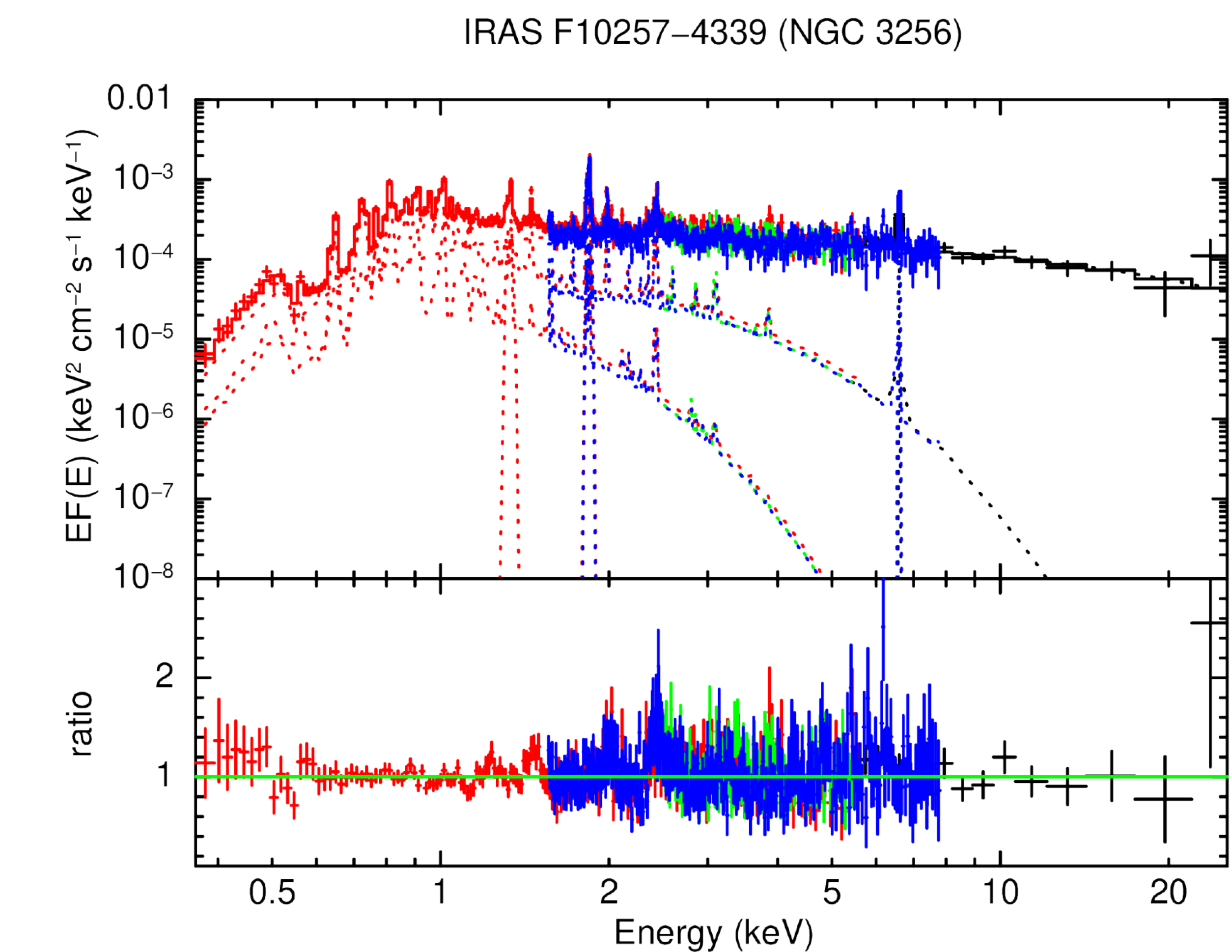}
        \caption{X-ray spectra, best-fit models, and ratios between the data and models of the starburst-dominant or hard X-ray undetected sources in 
        NGC~839, 
        MCG+08-18-013, 
        MCG--01-26-013, 
        NGC~3110,
        ESO~374-IG032,
        and NGC~3256.\\}
\label{C7-F}
\end{figure*}

\begin{figure*}
    \epsscale{1.15}
    \plottwo{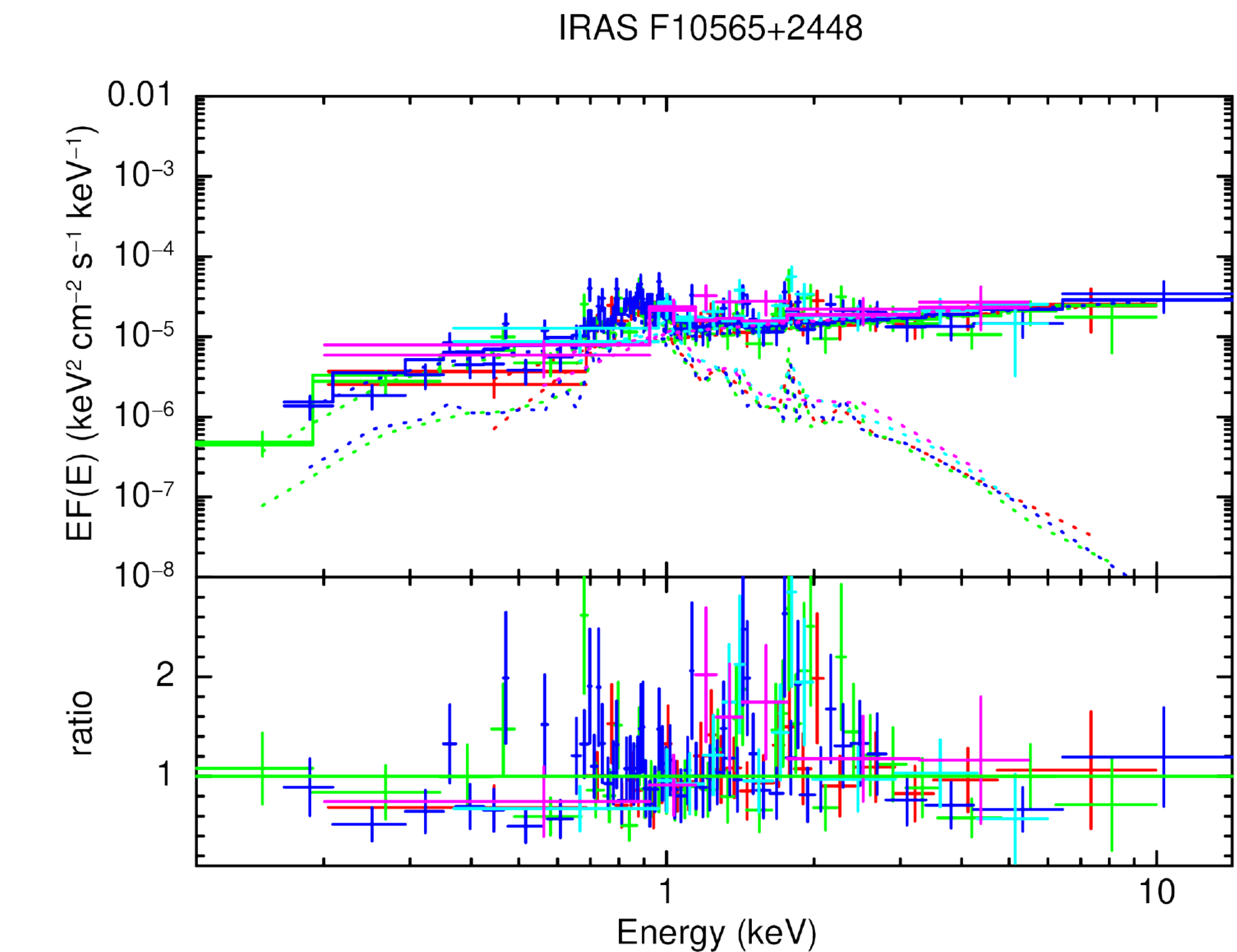}{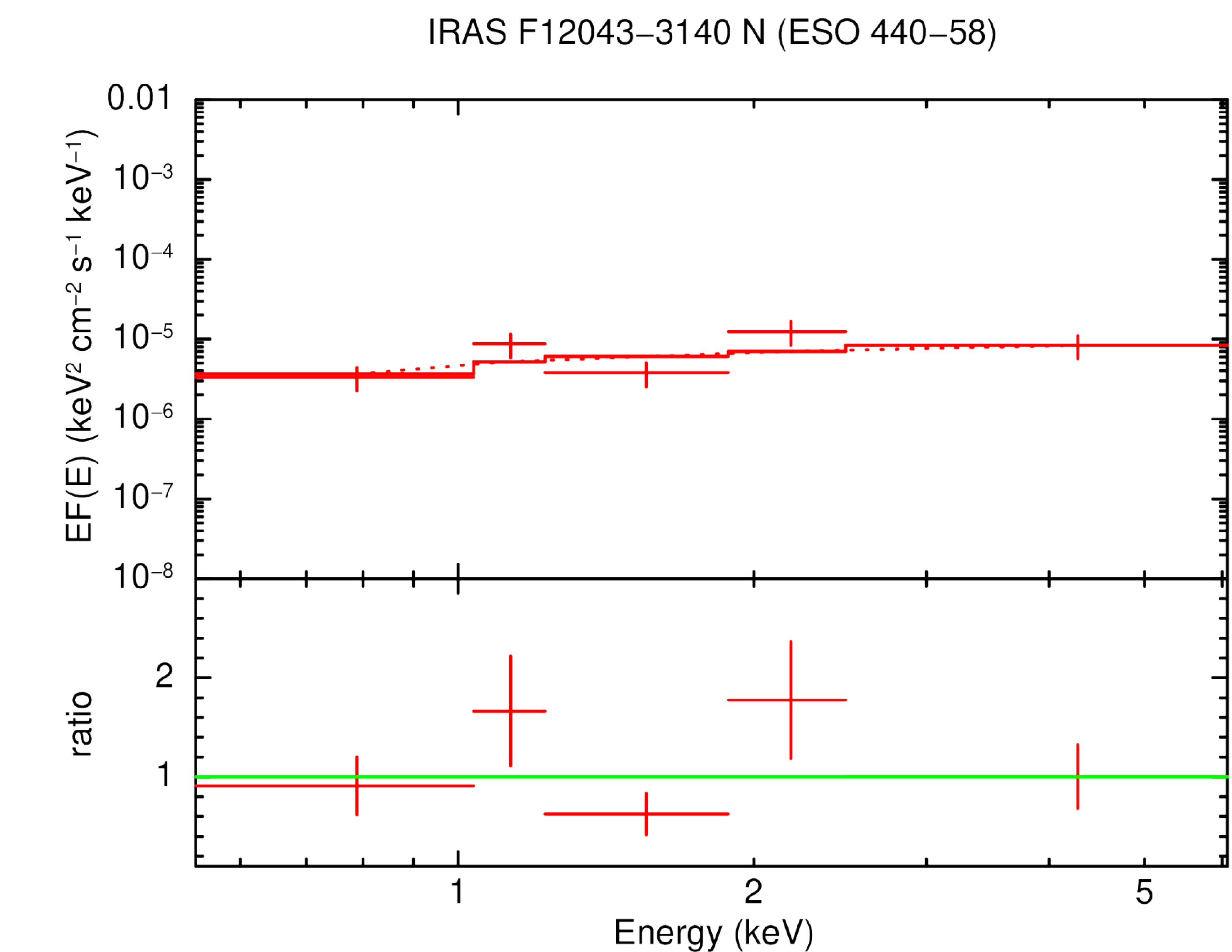}
    \plottwo{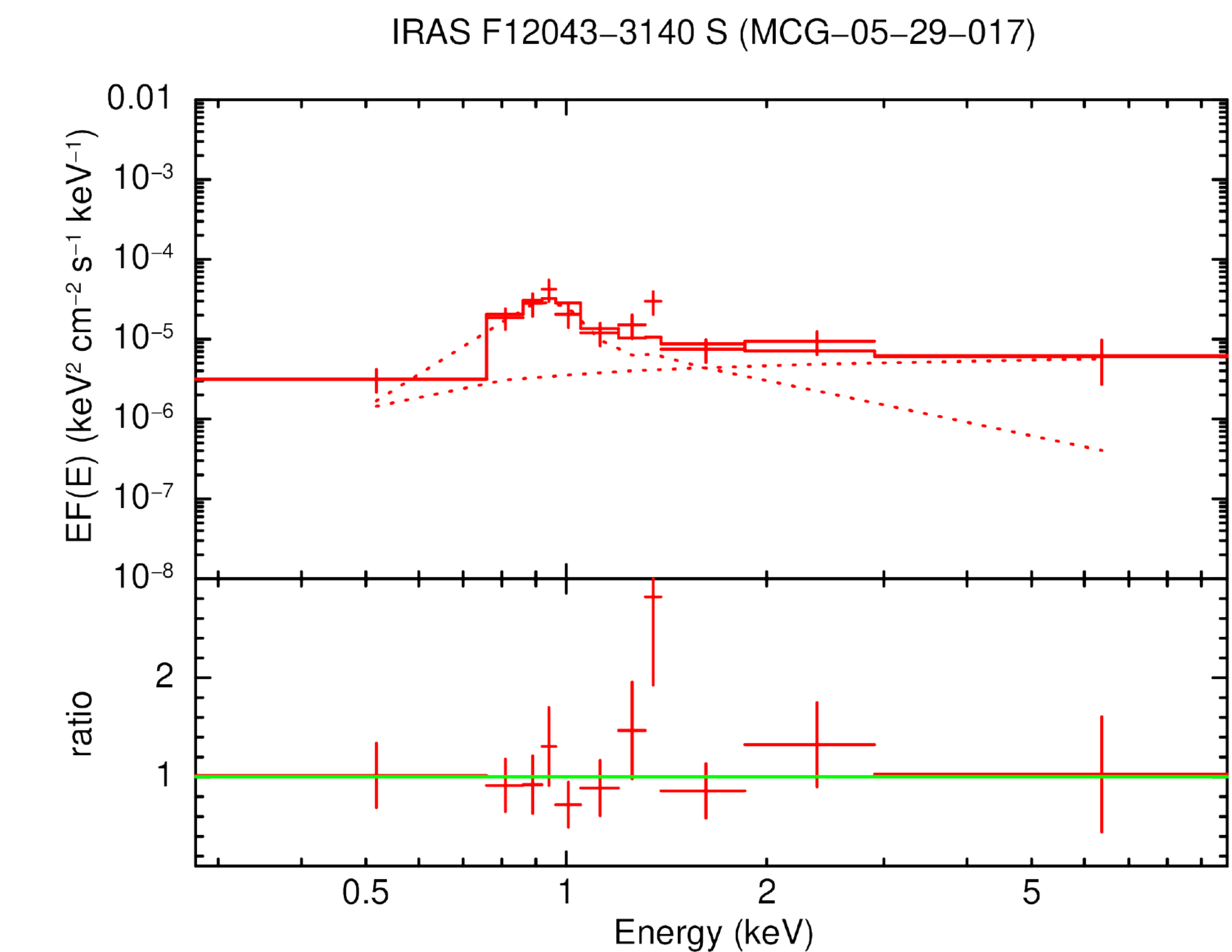}{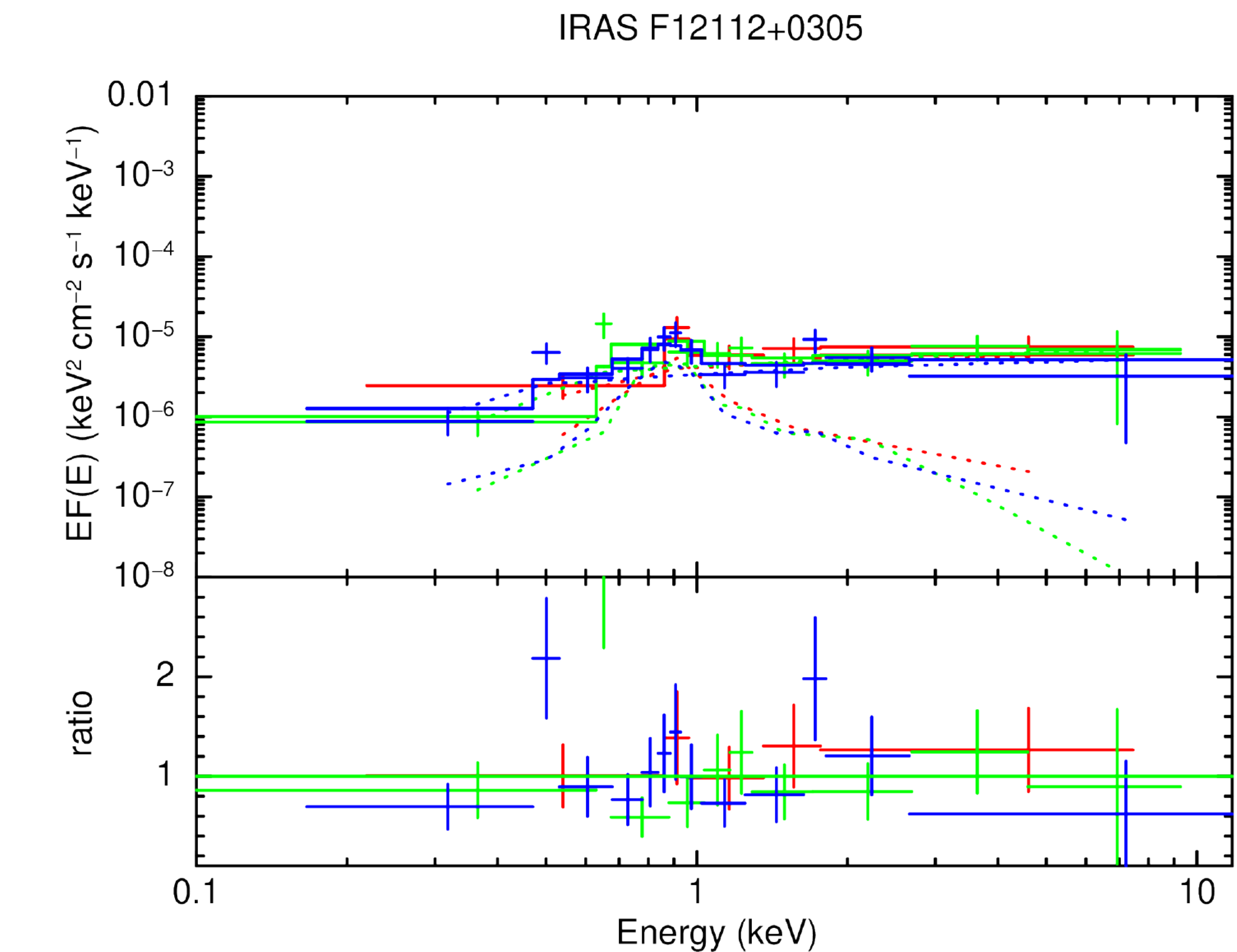}
    \plottwo{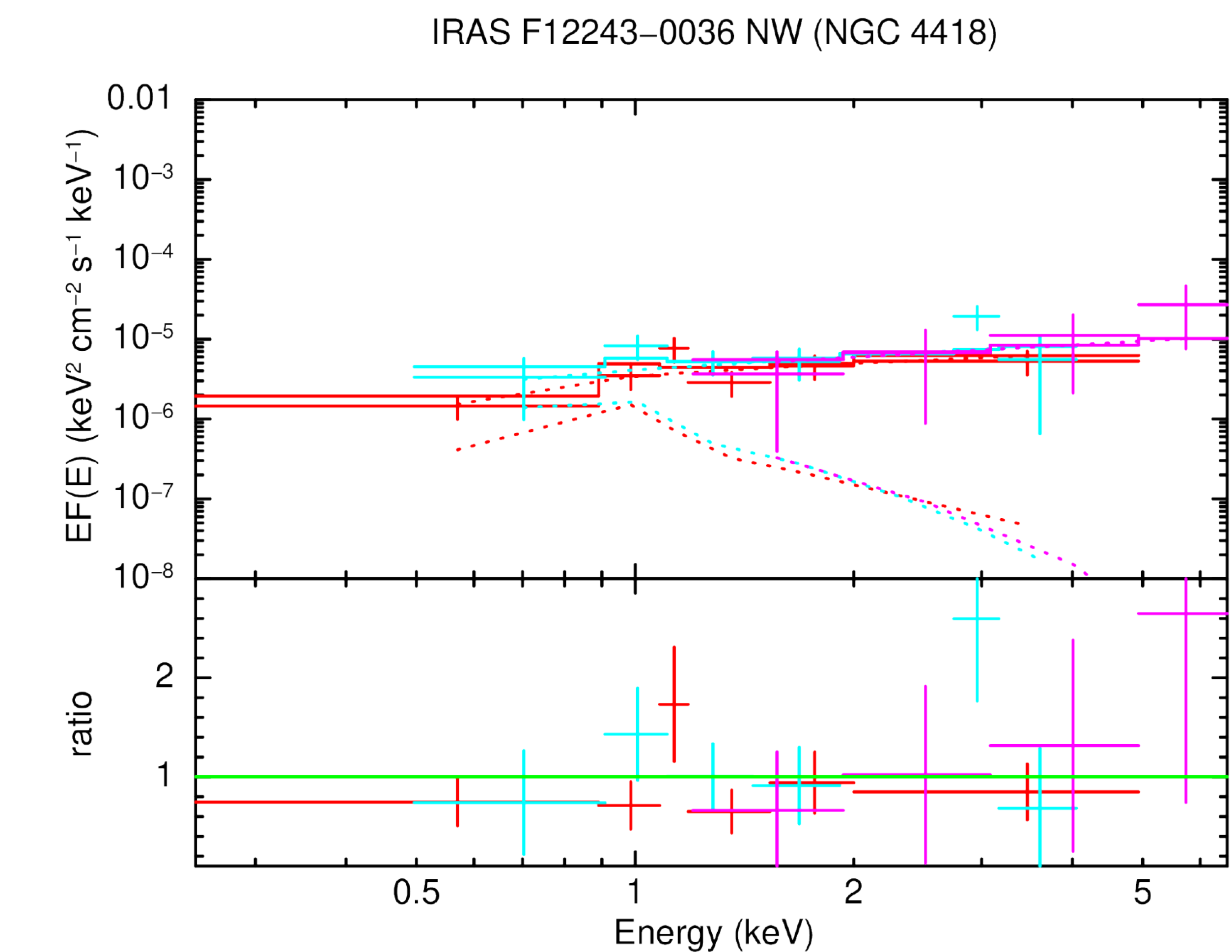}{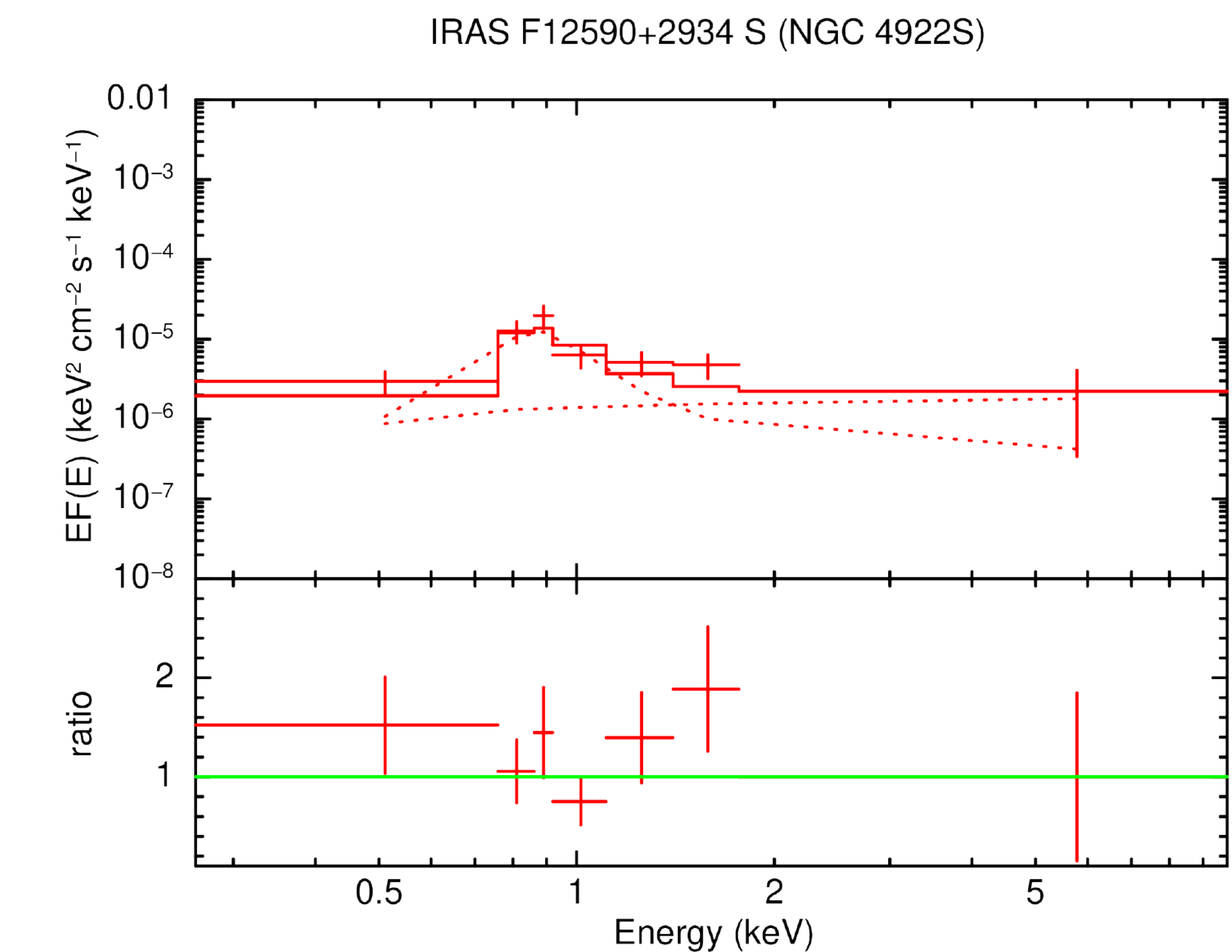}
        \caption{X-ray spectra, best-fit models, and ratios between the data and models of the starburst-dominant or hard X-ray undetected sources in
        IRAS F10565+2448,
        ESO~440-58, 
        MCG--05-29-017, 
        IRAS F12112+0305, 
        NGC~4418,
        and NGC~4922S.\\}
\label{C8-F}
\end{figure*}

\begin{figure*}
    \epsscale{1.15}
    \plottwo{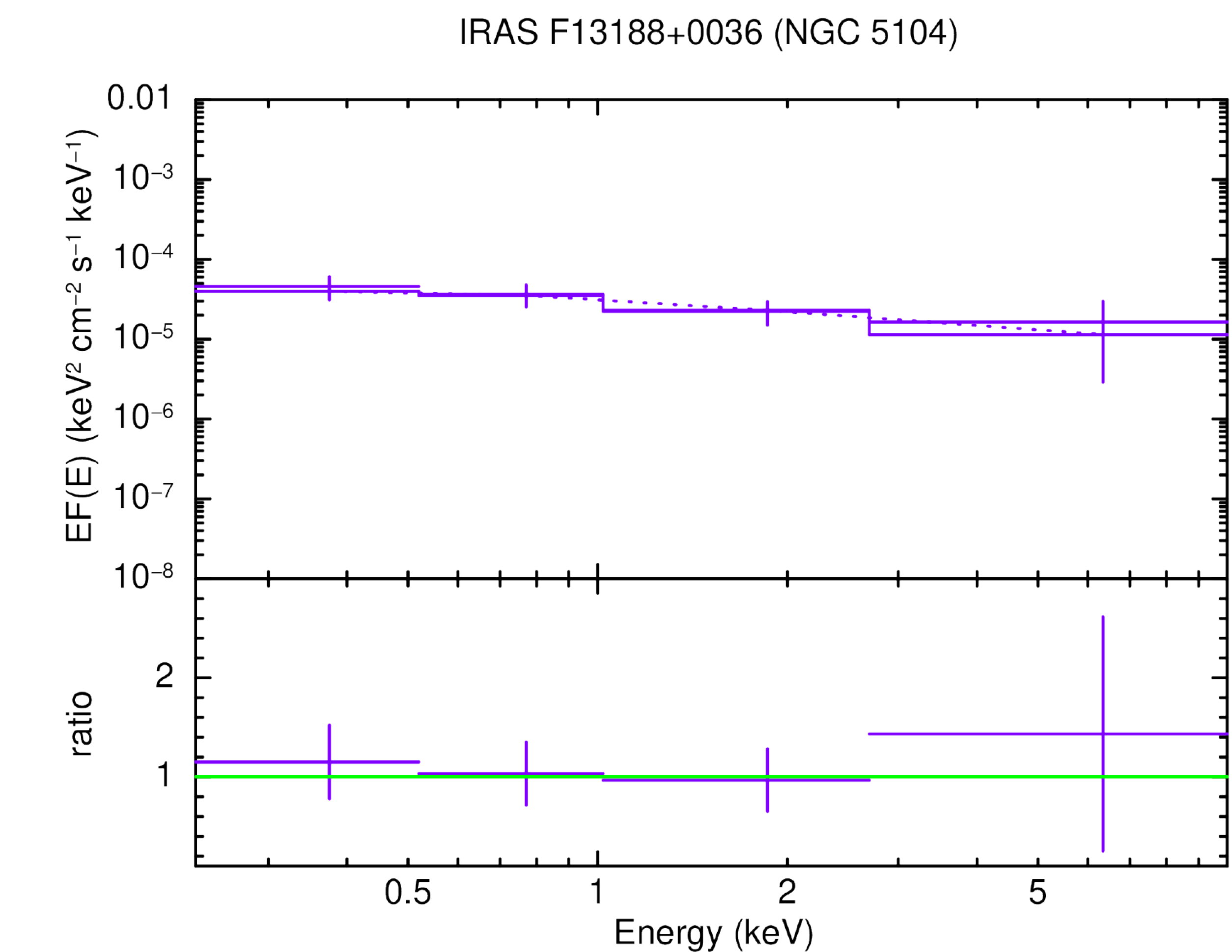}{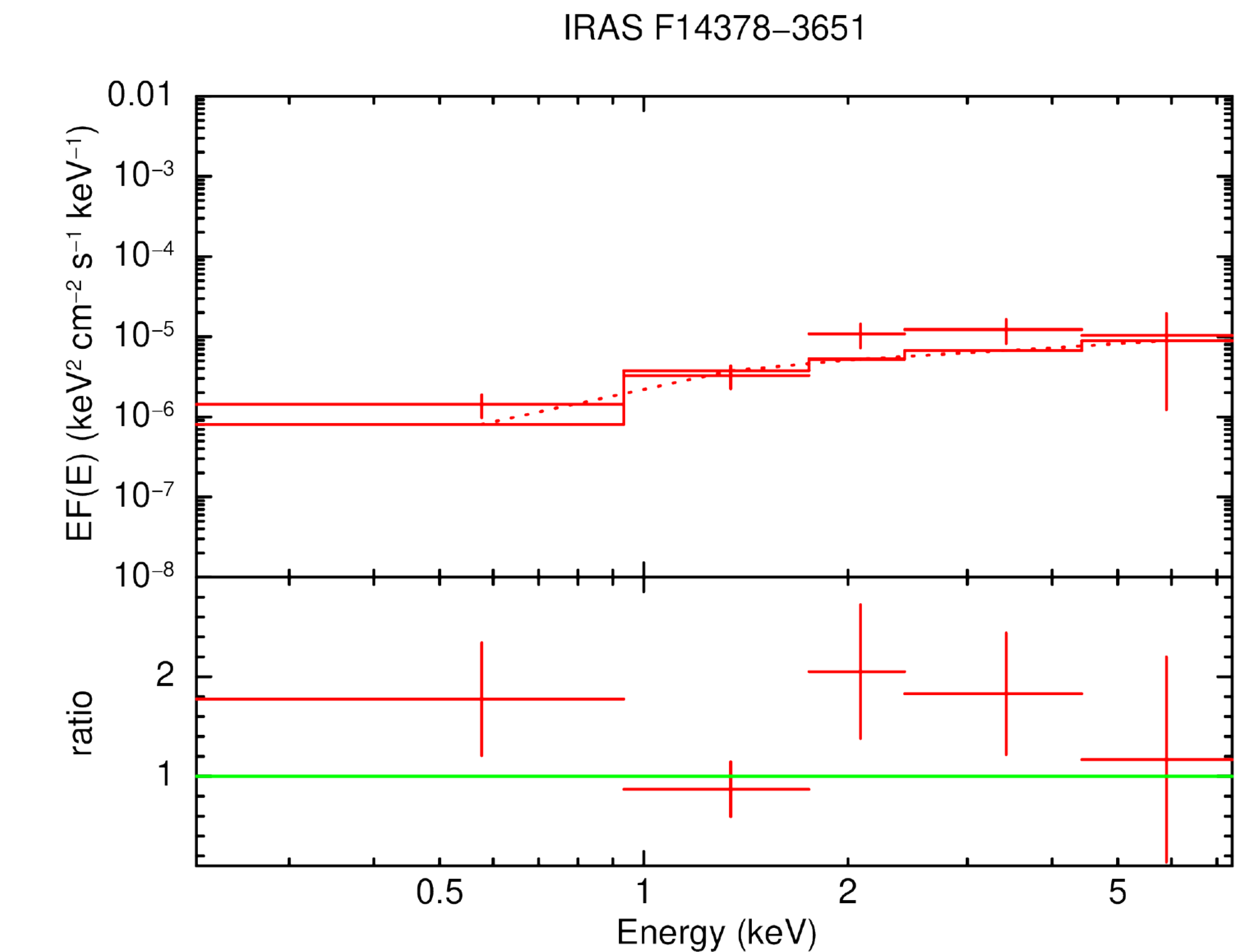}
    \plottwo{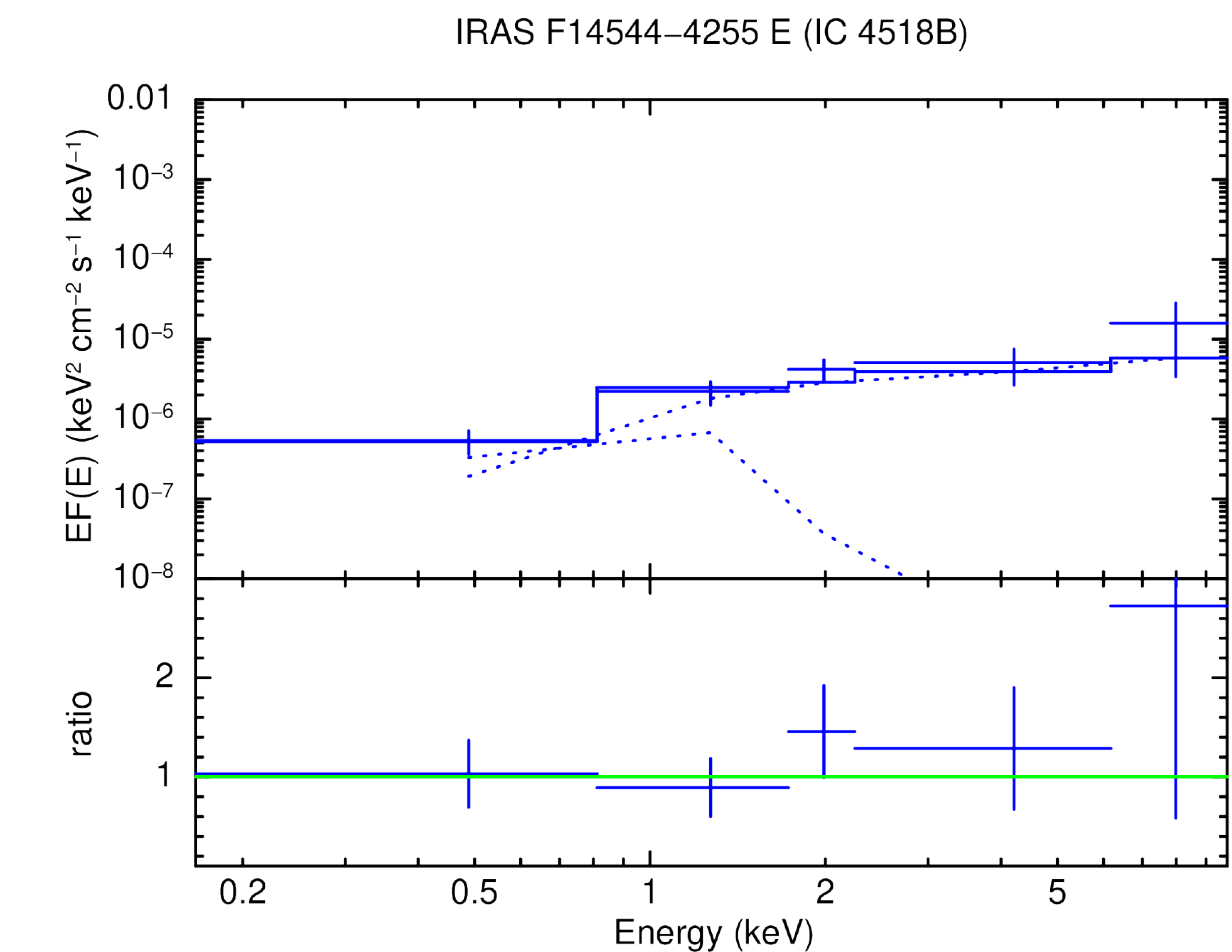}{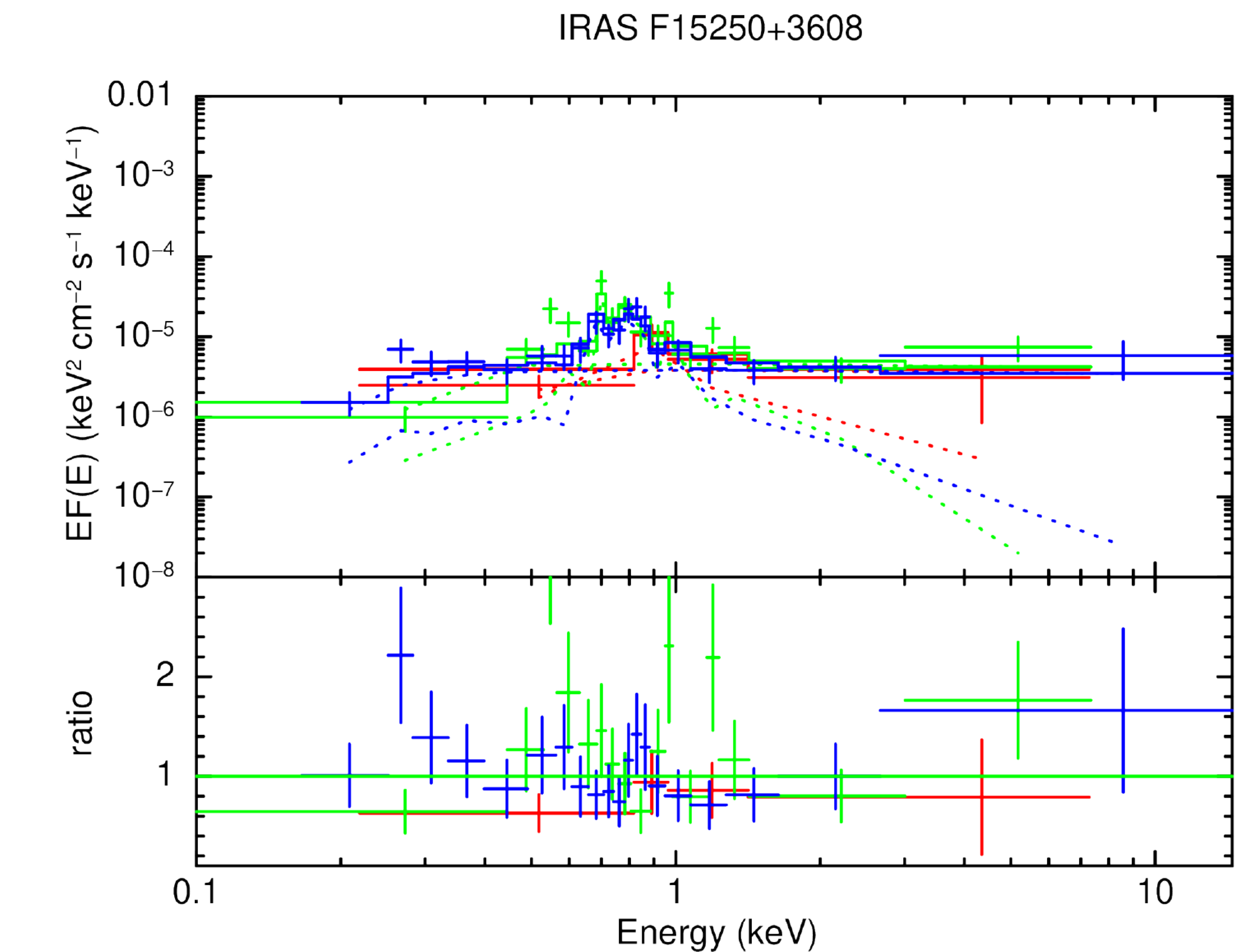}
    \plottwo{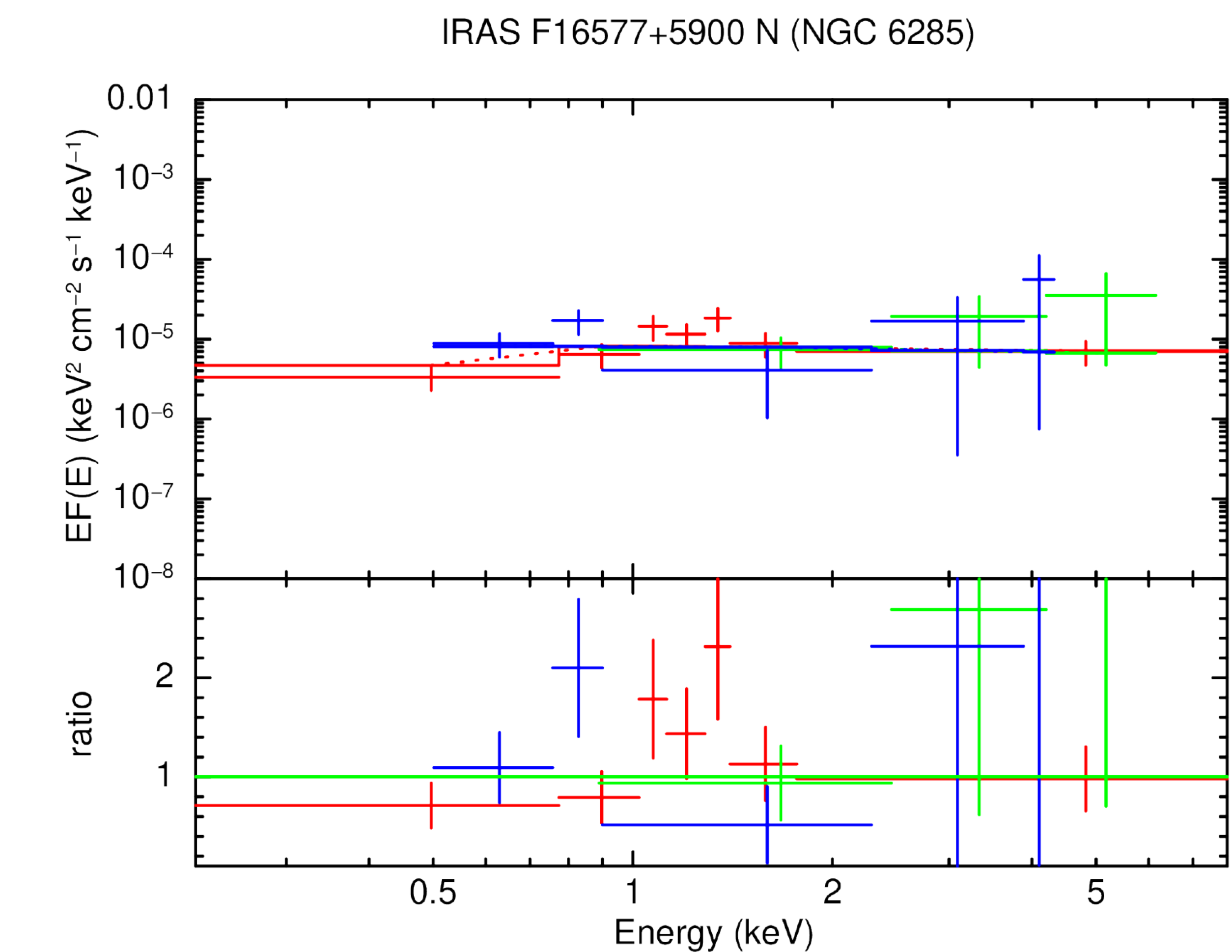}{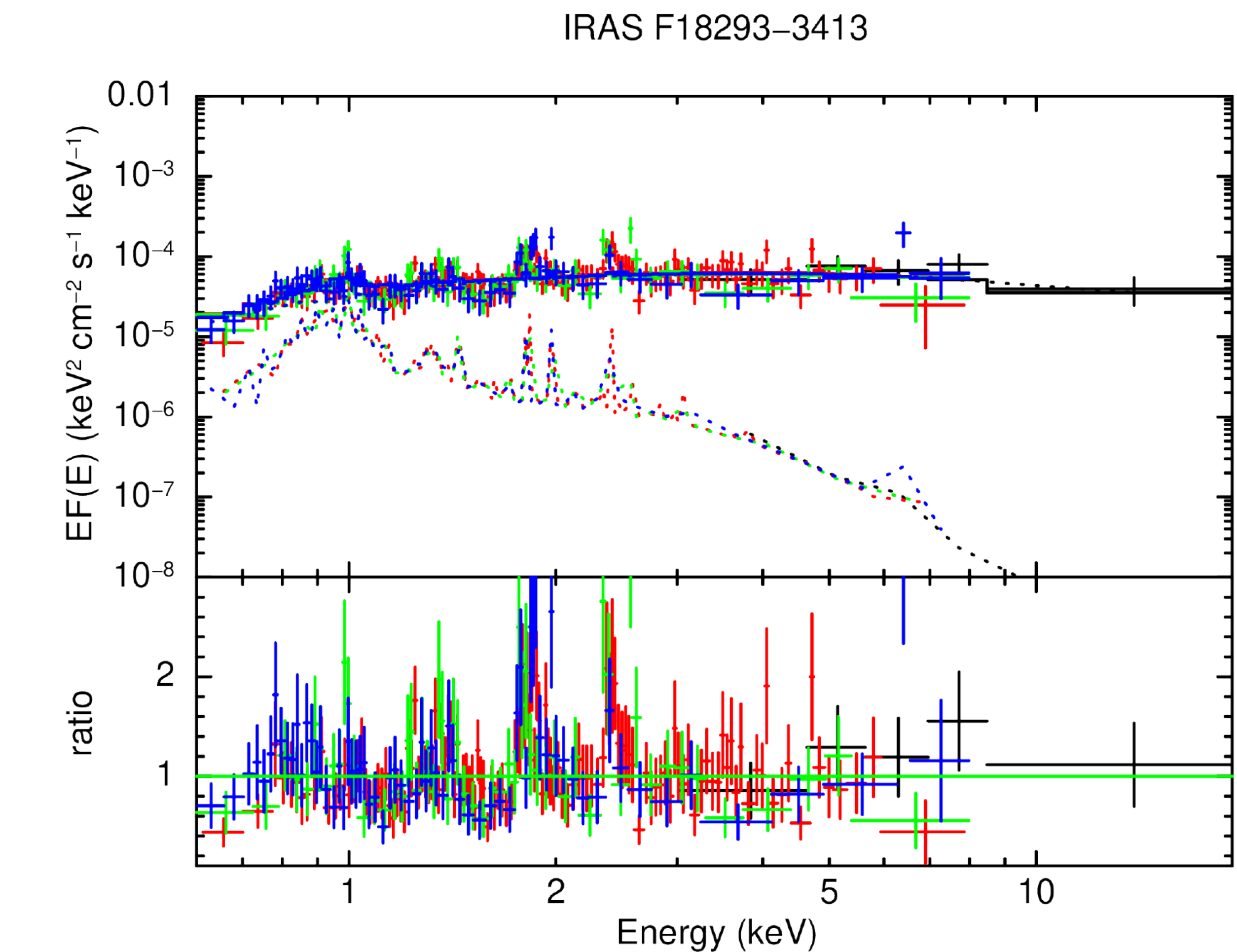}
        \caption{X-ray spectra, best-fit models, and ratios between the data and models of the starburst-dominant or hard X-ray undetected sources in
        NGC~5104,
        IRAS F14378--3651,
        IC~4518B,
        IRAS F15250+3608,
        NGC~6285,
        and IRAS F18293--3413.\\}
\label{C9-F}
\end{figure*}

\begin{figure*}
    \epsscale{1.15}
    \plottwo{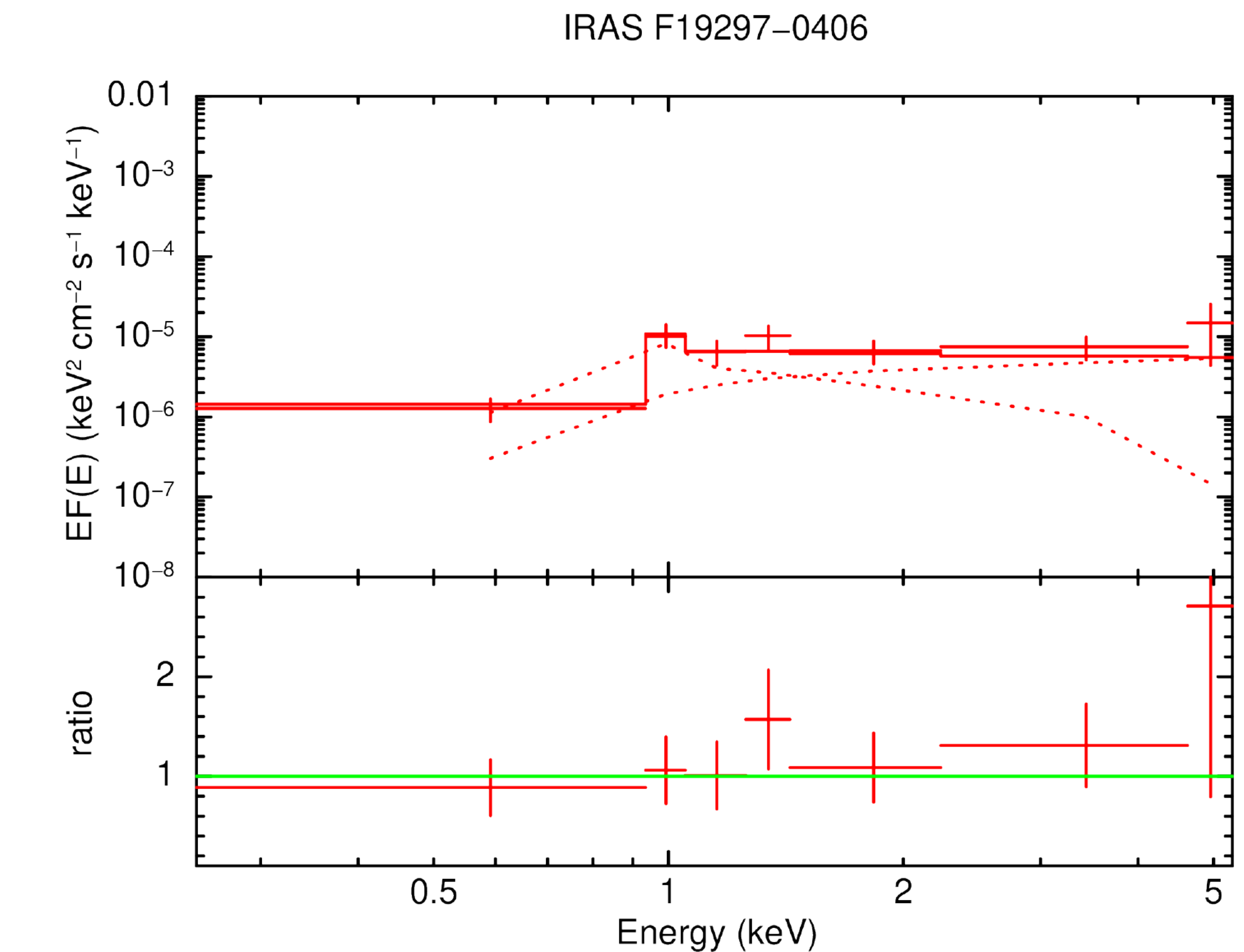}{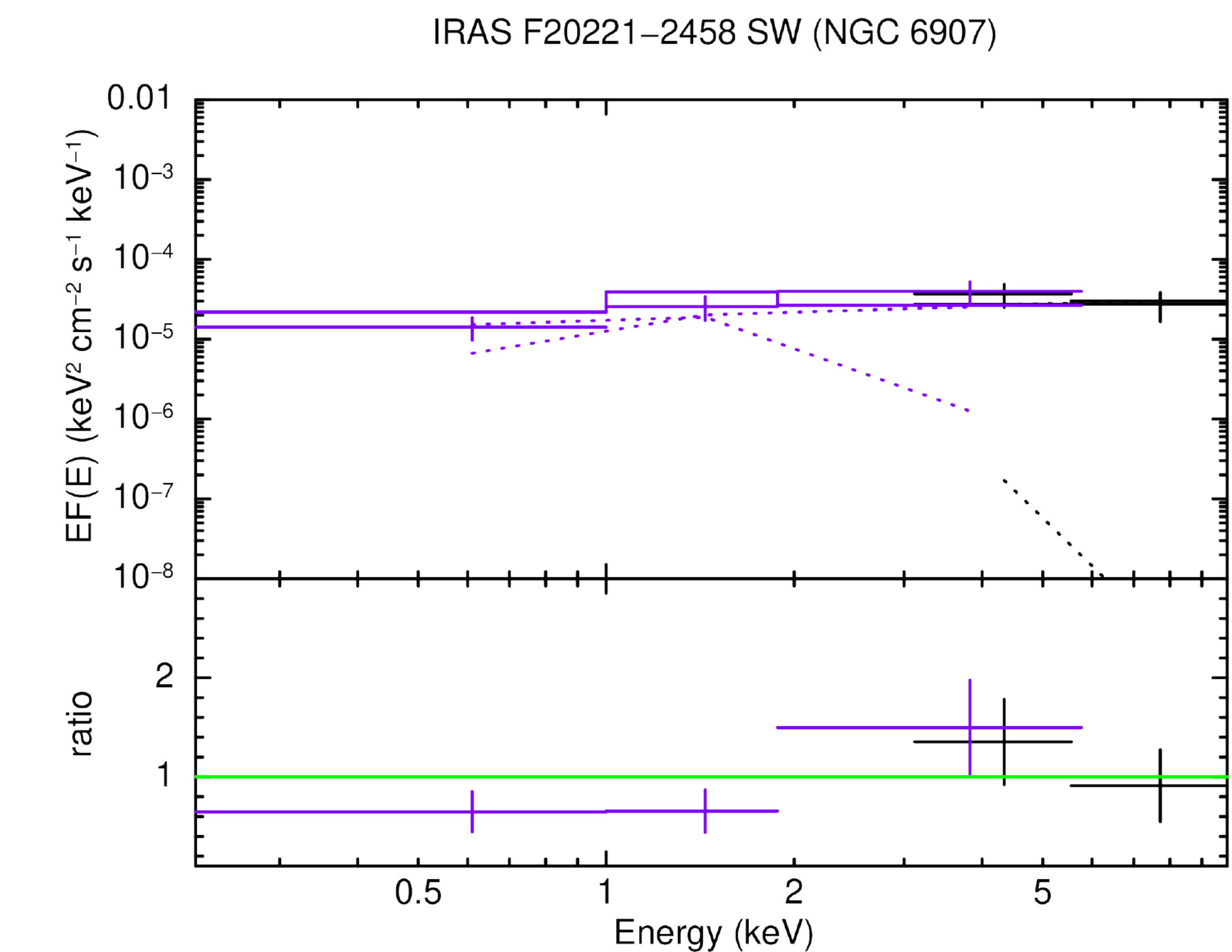}
    \plottwo{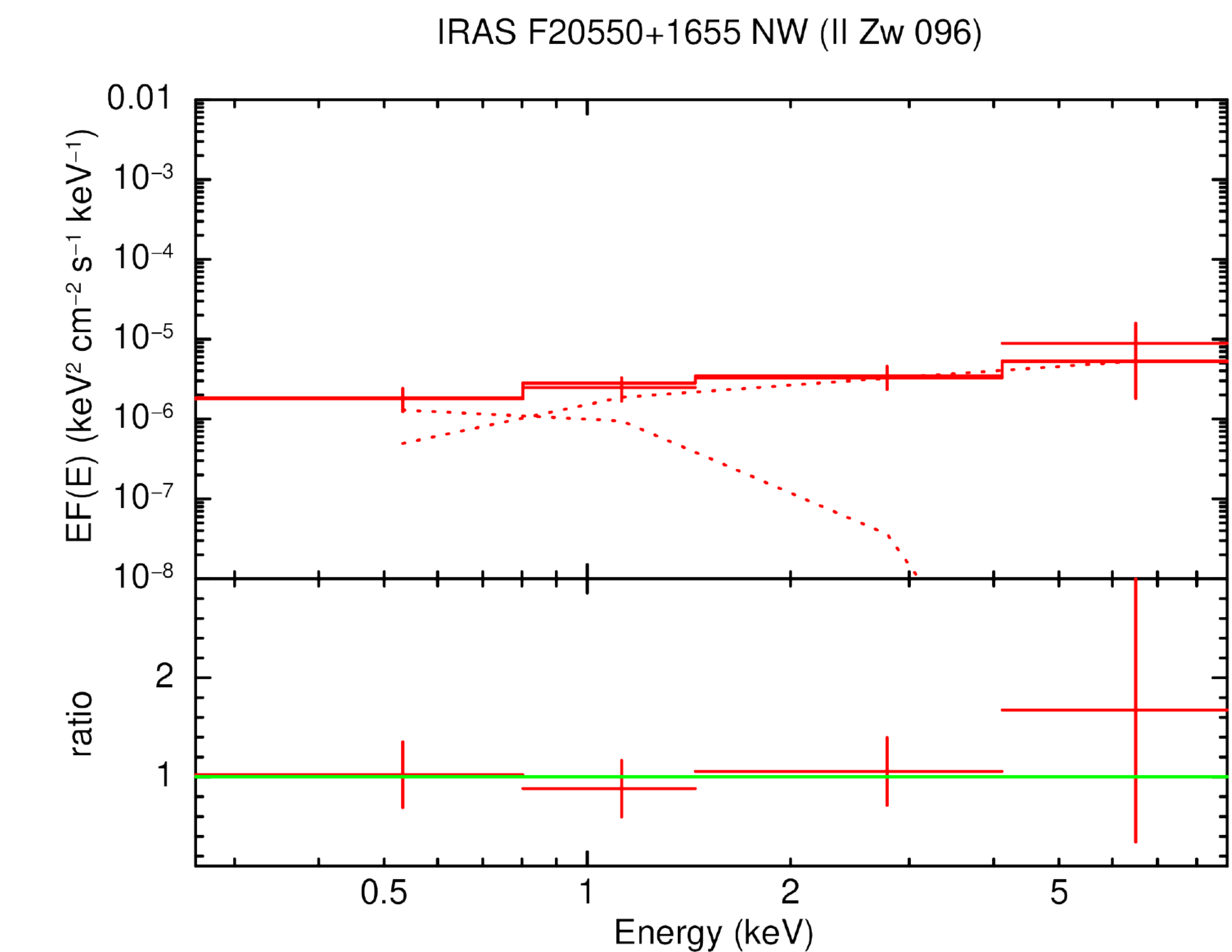}{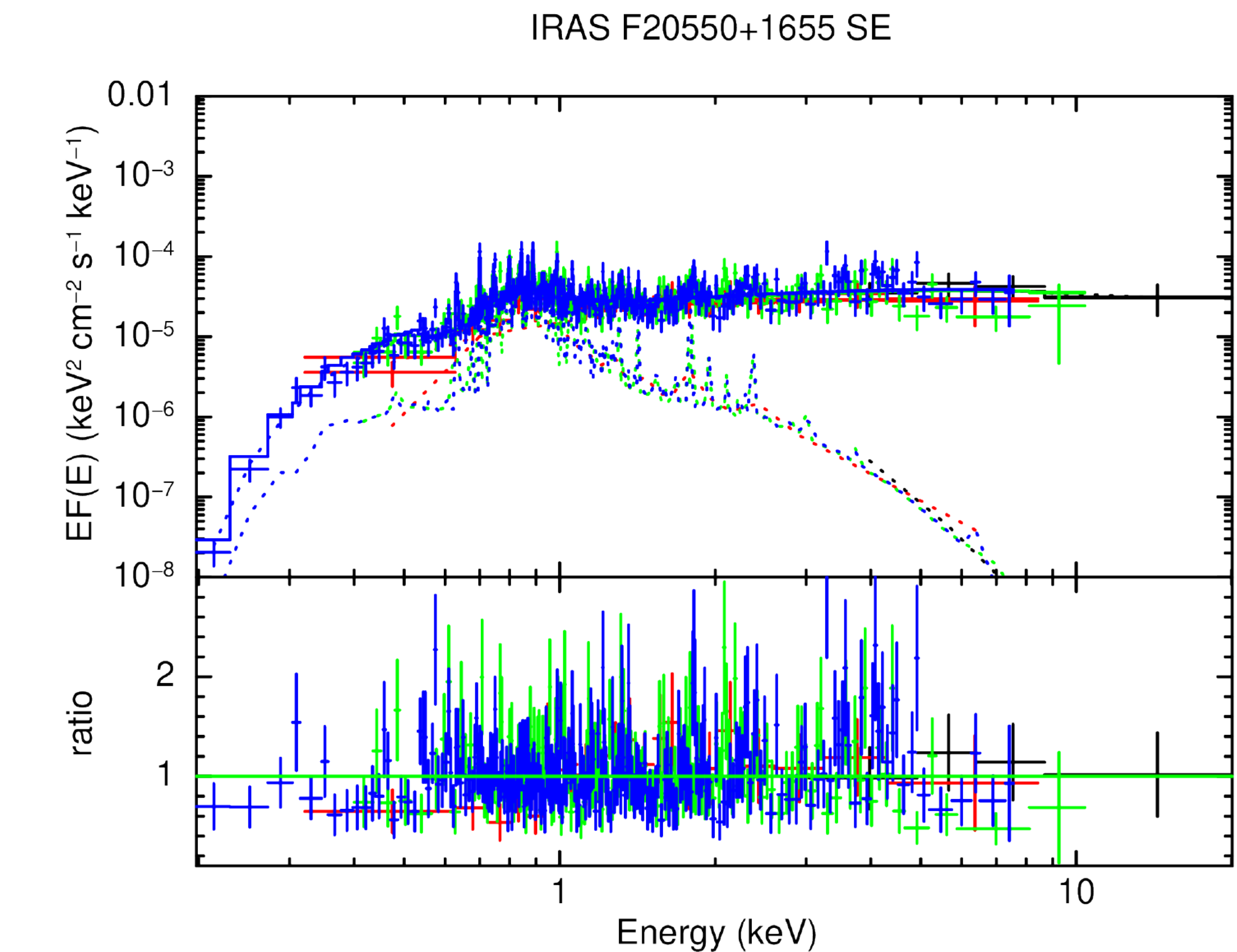}
    \includegraphics[width=0.5\textwidth]{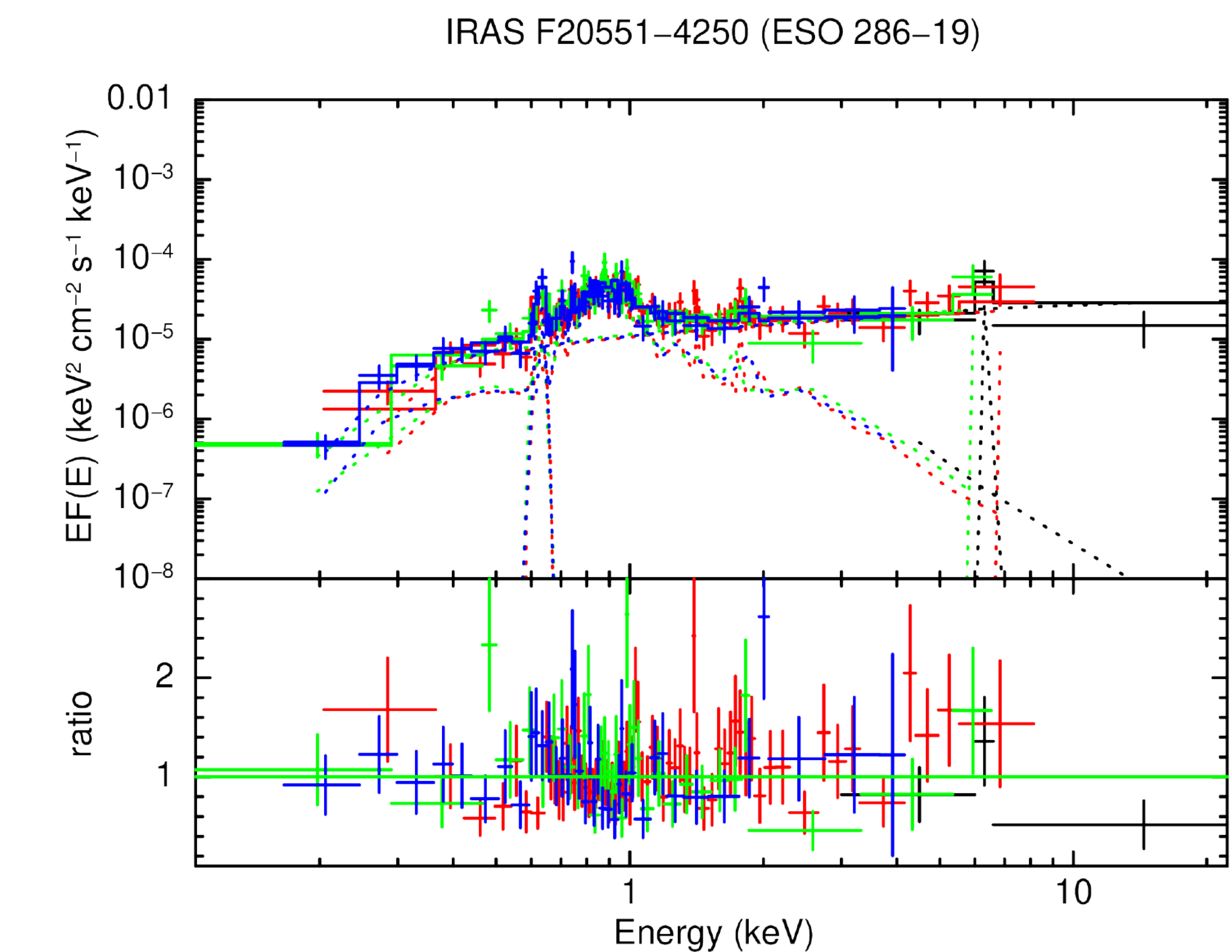}\hfill
        \caption{X-ray spectra, best-fit models, and ratios between the data and models of the starburst-dominant or hard X-ray undetected sources in 
        IRAS F19297--0406,
        NGC~6907,
        II~Zw~096,
        IRAS F20550+1655~SE,
        and ESO~286-19.\\}
\label{C10-F}
\end{figure*}

In this section, we display the X-ray spectra with the best-fit models
and ratios between the data and the models of all the sources in our sample (Figures~\ref{C1-F}--\ref{C5-F}: AGN
model, Figures~\ref{C6-F}--\ref{C10-F}: starburst-dominant model). The spectra are unfolded with
the energy responses and are presented in units of $E F(E)$, where
$F(E)$ is the energy flux at the energy $E$. When the Cash statistics
are adopted in the spectral fit, we merge the spectral bins to ensure a
significance of $>$3$\sigma$ per bin for viewing purposes. The colors of
the spectra classify the data of NuSTAR/FPM (black), Chandra/ACIS
(red), XMM-Newton/MOS (green), XMM-Newton/pn (blue), Suzaku/XIS-FI
(cyan), Suzaku/XIS-BI (magenta), Suzaku/HXD (orange), Swift/XRT
(purple), and Swift/BAT (sky blue).
\\
\\

\bibliography{reference}{}
\bibliographystyle{aasjournal}



\end{document}